\documentclass[12pt,a4paper]{report}
\usepackage[tbtags]{amsmath}
\usepackage{amssymb}
\usepackage{amsthm}
\usepackage{array}
\usepackage{xy}

\usepackage[iso-8859-7]{inputenc}
\usepackage{graphicx}

\def\<{\left\langle}
\def\>{\right\rangle}

\begin{document}
%
%
\newcommand{\g}{\greektext} 
\newcommand{\e}{\latintext}
\thispagestyle{empty}
\vspace*{5cm}
\normalfont
\begin{center}
Ph.D Thesis
\end{center}
\begin{center}
\textbf{\large Cosmological aspects of gauge mediated supersymmetry breakdown \normalsize}
\end{center}
\begin{center}
\vspace*{2cm}
{\textbf{Ioannis Dalianis} \\
Institute of Theoretical Physics, Faculty of  Physics \\
University of Warsaw,
ul. Ho\.za 69, Warsaw, Poland}
\end{center}
\begin{center}
\vspace*{4cm}
Supervisor: Zygmunt Lalak 
\end{center}
\begin{center}
\vspace*{5cm}
September 2011
\end{center}
\newpage
\tableofcontents

\newpage

\chapter{Introduction to the Thesis} 

\section{Overview}

\itshape Supersymmetry \normalfont is one of the most attractive candidates for a theory describing physics beyond the Standard Model. Until today no direct evidence of supersymmetry has been discovered. If it exists at the TeV scale, it stabilizes the hierarchy between the electroweak and the Planck scale, allows for gauge coupling unification with the minimal particle content and the lightest supersymmetric particle is a strong candidate for the dark matter of the universe. Despite of these strong motivations  a TeV supersymmetric theory is potentially accompanied with excessive flavour-changing and CP-violating effects, cosmological gravitino and moduli problems.  These problems are tightly related with the mechanism that breaks supersymmetry and mediates the breakdown to the low energy observable sector. The scenario of \itshape gauge mediation \normalfont  provides a compelling explanation for the absence of excessive flavour-changing phenomena. However, gauge mediation schemes are not free of cosmological problems. In this thesis, we study the details of some general cosmological problems of gauge mediation and we probe the supersymmetry breaking sector by cosmological arguments. We manifest that particular cosmological problems are naturally absent in the most general class of gauge mediation models without including additional ingredients or assumptions.

By construction the gauge mediation scenario is free of the Polonyi problem since the spurion that breaks the supersymmetry, i.e. the Goldstino superfield, is coupled with the observable sector with renormalizable interactions. In particular, in the ordinary gauge mediation schemes the spurion couples with the messenger fields with Yukawa couplings. Thereby its decay does not spoil the standard Big Bang Nucleosynthesis predictions that are in accord with the observations. Although these couplings of the spurion address the Polonyi cosmological problem they generically render the supersymmetry breaking minimum \itshape metastable \normalfont. The existence of many competing vacua, supersymmetric and non supersymmetric, poses the question of how natural it is for the system of fields to settle down into the phenomenological relevant vacuum with broken supersymmetry. We thoroughly investigate the problem and we conclude that the metastable supersymmetry broken vacuum becomes thermally selected for a wide range of the parameter space. The critical condition is the supersymmetry breaking sector to be sufficiently weakly coupled to the messenger fields. In fact this condition is expected to be fulfilled automatically in the general class of models that accept a microscopic interpretation. This result constitutes an essential improvement in the cosmological constraints applied to  the gauge mediation theories. Namely, the selection of the metastable vacuum does not impose any constraint on the maximal reheating temperature in the early universe. On the contrary, since the supersymmetry breaking vacuum is \itshape thermally attractive \normalfont this result suggests that high reheating temperatures were actually realized in the early universe. From the zero temperature point of view we can say that the gauge mediation supersymmetry breaking vacuum is thermally stable. 

It was recently shown \cite{Intriligator:2006dd} that the  supersymmetry breaking is a generic phenomenon in gauge theories. Supersymmetric Yang Mills theories can provide a microscopic description for the supersymmetry breaking sector and account for the dynamical generation of the scales. An additional, deeply fundamental problem is that of the vanishing \itshape cosmological constant \normalfont. Supersymmetry breaking and the vacuum energy are explicitly related. Moreover, supersymmetry being a space time symmetry  is tightly connected with gravity. However, the TeV supersymmetry breaking scale deprive supersymmetry from addressing the cosmological constant problem from the first place. Further structure has to be considered. The standard paradigm is adding a constant at the superpotential that renders the cosmological constant vanishing. In the thesis we take into account the gravity effects and we study the structure and the thermal behaviour of gauge mediation models supplemented with a constant term. Furthermore, we pursue an interpretation of this constant. \itshape String theory \normalfont provides a consistent unification of all fundamental forces of nature and provides a compelling microscopical completion of the supersymmetric field theories. In a stringy framework the constant term finds an elegant interpretation once the supersymmetry breaking sector is related with the sector that stabilizes the stringy moduli fields. This interrelation entails a connection between the supersymmetry breaking scale and the stabilization scale of the moduli. We consider the case of flux compactifications where the superpotential for the overall volume modulus accounts for the constant of the supersymmetry breaking sector. Although, the scale of the supersymmetry breaking in gauge mediation is relatively low we find that the basic stabilization schemes can be efficient. Hence, the early universe can be generically safe from high reheating temperature destabilization effects\footnote{A way to express in one word this stability of the supersymmetry breaking and the srtingy moduli vacua against thermal destabilization effects would be to name our universe 'pyrimahon' i.e. fireproof.}.

The presence of stringy \itshape moduli \normalfont fields that interact by Planck mass suppressed interactions and their mass is related with the scale of supersymmetry breaking implies an extra cosmological problem: the cosmological moduli problem. It is realized when moduli with mass of the order of the gravitino mass dominate the energy density of the early universe. This problem can be evaded in scenarios where the moduli fields become massive enough. Nevertheless, this problem 
is rather model dependent and can be circumvented e.g. by late entropy production contrary to the problem of the metastable vacuum selection and the problem of the modulus destabilization. 

Finally, a basic prediction of spontaneously broken supersymmetric theories is the massive \itshape gravitino \normalfont field. In the gauge mediation scenario it is the lightest supersymmetric particle and can account for the dark matter of the universe. It is also notorious for the cosmological problems that arise from its presence. We revisit the cosmological gravitino problem taking into account a basic feature of the supersymmetry breaking hidden sector: the fact that it possesses an approximate global $U(1)$ symmetry, the so called $R$-symmetry. Considering the underlying connection between the $R$-symmetry and the gaugino masses on the one hand and the temperature dependent variation of the $R$-symmetry breaking scale on the other hand, we claim that the production rate of the helicity $\pm \frac12$ component of the gravitino is actually suppressed at high temperatures thereby decreasing the value predicted for the gravitino abundance. This claim can significantly ameliorate the cosmological gravitino problem in gauge mediation. Moreover the resulting bounds on the reheating temperature are in a nice accordance with the temperatures necessary for a thermal selection of the supersymmetry breaking metastable vacuum.

\section{Organization of the Thesis}

The second chapter of the thesis is devoted to the review of the basic notions, formalism and properties of supersymmetric theories.  In chapter 3 we continue reviewing the mechanisms that implement the spontaneous breakdown of supersymmery. We comment on the attractive features of the gauge mediation and on the Polonyi cosmological problem.  In chapter 4 we exhibit the generality of the metastability in the supersymmetry breaking sector. We consider ordinary and direct gauge mediation models analysing in detail the basic models. In chapter 5 the thermal evolution of the supersymmetry breaking sector is presented and in chapter 6 the conditions that realize the selection of the phenomenologically relevant non-supersymmetric local vacuum are derived. In chapter 7 we briefly introduce some basics of string theory and focus on the phenomenological aspects of strings and in particular on the moduli fields. We couple the supersymmetry breaking sector with the sector that stabilizes the overall volume modulus and also consider alternative decoupled scenarios. In the chapter 8 the thermal effects on the modulus sector are studied and the related cosmological moduli problems are discussed leaving a more thorough and general discussion and presentation of the problem for the appendix. Finally, in the chapter 9 we discuss the gravitino cosmology and we revisit the gravitino production considering the implications of the $R$-symmetry thermal restoration. The results presented in the chapters  5 and 6 and some of the analysis of the chapter 4 have been published  at \cite{Dalianis:2010yk} and \cite{Dalianis:2010pq}.  The results of the chapter 9 can be also found at \cite{arXiv:1110.2072}. The material and the results of the chapters 7 and 8 are to appear in forthcoming publication (some early comments appeared at \cite{DL}).
The references are given in alphabetical order.

\section{Notation}
We gather here the symbols for some basic fields and quantities used in the text:
\begin{itemize}
	\item $X$ is the spurion (super)field that parametrizes the supersymmetry breaking hidden sector
\end{itemize}
\begin{itemize}
	\item $\phi$, $\bar{\phi}$ are the ordinary messenger (super)fields that mediate the supersymmetry breaking to the visible sector
\end{itemize}
\begin{itemize}
	\item $\lambda$ is the Yukawa coupling between the messenger and the spurion (super)fields
\end{itemize}
\begin{itemize}
  \item $\Lambda$ is an energy cut-off for the supersymmetry breaking sector and appears as an energy scale at the K\"ahler metric
\end{itemize}
\begin{itemize}
	\item $c$ is the constant added to the superpotential in order to cancel the cosmological constant at the vacuum
\end{itemize}
\begin{itemize}
	\item $M$ stands for the bare messenger mass (except if it is written otherwise) 
\end{itemize}
\begin{itemize}
	\item $\varphi$ collectively stands for scalar fields other than the ordinary messengers. For example it stands for gravitationally interacting (super)fields or for hidden degrees of freedom coupled to the spurion field $X$ with a Yukawa coupling $k$
\end{itemize}
\begin{itemize}
	\item $T_{\text{susy}}$ is the critical temperature that an approximately secord order phase transition towards the supersymmetric vacua takes place 
\end{itemize}
\begin{itemize}
	\item $T_X$ is the temperature that the supersymmetry breaking metastable minimum forms
\end{itemize}
\begin{itemize}
	\item $\rho$ is the overall volume modulus scalar field of type IIB string theory compactified in a Calabi-Yau Manifold
\end{itemize}
\begin{itemize}
	\item $\sigma$ is the imaginary part of the overall volume modulus scalar field, or the real part of the $\varrho$ field if $\varrho=i\rho$
\end{itemize}
\begin{itemize}
	\item $T_{de}$ is the temperature that the overall volume modulus minimum becomes an inflection point i.e. it is destabilized and $T_{ra}$, the temperature that it runs away to the decompactification - zero coupling limit
\end{itemize}
\begin{itemize}
   \item $m_\lambda$ or $m_{\tilde{g}}$ stands for the gaugino masses; the later in particular stands for the gluino mass
\end{itemize}
\begin{itemize}
   \item $\psi_\mu$ is the gravitino field
\end{itemize}
\begin{itemize}
	\item $M_P \simeq 2.4 \times 10^{18}$ GeV is the reduced Planck mass 
\end{itemize}

\section*{Acknowledgments}
\vspace*{.5cm}
\noindent 

First and foremost, I  would like to thank my advisor, Zygmunt Lalak, for his unwavering support, encouragement and guidance throughout the course of this thesis.

My fellows Misza Artymowski, Marcin Badziak, Anna Kami\'nska, Krzysztof Turzynski, Michalis Paraskevas for several discussions related to my research and Wojciech Kamyk Kami\'nski for many stimulating discussions about physics in general.

I would also like to thank David Langlois for hospitality at the APC Paris and collaboration in the first stages of this thesis; Alex Kehagias and George Zoupanos for always welcoming me at my home place university, NTU,Athens; and  CERN Theory Division  for hospitality.

Finally, I am thankful to the people of our Institute of Theoretical Physics at the University of Warsaw for the warm work enviroment and our secretery, Magda Mirecka, to whom I am grateful for her  generous assistance on countless bureaucratic issues.
 
This work was partially supported by the EC 6th FrameworkProgramme MRTN-CT-2006- 035863 and by Polish Ministry for Science and Education under grant N N202 091839. 

\chapter{Basics of Supersymmetry}

In this chapter we introduce the supersymmetry as a symmetry that stabilizes the electroweak scale and as a new spacetime symmetry. We discuss global and local supersymmetry and its breakdown. Some elements of the MSSM are presented.

Supersymmetry is a hypothetical space-time symmetry that transforms bosonic states into fermionic ones and vice versa. Bosons are mediators of interactions and fermions are the constituents of matter hence supersymmetry unifies matter and radiation. It plays an important role in modern particle physics, even though there is no direct experimental evidence for its existence. It is characterized by features such as vacuum stability ($E\geq 0$) and mild ultraviolet (UV) behaviour of the theory, that is a restricted form of divergences. Many theorists consider that these features are indications that supersymmetry is an essential ingredient of the ultimate unified theory of elementary particles, which perhaps is string theory.

Particle theorists have applied supersymmetry to the standard model of particle physics. Their main motivation is to protect the Higgs potential against quantum corrections. Because of very weak UV divergencies, one can naturally push the cut-off scale to an arbitrary high scale like the Planck scale ($\sim 10^{18}$ GeV).

\section{The Higgs Potential Cannot Stand Alone} 
The physical principle of \itshape naturalness \normalfont as articulated by 't Hooft states that a amall parameter is natural only when a symmetry is gained as it is set to zero \cite{'tHooft:1980xb}. In the context of quantum field theory, if a bare mass is unnaturally set to zero radiative corrections lead to a renormalized non-zero value. Therefore, if one wants a small renormalized value without a symmetry the bare value has to be fine tuned. The apparent violation of this principle in the value of the Higgs mass is known as "the gauge hierarchy problem"\footnote{A more dramatic fine tuning is that of the cosmological constant.}.

The Higgs potential triggers the spontaneous breaking of electroweak gauge symmetry;  \cite{Collins:1989kn, Binetruy:2006ad, Bailin:1994qt, Zee:2003mt, Kaku:1993ym}. A single elementary Higgs boson field can acount for all masses of quarks, leptons and gauge bosons through its interactions. To generate the symmetry breaking, we postulate a potential for the Higgs field 
\begin{equation}
V=\mu^2|h|^2+\lambda|h|^4.
\end{equation}
The assumption that $\mu^2<0$ is the complete explanation for electroweak symmetry breaking in the Standard Model. Since $\mu$ is a renormalizable of this theory, the value of $\mu$ cannot be computed from first principles and even its sign cannot be predicted. The parameter $\mu^2$ receives large additive radiative corrections from loop diagrams. For example the coupling of the Higgs field with itself and the top quark give quadratic ultraviolet divergencies. Applying a momentum cutoff $\Lambda$ the quantum contributions to the bare mass are
\begin{equation}
\mu^2=\mu^2_\text{bare}+\frac{\lambda}{8\pi^2}\Lambda^2-\frac{3y^2_t}{8\pi^2}\Lambda^2+...
\end{equation}
If we view standard model as an effective theory, $\Lambda$ should be taken to be the largest momentum scale at which this theory is still valid. The radiative corrections can easily change the sign of $\mu^2$ and the criterion $\mu^2<0$ is, hence, not a simple condition on the underlying parameters of the effective theory. Furthermore, if we try to embed the Standard Model in some larger GUT, such as $SU(5)$ for example, then the breakdown of this larger gauge summetry to $SU(3)_c\times U(1)_{em}$ requires two types of Higgs particles: $\Phi$ with $M_\Phi= {\cal O}(M_\text{GUT})$ in addition to the usual Higgs $h$ with $M_h= {\cal O}(M_\text{EW})$. In order to keep the $h$ scalar light while the $\Phi$ scalars heavy, we must ensure cancellations of the quadratic divergence to an accuracy 
\begin{equation}
\left(\frac{M_h}{M_\Phi}\right)^2 \sim \left(\frac{M_\text{EW}}{M_\Phi}\right)^2 \sim  10^{-24}
\end{equation}
if $M_\text{GUT} \sim 10^{14}$ GeV, which must be put in "by hand". The problem can be shown explicitly by considering a Higgs potential with two very different energy scales. If we are to have one set of Higgs fields $h$ with associated particles arising from vacuum expectation value $v \simeq 10^2$ GeV and another set of $\Phi$ with vev $v_0 \simeq 10^{14}$ GeV, then 
\begin {equation}
V(h, \Phi)=\lambda_1\left(|h|^2-v^2\right)^2+\lambda_2\left(|\Phi|^2-v^2_0\right)^2.
\end{equation}
Since both sets of Higgs fields interact with the gauge bosons, we get after renormalization corrections to the above potential of order $g^4 h^2 \Phi^2$ where $g$ is the gauge coupling. The minimum of the potential with respect to $h$ is shifted from $|h|=v$ to $h \simeq (g^4v^2_0/2\lambda_1)^{1/2} \simeq g^2 v_0$. For $\lambda_1 \sim 1$ the vev of the lower mass Higgs fields gets moved up within order of $g^2$ of the higher mass scale unless there are additional contributions to the potential adjusted to an accuracy $v^2/v^2_0 \sim 10^{-24}$ that cancel away these corrections. This problem is known as 'gauge hierarchy problem'.

Generally there are two different strategies to address this problem. One is to look for new strong-coupling dynamics at an energy scale of 1TeV or below. Then the Higgs field could be composite and its potential could be the result, for example, of pair condensation of fermion constituents. Higgs himself proposed that his field was a phenomenological description of a fermion pair condensation mechanism similar to that in superconductivity. Sometime later, Susskind and Weinberg proposed an explicit model of electroweak symmetry breaking by a new strong interaction called 'technicolor'. Today this approach is disfavoured because it typically leads to flavour changing neutral currents at an observable level and also conflicts the accurate agreement of precision electroweak theory with experiment.

The second strategy to solve the hierarchy problem (stabilize the scales) is to postulate that the electroweak symmetry is broken by a weakly coupled Higgs field, but that this field is part of a model in which the Higgs potential is computable. The Higgs mass term $\mu^2 |h|^2$ should be generated by well defined physics within the model and hence the $\mu^2$ term should not receive additive radiative corrections. This requires the presence of a symmetry that forbids the radiative corrections to induce a large mass for the Higgs in the Lagrangian. 

There are three ways to arrange a symmetry that forbids the term $\mu^2 |h|^2$. We can postulate a symmetry that shifts $h$ i.e. $\delta h= \epsilon v$, or a symmetry that connects $h$ to a gauge field whose mass can then be forbidden by gauge symmetry i.e. $\delta h= \epsilon \cdot A $ or, finally, a symmetry that connects $h$ to a fermion field, whose mass can then be forbidden by a chiral symmetry:
\begin{equation}
\delta h= \epsilon \cdot \psi.
\end{equation}
The first two options lead respectively to 'little Higgs' models and to models with extra space dimensions. The third option leads to supersymmetry.

\section{Supersymmetry: A new space-time symmetry} \normalsize
Supersymmetry is a symmetry that transforms bosonic states into fermionic ones and vice versa. 
\begin{equation}
Q_\alpha\left|b\right\rangle=\left|f\right\rangle.
\end{equation}
The generators of such a symmetry must carry a spinorial index, since they correspond to the transformation of an integer spin field into a spinor field. Therefore, they are not commuting with Lorentz transformations. In contrast, the SM's $SU(3)\times SU(2) \times U(1)$ are symmetries of internal degrees of freedom of the fields. In this sense, supersymmetry is necessarily a spacetime symmetry.

A new spacetime symmetry implies the presence of a new conserved charge. Noether procedure tells us that the charges may be interpreted as the generators of the transformations they are associated with. According to Coleman-Mandula theorem (1967) the conserved charges of any other symmetry apart from the Poincare group has to transform as a scalar under the Lorentz group. This "no-go" theorem shows that it is impossible to have a new space-time symmetry that mixes in a non trivial way with the Lorentz space-time symmetry.

The Poincare group is the sum of all space-time transformations that include translations, rotations and Lorentz boosts. Overall, this group has ten elements and has a representation in terms of vectors $P_\mu$ and symmetric $4 \times 4 $ matrices $M_{\mu\nu}$ that satisfy the commutation relations
\begin{equation}
[P_\mu, P_\nu]=0  \,\,\,\,
\end{equation}
\begin{equation}
[P_\mu, M_{\rho \sigma}]=i(\eta_{\mu \rho} P_\sigma - \eta_{\mu \sigma} P_\rho)
\end{equation}
\begin{equation}
[M_{\mu \nu}, M_{\rho \sigma}]= i(\eta_{\nu \rho} M_{\mu\sigma} - \eta_{\nu \sigma} M_{\mu \rho}-\eta_{\mu \rho} M_{\nu \sigma} + \eta_{\mu \sigma} M_{\nu \rho}) \end{equation}
\begin{equation}
[T_a, T_b]=iC_{abc}T_c.
\end{equation}
where $\eta=\text{diag}(-1, 1, 1, 1)$ the Minkowski metric and $T_a$ generators of an internal symmetry. According to the Coleman-Mandula theorem the $T_a$ are Lorentz scalars:
\begin{equation}
[T_a, P_\mu]=[T_a, M_{\mu \nu}]=0.
\end{equation}
If this was not the case then, as Witten (1981) showed, the two body scattering would look much different. In a two particle collision the conservation of $P_\mu$ and $M_{\mu \nu}$ leaves the scattering angle $\theta$ undetermined. If there was a Lie group that mixed with the Poincare group in a nontrivial way then these genertors would be associated with space-time, which would mean that the $\theta$ could only take discrete values. But the scatering amplitude is analytic in $\theta$ and so it would have to vanish for all $\theta$.

Supersymmetry escapes this "no-go" theorem because, in addition to the generators $P_\mu$, $M_{\mu \nu}$, $T_a$ that satisfy commutation relations, it involves fermionic generators that satisfy anticommutation relations. The supersymmetry algebra has a more general structure which is called "Graded Lie algebra" by the mathematicians. A deeper reason that supersymmetry although it is a (hypothetical) space-time symmetry escapes the Coleman-Mandula "no-go" theorem is that it involves a new type of coordinates that are grassmannian and correspond to the fermionic degrees of freedom. In this 'extended' superspace there is room for another space-time symmetry, the supersymmetry. 

We represent the supersymmetric generator $Q_\alpha$ as a Majorana spinor which is the simplest possible type of spinor with four real components. Since $Q_a$ is a spinor, it must satisfy 
\begin{equation}
[Q_\alpha, M_{\mu\nu}]=i{(\sigma_{\mu\nu})_{\alpha}}^\beta Q_\beta.
\end{equation}
This relation expresses the fact that the $Q_\alpha$ transforms as a spinor under the rotations generated by $M_{\mu\nu}$. The Jacobi identity of commutators 
\begin{equation}
[[Q_\alpha, P_\mu],P_\nu]+[[P_\nu, Q_\alpha],P_\mu]+[[P_\mu,P_\nu],Q_\alpha]=0
\end{equation}
requires that $Q_\alpha$ must be translationally invariant, that is 
\begin{equation} \label{2}
[Q_\alpha, P_\mu]=0
\end{equation}
Finally, the spinor $Q_\alpha$ satisfies the anticommutation relation
\begin{equation} \label{susyanti}
\{Q_\alpha, \bar{Q}_{\dot{\beta}}\}=2(\sigma^\mu)_{\alpha\dot{\beta}} P_\mu
\end{equation}
The translation invariance of the supersymmetric generator $Q_\alpha$ means that it also commutes with $P_\mu P^\mu$ which is the mass operator: $P_\mu P^\mu \left|b \right\rangle =m^2_b \left| b \right\rangle $ and $P_\mu P^\mu \left|f \right\rangle =m^2_f \left| f \right\rangle $. It is easy to see that 
\begin{equation} \label {susymass}
[P^\mu P_\mu, Q_\alpha]=(m^2_f - m^2_b) \left|f \right\rangle =0.
\end{equation}
Hence, supersymmetry implies that the bosons and fermions which are supersymmetric partners must have equal mass. However the observed particle spectrum is non-supersymmetric and supersymmetry is not a symmetry of the GeV scale particle physics. We will extensively discuss this fact in subsequent paragraphs and chapters, but here we are postponing this discussion presenting some more useful formalism for supersymmetry.

\subsection{Superspace}

At the anticommutation relation of the supercharges (\ref{susyanti}) dotted indices were introduced. It is due to the fact that the Lorentz algebra $SO(1,3)$ is locally isomorphic to $SU(2)\times SU(2)$. That is, a spinor representation of the Lorentz group is decomposed into irreducible representations of given chiralities: $4=2_L+2_R$. The dot was introduced in order to keep track of which $SU(2)$ we are refering. 
Also, as mentioned supersymmetry generates translations in superspace $\{x^{\mu}, \theta^\alpha, \bar{\theta}^{\dot{\beta}}\}$. Superspace includes both bosonic and fermionic coordinates, a notion invented by Salam and Strathdee. We can thus define a superfield $\Phi(x^{\mu}, \theta^a, \bar{\theta}^{\dot{\beta}})$ that lives in the superspace. The supercharges are represented as 
\begin{equation}
Q_\alpha= \frac{\partial}{\partial \theta^\alpha}-i (\sigma^\mu)_{\alpha \dot\alpha} \bar{\theta}^{\dot\alpha} \partial_\mu
\end{equation}
\begin{equation}
\bar{Q}_{\dot\beta}= - \frac{\partial}{\partial \bar{\theta}^{\dot{\beta}}}+i \theta^\beta (\sigma^\mu)_{\beta \dot{\beta}} \partial_\mu\, 
\end{equation}
Spinor derivatives on superfields orthogonal to $Q_\alpha$ and $\bar{Q}_{\dot{\beta}}$ that commute with supersymmetry transformations can be defined: $D_\alpha= \partial/\partial \theta^\alpha+i (\sigma^\mu)_{\alpha \dot\alpha} \bar{\theta}^{\dot\alpha} \partial_\mu$ and $\bar{D}_{\dot\beta}= - \partial/\partial \bar{\theta}^{\dot{\beta}}-i \theta^\beta (\sigma^\mu)_{\beta \dot{\beta}} \partial_\mu.$ Imposing the condition 
\begin{equation}
\bar{D}_{\dot{\beta}}\Phi =0
\end{equation}
on the superfield we find that this condition is also satisfied by the supersymmetric transformation of $\Phi$: $\bar{D}^{\dot{\alpha}}(\delta_s \Phi)=\delta_s (\bar{D}^{\dot{\alpha}} \Phi)=0$. Such a superfield is called as a \itshape chiral superfield \normalfont. Since the $\Phi$ depends on both ordinary and Grassmannian variable $\theta$ we can expand in power series in $\theta$. We find that the superfield $\Phi$ has to contain a Weyl fermion field $\Psi$ and two complex scalar fields $\phi$ and $F$. 

The above chiral superfields do not include gauge degrees of freedom. A vector field is a real field and is component of the vector superfield
\begin{equation}
V(x, \theta, \bar{\theta})=V^\dagger(x, \theta, \bar{\theta})
\end{equation}
A vector field is associated with a gauge transformation: $A_\mu(x) \rightarrow A_\mu(x) +\partial_\mu \alpha(x)$ in the abelian case. The supersymmetric form of this tranformation is $V \rightarrow V+i(\Lambda-\Lambda^\dagger)$ where $\Lambda$ a chiral superfield and its scalar component gives the gauge parameter $\alpha(x)$ of the transformation. This allows to set to zero components of the vector field $V$ when expanded in terms of the Grassmannian variables $\theta$, $\bar{\theta}$; this is the so called Wess-Zumino gauge. The gauge invariant degrees of freedom in the expansion of $V(x, \theta, \bar{\theta})$ denoted by $F_{\mu\nu}=\partial_\mu A_\nu -\partial_\nu A_\mu$, $\lambda(x)$ and $D(x)$ are the gauge field strength, the gaugino field and the auxiliary field $D(x)$. They turn out to be the components fields of the chiral superfield
\begin{equation}
W_\alpha =- \frac14 \bar{D}_{\dot{\alpha}} \bar{D}^{\bar{\alpha}} D_\alpha V 
\end{equation}
i.e. $\bar{D}^{\bar{\beta}} W_\alpha =0$ and gauge invariant. A supersymmetric action takes the form
\begin{equation}
S=\frac14 \int d^4x \, d^2\theta \, W^\alpha W_\alpha + \text{h.c.} \,= \, \int d^4 x \left[-\frac14 F^{\mu\nu}F_{\mu\nu}+i\lambda \sigma^\mu \partial_\mu \bar{\lambda}+\frac12 D^2 \right].
\end{equation}
When we couple matter to gauge fields we have to replace the gauge violating term $\Phi^\dagger \Phi$ with the gauge invariant $\Phi^\dagger e^{-2gqV}\Phi$. The $\Phi \rightarrow \Phi' = e^{2igq \Lambda}\Phi$ is the gauge transformation which, aslo preserves chirality. The action for a gauge invariant chiral superfield reads
\begin{equation} \nonumber
S =\int d^4 x d^2\theta d^2\bar{\theta} \Phi^\dagger e^{-2gqV}\Phi \,
\end{equation}
\begin{equation} \label{ca-1}
 = \, \int d^4 x\left[D^\mu \phi^* D_\mu \phi + i\Psi\sigma^\mu D_\mu \bar{\Psi} + F^* F + gq\left(D\phi^*\phi +\sqrt{2} \lambda\Psi\phi^*+\sqrt{2}\bar{\Psi} \phi \right)\right]
\end{equation}
where $D_\mu$ the covariant derivative. Therefore, in the superspace, the Lagrangian density has the form
\begin{equation}
{\cal L} = \int d^2 \theta \, d^2\bar{\theta}\, K(\Phi, \Phi^\dagger) + \int d^2\theta \left(f_a(\Phi)W^2_{\alpha a} + W(\Phi) \right).
\end{equation}
Finally, we note that apart from the chiral and the vector superfield there is also a third example: \itshape the linear superfield \normalfont. It describes the supermultiplet associated with the an antisymmetric tensor $b_{\mu\nu}$ known as Kalb-Ramond field \cite{Kalb:1974yc} and is present in string theory.
\\
\\
\itshape R-symmetry \normalfont
\\
The action (\ref{ca-1}) is also invariant under a non-trivial global $U(1)$ symmerty called $R$-symmetry that does not commute with supesymmetry. Such a symmetry is allowed in $N=1$ supersymmetry and its generator satisfies
\begin{equation}\nonumber
[Q_\alpha, R]=Q_\alpha
\end{equation}
\begin{equation}
[\bar{Q}_{\dot{\alpha}}, R]=- \bar{Q}_{\dot{\alpha}} \, .
\end{equation}
Since the $U(1)_R$ does not commute with supersymmetry it acts differently on the different components of the superfield and thus cannot leave the Grassmann variable $\theta$ invariant. For a chiral superfield
\begin{equation} \nonumber
R \Phi(x, \theta)=e^{ir\alpha} \Phi(x, e^{i\alpha}\theta)
\end{equation}
\begin{equation}
R \Phi^\dagger(x^*, \bar{\theta})=e^{-ir\alpha} \Phi^\dagger(x^*, e^{i\alpha}\bar{\theta})
\end{equation}
where $r$ is the $R$-charge of the supermultiplet. In terms of component fields has the value $r$, $r-1$ and $r-2$ for the component fields $\phi$, $\Psi$ and $F$ respectively. The Grassmannian variabe $R$-transforms like $\theta \rightarrow \theta'=e^{-i\alpha}\theta$ hence, the kinetic term $\int d^2 \theta \, d^2\bar{\theta} \Phi^\dagger \Phi$ is always $R$-invariant while the supepotential term $d^2 \theta W(\Phi)$ is invariant if it has $R$-charge $r=2$. This is the case if $W(\Phi)$ is a monomial in $\Phi$ and the $R$-charges add up to 2. 

Turning back to (\ref{ca-1}) the $W_\alpha$ has $R$-charge $r=+1$ and the $W^\alpha W_\alpha$ $r=+2$. At component level the gaugino $\lambda_\alpha$ has $R$-charge $r=+1$
\begin{equation}
R \lambda_\alpha (x) =e^{i\alpha} \lambda_\alpha (x)
\end{equation}
a fact that has important implications for the models of dynamical supersymmetry breaking that induce non-supersymmetric soft masses on gauginos as we will expose in the next chapters.
\\
\\
Another approach of constructing an action invariant under supersymmetry is that of J. Wess and B. Zumino (1974). This is the subject of the following section.
\section{Supersymmetric Lagrangian} \normalsize
In order to write down a supersymmetric Lagrangian we start with a theory with spin-0 bosons and spin-1/2 fermions. We consider the system of a complex scalar field $\phi(x)=(A(x)+iB(x))/\sqrt{2}$ and a Majorana spinor field $\Psi(x)$ which is called a chiral supermultiplet. Counting the degrees of freedom we see that the scalar field has two real whereas the Majorana spinor has four. The equation of motion projects out two of the  spinorial degrees of freedom however off-shell the Lagrangian is not supersymmetric. In order to have equal bosonic and fermionic degrees of freedom in the chiral supermultiplet we also include a complex scalar field $F(x)=(F_1(x)+iF_2(x))/\sqrt{2}$ which is called auxiliary field. The Lagrangian
\begin{equation} \label{L1}
L= \partial_\mu \phi^*\partial_\mu\phi+\frac{1}{2}\bar{\Psi}(i\gamma^\mu\partial_\mu-m)\Psi+F^*F+m(F\phi+F^*\phi^*)
\end{equation}
Since $F$ has no kinetic terms it is not a dynamical degree of freedom. The supersymmetric transformations read
\\
\begin{equation} \label{s-1}
\delta_s A=\bar{\epsilon}\Psi, \,\,\,\,\,\,  \delta_s B=i\bar{\epsilon}\gamma_5\Psi
\end{equation}
\begin{equation} \label{s-2} 
\delta_s \Psi =[-i\gamma^\mu\partial_\mu(A+iB\gamma_5)+F_1-iF_2\gamma_5]\epsilon,
\end{equation}
\begin{equation}  \label{s-3}
\delta_s F_1=-i\bar{\epsilon}\gamma^\mu\partial_\mu\Psi, \,\,\,\,\,\,  \delta_s F_2=-i\bar{\epsilon}\gamma_5\gamma^\mu\partial_\mu\Psi \, .
\end{equation}
We can add interactions at (\ref{L1}) between the chiral supermultiplet fields that do not spoil supersymmetry. The scalar interactions can be included in an analytic function $W(\phi)$ which is called superpotential
\begin{equation}
L=L_{KE}+FF^*+F\frac{\partial W}{\partial \phi}+F^*\frac{\partial W^*}{\partial \phi^*}- \frac{1}{2}\left(\frac{\partial^2 W}{\partial \phi^2} \bar{\Psi}_R \Psi_L+\frac{\partial^2 W^*}{\partial \phi^{*\,2}} \bar{\Psi}_L \Psi_R \right).
\end{equation}
Using the equation of motion $\partial L /\partial F$ which gives $F^*=-(m\phi^*+\lambda\phi^{*\,2})$ we can eliminate the auxiliary field $F$ and this yields the scalar potential $V(\phi)=\left|\partial W / \partial \phi\right|^2=|F|^2$. It is straightforward to generalize to the case of $n$ chiral supermultiplets $(\phi_i,\Psi_i)$ -with their corresponding auxiliary fields $F_i$- and a general superpotential $W(\phi_i)$. Solving again for $F_i$ yields the scalar potential 
\begin{equation}
V(\phi_i)=\sum_i\left| \frac{\partial W}{\partial \phi_i}\right|^2=\sum_i|F_i|^2.
\end{equation}
A combination of a susy theory with gauge theory is clearly necessary if these ideas are to make contact with the real world. In addition to the chiral multiplets we must include the "gauge" supermultiplets
\begin{equation}
(A^a_\mu, \lambda^a), \,\,\,\,\, a=1,2,..., N^2-1
\end{equation} 
where $A^a_\mu$ are the spin-1 gauge bosons of the gauge group G which is taken to be SU(N). The $\lambda^a$ are their Majorana superpartners, the so called "gauginos". These boson-fermion pairs, which in the absence of gauge symmetry breaking are assumed to be massless, belong to the adjoint representation of the gauge group. Here we also add a real auxiliary pseudoscalar field $D$. This makes the 3+1 bosonic degrees of freedom to match the 4 fermionic degrees of freedom in the off-shell formulation. A supersymmetric and gauge invariant Lagrangian reads
\begin{equation}
L_G=-\frac{1}{4}F^a_{\mu\nu}F^{a\mu\nu}+\frac{1}{2}\bar{\lambda^a}i \gamma^\mu D_\mu\lambda^a+\frac{1}{2}D^aD^a
\end{equation}
where $F_{\mu\nu}$ is the covariant field strength and $D_\mu$ the covariant derivative of the gaugino field. The $L_G$ is invariant under the supersymmetric transformations
\begin{equation}
\delta_s A^\mu_a=\epsilon \gamma^\mu \gamma_5\lambda^a
\end{equation}
\begin{equation}
\delta_s \lambda^a_r=-D^a \epsilon_r+\frac{1}{2}(\sigma^{\mu\nu}\gamma_5 \epsilon)_r F^a_{\mu\nu}
\end{equation}
\begin{equation}
\delta_s D^a= -i \bar{\epsilon}\gamma^\mu\gamma_5\partial_\mu \lambda^a.
\end{equation}
This pure gauge Lagrangian the equation of motion, $\partial L_G/ \partial D^a=0$, implies $D^a=0$; it will become non-zero when the chiral fields are coupled in. To include the chiral fields $\phi^i, \Psi^i_L$ we add $L_\text{chiral}$ of ($\ref{L1}$) but substitute the covariant derivative $D_\mu$ for $\partial_\mu$ in the kinetic energy terms, i.e. 
\begin{equation}
\partial_\mu \rightarrow D_\mu=\partial_\mu+ i g_G T^a A^a_\mu
\end{equation} 
where $T^a$ are the matrices representing the generators of the gauge group in the representation to which $(\phi_i, \Psi_i)$ belong. To ensure the supersymmetry of the combined "chiral-gauge" Lagrangian we must include two further terms 
\begin{equation}
L=L_\text{chiral}+L_G-g_G\phi^*_i(T_a)_{ij}\phi_jD^a+[\sqrt{2}g_G\phi^*_i\bar{\lambda}(T^a)_{ij}P_L\Psi_j + \text{h.c.}]
\end{equation}
where $P_L\equiv \frac{1}{2}(1-\gamma^5)$. Using $\partial L_G/ \partial D^a=0$ to eliminate the auxiliary fields gives 
\begin{equation}
D^a=g_G\phi^{*i}(T^a)^j_i\phi_j.
\end{equation}
Therefore, the total potential for the scalar field is given by
\begin{equation}
V(\phi, \phi^{*i})=\sum_iF^{*i}F_i+\frac{1}{2}\sum_a(D^a)^2+\frac{1}{2}\sum_m(D^m)^2=
\end{equation}
\begin{equation} \label{Vtot1}
=\sum_i\left|\frac{\partial W}{\partial \phi_i}\right|^2+ \frac{1}{2}\sum_a g^2_a[\phi^{*i}(T^a)^j_i\phi_j]^2+\frac{1}{2}\sum_m\left[ g_m\sum_iq^m_i\phi^{*i}\phi_i-\xi^m\right]^2
\end{equation}
In the last term we wrote seperately the contribution to the scalar potential coming from an $m$ number of abelian $U(1)_m$ gauge groups. But the novel terms are the so-called Fayet-Iliopoulos terms, labeled $\xi^m$ for $U(1)_m$. These terms are allowed in the Lagrangian because they do not spoil the supersymmetry. The supersymmetry transformation for the auxiliary field $D^m$ is a total derivative and therefore the term $L^m_\text{FI}=-\xi^m D^m$ preserves supersymmetry. However, the Fayet-Iliopoulos term is not allowed for a nonabelian symmetry because it is not gauge invariant.

\section{Departing from Exact Supersymmetry} \normalsize

The terms that we have written so far preserve exact supersymmetry but the particles observed in nature show no sign whatsoever of a degeneracy between fermions and bosons, see (\ref{susymass}). A fully supersymmetric model would contain a massless fermionic partner of the photon and a charged scalar particle with the mass of the electron. These particles manifestly do not exist. Supersymmetry if it is relevant to nature, must be broken.

The breaking could be either explicit or spontaneous. In any case it has to be soft i.e. small enough to preserve the good features of supersymmetry and yet large enough to push the supersymmetric partners out of the reach of current experiments. Explicit breaking would be, in general sense, ad hoc. The susy generators would no longer commute with the Hamiltonian 
\begin{equation}
[Q_\alpha, H]\neq 0.
\end{equation}
However, we would inevitably lose the nice renormalization theorems and any attempt ro embrace gravity via local susy would be prohibited. So instead we prefer the spontaneous breaking of supersymmetry. We assume that the Lagrangian is supersymmetric but the vacuum state is not, that is the physical vacuum state $\left|0 \right\rangle$ is not invariant under a supersymmetry transformation i.e.
\begin{equation}
[Q_\alpha, H]= 0, \,\,\,\,\,\, \text{but} \,\,\,\,\,\,  Q_\alpha \left|0\right\rangle \neq 0 \,\,\,\,\, \text{or} \,\,\,\,\,\,  \bar{Q}_{\dot{\alpha}} \left|0\right\rangle \neq 0.
\end{equation}
This has important implications for the energy of the ground state. The anticommutator relation of the susy generators (\ref{susyanti}) can be related to the Hamiltonian with the aid of the identity
\begin{equation}
Tr(\sigma^{\mu}\bar{\sigma}^\nu)=\sigma^\mu_{\alpha\dot{\beta}}(\bar{\sigma}^\nu)^{\dot{\beta}\alpha}=2 \eta^{\mu\nu}.
\end{equation}
Then the (\ref{susyanti}) can be inverted to obtain
\begin{equation}
P^\nu=\frac{1}{4}(\bar{\sigma}^\nu)^{\dot{\beta}\alpha}\{Q_\alpha, \bar{Q}_{\dot{\beta}}\}
\end{equation}
and the Hamiltonian reads
\begin{equation}
H=P^0=\frac{1}{4} \left( Q_1\bar{Q}_{\dot{1}}+\bar{Q}_{\dot{1}}Q_1+Q_2\bar{Q}_{\dot{2}}+\bar{Q}_{\dot{2}}Q_2 \right).
\end{equation}
Hence, the energy of the vacuum cannot be negative. When supersymmetry is unbroken in the vacuum state this state has zero energy and when supersymmetry is spontaneously broken in the vacuum it has positive energy. As a result, whenever a supersymmetric vacuum state exists then it is the global minimum of the potential and when if there are more than one then it is degenerate with the other supersymmetric states. A non-supersymmetric state can be the global minimum only if the potential possesses no supersymmetric minimum. 

The spontaneous supersymmetry breaking can arise from some fields in the theory that acquire vacuum expectation values that are not invariant under supersymmetry transformations. In a theory with chiral superfields the only one of the supersymmetry transformation laws  whose expectation value can have a non-zero right hand side without breaking Lorentz invariance is 
\begin{equation}
\left\langle 0 \left| \delta_s \Psi_i \right| 0 \right\rangle = \left\langle 0 \left| F_i \right| 0 \right\rangle \epsilon. \,\,\,\,\,\,\,
\end{equation}
Hence, when one of the auxiliary fields of some chiral supermultiplet $\Phi_i$ has non-zero vacuum expectation value i.e. $ \left\langle 0 \left| F_i \right| 0 \right\rangle \neq 0 $ the supersymmetry breaks spontaneously.  

In a theory that a vector superfield is present there is the possibility that the field $D^a(x)$ has a non-zero vev $ \left\langle 0 \left| D^a \right| 0 \right\rangle \neq 0$ and the susy transformation
\begin{equation}
\left\langle 0 \left| \delta_s \lambda^a_r \right| 0 \right\rangle = -\left\langle 0 \left| D^a \right| 0 \right\rangle \epsilon_r 
\end{equation}
signals the spontaneous breaking, again here, of supersymmetry without violating the Lorentz invariance.

Once spontaneous supersymmetry breaking has occured a massless Goldstone fermion is expected to appear because the supersymmetry generator is fermionic, mush as a Goldstone boson appears when ordinary global symmetries are spontaneously broken. When a single auxiliary field $F_i$ acquires a vev the Goldstone fermion will be the spinor $\Psi_i$ in the supermultiplet $\Phi_i$ to which $F_i$ belongs. Similarly, for the case that an auxiliary field $D^a$ acquires a vev then a massless gaugino $\lambda^a_r$ that belongs at the same vector supermultiplet with the $D^a$ will be the Goldstone fermion associated with the spontaneous breaking of global supersymmetry. However, when global supersymmetry becomes local in supergravity theories the Goldstone fermion disappears from the spectrum. 

\section{Soft Supersymmetry Breaking} \normalsize

Supersymmetry explains the stabilization of the electroweak scale in a compelling way. The radiative corrections to the Higgs mass that destroy the hierarchy vanish in an exact supersymmetric theory. Since nature is not supersymmetric at the GeV scale it is of central importance to see whether a broken supersymmetry can still protect the Higgs potential from quadratic divergencies. Otherwise the main \itshape raison d' \^etre \normalfont of supersymmetry is lost. Fortunately, when supersymmetry is softly broken the quadratic divergencies are under control.

In the Wess-Zumino model, for example, we can demonstrate the breaking of supersymmetry by modifying the mass squared of the components $A$ and $B$ of the complex scalar in the chiral supermultiplet ($\phi, \Psi$). Assuming initially exact supersymmetry then all the fields in the supermultriplet have the same mass. It can be shown that the quadratically divergent self-energy diagrams for the scalars cancel due to the opposite contribution of the fermion. The only divergencies that are present are the logarithmic ones in the wave function renormalization. Let us now assume that the scalar $A$ and the pseudoscalar $B$ have masses squared $m^2+\delta m^2_A$ and $m^2+\delta m^2_B$ respectively and the fermion $m_\Psi=m$. Then the self-energy diagram for the scalar $A$ at one loop coming from the interaction with itself is
\begin{equation} \label{Aloop-1}
\sim \lambda^2 \int \frac{d^4 k}{(2\pi)^4}\frac{1}{k^2-m^2-\delta m^2_A} = \int \frac{d^4 k}{(2\pi)^4}\frac{1}{k^2-m^2}\left( 1+ \frac{\delta m^2_A}{k^2-m^2}\right) +\text{finite};
\end{equation}
the contribution of the pseudoscalar $B$ is also of the same form. It is apparent from (\ref{Aloop-1}) that quadratic divergencies cancel again and only a logarithmic divergence survives.

This splitting in the masses of complex scalar is represented as 
\begin{equation}
\delta {\cal L} _\text{soft} = - M^2 \phi^* \phi - \delta M^2(\phi^2+\phi^{*2})
\end{equation}
where $M^2$ is a supersymmetric mass squared and the $\delta M^2$ is the one that breaks supersymmetry "softly". On the contrary, one can not shift softly the mass of the fermion in the chiral supermultiplet because this would generate quadratic divergencies. Hence, a soft mass for matter fermion does not exist; such a mass would correspond to a hard breaking of supersymmetry. 

Another soft supersymmetry breaking term comes from the modification of the couplings of the form 
\begin{equation}
\delta L_\text{soft} = - A\lambda (\phi^3+\phi^{*3})
\end{equation}
If the term $\phi^3$ is allowed by the gauge symmetry, then in all generality, it is also allowed in the superpotential. Such a term provides only cubic scalar interactions which cannot lead to quadratically divergent two-point functions.

The last possibility of soft breaking exists in the vector supermultiplet. This is a mass for the supersymmetric partners of the gauge fields, the gauginos
\begin{equation}
\delta L_\text{soft}= - \frac{1}{2} M_\lambda \bar{\lambda}\lambda.
\end{equation} 
The above soft susy breaking terms seems to break explicitly supersymmetry. However, they should be considered as part of the low energy effective Lagrangian that does not include the sector responsible for the spontaneous breaking of supersymmetry.

\section{Local Supersymmetry Embraces Gravity}

In the previous sections we constucted a supersymmetric Lagrangian which yields an action that is invariant under supersymmetric transformations e.g. (\ref{s-1})-(\ref{s-3}) for a theory with a chiral superfield
\begin{equation} \label{}
\delta \phi \sim \bar{\epsilon} \Psi, \, \, \, \delta \Psi \sim \epsilon\gamma^{\mu}\partial_{\mu} \phi, \, \, \, \delta F \sim \bar{\epsilon}\gamma^{\mu}\partial_{\mu} \Psi.  
\end{equation}    
The $\epsilon$ is a spinorial "small" parameter that did not depend on spacetime. This symmetry is global. If we now allow $\epsilon=\epsilon(x)$, hence local supersymmetry, then new terms appear that spoil the supersymmetry of the theory. We can ensure local susy invariance by adding a term whose transformations will compensate the unwanted terms. This term must transform according to 
\begin{equation}
\delta\psi_{\mu r} = \frac{2}{\kappa}  \partial_{\mu} \epsilon_r 
\end{equation} 
Therefore, the new term that we introduce, $\psi_{\mu r}$, has both spinor and vector indices. It describes a spin-$3/2$ particle. Under susy transformations, this new term cancels at first place the unwanted term coming from the spacetime dependence of $\epsilon(x)$ however, it introduces new terms that render the action non-supersymmetric. We must now add in the Lagrangian an extra term 
\begin{equation} \label {grav1}
{\cal L}_g=-g_{\mu\nu}T^{\mu\nu}.
\end{equation} 
We introduced a tensor that we identify with the metric tensor of Einstein gravity and thus, the (\ref{grav1}) is the contribution of the scalar field to the Lagrangian of general relativity. In conclusion, it is possible to construct a locally supersymmetric Lagrangian with the fields $\phi$, $\Psi$, $\psi_{\mu}$ and $g_{\mu\nu}$. A massless spin-$2$ particle together with its susy partner, the massless spin-$3/2$ were introduced. The constant $\kappa$ is the gravitational coupling. Hence
\begin{equation}
\kappa^{-1}=\sqrt{\frac{\hbar c}{8 \pi G_n}}=\frac{m_{Pl}}{\sqrt{8 \pi}}\equiv M_P=2.4 \times 10^{18} \,\text{GeV}.
\end{equation}
Just as the local $U(1)$ phase invariance allowed us to "derive" Maxwell's theory of electromagnetism, so the requirement of local susy allows us to deduce classical gravity (general relativity) and gravitational interactions. 

The simplest possible supersymmetric gravity (sugra) model is that with only the gravity multiplet which consists of a spin-2 graviton, $g_{\mu\nu}$, and a spin-3/2 Majorana gravitino, $\psi_{\mu r}$. The Einstein Lagrangian is 
\begin{equation}
{\cal L}^{(2)}=-\frac{M^2_P}{2}eR
\end{equation}
where $e\equiv \text{det}({e^m}_\lambda)=[\text{det}({e_m}^\lambda)]^{-1}=[-\text{det}(g_{\mu\nu})]^{1/2}$ and ${e^m}_\lambda$ is the vierbein field which may be regarded as the square root of the metric. The Lagrangian that describes the massless spin-3/2 field $\psi_\mu$ is described by the Rarita-Schwinger Lagrangian
\begin{equation}
{\cal L}^{(3/2)}=-\frac{1}{2}\epsilon^{\mu\nu\rho\sigma}\bar{\psi}_\mu\gamma_5\gamma_\nu\partial_\rho\psi_\sigma.
\end{equation} 
A first guess for a sugra Lagrangian is the sum of the ${\cal L}^{(2)}$ and ${\cal L}^{(3/2)}$ 
\begin{equation}
{\cal L}={\cal L}^{(2)}+{\cal L}^{(3/2)}
\end{equation} 
which includes covariant derivatives instead of the $\partial_\rho$ since we work in a curved spacetime. The simple gravity multiplet (${\epsilon^m}_\lambda$, $\psi_{\mu r}$) is adequate "on-shell" but it must be supplemented by additional auxillary fields in order that the "off-shell" algebra is closed. Counting the degrees of freedom we find that we have 12 fermionic and 6 bosonic degrees of freedom. This mismatch means we need at least six auxiliary bosonic fields. The off-shel algebra closes if we introduce a scalar $S$, a pseudoscalar $P$ and an axial vector $A_\mu$ with a sugra Lagrangian of the form 
\begin{equation}
{\cal L}={\cal L}^{(2)}+{\cal L}^{(3/2)}-\frac{e}{3}(S^2+P^2-A^2_\mu)
\end{equation} 
where the auxiliary fields are eliminated by the equation of motion. 

To obtain a local supersymmetric Yang-Mills theory we must couple the "pure" sugra Lagrangian to a Lagrangian that describes the gauge and matter fields. We construct covariant derivatives $\partial_\rho\rightarrow D_\rho$ which enable the graviton to couple to the spin-1, spin-1/2 and spin-0 fields. We next add to the Lagrangian the terms that are required to ensure local susy. We then eliminate the auxiliary fields in favour of the dynamical fields. The final Lagrangian (keeping only few essential terms) has the form 
\begin{equation} \label{fullsugra}
\frac{1}{e}{\cal L}=-\frac{1}{2}R+G^j_i\partial_\mu\phi^i\partial^{\mu}\phi^*_j-e^G[G_iG_j(G^{-1})^i_j-3]+V_D-\frac{1}{4}\text{Re}(f_{ab})F^a_{\mu\nu}F^{\mu\nu b}+e^{G/2}\bar{\psi}_\mu\sigma^{\mu\nu}\psi_\nu+...
\end{equation} 
where the gravitational coupling (or the reduced planck mass) was set equal to unity $M_P=1$ and where
\begin{equation}
G_i\equiv \frac{\partial G}{\partial\phi^i}, \  \  \ G^j\equiv \frac{\partial G}{\partial\phi^*_j}, \ \ \ G^j_i\equiv \frac{\partial^2 G}{\partial\phi^i\partial\phi^*_j}.
\end{equation} 
The full Lagrangian is characterized by two arbitrary functions of the scalar fields: a real function $G(\phi^i,\phi^*_i)$ called generalized K\"ahler potential and an analytic function $f_{ab}(\phi^i)$. These functions determine the general forms allowed for the kinetic energy terms of the scalar fields $\phi^i$ and of the gauge fields $A^a_\mu$ respectively. The scalar kinetic-energy term demonstrates that $G^j_i$ plays the role of the metric in the space spanned by the scalar fields and is referred to as a K\"ahler metric. In the absence of gravity $G^j_i\rightarrow\delta ^j_i$ and $f_{ab}\rightarrow\delta_{ab}$. 

The Lagrangian (\ref{fullsugra}) contains a scalar potential of the form  
\begin{equation} \label{sugrascalar-1}
V(\phi,\phi^*)=V_D+e^G[G_iG_j(G^{-1})^i_j-3].
\end{equation} 
The $V_D$ is the part of the potential stemming from the $D^a$ auxilary fields. The origin of the first term in the square brackets is related to the elimination of the auxiliary field term $|F|^2$ including now the K\"ahler metric
\begin{equation} \label{}
|F|^2\rightarrow \left|\frac{\partial W}{\partial \phi^i}\right|^2(G^{-1})^i_j.
\end{equation} 
The second term in the bracket comes from the elimination of the auxiliary scalar field terms, $-|S+iP|^2$ in the sugra part of the Lagrangian. The negative sign has considerable importance. The $e^G$ factor arises from the Weyl rescaling of the $e^\mu_\lambda$ fields required to bring the first term in the sugra Lagrangian (\ref{fullsugra}) into the canonical Einstein form, $-R/2$. This rescaling implies a redefinition of the fermion fields and hence a factor $e^{G/2}$ in the last part of the Lagrangian. Owing to this term, when the local susy is spontaneously broken the gravitino acquires a mass  
\begin{equation} \label{}
m_{3/2}=e^{G/2},
\end{equation} 
where $G$ evaluated at the minimum of the potential (\ref{sugrascalar-1}).

In general, in the presence of gravity there is no requirement that the Lagrangian should be renormalizable and there is no reason the $G(\phi, \phi^*)$ to correspond to only renormalizable kinetic terms. However, the function $G$ has to satisfy certain conditions for the theory to be well defined. We require $G^i_j>0$ so that the kinetic terms of the scalar fields have the correct sign. A special choice is
\begin{equation} \label{}
G= \frac{\phi^i \phi^*_i}{M^2_P}+\text{log}\left|\frac{W}{M^3_P}\right|^2
\end{equation}  
where we reintroduced the Planck mass. With this choice of the K\"ahler potential the sugra scalar potential (\ref{sugrascalar-1}) is written
\begin{equation} \label{sugrascalar2-1}
V(\phi,\phi^*)=e^K(W_i+M^{-2}_P K_iW)(K^{-1})^i_j(W^{\dagger j}+M^{-2}_P K^j W^\dagger)-3e^K  M^{-2}_P|W|^2+ V_D
\end{equation} 
where $V_D=1/2 g^2 G^i(T_a)_{ij}\phi_j G^k(T_a)_{kl}\phi_l$. 
In the limit of vanishing gravity, $M_P\rightarrow \infty$, the scalar potential reduces to the potential of the global supersymmetry $V=\sum_i|\partial W/\partial\phi^i|^2+\frac{1}{2}g^2\sum_a(\phi^*T^a\phi)^2$.

\section{Local Supersymmetry Breaking}

In the case of globally supersymmetric theories a supersymmetric vacuum state has zero energy, and if supersymmetry was spontaneously broken in the vacuum state has positive energy. When gravity is included the vacuum energy is no longer positive definite. At the scalar potential (\ref{sugrascalar-1}) we see the negative term $V=-3e^G M^4_P= -3 e^K |W|^2 M^4_P$.
There is the possibility here of a ground state with negative energy. 

In the local supersymmetric case spontaneous breaking also occurs when at least one of the fields in the theory has a vev that is not invariant under supersymmetry transformations. Asking for Lorentz invariance the only local supersymmetry transformation laws that can be non-zero are 
\begin{equation} \label{locF-1}
\left\langle 0 \left| \delta \Psi_i \right| 0 \right\rangle = - e^{G/2}(G^{-1})^i_j G_j \epsilon
\end{equation}
and 
\begin{equation} \label{locD-1}
\left\langle 0\left| \delta \lambda_a \right| 0 \right\rangle = \frac{i}{2} g Re f^{-1}_{ab} G^i(T_b)_{ij} \phi_j \epsilon 
\end{equation}
where expectation values of the fields is assumed. For a theory with a gauge non-singlet chiral superfield $\Phi_i$ with canonical kinetic terms the right hand sides of (\ref{locF-1}) and (\ref{locD-1}) can be non-zero. In particular in the former case
\begin{equation}
\frac{\partial W}{\partial \phi_i}+\phi^*_i W \neq 0
\end{equation}
for some values of the fields, generalizing $F$-term supersymmetry breaking to supergravity. In the later case
\begin{equation}
G^i(T_a)_{ij}\phi_j \neq 0 
\end{equation}
generalizing $D$-term breaking in supergravity.

Another important point is the value of the vacuum energy. In principle, a supersymmetry breaking minimum can have negative, positive or zero value. For phenomenological reasons it is desirable to tune its value to zero. Apart from the cosmological constant problem it is also the definition of the mass which is subtle when the spacetime is anti-de Sitter. If the minimum is anti-de Sitter with energy density $-V_0$ it is possible to achieve cancelation of this vacuum energy thanks to a positive contribution that can be tuned to be $+V_0$.
Then, from the sugra Lagrangian (\ref{fullsugra}) the term $e^{G/2}M_P$ corresponds to $m_{3/2}$ since, $\bar{\psi}_\mu \sigma^{\mu\nu}\psi_\nu = 1/2 \bar{\psi}^\mu \psi_\mu$. Hence we see that when supersymmetry is broken the gravitino has mass
\begin{equation} \label{m3/2}
m^2_{3/2}  \equiv e^G M^2_P = M^{-2}_P\frac{V_0}{3}.
\end{equation}
The breaking is now demonstrated by a non-zero gravitino mass (which can easily recognized in the Lagrangian thanks to the Minkowski background). Since the graviton stays massless, a massive gravitino is a clear sign of supersymmetry breaking in the spectrum. 

The last statement is tightly connected to the appearence of the massless Goldstino due to the spontaneous breaking of supersymmetry. In locally supersymmetric theories the Goldstino field is 'eaten' by the gauge field of local supersymmetry, namely the gravitino, and in this way the gravitino acquires a mass. Similar to the Higgs mechanism, there is a choice of supersymmetric gauge in which the Goldstino disappears from the physical spectrum providing the helicity $\pm 1/2$ degrees of freedom to the gauge transformed gravitino $\psi_{\mu r}$.  This is the so called  "super-Higgs" mechanism. Hence, local supersymmetric theories have the attractive features that do not suffer from massless Goldstino fields and can give a Minkowski supersymmetry breaking vacuum as presented above. 

\section{The Minimal Supersymmetric Standard Model}

Here we briefly review the minimal version of a supersymmetric Standard Model (MSSM). The observable fields are considered to be part of superfields. The missing components of the supermultiplets have to be massive enough in order to evade experimental limits. Their masses should be soft i.e. they are at most corrected logarithmically. The MSSM soft masses are listed below:
\begin{enumerate}
	\item  Mass terms for squarks, sleptons and Highs fields:
	\begin{equation}\nonumber
	{\cal L}_\text{scalars}= Q^*m^2_Q Q+ \bar{U}^*m^2_U \bar{U} +\bar{D}^* m^2_D \bar{D}+L^*m^2_L L + \bar{E}^* m^2_E \bar{E}
	\end{equation}
	\begin{equation}
	m^2_{H_U}|H_U|^2+m^2_{H_D}|H_D|^2+ B_\mu H_U H_D + \text{c.c.}
	\end{equation}
	The $m^2_Q$, $m^2_U$ etc. are hermitian matrices in the space of flavours. The first five matrices are $3\times 3$ Hermitian matrices (45 parameters); the Higgs mass terms add an additional, for a total of 49 parameters.
  \item  Cubic couplings of the scalars:
   \begin{equation} 
  {\cal L}_A= H_U Q A_U \bar{U}+ H_D Q A_D \bar{D}+ H_D L A_E \bar{E}+ c.c.
  \end{equation}
  The matrices $A_U$, $A_D$ and $A_E$ are complex matrices, each with 18 real entries so we have additional 54 parameters.
  \item Mass terms for $b$, $w$, and $\lambda$ gauginos of the gauge groups  $U(1)$, $SU(2)$ and $SU(3)$ respectively:
  \begin{equation}
  m_1 bb + m_2 ww + m_3 \lambda \lambda
  \end{equation}
  These are three complex quantities, making six additional parameters
  \item The $\mu$ term for the Higgs field, 
  \begin{equation}
  W_\mu=\mu H_U H_D
  \end{equation}
  which accounts for two additional parameters.
\end{enumerate}
Counting the numbers of the MSSM parametrs beyond those of the Standard Model it appears to be 111 new parameters.
However, the MSSM lagrangian has symmetries which are broken by the general soft breaking terms that include a "Peccei-Quinn" symmetry under which the $H_U$ and $H_D$ rotate by the same phase and accordingly the quarks and leptons; a continuous $R$-symmetry. Redefining the fields reduces the number of parameters to 105.

Over the years, there have been extensive searches for superpartners of ordinary particles and these severely constrain the spectrum. These translate to constraints on the construction of supersymmetric models including models of Dynamical Supersymmetry Breaking. The phenomenological requirements to be fulfilled are basically 
\begin{itemize}
\item the $Z$ boson mass, $m_Z=91$ GeV
\item the Higgs boson mass constraint, $m_h> 114$ GeV \cite{Barate:2003sz}
\item the Higgsino mass constraint, $m_{\tilde{H}} > 94$ GeV \cite{Abdallah:2003xe}
\item the gaugino mass constraint, $m_\lambda > {\cal O}(100)$ GeV \cite{Abdallah:2003xe}, however, one of the neutralinos can be very light \cite{Hooper:2002nq, Belanger:2003wb, Dreiner:2009yk}
\item the top quark mass, $m_t=173$ GeV \cite{:2009ec}
\item and the bound on the electric  dipole moments of the neutron, $d_n< 3 \times 10^{26}e \,cm$  \cite {Baker:2006ts}
\end{itemize}
These bounds come from direct searches of superpartners but also from the absence of Flavour Changing Neutral Currents (suppression of $K\leftrightarrow \bar{K}$, $D\leftrightarrow \bar{D}$ mixing; $B \rightarrow s+ \gamma$, $\mu \rightarrow e+\gamma$,... ) \cite{Masiero:2002xj} and suppression of the $CP$ violation ($d_n$, phases in $K \bar{K}$ mixing).

These imply a spectrum that is highly degenerate and $CP$ violating phases in the soft breaking Lagrangian are suppressed. This happens in many gauge mediated models (see next chapter) and in special regions of some superstring moduli space \cite{Kaplunovsky:1993rd}. Other possible explanations include flavour symmetries \cite{Banks:1994yg}.

\subsection{The interference with Cosmology}

An important observational output from cosmology is the certainty of the dark matter existence. The dark matter relic density has been determined with good precision. Upcoming observations, such as by the Planck satellite are likely to reduce the uncertainties in the relic density determination to $1\%$ level, given the standard cosmological assumptions. Astrophysical experiments may also detect dark matter either directly through its interactions with ordinary matter or indirectly through its annihilation decay products. Such data, combined with astrophysical inputs such as the dark matter halo profile and the local density provide information about the strength $\chi N$ scattering and $\chi\chi$ annihilation, where $\chi$ the dark matter particle and $N$ an ordinary nucleus.

Until today, the cosmological observations and the dark matter detection experiments cannot discover supersymmetry. In particular, the properties of dark matter are loosely constrained. If dark matter is discovered in direct or indirect detection experiments, its mass and interaction strengths will be bounded but only very roughly at first. Moreover, the microscopic implications of such experiments are clouded by significant astrophysical ambiguities, such as the dark matter velocity distribution, halo profiles etc.

\subsection{The Supersymmetric Fine Tuning Problem}

The experimental lower limit bound on the Higgs boson mass from LEP-II, $m_h>114$ GeV \cite{Barate:2003sz} has put a threat on supersymmetric models. 
In minimal supersymmetry the Higgs field $h$ is a linear combination of the two Higgs doublets, $H_U$ and $H_D$. The potential for $h$ is given by \cite{Kitano:2006gv}
\begin{equation}
V=m^2_h |h|^2+\frac{\lambda_h}{4}|h|^4,
\end{equation}
where $m^2_h$ is negative and $\lambda_h$ is positive. By minimizing the potential we obtain $\left\langle h \right\rangle^2 =-2 m^2_h/\lambda_h$ and the physical Higgs boson mass is 
\begin{equation}
M^2_\text{Higgs}=\lambda_h \left\langle h \right\rangle^2=-2 m^2_h
\end{equation}
For moderately large $\text{tan} \beta \equiv \left\langle  H_U \right\rangle / \left\langle H_D \right\rangle$ e.g. $\tan \beta \gtrsim 2$, $m_h$ can be written as 
\begin{equation}
\left. \left. m^2_h=|\mu^2|+m^2_{H_U}\right|_\text{tree} + m^2_{H_U} \right|_\text{rad}\, ,
\end{equation}
where $\mu$ is the supersymmetric mass for the Higgs doublets and $\left. m^2_{H_U}\right|_\text{tree}$ and $\left. m^2_{H_U} \right|_\text{rad} $ represent the tree-level and radiative corrections to the soft supersymmetry-breaking mass squared for $H_U$.  The dominant contribution to $\left. m^2_{H_U} \right|_\text{rad} $ arises from top-stop loop:
\begin{equation}
\left. m^2_{H_U} \right|_\text{rad} \simeq -\frac{6 y_t}{(4\pi^2)} \left(m^2_{Q_3} + m^2_{U_3}+|A_t|^2  \right) \ln \left(\frac{M_\text{mess}}{m_{\tilde{t}}} \right)\, ,
\end{equation}
where $y_t$ is the top Yukawa coupling, $m_{Q_3}$ and $m_{U_3}$ soft supersymmetry breaking masses for the third-generation doublet quark, $Q_3$ and singlet up-type quark $U_3$, and $A_t$ the trilinear scalar interaction term for the top squarks (a product of the $A$ parameter and the top Yukawa coupling i.e. ${\cal L}= -y_t A_t Q_3 U_3 H_U + \text{h.c.}$). The $M_\text{mess}$ represents the scale at which squark and sleptons are generated.
 
In order to satisfy the experimental bound, we need either a heavy scalar top quark (stop) or a large $A_t$-term (the stop-stop-Higgs coupling) since a significant one-loop contribution to $m_h$ is necessary. On the other hand, once we have large $m_{\tilde{t}}$ or $A_t$ it induces a large one-loop contribution to the soft mass term $m^2_{H_U}$. This immediately means that there is a fine tuning in the electroweak symmetry breaking, namely
\begin{equation}
\left. \left. \frac{M^2_Z}{2} \simeq -|\mu|^2 -m^2_{H_U}\right|_\text{tree}- m^2_{H_U}\right|_\text{rad}
\end{equation}
$M_Z$ is the $Z$ boson mass. If $\left. m^2_{H_U}\right|_\text{rad}\gg M^2_Z$ we need cancelation between $\left. m^2_{H_U}\right|_\text{rad}$ and either $\mu^2$ or $\left. m^2_{H_U}\right|_\text{tree}$ to reproduce the correct value of the $Z$ boson mass. A cancellation of at least ${\cal O}(1-5\%)$ is necessary to satisfy the bound on the Higgs boson mass in a generic gravity or gauge mediation modes \cite{Kitano:2006gv}.

A variety of solutions have been proposed to this problem. Models with an additional singlet beyond the MSSM fields are known collectively as the "Next to Minimal Supersymmetric Standard Model" (NMSSM).

\chapter{Schemes of Supersymmetry Breaking}

Firstly, $F$-term and $D$-term paradigms of spontaneous supersymmetry breaking are presented. The necessity for a hidden sector and its features are discussed. Secondly, the gravity, anomaly and gauge mediation schemes of supersymmetry breaking to the visible sector are described. The gauge mediation is presented in detail.  

\section{F-term Supersymmetry Breaking}
In the first chapter it was shown that supersymmetry breaks spontaneously when one of the auxiliary fields $F_i$ or $D^a$, of a chiral and a vector supermultiplet respectively, acquires a vev. This can happen dynamically by constructing models that implement this spontaneous breaking. A simple example is model which has a non-supersymmetric ground state. In the global limit, $M_P \rightarrow \infty$, this is a state of positive energy. This breaking of supersymmetry can be realized in a theory with chiral superfields $\Phi_i$. Working from now on in the global limit, the potential reads $V=\sum_i |\partial W /\partial \varphi_i|$. Hence all the information -the interactions- is included in the superpotential which for a renormalizable theory takes the general form 
\begin{equation}
W(\Phi_i)=f_i \Phi_i + \frac{1}{2}m_{ij} \Phi_i \Phi_j + \frac{1}{3} \lambda_{ijk}\Phi_i\Phi_j\Phi_k.
\end{equation}
The simplest example of a model without any supersymmetric minima is the O'Raifeartaigh model \cite {O'Raifeartaigh:1975pr} has three chiral superfields $\Phi_1$, $\Phi_2$ and $\Phi_3$ ($\varphi_1, \varphi_2, \varphi_3$, the dynamical scalar degrees of freedom) with superpotential
\begin{equation}
W(\Phi_1, \Phi_2, \Phi_3)=\lambda \Phi_1(\Phi^2_3-\mu^2)+M \Phi_2 \Phi_3.
\end{equation}
For $\mu^2<M^2/2\lambda^2_1$ the absolute minimum of the potential 
\begin{equation}
V=\sum^3_{i=1} |F_i|^2 = \lambda^2 |\varphi^2_3-\mu^2|^2+M^2|\varphi_3|^2+|M\phi_2 + 2\lambda\varphi_1\varphi_3|^2
\end{equation}
occurs at $\left\langle \varphi_2 \right\rangle= \left\langle \varphi_3 \right\rangle =0$ and $\left\langle \varphi_1 \right\rangle$ is undetermined. In a first view $\left\langle \varphi_1 \right\rangle$ seems to be a flat direction. However, since it is a supersymmetry breaking direction, the interctions of the $\varphi_1$ with the rest of the fields induce a potential at 1-loop approximation for the $\varphi_1$ and the degeneracy gets lifted. Thus it is usually called pseudo-flat direction. At this absolute minimum 
\begin{equation}
F^\dagger_1=\lambda \mu^2,  \,\,\,\ F^\dagger_2=F^\dagger_3=0
\end{equation}
and
\begin{equation}
V=\lambda^2 \mu^4>0.
\end{equation}
Because $F_1$ is non-zero we expect that the fermion of the $\Phi_1$ chiral supermultiplet, $\Psi_1$, to be the Goldstone fermion associated with the spontaneous breaking. This may be verified by looking at the fermion mass matrix $\partial^2 W/ \partial \varphi_i\partial \varphi_j $. Hence at tree level the $\Phi_1$ supermultiplet has supersymmetric masses. On the other hand the masses of $\Phi_2$ and $\Phi_3$ supermultiplets are no longer supersymmetric. One finds $\sqrt{M^2-2\lambda F_1}$ for the Re($\varphi_2$), $\sqrt{M^2+2\lambda F_1}$ for the Im($\varphi_2$) and $M$ for $\varphi_3$. The off-diagonal structure of the fermion mass matrix signals that the two Weyl spinors $\Psi_2$ and $\Psi_3$ can combine to give a single Dirac fermion with mass $\pm M$ (the sign, as in any fermion mass, can be redefined through phase redifinitions). The particular mass squared splitting $\pm 2\lambda F$ in the components of the complex scalar $\varphi_2$ is due to its direct coupling to the $\Phi_1$ which contains the Goldstone fermion.

One may observe that in the above theory the sum squared masses of the bosons is twice the sum of the mass squared of the fermions. This is a tree level feature of theories of chiral superfields
\begin{equation} \label{Str2}
\text{STr} M^2 \equiv \sum^{1/2}_{J=0} (-1)^{2J}(2J+1)M^2_J= \sum M^2_\text{boson}- 2\sum M^2_\text{fermion}
\end{equation} 
where $\text{STr} M^2$ and is called the supertrace. In the presence of supersymmetry breaking equation (\ref{Str2}) is modified by radiative corrections, though it does not receive divergent contributions (section 1.6). It is also modified in theories containing vector superfields if $D$-terms develop vevs. Nevertheless, such a relation poses severe phenomenological problems: it means that the average boson mass squared coincides with the average fermion mass squared which is not supported experimentally. For the electron, for example, this would imply that one of the two scalar selectrons must have a mass less than or equal to that of the electron. However, limits on the supersymmetric particles tend to show that on average, bosons are much heavier (not discovered actually (2011)) than the fermions in the hypothetical chiral supermultiplets. 

\section{D-term Supersymmetry Breaking}

In gauge theories it is possible to break supersymmetry utilizing the auxiliary $D(x)$ field of a vector superfield. Let us assume a $U(1)$ gauge theory interacting with chiral superfields $\Phi_i$ having charges $e_i$. This yields a $D$-term 
\begin{equation} \label{Dterm2}
D=- \sum_i e_i \varphi^\dagger_i\varphi_i.
\end{equation}
The vev of the $D$ field can be non-zero if the scalars $\varphi_i$ have non-zero vevs. This last depends on the way the scalars interact, hence, on the superpotential $W(\Phi)$. However, it is possible to realize a $D$-term supersymmetry breaking without relying on the form of the superpotential. From (\ref{Vtot1}) we see that for $U(1)$ gauge theories there is an additional gauge invariant supersymmetric term: the Fayet-Iliopoulos $D$-term \cite{Fayet:1974jb}
 and the (\ref{Dterm2}) reads 
\begin{equation} \label{DFIterm2}
D=- \left( \xi  + \sum_i e_i \varphi^\dagger_i\varphi_i\right).
\end{equation}
It is then possible to break supersymmetry in a gauge theory with just a single scalar chiral field $\Phi$ having charge $e$. The Lagrangian is 
\begin{equation}
{\cal L}= \frac{1}{32}(W^\alpha W_\alpha)_F+(\Phi^\dagger e^{2eV} \Phi)_D+ \xi(V)_D
\end{equation}
where $W_\alpha$ the field strength (chiral) superfield and $V$ the vector superfield. The $U(1)$ gauge invariance forces the superpotential, and therefore the $F$-terms, to vanish. The $D$-term reads $D=- \left( \xi  +  e \varphi^\dagger\varphi\right)$ and if $\xi e > 0$ then the scalar potential $V(\varphi)=D^2/2$ cannot vanish and it gets minimized for a zero vev of $\varphi$. Thus the $U(1)$ gauge invariance is unbroken but since $\left\langle  D \right\rangle =- \xi \neq 0$ the ground state of the potential has positive energy $\xi^2/2$ and supersymmetry is spontaneously broken. The scalar field $\varphi$ acquires a non-zero mass $m^2_\varphi= e \xi$ while its fermionic superpartner $\Psi$ remains massless. The unbroken gauge invariance ensures that the gauge field $A_\mu$ remains massless. The gaugino field $\lambda$ also remains massless for it is the Goldstone fermion associated with the spontaneous breaking of the global supersymmetry.

Models of $D$-term breaking that include more fields can be constructed. Generally, the Fayet-Iliopoulos term affects the scalar sector while the fermion sector does not feel the broken supersymmetry. Moreover, an abelian gauge symmetry has potential problems with quantum anomalies (triangle anomaly).

As mentioned it is also possible the $D$-term to be non-zero without the presence of a Fayet-Iliopoulos term. For example, in an $SU(3)$ supersymmetric Yang-Mills theory 
\begin{equation}
V_D=\frac{1}{2}\left(\sum_3 \varphi^\dagger T^a \varphi - \sum_{\bar{3}} \bar{\varphi} T^a \bar{\varphi}^\dagger \right)^2.
\end{equation}
Generally the terms in the parenthesis can sum to zero. However, one can arrange a superpotential such that the solutions of the $F_i=0$ conditions do not coincide with the solutions of the $D=0$ conditions. This can lead to a $D$-term breaking again with the sum rule (\ref{Str2}) valid at tree level.

\section {The Hidden Sector Hypothesis} \normalsize
The conclusion of the last two sections is that these models presented cannot be straightly applied to the Standard Model and explain the mass spectrum of the particles we observe. The problem is rooted in the sum rule (\ref{Str2}) which is, however, valid only at tree level. Hence, the phenomenological problems encountered by both $F$ and $D$ type breaking can be evaded once we isolate the supersymmetry breaking sector from the sector of quarks, leptons and their supersymmetric partners.

Isolation of the supersymmetry breakig sector means that there are no tree level couplings between the superfield that breaks supersymmetry (via its $F$ or $D$ auxiliary fields) with the superfields of the Supersymmetric Standard Model (SSM). The first step on the way of constructing (or discovering from another point of view) this, so called, \itshape hidden sector \normalfont is to identify its energy scale. Heuristically we can say that the splitting in the supermultiplets has to be proportional to the vev of the $F$ or $D$ field (i.e. the scale of supersymmetry breaking) and to the strength of the interaction that mediates this breaking. As explained, there must not be any tree level couplings, hence the coupling between the supermultiplet that contains the Goldstone fermion and the observable fields has to be either loop suppressed or gravitational. Taking into account that the splitting in the supermultiplets is experimentally constrained to be at least of the order of ${\cal O} (100)$ GeV this implies that the non-zero vev of the $F$ field is much larger than the masses of the observed fields. Hence, the hidden sector is characterized by a new mass scale well above the electroweak scale.

We are thus led to the following picture, which provides a phenomenologically reasonable supersymmetric extension of the Standard Model. Apart from extending the Standard Model fields to superfields we also introduced a hidden sector with no direct coupling to quarks, leptons and Standard Model gauge bosons. Supersymmetry is spontaneously broken in this hidden sector. A weak interaction couples the two sectors and induces a supersymmetry-breaking effective interaction for the Standard Model particles and their superpartners. If $M_\text{HS}$ is the mass scale of the hidden sector, the supersymmetry breaking mass terms induced for the Standard Model sector are of the order of 
\begin{equation} \label{m2}
m_{SB} \sim \frac{\left\langle F \right\rangle}{M_\text{mess}} \sim \frac{M^2_\text{HS}}{M_\text{mess}}
\end{equation}
where $M_\text{mess}$ is the mass of the particle responsible for the weak connection between the two sectors. The mass $M_\text{mess}$ is called the \itshape messenger scale \normalfont. The universal gravity interaction is always present. Hence \itshape supergravity \normalfont is a messenger by default; if it is aslo the only one then $M=M_{P}$ and then $M_\text{HS} \sim 10^{11}$ GeV. In this scenario the superpartners acquire mass of the order of the parameter $m_{SB}$ in (\ref{m2}). It is also very reasonable to assume that the hidden sector is charged under the standard model gauge group. In that case the mass $M_\text{mess}$ can be much lower than the $M_P$. This implies that the scale of supersymmetry breaking $M_\text{HS}$ has to be lower here in order to keep the mass of the superpartners $m$ at the same scale. This scenario is called \itshape gauge mediation \normalfont which is also the direction of research of this thesis. Finally, there is an alternative to supergravity and gauge mediation scenarios the so called \slshape anomaly mediation \normalfont.

The breaking of supersymmetry results in a splitting in the supermultiplets. Still, the quarks, the leptons and the gauge bosons cannot obtain mass until the $SU(2)_L \times U(1)_Y$ is broken. Nonetheless, it is attractive consider that the symmetry-breaking terms that give mass to the superpartners cause the $SU(2)_L \times U(1)_Y$ to be spontaneously broken, at more or less the same scale.

\section{Supergravity Mediation} \normalsize

A straightforward way to hide a sector it is to assume that it has no any Standard Model charges. Hence, the supersymmetry breaking effects will be communicated by gravity which is always present.

\subsection{A Supergravity Application: The Polonyi Model} \normalsize

The simplest example of gravity mediated supersymmetry breaking to the observable sector is the Polonyi model. We assume that we have only one chiral gauge-singlet supermultiplet for which the superpotential is 
\begin{equation} \label{}
W({\cal X})=m^2({\cal X}+\beta)
\end{equation} 
where ${\cal X}$ denotes the chiral superfield in the hidden sector\footnote{In the following we will adopt the convenient notation where $X$ stands for both the superfield and the scalar component of the multiplet.}. From ($\ref{sugrascalar2-1}$) the scalar potential for the scalar component $X$ of ${\cal X }$ and for canonical K\"ahler becomes 
\begin{equation} \label{Pol-2}
V=e^{X^{\dagger}X}\left(\left|(m^2+\frac{m^2(X+\beta)X^\dagger}{M^2_P}\right|^2-3\frac{m^4}{M^2_P}\left|X+\beta\right|^2\right).
\end{equation} 
At the minimum $\partial V/\partial X=\partial V/\partial X^\dagger=0$ we impose the condition of zero cosmological constant $V=0$ and we get
\begin{equation} \label{}
X_{min}=(\sqrt{3}-1)M_P,  \ \  \  \ \beta=(2-\sqrt{3})M_P.
\end{equation} 
The gravitino mass at this ground state is 
\begin{equation} \label{}
m_{3/2}=M_P\, e^{G/2}=\frac{|W|}{M^2_P}\,\, e^{X^\dagger X/2M^2_P}=\frac{m^2}{M_P}e^{2-\sqrt{3}}
\end{equation}  
and the scale of supersymmetry breaking defined by 
\begin{equation}
M^2_\text{SB} =e^{G/2} (G^{-1})^j_i G_j
\end{equation}
is $M^2_\text{SB} = \sqrt{3} m_{3/2}M_P$. 
As we will see below, the gravitino mass is tightly related to the soft masses. We want to have $m_{3/2}\sim M_W\sim 1$ TeV. Hence, the parameter $m$ in the hidden sector must be an intermediate scale
\begin{equation} \label{}
m\simeq \sqrt{M_Pm_{3/2}}\sim \sqrt{M_PM_W}\sim 10^{11} \text{GeV}.
\end{equation}  

\begin{figure}[htbp]\begin{center}
\includegraphics[width=0.495\linewidth]{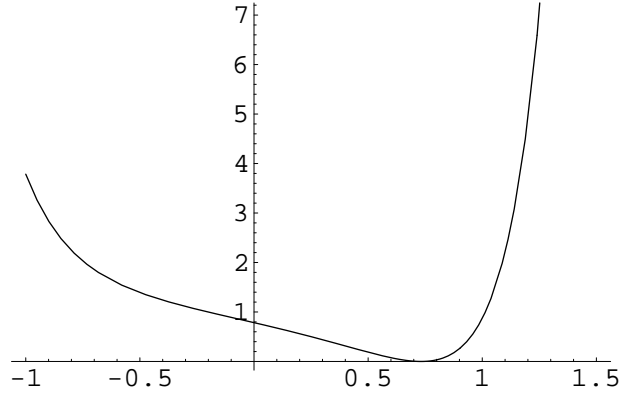}
\caption{ The shape of the Polonyi potential. The horizontal axis is the Polonyi field $X$ value in $M_P=1$ units. The vertical axis is the potential $V(X)$ density in $m^4$ units. The minimum is close to $M_P$, it is non-supersymmetric and the cosmological constant there (the Polonyi field potential energy density) is zero.}
\end{center}\end{figure}

In order the information of supersymmetry breaking to reach the observable sector we couple the hidden to the low-energy observable multiplets assuming that the superpotential is 
\begin{equation} \label{}
W({\cal X},\Phi^i)=W_h({\cal X})+W_o(\Phi^i).
\end{equation} 
In the limit of global supersymmetry  the scalar potential reads
\begin{equation}
V(X, \phi^i)= \left| \frac{dW_h}{dX}\right|^2+ \sum_i\left|\frac{\partial W_o}{\partial \phi^i} \right|^2
\end{equation}
and the two sectors are completely decoupled. Restoring the $M_P$ we see that the two sectors are coupled only gravitationally. Assuming also here a flat K\"ahler metric we can calculate the scalar potential asking again for zero cosmological constant at the minimum. After parametrizing the minimum as $X=aM_P$ and the hidden sectror superpotential as $W_h=bm^2M_P$ the gravitino mass reads $m_{3/2}\simeq bm^2e^{a^2/2}/M_P$.  
The low energy effective theory can be obtained in the limit $M_P\rightarrow \infty$ and holding at the same time the $m_{3/2}$ fixed. Then the low energy effective potential takes the form 
\begin{equation} \label{Veff-2}
V_\text{eff}=m^4\left[(1+ab)^2-3b^2\right]e^{a^2}+\left|\frac{\partial \hat{W}_o}{\partial \phi ^i}\right|^2+ V_\text{soft}+ V_D
\end{equation}  
where 
\begin{equation} \label{Vsoft-2}
V_\text{soft}=m^2_{3/2}|\phi^i|^2+m_{3/2}\left[\frac{\partial \hat{W}_o}{\partial \phi ^i}\phi^i+(A-3)\hat{W}_o+\text{h.c.}\right]
\end{equation} 
and $\hat{W}_o\equiv e^{a^2/2}W_o$, $A\equiv a(1/b+a)$. The first term in the (\ref{Veff-2}) is the cosmological constant of the theory which can be set to zero by the choice $(1+ab)^2=3b^2$. Actually, there will be corrections by constant terms of order $m^2_{3/2}/M^2_P$ and the cosmological constant can be fine tuned to zero only after taking into account any symmetry breaking in the light sector. The second term is of the standard form for unbroken global supersymmetry with superpotential $\hat{W}$. The $V_\text{soft}$ are the supersymmetry breaking terms. The last term are the $D$-terms which, in general, the low energy effective potential may contain. They read $V_D=\text{Re} f^{-1}_{ab} D_a D_b$ with $D_a=g\phi^{*i} (T_a)^j_{i}\phi_j$.

The part of the low-energy Lagrangian involving chiral spin-$1/2$ \itshape fermions \normalfont is as for unbroken global supersymmetry and hence, all the supersymmetry breaking terms are soft. This can be seen from the complete supergravity Lagrangian for chiral supermultiplets \cite{Bailin:1994qt}. Therefore, $m_{3/2}$ is the universal supersymmetry breaking mass splitting between bosons and fermions in the same chiral supermultiplet. It is possible to obtain an effective potential in which the  \itshape universality \normalfont is absent by taking the kinetic terms for the chiral superfields to be non-minimal. However, universal supersymmetry breaking scalar masses are desirable in order to avoid the flavour-changing neutral currents in the observable sector.

The parameters in the (\ref{Veff-2}) should be understood as being defined at the Planck scale. Thus to employ the effective potential at the electroweak scale it will be necessary to run the parameters between the two energy scales by means of the renormalizaion group equations. According to the susy nonrenormalization theorems the superpotential parameters are not subject to renormalization however, a nontrivial renormalization occurs in the soft-susy breaking parameters. In this way a negative mass squared for the Higgs boson at low energies can be achieved despite its positive soft mass squared at the Planck scale.

At the $V_\text{soft}$ (\ref{Vsoft-2}) we recognize the soft scalar masses of order $m_{3/2}$ and the $A$-terms. The last, for $W(\phi)=\lambda \phi^3$ simply reads $Am_{3/2} \lambda (\phi^3+\phi^{*3})$. However, the last type of soft terms, the \itshape gaugino masses\normalfont, is missing. It turns out that they can be generated at tree level only if non-minimal gauge kinetic terms are present with $f_{ab}$ being a non-trivial function of chiral superfields. From the full supergravity Lagrangian we see that
\begin{equation} \label{gaugino-2}
{\cal L} \, \supset \, \text{Re} [f_{ab}(X, \phi^i)] \, \frac{i}{4} \bar{\lambda}_a \gamma^{\mu}\partial_\mu\lambda_b -\frac{1}{4} \frac{\partial f_{ab}}{\partial X} F_X \bar{\lambda}_{Ra}\lambda_{Lb} + \, \text{h.c.}
\end{equation}
The $\left\langle  F _X \right\rangle$ is of order $m_{3/2}M_P$ and hence, in the case that the gauge kinetic function depends on the Polonyi field $X$ the gaugino masses are also of $m_{3/2}$ order. In (\ref{gaugino-2}) we explicitly included the possibility of non-universal gaugino masses, although the simple Polonyi model, based on the flavour-blind gravity, yields universal supersymmetry breaking terms. However, more complete models that unify gravity with standard gauge interactions exhibit non-universalities since flavour-blindness is lost.

\subsection{Gaugino Condensation and Universality} \normalsize

Another possible mechanism for supersymmetry breaking in a supergravity theory would be for a product of two fermionic gaugino fields to develop a vev. Including gaugino terms there is the local supersymmetric transformation 
\begin{equation}
\delta_s \Psi_i = -\frac{1}{8} f_{abj}(G^{-1})^j_i \lambda_a \lambda_b + \text{other terms},
\end{equation}
with $f_{abj}=\partial f_{ab}/\partial \phi^{*j}$. The $f_{ab}$ is the coefficient of the gauge field strength term. An expectation value for $\lambda^a\lambda^b$ can break supersymmetry by making the expectation value of $\Psi_i$ non-invariant under a supersymmetry transformation. For this to occur it is necessary for some components of $f_{abj}$ to be non-zero, which requires non-minimal gauge field kinetic terms.  This condition is fulfilled in a rather model-indpendent way in the framework of superstring inspired models. Assuming $f_{ab}=S\delta_{ab}/M_P$ in order to achieve universality the gauge kinetic terms have the form 
\begin{equation}
{\cal L}=  -\frac{1}{4} \frac{\text{Re} \, S}{M_P} F^{a\mu\nu} F^{a}_{\mu\nu}+\frac{1}{4} \frac{\text{Im} \, S}{M_P} F^{a\mu\nu} \tilde{F}^a_{\mu\nu}
\end{equation}
where $\tilde{F}_{a\mu\nu}=\epsilon_{\mu\nu\rho\sigma}F_a^{\rho\sigma}$ is the dual of the gauge field strength $F_a^{\rho\sigma}$ and the $S$ field can be identified as the string dilaton. The gauge fermions $\lambda^a$ of the vector superfield can condense in analogy to QCD, which leads to quark-antiquark condensates. The vector superfield must belong to the hidden sector and hence we assume a new asymptotically free gauge interaction of group $G_h$, with gauge coupling $g$, under which the observable fields are neutral. The gauge coupling explodes at a scale which is approximately given by
\begin{equation}
\Lambda_c = M e^{-8\pi^2/(b_0 g^2)}
\end{equation}
where $b_0$ is the one-loop beta function coefficient associated with the hidden sector gauge symmetry. At this scale, the gauginos are strongly interacting and it is expected to condense with $|\left\langle \bar{\lambda}^a \lambda^b \right\rangle| \sim \Lambda^3_c$. Since the $S$ field is identified with the string dilaton then the real part of its vev should fix the values of the gauge coupling
\begin{equation}
\left\langle \text{Re}\, S \right\rangle = \frac{1}{g^2}.
\end{equation}
Then the scale $\Lambda_c$ is written in terms of the vev of the dilaton field
\begin{equation}
|\left\langle \bar{\lambda}^a, \lambda^b \right\rangle| \sim \Lambda^3_c \sim M^3e^{-12 \pi^2 \left\langle S+\bar{S} \right\rangle/b_0}.
\end{equation}
Such a dependence results in an effective potential for the dilaton which is, typically, exponentially decreasing and induced by the four gaugino term present in the supergravity Lagrangian. This lifts the initially flat $S$-direction and yields in the superpotential an extra term 
\begin{equation}
W(S) \sim e^{-24 \pi^2 S/b_0}
\end{equation}
and for finite values of the $S$-field the $F_S \sim e^{24\pi^2 S/b_0}$ breaks supersymmetry. However, the potential is a runaway towards $S\rightarrow \infty$ where supersymmetry is restored. There have been proposed several ways to stabilize the dilaton. One is via a multiple gaugino condensation considering more than one gauge groups, a model know as the racetrack model. A second way is to advocate the presence of a constant in the effective potential.  Another example of stabilization mechanism is to use a nontrivial K\"ahler function for the dilaton. In string theories, the simple dependence $K=-\text{ln}(S+\bar{S})$ receives non-perturbative corrections which may lead to the stabilization of the $S$ field. Hence, it is possible to stabilize the dilaton $S$ at a finite value breaking also supersymmetry at this vacuum.  

The next step towards the construction of a viable phenomenological model is to analyze how the breakdown of supersymmetry is felt in the observable sector. In gravity mediation this can be rather model dependent. For example in the no-scale models, all soft supersymmetry breaking terms vanish at tree level. Although it is possible to generate nonzero soft masses the universality is usually not expected to hold. 

Let us start with the case of the gaugino masses. The boundary values for the soft supersymmetry breaking terms do not exhibit universality by default. The general expression for gauginos, as we wrote at (\ref{gaugino-2}), reads 
\begin{equation} \label{gauginoS-2}
{\cal L} \, \supset \,  -\frac{1}{4} \frac{ F_{X_{ab}}}{M_P} \bar{\lambda}_{Ra}\lambda_{Lb} + \, \text{h.c.}
\end{equation}
which yields non-universal masses. In context of string theories, this dependence on the flavour indices is due to corrections from superheavy thresholds that the dilaton coupling receives that result in a gauge kinetic term of the form $-1/4[\text{Re} \, S + \Delta_a(T, \bar{T})]F^{a\mu\nu}F^a_{\mu\nu}$ where $\Delta_a$ is a function of a generic modulus field $T$. 

For the case of sfermion masses similar problems are encountered. Assuming that the $F$-term of a Standard Model gauge singlet chiral superfield $X$ breaks supersymmetry. The K\"ahler potential is generally expected to have the form
\begin{equation}
K= - \frac{X^\dagger X \Phi^\dagger \Phi}{M^2_{P}}+ \left(\frac{X \Phi^\dagger\Phi}{M_P} +\, \text{h.c.} \right)+...
\end{equation}  
where $\Phi$ represents the quark and lepton superfields in the SSM and coefficients of ${\cal O}(1)$ were omitted. Since there is no reason for the alignment of the flavour structure too large flavour changing processes are predicted. This Lagrangian will yield flavour mixing of ${\cal O}(1)$ which are unacceptable unless the sfermion masses are of ${\cal O}(10)$ TeV or heavier.

It seems gravity mediation has a flavour problem because of the above non-renormalizable operators that can induce flavour changing neutral current in the observable sector. These Planck suppressed operators render the soft terms sensitive to the ultraviolet physics all the way to the Planck or compactification scale and so to flavour violation at all scales. Postulating precise relations among squark masses at high scale, such as universality or proportionality do not address the problem. These relations are detuned under renormalization group evolution. For example, in grand unified theories, large flavour violation can be induced by running between the Planck and the GUT scales \cite{Dimopoulos:1996yq}.   However, there are schemes that avoid these phenomenological obstacles. For example, in string theory, discrete family symmetries and/or universal mediation schemes such as dilaton domination in the heterotic string can avoid the flavour problem, see e.g. \cite{Nilles:2008pt}.

Contrary to the flavour problem, gravity provides a natural explanation of the scale of the supersymmetric $\mu$ parameter in the superpotential. In the case that there is a hidden sector field $X$ with an $F$-component, $F_X= {\cal O}(M^2_\text{susy})$ then it generates the $\mu$ parameter in the low energy superpotential at the correct order of magnitude. This mechanism  \cite{Giudice:1988yz} relies on the operator
\begin{equation} \label{GM-2}
\int d^4 \theta \frac{X^\dagger}{M_P} H_U H_D
\end{equation}
and produces $\mu= {\cal O} (M^2_\text{susy}/M_P)$ which is appropriately of the order of the weak scale.

\subsection{The Cosmological Polonyi/Moduli Problem}

The Polonyi field mass is $m^2_X = {\cal O}(m^2/M_P)$, i.e. of the order of the gravitino mass, and its interactions are $M_P$ suppressed. It is a light field and decoupled from the thermal plasma and its decay width is 
\begin{equation} \label{de-2}
\Gamma_X = \frac{c}{4\pi}\frac{m^3_X}{M^2_P}.
\end{equation}
and decays after the Big Bang Nucleosynthesis (BBN) has been completed. It reheats the universe with a temperature
\begin{equation}
T_X= \left(\frac{\pi^2 g_*}{90} \right)^{-1/4} \sqrt{M_P \Gamma_X} \simeq \, 5.5 \times 10^{-3} \text{MeV} \times \sqrt{c}
\left( \frac{m_X}{1 \text{ TeV}}\right)
\end{equation}
An energy, stored in its oscillations,  released at such a late time causes a cosmological disaster \cite{Coughlan:1983ci}. A similar problem, caused by a fermion this time, is the gravitino decay that spoils the BBN \cite{Ellis:1984eq}. We note that in the gravity mediation scenarios the gravitino is an unstable particle.

The Polonyi field is a scalar and during an inflationary phase it is expected to be displaced from the minimum of the potential (\ref{Pol-2}). The details of the displacement are rather model dependent and we refer the reader to the appendix. Here, we review the cosmological problems caused by scalars with gravitational interactions, calling them collectively moduli -however, they should no be identified with the stringy moduli fields. The conclusion is that \itshape the  mass of the Polonyi field lies in a range that can hardly be compatible with the standard cosmology. \normalfont
\\

There are two potential dangers with the gravitational relics, $\varphi$, (moduli fields). If their lifetime is smaller than the age of the universe, their decay might have released a very large amount of entropy in the universe and diluted its content. If their lifetime is larger than the age of our universe, they might presently still be oscillating around their minimum and the energy stored in these oscillations may overclose the universe. The moduli fields have to be either superlight or superheavy in order to be compatible with the standard cosmology.
 
From the (\ref{de-2}) one thus deduces that the modulus will decay at present times if $\Gamma_{\varphi}\sim H_0$, that is if its mass $m_{\varphi}$ is of order $(H_0M_P^2)^{1/3} \sim 20$ MeV. 

The other relevant quantity is the initial value $f_{\varphi}$ with respect to its ground state value $\varphi_0$. Presumably at very high energy (that is above the phase transition associated with dynamical supersymmetry breaking) where the flat direction is restored, one expects generically that $f_\varphi \sim M_P$ since this is the only scale available (see apendix for specific analysis)
\\
\\
\itshape Light Moduli \normalfont
\\
The first case to be considered is that of a light modlus, i.e. $m_{\varphi}<20MeV$. In this case the field has not decayed at the present time. The equation of evolution for the field $\varphi$ reads
\begin{equation} \label{em1} 
\ddot{\varphi}+3H\dot{\varphi}+V'(\varphi)=-\Gamma_{\varphi}\dot{\varphi}
\end{equation}
where the last term accounts from particle creation due to the time variations of $\varphi$.
If we assume that the universe is initially radiation dominated, then $H \sim T^2/M_P$. As long as $H>m_{\varphi}$ the friction term $3H\dot{\varphi}$ dominates in the equation of motion and the field $\varphi$ remains frozen at its initial value $f_\varphi$. When $H \sim m_\varphi$, i.e. $T_I \sim (m_\varphi M_P)^{1/2}$, the field $\varphi$ starts oscillating around the minimum of the potential which we approximate as
\begin{equation} 
V(\varphi)=\frac{1}{2}m_\varphi ^2(\varphi-\varphi_0)^2+O[(\varphi-\varphi_0)^3].
\end{equation}
A solution is of the form 
\begin{equation} 
\varphi={\varphi}_0+A(t)\cos(m_\varphi t),
\end{equation}
with $|\dot{A}/A| \ll m_\varphi$ and initial amplitude $A(t_i) \sim f_\varphi$. Since, $\Gamma_\varphi < H_0 <H$ one may neglect the decay term of ($\ref{em1}$) and finds that the energy density stored in the field $\varphi$ behaves as non-relativistic matter. Since $a\propto T^{-1}$ the energy density, due to the coherent oscilliation, at temprature $T$ is 
\begin{equation} \label{rhomod}
\rho_{\varphi}(T)=\rho_{\varphi}(T_i)\left(\frac{T}{T_i}\right)^3 \sim m_\varphi^2 f_\varphi^2\left(\frac{T}{T_i}\right)^3.
\end{equation}
\\
Since the radiation energy $\rho_R(T)$ behaves as $T^4$, $\rho_\varphi / \rho_R$ increases as the temprature of the universe decreases and one reaches the time where the energy of the scalar field oscillations dominates the energy density of the universe. One should then make sure that $\rho_\varphi(T_0)<\rho_c$. Using ($\ref{rhomod}$) and $T_i=(m_\varphi m_P)^{1/2}$ one finds the constraint for the modulus field mass
\begin{equation} 
m_\varphi <M_P\left(\frac{\rho_c m_P}{f_\varphi ^2T_0 ^3}\right)^2 \sim 10^{-26} eV
\end{equation}
for the natural value $f_\varphi \sim M_P$. Thus if 
\begin{equation} 
10^{-26}\text{eV} <m_\varphi<20 \text{MeV}
\end{equation}
there is too much energy stored in the $\varphi$ field, which has not decayed in present times.
\\
\\
\itshape Heavy Moduli \normalfont
\\
The second case where $m_\varphi > 20MeV$ that is the scalar field has decayed at present times. Decay occurs at a temprature $T_D \sim \Gamma _\varphi$ and assumming that at that time the $\rho_\varphi$ dominates the energy density we have    
\begin{equation} 
\Gamma_\varphi ^2 \sim \frac{\rho_\varphi(T_D)}{M_P ^2}=\frac{\rho _\varphi(T_i)}{M_P ^2}\left(\frac{T_D}{T_i}\right)^3
\end{equation}
At decay, all energy density is transfered into radiation. Thus the reheating temprature $T_{rh}$, that is the tempature of radiation issued from the decay, is given by the condition
\begin{equation} 
\rho_\varphi(T_D) \sim g_* T_{rh} ^4
\end{equation}
From the above equations we find for the reheating temprature 
\begin{equation} \label{rhomod}
T_{rh} \sim M_P^{1/2}\Gamma_\varphi ^{1/2} \sim \frac{m_\varphi ^{3/2}}{M_P ^{1/2}}.
\end{equation}
The entropy release is 
\begin{equation} \label{entropyrel}
\sigma\equiv\frac{S_{rh}}{S_D}=\left(\frac{T_{rh}}{T_D}\right)^3 \sim \frac{1}{m_P^{1/2}\Gamma_\varphi^{1/2}}\frac{\rho_\varphi(T_i)}{T_i ^3}
\end{equation}
\\
This gives, using $T_i \sim m_\varphi^{1/2} M_P^{1/2}$ and $\rho_\varphi(T_i) \sim M_\varphi ^2 f_\varphi^2$, $\sigma \sim f_\varphi ^2 /(m_\varphi M_P)$. With initial field value of order mass Planck, $f_\varphi \sim M_P$ this gives a very large entropy release as long as the modulus remains much smaller than the planck scale. This energy release must necessarily precede nucleosynthesis since otherwise it would dilute away its effects. This condition, namely $T_{rh} >1$MeV, gives $m_\varphi >10$TeV. Thus for $20\text{MeV} < m_\varphi < 10$TeV the entropy release following the decay of the modulus field is too large to be consistent with present observations. 
In the absence of other effects, we are left with only superlight moduli fields ($m_\varphi <10^{-26}eV$) or heavy ones ($m_\varphi >10$TeV).    
\\
\\
The \itshape summary  \normalfont of this section is that, although gravity mediation is a natural way to mediate the breakdown of supersymmetry to the visible sector it is usually accompanied by severe problems: strong constrains from the FCNC in the visible sector and a gravitationally interacting Polonyi field that spoils the success of the BBN predictions. An additional gravitational problem is the decay of the gravitino that restricts the reheating temperature to values $T_{rh}<10^9$
 GeV and a leptogenesis scenario cannot be realized. Except for the gravity mediation scenario there are also different mechanisms.  Two popular alternatives are the anomaly and the gauge mediation that we discuss below. Gauge mediation, in particular, is the central scenario studied in this thesis.

\section{Anomaly Mediation} \normalsize

In the supergravity theories there is a further contribution to the soft masses called anomaly mediation \cite{Randall:1998uk, Giudice:1998xp}. It originates from the regularization of the integrals present in the loop calculations. Aslo, as it was mentioned \cite{Dine:1997nq} this particular phenomenon is common in Quantum Field Theory and occurs also in supersymmetric theories without gravity.
The regulator necesserily breaks supersymmetry in models of supersymmetry breaking, i.e. the cut-off induces an extra mass spliting in the observable multiplets and hence, contributes to the gaugino masses and the $A$-terms.  Such a contribution is always present and gives the leading effect in models where there are no singlet fields present in the hidden sector, or models in which the singlets do not couple to the gauge fields in the required way.  
Assuming a cut-off of a theory the energy scale $\Lambda$ then this can be interpretted in terms of a field. Then the cut-off of the momenta in the integral can be associated with an effective F-term $\ F_\Lambda= \Lambda m_{3/2}$. Including the one-loop renormalization effects the gauge kinetic term of the observable sector reads
\begin{equation}
{\cal L} =- \frac{1}{4} \left( \frac{1}{g^2}+ \frac{b^{(1)}}{8\pi^2} \log\frac{\Lambda}{\mu}\right) F_{\mu\nu} F^{\mu\nu}.
\end{equation}
Hence the gaugino masses are non-zero
\begin{equation}
{\cal L'}= \frac{1}{4} \frac{b^{(1)}}{8\pi^2} \frac{1}{\Lambda} F_\Lambda \bar{\lambda}_R \lambda_L +\text{h.c} = \frac{1}{4} \frac{b^{(1)}}{8\pi^2} m_{3/2} \bar{\lambda}_R \lambda_L + \text{h.c}
\end{equation}
and equal to 
\begin{equation}
m_{\lambda}= -\frac{b^{(1)}g^2}{16 \pi^2} m_{3/2}
\end{equation}
where the non-canonical form of the kinetic terms (\ref{gaugino-2}) was taken into account. In the case that the anomaly (actually, the should be called super conformal)  mediated contributions dominate the prediction for gaugino masses is strikingly different from gravity mediation. For the minimal supersymmetric standard model one finds that the wino, $\tilde{A}^3$ is lighter than the bino $\tilde{B}$ and vecomes the LSP. The wino annihilation is very efficient and hence, the predicted thermal relic density of such a LSP is insufficient to account for dark matter. 

Similarly, $A-terms$ are generated 
\begin{equation}
A_{ijk}= \frac{1}{2} (\gamma_i+\gamma_j+\gamma_k) m_{3/2}
\end{equation}
where the $\gamma_i=\mu (d\ln Z_i/d\mu)$, with $\mu$ the renormalization scale, are the anomalous dimensions of the corresponding cubic term $\lambda_{ijk} \phi^i \phi^j\phi^k$ in the superpotential. The $\gamma$'s are functions of the gauge couplings $g_j$ and Yukawa couplings $\lambda_k$. 

The scalar masses squared receive contribution 
\begin{equation}
\tilde{m}^2_i=-\frac{1}{4}\frac{d\gamma_i}{d\log\mu} m^2_{3/2}
\end{equation}
at the renormalization scale $\mu$. For the sleptons case the above contribution is negative and the sleptons are tachyons. Unless there is an extra contribution from the hidden sector to the soft scalar masses anomaly maediation is characterized by this instability that cause a spontaneous breaking of $U(1)_{em}$ and makes the photon massive. Hence, for phenomenological purposes one usually introduces a universal contribution $m_0$ in addition to the anomaly mediated ones.

For having $m_{\lambda}= {\cal O}$(100) GeV, a large gravitino mass $m_{3/2} \sim 10- 100$ TeV is needed. Because of the flavour blindness of the mediation mechanism, there is no supersymmetric flavour problem. The large value of $m_{3/2}$ enhances the decay rate of the gravitino. This makes the gravitino cosmologically harmless as it decays before the BBN starts. The cosmological moduli problem (or Polonyi) problem is also absent. The field responsible for the supersymmetry breaking can have any conserving charges, and thus it is reasonable to assume that it is attracted by the symmetry enhanced points during and after inflation so that there is no large initial amplitude. Another phenomenological puzzle of the anomaly mediation is the derivation of a correct order of magnitude for the $\mu$ and $B\mu$ parameters since the absence of a gauge-singlet field in the Hidden sector forbids the operator (\ref{GM-2}).

\section{Gauge Mediation} \normalsize

If the messenger scale is well below the Planck scale, it is likely that the usual gauge interactions of the Standard Model play some role in the messenger sector. This is because Standard Model gauginos couple at the renormalizable level only through gauge interactions. If the Higgs bosons received masses predominantly from non-gauge interactions in the messenger sector, with only a small contribution from gauge interactions, the Standard Model gauginos would be unacceptably lighter than the electroweak scale. It is therefore interesting to consider theories in which the Standard Model gauge interactions act as messengers of supersymmetry breaking \cite{Dine:1981za,Dimopoulos:1981au,Dine:1981gu,AlvarezGaume:1981wy,Dine:1982zb,Dimopoulos:1982gm,Nappi:1982hm, Dine:1993yw,Dine:1994vc,Dine:1995ag,Giudice:1998bp} and some standard reviews \cite{Giudice:1998bp, Dimopoulos:1996yq}. This mechanism occurs if supersymmetry is realized non-lineraly in some sector which transforms under the Standard Model gauge group. Supersymmetry breaking in the visible sector spectrum then arises as a radiative correction.

Low energy supersymmetry removes power low sensitivity to ultraviolet physics. Although the parameters of the low energy theory are renormalized they are only logarithmically sensitive to effects in the ultraviolet. Hence, if the messenger sector for supersymmetry breaking is well below the scale the Yukawa hierarchies are generated, the soft terms can be insensitive to the flavour sector. Naturally small flavour violation can result without specially designed symmetries.

Standard model gauge interactions act as messengers of supersymmetry breaking if fields within the supersymmetry breaking sector transform under the Standard Model gauge group. Integrating out the messenger sector fields gives rise to radiatively generated soft terms within the visible sector. Supersymmetry breaking entails the presence of a Goldstino superfield  that we assume it overlaps with a chiral superfield $X$, which acquires a vev along the scalar\footnote{For the sake of economy, we use the same notation for the superfield  $X$ and its scalar component.} and auxiliary components
\begin{equation} \label{X-2}
\left\langle X \right\rangle_\text{superfield}=\left\langle X \right\rangle +\theta^2 F.
\end{equation}
The parameters $\left\langle X \right\rangle$ and $\sqrt{F}$ which are the fundamental scales in the theory and  can vary from severals tens of TeV to almost the GUT scale. The $X$ can coincide with the Goldstino superfield. This is the case considered here; however, there are hidden sectors in which the Goldstino is a linear combination of deifferent fields. The mass scale $\sqrt{F}$ is the measure of supersymmetry breaking in the messenger sector that consists of the chiral messenger superfields $\phi$, $\bar{\phi}$.  The messenger mass eigenvectors are $(\phi+\bar{\phi}^\dagger)/\sqrt{2}$ with squared-mass eigenvalues
\begin{equation}
m^2_{\pm} =\lambda^2 \left\langle X \right\rangle^2 \pm \lambda F\, .
\end{equation}
Ordinary particle supermultiplets are degenerate at the tree level because of gauge interactions between observable and messenger fields. The positivity of the messenger squared masses requires $F<\lambda \left\langle X \right\rangle^2$ or simply $F< \left\langle X \right\rangle^2$. Assuming $F\ll \left\langle X \right\rangle^2$ the effective field theory below the messenger scale $\left\langle X \right\rangle$  supersymmetry breaking can be treated as small effect.  For energies below the mass $\left\langle X \right\rangle$, an effective theory can be defined by integrating out the heavy messenger fields and soft supersymmetry breaking terms arise from 
\begin{equation} \label{Kf-2}
K_\text{gauge} = -\frac{4g^4 N_\text{mess}}{(4\pi)^4} C_2(R) (\text{ln}|X|)^2 Q^\dagger Q \, , \quad\quad f_\text{gauge} = \frac{1}{2}\left(\frac{1}{g^2}- \frac{2N_\text{mess}}{(4\pi)^2} \text{ln} X \right) W^\alpha W_\alpha.
\end{equation}
where $C_2(R)$ is the quadratic Casimir factors for the observable fields $Q$ and $N_\text{mess}$ are the messenger fields. Hence the gaugino masses arise at one-loop and the sfermions at two loops respectively from the operators 
\begin{equation}
\int d^2\theta \ln X W^\alpha W_\alpha + \text{h.c.}\, , \quad\quad
\int d^4 \theta \ln (X^\dagger X) Q^\dagger e^V Q
\end{equation}

A different way to derive the soft masses is the wave function renormalization procedure \cite{Giudice:1997ni}. The renormalizable terms are those not suppressed by powers of $\left\langle X \right\rangle$. Due to the non-renormalization of the superpotential, all the relevant $\left\langle X \right\rangle$ dependence of the low energy effective theory is contained in the gauge and matter wave-function renormalizations $S$ and $Z_Q$. In the presence of a single mass scale $\left\langle X \right\rangle$, this dependence is logarithmic and can be calculated by solving the renormalization-group (RG) equations in the exact supersymmetric theory. At the 
end  the mass parameter $\left\langle X \right\rangle$ can be analytically continued in superspace into a chiral superfield $X$.  Holomorphy dictates the correct analytic continuation $\left\langle X \right\rangle \rightarrow X$ to be performed in $S$. For the $Z_Q$ the only substitution that is consistent with the chiral reparametrization $X\rightarrow e^{i\varphi}X$ is given by $\left\langle X \right\rangle \rightarrow \sqrt{X X^\dagger}$. By replacing the $X$ with its background value $\left\langle X \right\rangle_\text{superfield}$ (\ref{X-2}) all the relevant supersymmetry breaking effects can be derived. The supersymmetry breaking gaugino masses, squark and slepton masses and coefficients of the trilinear $A$-terms, defined by 
\begin{equation}
{\cal L} = -\frac12 \left(m_\lambda \lambda\lambda +\text{h.c}\right) - m^2_{\tilde{Q}} Q^\dagger Q -\left(\sum_i A_iQ_i\partial_{Q_i} W(Q)+\text{h.c.} \right) 
\end{equation}
have the general expressions \cite{Giudice:1997ni}
\begin{equation}
\left. m_\lambda(t)= -\frac12 \frac{\partial S(X, t)}{\partial \ln X}\right|_{X=\left\langle X \right\rangle} \frac{F}{\left\langle X \right\rangle}
\end{equation}
\begin{equation}
\left. m^2_{\tilde{Q}}(t)= - \frac{\partial^2 Z_Q(X,X^\dagger, t)}{\partial \ln X \partial\ln X^\dagger}\right|_{X=\left\langle X \right\rangle} \frac{FF^\dagger}{\left\langle X \right\rangle \left\langle X \right\rangle^\dagger}
\end{equation}
\begin{equation}
\left. A_i(t)=  \frac{\partial \ln Z_{Q_i}(X,X^\dagger, t)}{\partial \ln X}\right|_{X=\left\langle X \right\rangle} \frac{F}{\left\langle X \right\rangle} \,\,.
\end{equation}
The $t=\ln \left\langle X \right\rangle^2 /\mu^2$ and $\mu$ is the low-energy scale at which the soft masses are defined. The gauge and chiral wave function renormalizations $S$ and $Z_Q$ are obtained by integrating the Renormalization Group (RG) differential equations in the supersymmetric limit.

\subsection{Soft Masses in Gauge Mediation}

The minimal model can be illustrated by the chiral field that has a vev (\ref{X-2}) and is coupled to a vector-like sets of fields, transforming as a single $\bold{5}$  and $\bold{\bar{5}}$ of $SU(5)$ i.e. messengers
\begin{equation}
W=X\left(\lambda_\ell \ell \bar{\ell} +\lambda_q q \bar{q} \right).
\end{equation}
For $\lambda_{q, \, \ell}F< \lambda^2_{q, \, \ell} \left\langle X \right\rangle^2$ the $\ell$, $\bar{\ell}$, $q$ and $\bar{q}$ are massive with supersymmetry breaking splittings of order $F$. The fermion masses are given by 
\begin{equation}
m_q= \lambda_q \left\langle X \right\rangle ,\,\,\,\,\,\,\,\,\,\,\,\,\, m_\ell = \lambda_\ell \left\langle X \right\rangle
\end{equation}
while the scalar splittings are 
\begin{equation}
\Delta m^2_q=\lambda_q F, \,\,\,\,\,\,\,\,\,\,\,\,\, \Delta m^2_\ell = \lambda_\ell F
\end{equation}
In such a model masses for gauginos are generated at one loop and for scalars at two loops. 
The result for the gaugino masses is 
\begin{equation}
m_{\lambda_i}= \frac{\alpha_i}{4\pi}\Lambda_0
\end{equation}
while for the squark and slepton masses it is 
\begin{equation}
\tilde{m}^2= 2\left[C_3\left(\frac{\alpha_3}{4\pi} \right)^2+C_2\left(\frac{\alpha_2}{4\pi} \right)^2 +\frac{5}{3}\left(\frac{Y}{2}\right)^2\left(\frac{\alpha_1}{4\pi} \right)^2 \right] \Lambda^2_0
\end{equation}
where $\Lambda_0 \equiv F_X/\left\langle X \right\rangle$. The $C_3=4/3$ for color triplets and zero for singlets; $C_2=3/4$ for weak doublets and zero for singlets. 

The spectrum has a number of notable features. Firstly, there is only one parameter that describes the masses of the three gauginos and the squark and the sleptons. Secondly , the Flavour Changing Neutral Currents are automatically suppressed: each of the matrices $m^2_Q$ etc is proportional to the unit matrix; the $A$-terms are highly suppressed since they receive no contributions before the three loop order. Thirdly, the CP conservation is automatic. Finally, this model cannot generate a $\mu$-term that is protected by symmetries and some further structure is necessary.

\subsection{General Gauge Mediation / Departures from MGM}

The assumption of a single set of messengers and one singlet responsible for supersymmetry and $R$-symmetry breaking is the minimal one. Formulating the problem in a general way leads to the so called "General Gauge Mediation" \cite{Meade:2008wd}. There, the problem was studied in terms of correlation functions of gauge supercurrents. Analyzing the restrictions imposed by Lorentz invariance and supersymmetry on these correlation functions it is was found that the general gauge mediation spectrum is described by three complex parameters and three real parameters.

It is easy to see how one can obtain a larger set of parameters in simple weakly coupled models. Taking a model with a set of singlets the supepotential of gauge mediation reads
\begin{equation}
W=\lambda^q_i X_i q\bar{q} + \lambda^l_i X_i \ell \bar{\ell}.
\end{equation}
Here, unlike the case of minimal gauge mediation, the ratio of the splittings in the multiplets to the average (i.e. fermion) masses is not the same for $q$, $ \bar{q}$ and $\ell$, $\bar{\ell}$. For the fermion masses 
\begin{equation}
m_q= \sum \lambda^q_i \left\langle X_i \right\rangle, \,\,\,\,\,\,\,\,\,\,\,\, m_\ell =\sum \lambda^\ell_i \left\langle X_i \right\rangle
\end{equation}
while the scalar splittings are
\begin{equation}
\Delta m^2_q = \sum \lambda^q_i F_i  \,\,\,\,\,\,\,\,\,\,\,\, \Delta m^2_\ell = \sum \lambda^\ell_i F_i
\end{equation}
In the case of minimal gauge mediation, the one loop contributions for fields carrying color were proportional to $\Delta m^2_q/m^2_q$ while those contributing to fields carrying weak charge were proportional to $\Delta m^2_q/m^2_q$. Generalizing the previous computations the masses for the gauginos are found to be
\begin{equation}
m_g= \frac{\alpha_3}{4\pi} \Lambda_q, \,\,\,\,\,\, m_w= \frac{\alpha_2}{4\pi} \Lambda_\ell, \,\,\,\,\,\, m_b= \frac{\alpha_1}{4\pi} \left(\frac{2}{3}\Lambda_q+\Lambda_\ell \right). 
\end{equation}
where
\begin{equation}
\Lambda_q= \frac{\lambda^i_q F_i}{\lambda^j_q \left\langle X_j \right\rangle},\,\,\,\,\,\,\, \Lambda_\ell= \frac{\lambda^i_\ell F_i}{\lambda^j_\ell \left\langle X_j \right\rangle}
\end{equation}
and summation over the indices $i$ and $j$ is understood. For the mass of squarks and sleptons one gets
\begin{equation}
\tilde{m}^2= 2\left[C_3\left(\frac{\alpha_3}{4\pi} \right)^2\Lambda^2_q+C_2\left(\frac{\alpha_2}{4\pi} \right)^2\Lambda^2_\ell +\left(\frac{Y}{2}\right)^2\left(\frac{\alpha_1}{4\pi} \right)^2 \left(\frac{2}{3}\Lambda^2_q+\Lambda^2_\ell \right)\right]
\end{equation}
In the pesent example there are two complex parameters the $\Lambda_q$ and $\Lambda_\ell$.  But generally, we have three independent parameters for each Majorana gaugino and for the scalars one can inroduce a real parameter $\Lambda^2_c$ for the contribution from $SU(3)$ gauge fields, $\Lambda^2_w$ for those from $SU(2)$ gauge fields and $\Lambda^2_Y$ for those from hypercharge gauge fields. Hence, the general sfermion mass matrix reads
\begin{equation}
\tilde{m}^2= 2\left[C_3\left(\frac{\alpha_3}{4\pi} \right)^2\Lambda^2_c+C_2\left(\frac{\alpha_2}{4\pi} \right)^2\Lambda^2_w +\frac{5}{3}\left(\frac{Y}{2}\right)^2\left(\frac{\alpha_1}{4\pi} \right)^2 \Lambda^2_Y \right].
\end{equation}
One can construct models which exhibit the full set of parameters. One feature of minimal gayge mediation that is not immediatelly inherited by the general gauge mediation is the suppression of new sources of CP violation. Because the gaugino masses are independent parameters they introduce additional phases which are CP violating. This fact is one of the challenges (or obstacles) of general gauge mediation model building \cite{Dine:2009gy}.

\subsection{The Decay of the Goldstino Scalar in Gauge Mediation}

Contrary to the gravity mediation the scalar component of the Goldstino superfield is not associated with cosmological problems see e.g. \cite{Ibe:2006rc}. The spurion field $X$ is coupled directly to the messenegrs $\phi$, $\bar{\phi}$ and to the MSSM through loop diagrams with the messenger particles. We list the various coupling below.
\begin{itemize}
\item \itshape Gauge Bosons \normalfont
\end{itemize} 
The low energy values of the gauge coupling constants depend on the $X$ through $\log X$ because it changes the scale at which the $\phi$ and $\bar{\phi}$ are integrated out. The kinetic term of the gauge bosons are 
\begin{equation}
{\cal L} =- \frac{1}{8 g^2(X)} F_{\mu\nu} F^{\mu\nu} +\text{h.c.}
\end{equation}
The gauge constant is given by
\begin{equation}
\frac{1}{g^2(X)} = \frac{1}{g^2_0} - \frac{2(b_L+N)}{(4\pi^2)} \ln \frac{\lambda X}{\tilde{\Lambda}}-\frac{2b_L}{(4\pi)^2}\ln\frac{\mu_R}{\lambda X} 
\end{equation}
where $g_0$ the gauge coupling constant at a scale $\tilde{\Lambda}$, the $b_L$ is the beta function coefficient below the messenger scale and $\mu_R$ is the renormalization scale. After canonically normalizing the kinetic terms of gauge bosons the interaction term with the gauge bosons is obtained
\begin{equation} \label{gb-2}
{\cal L} = \frac{g^2 N}{(4\pi)^2} \frac{1}{\left\langle X \right\rangle}\frac14 X F_{\mu\nu}F^{\mu\nu} +\text{h.c.}
\end{equation}
\begin{itemize}
\item \itshape Gauginos \normalfont
\end{itemize} 
Similarly, one obtains the interaction term with the gauginos
\begin{equation}
{\cal L}_\text{gaugino}= -\frac{1}{2}m_{\lambda}(X) \lambda\lambda +\text{h.c.}.
\end{equation}
The gaugino mass is given by the formula
\begin{equation}
m_{\lambda} (\left\langle X \right\rangle)= \frac{g^2 N}{(4\pi)^2} \frac{F}{\left\langle X \right\rangle }
\end{equation}
Considering excitatins of the $X$ about the vacuum value $\left\langle X \right\rangle$ then the $(\left\langle X \right\rangle +X)^{-1} \simeq \left\langle X \right\rangle^{-1} (1+X/\left\langle X \right\rangle)$ and the interaction term of the spurion with the gaugino reads
\begin{equation} \label{gau-2}
{\cal L}_\lambda= \frac12 \frac{m_\lambda}{\left\langle X \right\rangle} X\, \lambda\lambda +\text{h.c.}
\end{equation}
Comparing the interaction of $X$ with the gauginos with the one with the gauge bosons ones sees that (\ref{gau-2}) gives larger contribution to the decay width of $X$ by one loop factor when $m_X \sim m_\lambda$.
\begin{itemize}
\item \itshape Sfermions \normalfont
\end{itemize} 
The soft sfermion mass term in the Lagrangian is
\begin{equation}
{\cal L}_{\text{scalar}} = - m^2_{\tilde{f}}(X) \tilde{f}^\dagger \tilde{f} 
\end{equation}
and with a similar reasoning one finds
\begin{equation}
{\cal L}_{\tilde{f}} = \frac{m^2_{\tilde{f}}}{\left\langle X \right\rangle}\, X \, \tilde{f}^\dagger \tilde{f}  +\text{h.c.}
\end{equation}
This interaction yields a decay width that is one-loop order of magnitude larger that the (\ref{gb-2}). Analogously, the interaction Lagrangian with the Higgs fields can be found. We note that under certain limits direct couplings of the Higgs superfields with the $X$ are allowed in the superpotential; also, a general coupling at the K\"ahler potential often appears in the models. Hence, for the decay width of the spurion to Higgses may have an another contribution apart from that through messengers.
\begin{itemize}
\item \itshape Gravitino \normalfont
\end{itemize} 
Finally, the scalar Goldstino $X$ can decay to gravitinos. The largest contribution for the $X$ decay comes from the coupling to the longitudinal mode of the gravitino which is the goldstino field $\tilde{x}$, i.e. the fermionic component of the $X$ superfield. In the case of a quatric correction to the K\"ahler potential of the form $|X|^4/\Lambda^2$ the coupling is
\begin{equation}
{\cal L}_{3/2}= -\frac{2F^\dagger_X}{\Lambda^2}\, X^\dagger \, \tilde{x} x + \text{h.c.} 
\end{equation}
that can be written in terms of the longitudinal component $\psi$ of gravitino field $\psi_\mu$
\begin{equation}
{\cal L}_{3/2}=-\frac12 \frac{m_{3/2}}{\left\langle X \right\rangle}\, X^\dagger\, \bar{\psi} \psi +\text{h.c.}
\end{equation}
The decay width of the gravitino is supressed by ${\cal O} ((m_{3/2}/m_\lambda)^2)$ compared to the gaugino modes. The same also is the suppression compared to  the sfermion modes. The decay width to gravitinos compared to the gauge bosons is larger but still supressed by ${\cal O} ((m_{3/2}/m_X))$.

We note that due to kinimatic reasons the $X$ does not decay to messengers since $m_X<m_\phi$.
\\
\\
\itshape Conclusions \normalfont
\\
The above interaction terms suggest that the spurion in gauge mediation does not cause Polonyi-like cosmological problems. The decay of $X$ takes place before the BBN and reheats the universe to a temperature $T_{rh}= {\cal O}(1-100)$ MeV. It is also possible that the oscillations of $X$ field in gauge mediation never dominate the energy density of the universe and its decay does not cause an extra entropy production. This is, actually, one of the main results presented in this thesis.

\chapter{Metastable Dynamical Supersymmetry Breaking Models}

In this chapter the basic paradigms of $F$-term supersytmmetry breaking schemes are presented. The metastability is demonstrated to be a generic feature. A thorough presentation is required for the study of the thermal behaviour of these models carried out at \cite{Dalianis:2010yk} and \cite{Dalianis:2010pq}.  
\\
\\
The introduction of soft breaking parameters cannot be satisfactory from many points of views. First of all, the theory is not complete in the ultraviolet due to the logarithmic divergencies in the soft breaking parameters. Moreover, all the soft breaking parameters appear to be independent parameters. Finally, the scales seem to be arbitrary and an explanation for the hierarchies is desirable. In this chapter models and that yield 
\begin{equation}
\left\langle X \right\rangle = X + \theta^2 F
\end{equation}
i.e. supersymmetry breaking will be demonstrated. From the low energy point of view this might look enough but one should provide an explanation for the values of the parameters $F$ and $X$. Weakly coupled models and relations with strong coupling dynamics will be aslo discussed. A connection of low energy supersymmetry to some underlying structure, in particular string theory, is an extra motivation for such models and will be further pursued at the chapter 7.

The prototype model is the O'Raifeartaigh paradigm \cite{O'Raifeartaigh:1975pr}. It can account for the a calculable low-energy description of possible strong gauge dynamics that have been integrated out. The K\"ahler potential is expected to be non-canonical, however complete models with canonical K\"ahler have been constructed \cite{Intriligator:2006dd}. The perturbative O'Raifeartaigh model corresponds to the case of a generic superpoptential with an $R$-symmetry and with non-compact pseudoflat direction (pseudomodulus). Pseudoflat means that it is flat only in the classical limit and perturbative corrections lift the flatness. This is possible because here the supersymmetry is broken i.e. it is a direction with constant positive energy classically. Everywhere along the pseudoflat direction away from the origin the $R$-symmetry is spontaneously broken apart from the origin where the $R$ symmetry is preserved. The $R$-symmetry, as will be explained in the next section, is an essential feature of the dynamical supersymmetry breaking models.

\section{The R-symmetry, SUSY Breaking and Metastability}

It was pointed out at the reference \cite{Nelson:1993nf} that there is a deep connection between $R$-symmetry and supersymmetry breaking. For example a Wess-Zumino model the presence of the $R$-symmetry is a necessary condition for supersymmetry breaking if the superpotential is a generic function of the fields i.e. if it contains enough terms. Theories with non-generic superpotentials, which typically have some interaction or mass terms set to zero or fine tuned, are unstable under small variations of the couplings. 

A Wess-Zumino model and the scales involved can be the low-energy description  of some strong dynamics that have been integrated out. Typically the K\"ahler potential is non-canonical.  An effective theory described by a Wess-Zumino model with $n$-chiral superfields $\varphi_i$, $i=1,...,n$  preserves supersymmetry when the $n$ complex equations $\partial_{\phi_i} W$ with $n$ complex unknowns have a solution. For a generic superpotential with no symmetry requirements there is always a solution. Considering a theory with an $R$-symmetry which is spontaneously broken say, by the vev of the field $\varphi_1$. One can parametrize the fields by $\varphi_1$ and by the $R$-invariants $\chi_i= \varphi^{1/r_1}_i/\varphi^{1/r_1}_1$, for $i=2,...,n$ where $r_i$ is the charge of $\phi_i$. Since the $W$ has charge 2 it will be of the form $W=\varphi^{2/r_1}_1 \Omega(\chi)$. By using $\chi \neq 0$ supersymmetric condition is 
\begin{equation}
\partial_{\varphi_1} W= \frac{2}{r_1} \varphi^{-1+2/r_1} \, \Omega(\chi_i) =0 \quad \quad \text{i.e.} \quad \quad \Omega(\chi_i)=0
\end{equation}
\begin{equation}
\partial_{\chi_i} W= \varphi^{2/r_1}_1 \partial_{\chi_i} \Omega(\chi_i)=0 \quad \quad \text{i.e.} \quad \quad \partial_{\chi_i}\Omega(\chi_i)=0
\end{equation}
The above system of equations correspond to $n$ equations with $n-1$ variables $\chi_i$ which cannot be solved for generic $\Omega$ and hence, supersymmetry is broken. It can be further shown that the presence of an $R$-symmetry is a necessary condition for supersymmetry breaking and a spontaneously broken $R$ symmetry provides a sufficient condition \cite{Nelson:1993nf}. We note that in case of singular points in K\"ahler metric a specific study is needed. Also, it is possible the effective potential generated by the non-perturbative dynamics can be non-generic and dynamical supersymmetry breaking without an $R$-symmetry can occur.

The connection between spontaneously broken $R$-symmetry and dynamical supersymmetry breaking may be a phenomenological problem. The gluino mass has  $R$-charge 2 this implies that the $R$-Goldstone boson (R-axion) couples to the QCD anomaly and since its scale $f_a$ is typically low ${\cal O}(10-100)$ TeV is phenomenologically problematic. However, there are sources of explicit $R$-breaking which are small enough not to restore supersymmetry and large enough to give the axion a mass that renders it phenomenologically harmless. The breaking of the $R$-symmetry restores supersymmetry but not in a nearby vacuum. An example is the modification of the O'Raifeartaigh model with a small parameter $\epsilon$ \cite{Intriligator:2007py}
\begin{equation} \label{We-3}
W=W_{O'R}(X, \varphi_1, \varphi_2)+\delta W(\epsilon, \varphi_1, \varphi_2)= FX+ \frac12 h X \varphi_1^2+m\varphi_1\varphi_2  + \epsilon m \varphi^2_2
\end{equation}
with $\epsilon \ll 1$. There are two supersymmetric vacua at
\begin{equation} \label{eps-3}
\left\langle \varphi_1 \right\rangle_{susy} = \pm \sqrt{-2F/h}\, ,  \quad \left\langle \varphi_2 \right\rangle_{susy} = \mp \frac1\epsilon\sqrt{-2F/h}\, , \quad \left\langle X \right\rangle_{susy} = \frac{m}{h\epsilon}
\end{equation}
For small $\epsilon$, the supersymmetric (\ref{eps-3}) vacua have $X$ far from the origin. As the parameter $\epsilon \rightarrow 0$ these supersymmetric vacua are pushed to infinity and the local analysis of the supersymmetry-breaking vacuum is expected to be unaffected. Moreover, at \cite{Nelson:1993nf} it was shown that in some models the $1/M_P$ suppressed dimension-five operators also break $R$-symmetry explicitly. In addition, the cancellation of the cosmological constant requires the presence of a constant in the supepotential that breaks the $R$-symmetry at the same scale as supersymmetry and this gives to the $R$-axion an acceptably large mass \cite{Bagger:1994hh}. There models, as well,  that $R$ is broken at a higher scale than supersymmetry. 

The explicit breaking of the $R$-symmetry is also welcome and expected from a different reason. Although a generic dynamical breaking of supersymmetry is closely associated with a global $R$-symmetry it is unlikely that any fundamental theory exhibits continuous symmetries. The rough argument is that one needs to rotate fields all over space-time at once. This is at odds with the "spirit" of relativity. There is actually a theorem in string theory that there are no global symmetries \cite{Banks:1988yz}.  The $R$-symmetry of models like the O'Raifeartaigh should be approximate. The $W_{O'R}$ of (\ref{We-3}) has an $R$-symmetry with $R(X)=2$ , $R(\varphi_2)=2$ and $R(\varphi_1)=0$ that might be a consequence of a discrete $R$-symmetry under which 
\begin{equation}
X \rightarrow e^{2\pi i/N}X \quad \quad \varphi_2 \rightarrow e^{2\pi i/N} \varphi_2 \quad \quad W_{O'R} \rightarrow e^{2\pi i/N}W_{O'R} 
\end{equation} 
Along with the $\varphi_2 \rightarrow - \varphi_2$, $\varphi_1 \rightarrow \varphi_1$ symmetry this accounts for the structure of the Lagrangian at the renormalizable level. Nevertheless, couplings like $1/M_*^{N-2}X^{N+1}$ are allowed by the symmetry. As a result the zero enegrgy condition
\begin{equation}
\frac{\partial W}{\partial X}=0
\end{equation}
has a solution with $X \sim (M_*^{N-2}F)^{1/N}$ i.e. large enough not to cancel the susy breaking minimum about the origin.

We can summarize saying that the dynamical supersymmetry breaking of generic models is realized in metastable vacua; this is the essential lesson of this section.

\section{Models of Ordinary Gauge Mediation with R-Symmetry}

The most general renormalizable, gauge invariant superpotential describing the couplings between the messengers and any number of singlets $X_k$ is the following
\begin{equation}
W=\left(\lambda^{(k)}_{ij}X_k +M_{ij}  \right) \phi_i \bar{\phi}_j= \left(\lambda_{\ell ij}^{(k)} X_k +M_{\ell ij}  \right) \ell_i \bar{\ell}_j+ \left(\lambda_{q ij}^{(k)} X_k +M_{q ij}  \right) q_i \bar{q}_j
\end{equation} 
where in the right hand side of the equation we have decomposed the $\phi$, $\bar{\phi}$ into their $SU(2)$ doublet and $SU(3)$ triplet components,  $\ell$, $\bar{\ell}$ and $q_i$ and $\bar{q}_i$ respectively. This douplet/triplet splitting is similar in spirit to the doublet/triplet splitting that already happens in susy GUT embeddings of the MSSM.  This model, can be reduced to a model with only one singlet thanks to a a unitary transformation that rotates the singlet fields so that only one of them acquires a susy-breaking F-component vev. Let us call this singlet simply as $X$. The remaining singlets only have scalar components vevs $\left\langle X_k\right\rangle =X_k$. Hence, they induce tree level masses for the messengers and the superpotential reduces to the form
\begin{equation} \label{eogm1-3}
W=\left(\lambda_{ij}X +m_{ij}  \right) \phi_i \bar{\phi}_j= \left(\lambda_{\ell ij} X +m_{\ell ij}  \right) \ell_i \bar{\ell}_j+ \left(\lambda_{q ij} X +M_{q ij}  \right) q_i \bar{q}_j\, .
\end{equation}
The above superpotential means that the simple Ordinary Gauge Mediation model can be extended with arbitrary supersymmetric mass terms for the messengers.  The models (\ref{eogm1-3}) can be readily completed into generalized O'Raifeartaigh models simply by adding the $\delta W= FX$ term
\begin{equation} \label{eogm2-3}
W=\lambda_{ij} X \phi_i \bar{\phi}_j + m_{ij} \phi \bar{\phi}_j +FX \equiv {\cal M}(X)_{ij} \phi_i \bar{\phi}_j
\end{equation}
that preserves the $R$-symmetry and ${\cal M}$ the messenger mass matrix. The superpotential must always have definite $R$-charge $R(W)=2$.
However, as we will explain below,  cases that the $R$-symmetry is broken spontaneously or explicitly will be pursued in order the models to be phenomenologically viable. We note that the models always possess a trivial $R$-symmetry under which $R(X)=0$. 

\subsection{Coleman-Weinberg Potential}

The interaction between the spurion $X$ and the messenger fields $\phi_i$, $\bar{\phi}_i$ generate at one-loop a Coleman-Weinberg potential on the pseudomoduli space. This one-loop correction is rather important because the $X$-field value is undetermined i.e. it a pseudo-modulus or pseudo-flat-direction . We refer that, contrary to the goldstone bosons, the different values for the $X$ are not physically equivalent: the vev of $X$ controls the mass for the messenger fields. The minima of the Coleman-Weinberg, if they exist, are susy-breaking vacua of the theory.  On the other hand the one-loop correction to the messenger direction is of a minor importance since the messengers (\ref{eogm2-3}) have a tree level mass\footnote{The one-loop at the messenger direction would be of a special interest in the hypothetical case that the coupling between the $X$ and the messengers at the potential could turn negative. However, in the Lagrangian appears the absolute value: $\delta {\cal L}=-|\lambda|^2 |X|^2|\phi^2| $.}.
\begin{equation} \label{CW-3}
V^{(1)}_\text{eff} = \frac{1}{64\pi^2}\, \text{STr} \tilde{\cal M}^4\ln \frac{\tilde{\cal M}^2}{M^2_{cutoff}}=\frac{1}{64\pi^2} \text{STr }\left(\text{Tr } m^4_B \ln \frac{m^2_B}{M^2_{cutoff}}- \text{Tr }m^4_F \ln \frac{m^2_F}{M^2_{cutoff}} \right),
\end{equation}
where $m^2_B$ and $m^2_F$ are the tree-level boson and fermion masses as a function of the expectation value of the pseudo-modulus $X$. The $\text{STr} \tilde{\cal M}^4$ is independent of the pseudo-moduli. The ultraviolet cut-off $M^2_{cutoff}$ can be absorbed into the renormalization of the coupling constants appearing in the tree level vacuum energy $V_0$. The quantum correction (\ref{CW-3}) generates a non-hierarchical phase and in the case that the O'Raifeartaigh model is augmented with a gauge symmetry under which the pseudo-modulus is charged then an "inverted hierarchy" with $\left \langle X \right\rangle$ can be generated \cite{Witten:1981nf}. In the non-hierarchical case that we are interested here the (\ref{CW-3}) generates an effective potential around the $X=0$ of the form
\begin{equation}
V^{(1)}_\text{eff} =V_0 + m^2_X |X|^2+ {\cal O}\left( |X|^4 \right)
\end{equation}
The sign of the $m^2_X$ determines whether or not the $R$-symmetry is spontaneously broken. We note that for the O'Raifeartaigh model (\ref{We-3}) the potential grows quadratically only near the origin; for $X \gg m$ it grows logarithmically.

\subsection{ Metastability and Gaugino Masses}

Coming back to the case of $R$-symmetry it was shown at \cite{Cheung:2007es} for the models (\ref{eogm2-3}) that there is an essential relation between the $R$-charges, the structure of the messenger mass matrix ${\cal M}$ and the gaugino masses. The condition $R(W)=2$ implies that 
\begin{equation} \label{l-3}
\lambda_{ij} \neq 0 \,\,\,\,\,\,\, \text{only if} \,\,\,\,\,\,\, R(\phi)+R(\bar{\phi})=2-R(X)
\end{equation}
\begin{equation} \label{m-3}
m_{ij} \neq 0 \,\,\,\,\,\,\, \text{only if} \,\,\,\,\,\,\, R(\phi)+R(\bar{\phi})=2.
\end{equation}
The $R$-symmetry constrains the structure of the superpotential. One sees from (\ref{l-3}) and (\ref{m-3}) that an interaction term $\lambda_{ii} X \phi_i\bar{\phi}_i$ and a messenger mass $m_{ii}\phi_i\bar{\phi}_i$ cannot both exist. A direct implication of this fact, i.e. that not all the messengers can have bare mass and be coupled to the spurion $X$, is that \itshape the messengers either cannot be stable at $X=0$ or they are stable but deficiently coupled to the Goldstino superfield. \normalfont We analyse this connection between the $R$-symmetry, the shape of the potential and the phenomenology in the following.

A formula for the running of gaugino and sfermion soft masses at the messenger scale can be given by generalizing the wavefunction renormalization technique of \cite{Giudice:1997ni}. The gaugino masses are given by
\begin{equation} \label{mg-3}
m_{\lambda_i} = \frac{\alpha_i}{4\pi} \Lambda_G, \,\,\,\,\,\,\, \Lambda_G= F \partial_X \ln \det {\cal M} =\frac{nF}{X}
\end{equation}
and the sfermion masses
\begin{equation}\label{mf-3}
m^2_{\tilde{f}}= 2 \sum^3_i  2 C_{\tilde{f},i} \left( \frac{\alpha_r}{4\pi}\right)^2 \Lambda^2_S, \,\,\,\,\,\,\, \Lambda^2_S =\frac{1}{2}|F|^2 \frac{\partial^2}{\partial X \partial X^*} \sum^{N}_{j=1} \left(\ln |{\cal M}_j|^2\right)^2
\end{equation}
The key point here is the determinant of the mass matrix ${\cal M}$. It was proven at (\ref{eogm2-3}) that the messenger mass matrix is written as product of a power of the field $X$ and a function $G(m,\lambda)$ of the couplings:
\begin{equation} \label{det-3}
\det {\cal M} =X^n G(m, \lambda), \,\,\,\,\,\,\,\,\,\, n= \frac{1}{R(X)} \sum^N_{i=1}\left(2-R(\phi_i) - R(\bar{\phi}_i) \right).
\end{equation}
This parametrization of the determinant and the relations between the $R$-charges and the matrices of couplings $\lambda_{ij}$ (\ref{l-3}) and masses $m_{ij}$ (\ref{m-3}) lead to an important classification of models. The 
\begin{itemize}
	\item \itshape Type I models \normalfont
\end{itemize}
which have $\det m \neq 0$. Via a bi-unitary transformation it is convenient to go to a basis where the $m_{ij}$ is diagonal and in this basis the fields come in pairs with $R$-charges $R(\phi)+R(\bar{\phi}_i)=2$. According to (\ref{det-3}) 
\begin{equation}
n=0 \,\,\,\,\,\,\,\, \text{and} \,\,\,\,\,\,\,\, \det {\cal M}= \det m
\end{equation}
which necessary implies that  $\det \lambda =0$. Since these models have $\det m \neq 0$ and $\det \lambda =0$ the messengers are all stable in the region about $X=0$. At large values of $X$ the messenger may become tachyonic. Thus these models have a stable messenger sector only for
\begin{equation}
|X|<X_\text{max}
\end{equation}
An example of such a type of models with $N$ messenger pairs is the following:
\begin{equation}
W=FX+m_{i} \phi_i \bar{\phi}_i+\lambda_{i}X\phi_i \bar{\phi}_{i+1}
\end{equation}
where $i$ runs over the values $1...N$. 
For the minimal case i.e. $W=FX+m \phi_1 \bar{\phi}_1 + m\phi_2\bar{\phi}_2 +\lambda X \phi_1 \bar{\phi}_2$ the $X_\text{min}$ the messengers can be stable for $m^2> \lambda F$ for all the moduli space $X$.

The fact that $\det {\cal M} = \det m$ means that the gaugino masses all vanish at leading order in $F$ (\ref{mg-3}) and these models are pathological phenomenologically. This leads to a large hierarchy between the gaugino and the squark masses which in turn exacerbates the fine-tuning problems of gauge mediation.
\begin{itemize}
	\item \itshape Type II models \normalfont
\end{itemize}
which have $\det \lambda \neq 0$. Correspondingly, a bi-unitary transformation can diagonalize the $\lambda_{ij}$ and then the fields come in pairs with $R(\phi)+R(\bar{\phi}_i)=0$. According to (\ref{det-3}) 
\begin{equation}
n=N \,\,\,\,\,\,\,\,\,\, \text{and} \,\,\,\,\,\,\,\,\, \det{\cal M} =X^N \det \lambda\, .
\end{equation}
The type II models have a stable messenger spectrum for
\begin{equation}
|X|>X_\text{min}
\end{equation} 
for some $X_\text{min}$. A minimal example of type II models that we will invoke later has superpotential
\begin{equation} \label{Ws-3}
W=FX +\lambda_1 X \phi_1 \bar{\phi}_1 +\lambda_2 X \phi_2 \bar{\phi}_2 +m\phi_1\bar{\phi}_2
\end{equation}
and a simple choice is the $\lambda_1=\lambda_2 \equiv \lambda$. We see that the messengers  are not protected by a bare mass term.  Let us exhibit the tachyonic behaviour of the scalar messenger masses explicitly by considering the masses of the scalar messengers $\phi_1$ and $\bar{\phi}_1$ about the origin.  The scalar matrix is given by the derivatives of the superpotential
\begin{equation}
\left(m^2_0\right)_{a b} =\left(\begin{array}{cl}
W^{\dagger a c} W_{cb} &  W^{\dagger abc} W_c \\
W_{abc} W^{\dagger c} &  W_{ac} W^{\dagger cb} \\
\end{array} \right). 
\end{equation}
This is a 10 by 10 matrix for the five fields of (\ref{Ws-3}). Focusing e.g. on the derivatives with respect to the fields $\phi_1$ and $\bar{\phi}_1$ we find a four by four submatrix that in the vicinity of the origin $\phi_i$, $\bar{\phi} \simeq 0$ and in terms of the singlet $X$ reads
\begin{equation}
\left(m^2_0\right)_{\phi_1\, \bar{\phi}_1} =\left(\begin{array}{ccll}
|\lambda|^2|X|^2+m^2 &0 & 0 & \lambda^\dagger F \\
0  & |\lambda|^2|X|^2+m^2 &  \lambda F^\dagger & 0 \\
0  &   \lambda^\dagger F &  |\lambda|^2 |X|^2 & 0 \\
\lambda F^\dagger & 0 &0 & |\lambda|^2 |X|^2
\end{array} \right).
\end{equation}
Taking real values for the $F$, $\lambda$ then the four eigenvalues are by two degenerated and read 
\begin{equation} \label{mas-3}
m^2_{\pm}= \lambda^2|X|^2 +\frac{m^2}{2} \pm \frac{1}{2}\sqrt{m^4+4 \lambda^2 F^2} 
\end{equation}
For $X= 0$ the $m^2_-$ eigenvalues of (\ref{mas-3})  are negative and thus tachyonic around the origin. They become positive only for a non-zero value $X_\text{min}$ for the spurion. For vanishing mass $m$ the tachyonic region is $|X|<\sqrt{F/\lambda}$. Obviously, if there was a mass term $m_{ii} \phi_i\bar{\phi}_i$ it could prevent $m^2_-$ from becoming tachyonic at $X\sim 0$; this is what happens at type-I models.

These models do not suffer from the vanishing gaugino masses problem. Indeed, since $n \neq 0$ and the $\det {\cal M}$ depends on the $X$ field the gaugino masses are nonzero at leading order in $F/M^2_\text{mess}$. They are phenomenologically rather attractive generating gaugino and sfermion masses at the same scale parametrically. This is the main reason that a further study of these kind of models is well motivated. In particular, it is challenging to examine whether such phenomenologically metastable vacua can be also "cosmologically viable". This is, partly, a direction of this dissertation.

\begin{itemize}
	\item \itshape Type III models \normalfont
\end{itemize}
these models correspond to $\det \lambda =\det m=0$ that combine features of the type I and type II. The $n \neq 0$ and the gaugino masses are non-vanishing at leading order as in type II models. These models have a stable messenger sector only for 
\begin{equation}
X_\text{min} < |X| <X_\text{max}
\end{equation}

\section{Non-Vanishing Gaugino Masses and the Vacuum Structure}

The features of the vacuum metastability are related with the all-important mass scale of the gauginos via the relation
\begin{equation} \label{gau-3}
m_{\lambda} \sim F^\dagger \frac{\partial}{\partial X}\log\det(\lambda X+m)
\end{equation}
where $(\lambda X+m)\equiv {\cal M}(X)$. This can be demonstarted by looking into the zero energy conditions for the superpotential $W=FX+\frac12 {\cal M}^{ij}(X)\phi_i\phi_j $ :
\begin{equation}
F_X=F +\frac12 {\cal M}'(X)^{ij} \phi_i \phi_j \quad \quad F_{\phi_i}= {\cal M}(X)^{ij}\phi_j.
\end{equation}
If the $\det{\cal M}(X)$ depends on $X$, then there will necessarily be values $X=X_0$ where it vanishes and then the $F_{\phi_i}=0$ has a solution for non-vanishing $\phi_{i, 0}$. The $X_0$, $\phi_{i,0}$ are supersymmetric vacua because there exists a $\phi_i$ that satisfies the $F_X=0$ condition since the solutions of the $F_{\phi_i}=0$ are infinite at $X=X_0$. Let us consider the case where $\det(\lambda X+m)=\det(\lambda X)$. The determinant vanishes for $X=0$. There, according to the previous line, there are supersymmetric solutions for non zero $\phi_i$. Hence, the origin $X=0$ has non-zero energy $ F^2$ but it is tachyonic towards the supersymmetric vacua $X_0=0, \phi_{i,0} \neq 0$. The origin is unstable and a supersymmetry breaking vacuum can exist only at larger vevs for the $X$ field. 

On the other hand if $\det {\cal M}$ is a non-zero, $X$-independent constant (n=0 of (\ref{det-3})) then the unique solution of $F_{\phi_i}=0$ is the $\phi_i=0$. For $X=0$ this is a locally stable supersymmetry breaking vacuum.  However, the (\ref{gau-3}) says that the gaugino masses vanish at these models. Since there is no such cancellation for the sfermion masses, this generally implies that the gauginos are much lighter than the sfermions in such models. This happens even if the $R$-symmetry is spontaneously broken at the vacuum. Hence, $R$-symmetry breaking is a \itshape necessary but not sufficient \normalfont condition for non-zero gaugino masses. This simple result suggests that this general class of models characterized by a susy-breaking vacuum which is the ground state in the low energy renormalizable approximation i.e. O'Raifeartaigh-like models with no tachyonic messenger masses at the origin, cannot be phenomenologically viable, unless one is prepared to accept an exacerbated little hierarchy problem and the attendant fine tuning coming from very heavy sfermions. A remark on these models is that the field responible for the susy breakdown (Goldstino superfield) does not couple with all the messengers in the superpotential. This feature is characteristic of the \itshape direct mediation \normalfont models. There, a subgroup of the flavour group is gauged and  identified with the Standard Model a fact that enables some messengers to participate at the supersymmetry breaking.

In the case that $\det(\lambda X+m)=\det\lambda X$ the gaugino masses do not vanish. These phenomenologically attractive models have a superpotential where the the Goldstino superfield couples with all the messengers via $\delta W= \lambda X \phi\bar{\phi}$ so they always become tachyonic around $X=0$ except if a bare $R$-violating "bare" mass term $M \phi\bar{\phi}$ also exists (then, the tachyonic directions are simply shifted from the origin). The conclusion is that the \itshape gauginos are non zero at leading order because the pseudomoduli space is not locally stable everywhere \normalfont \cite{Komargodski:2009jf, Abel:2009ze}. This class of models with explicit messengers are collectively called \itshape ordinary gauge mediation \normalfont models and predict a universal form for the gaugino and sfermion masses such that they are of the same order.

\section{Minimal Models with Non-Vanishing Gaugino Masses}

\subsection{Canonical K\"ahler and Spontaneous R-Symmetry Breaking}

The above suggest that we should build models with $\det(\lambda X+m)=X^n G(m, \lambda)$ with $n\neq 0$. The obvious question is the mechanism that stabilizes the spurion at $X \neq 0$ outside the tachyonic region. One strategy is to keep the $R$-symmetry at the low energy effective theories. One interesting model that exhibits supersymmetry breaking and breaks spontaneously the $R$-symmetry   was constructed by Shih \cite{Shih:2007av} and is the (\ref{Ws-3}):
\begin{equation} 
W=FX +\lambda_1 X \phi_1 \bar{\phi}_1 +\lambda_2 X \phi_2 \bar{\phi}_2 +m\phi_1\bar{\phi}_2\, .
\end{equation}
The particular feature of this model is that for $R$-charge assignments $R(\phi_1)=-2$, $R(\bar{\phi}_1)=2$, $R(\phi_2)=-4$ and $R(\bar{\phi_2})=4$ the Coleman-Weinberg (CW) potential has a susy and $R$-symmetry breaking minimum at $X\neq 0$ in some regime of parameters. The CW potential to leading order in $F^2$ is given by \cite{Cheung:2007es}
\begin{equation}
V_\text{CW} = \frac{5\lambda^2 F^2}{32\pi^2} V(x) 
\end{equation}
where $V(x)$ an non-polynomial function of $x=\lambda X/m$. The minimum lies at vevs of the order of $x={\cal O}(1)$. In this example, the messengers participate directly in the susy breaking dynamics, in that their radiative effects generate the Coleman-Weinberg potential. Hence, this constitutes a model of ordinary direct mediation.

\subsection{K\"ahler Stabilization Models with explicit R-symmetry Breaking}

\subsubsection{Non-Minimal K\"ahler from Integrating Out Heavy Fields}

The low energy effective theory (\ref{Kf-2}) is valid up to the messenger scale $\lambda \left\langle X \right\rangle$.  This theory is UV completed once the messenger particles are introduced via the interaction $\delta W = \lambda X \phi \bar{\phi}$ and e.g. a canonical K\"ahler. This represent the full model instead of terms involving $\log X$. 
Nevertheless, above a mass scale $\Lambda$ we may need a further UV completion. The simplest model is the O'Raifeartaigh
\begin{equation}
K=X^\dagger X + A^\dagger A +B^\dagger B
\end{equation}
and
\begin{equation}
W_X= m^2X+\frac{k}{2}XA^2 +M_{AB} AB
\end{equation}
where $k$ and $M_{AB}$ are a coupling constant and a mass for $A$ and $B$ respectively. There is an approximate Peccei-Quinn (PQ) symmetry with charges $PQ(A)=1$ and $PQ(B)=1$. By integrating out the massive fields $A$ and $B$ we obtain the K\"ahler term 
\begin{equation} \label{nmK-3}
\delta K= -\frac{(X^\dagger X)^2}{\Lambda^2}
\end{equation}
with 
\begin{equation}
\frac{1}{\Lambda^2} = \frac{|k|^4}{12(4\pi)^2}\frac{1}{M^2_{AB}}
\end{equation}
at one loop level. The O'Raifeartaigh model itself may be an effective theory of some dynamical breaking models (e.g. this is the case when the fields $A$, $B$ generically can have couplings to other fields as well, that are not small.) 

\subsubsection{Explicit R-Symmetry Breaking Mass for the Messenger Fields}

At low energies $U(1)_R$ violating effects in the supersymmetry breaking sector arise from higher dimension operators and are suppressed by powers of the cut-off scale. According to Nelson-Seiberg argument such a theory would not break supersymmetry. Yet, it may have a local supersymmetry breaking minimum. The \cite{Murayama:2006yf} introduced a "bare" for messengers (contrary to the (\ref{eogm1-3})) that violate the $R$-symmetry. Integrating out the messengers about the supersymmetry breaking vacuum the $R$-symmetry is restored and appears as an accidental symmetry. Their proposal was to supplement the ISS model (see section 3.7) i.e. an $SU(N)$ QCD with massive vector-like quarks $Q^i$ and $\bar{Q}^i$ $(i=1,..., N_f)$ with massive messengers $\phi$ and $\bar{\phi}$:
\begin{equation} \label{mur-3}
W_\text{tree} = m_{ij}\bar{Q}^i Q^j +\frac{\lambda_{ij}}{M_P} \bar{Q}^iQ^j \phi \bar{\phi} + M \phi\bar{\phi}.
\end{equation}
According to ISS \cite{Intriligator:2006dd} the $SU(N)$  SQCD theory in the free magnetic phase, $N+1 \leq N_f \leq \frac32 N$ breaks supersymmetry on a metastable minimum if the quark masses $m_{ij}$ are much smaller than the dynamical scale $\Lambda$. In particular, the ISS model discovered the supersymmetry breaking vacua in the non-calculable regime of the electric $SU(N)$  SQCD theory that generally possesses $N$ supersymmetric vacua. The ISS model has $U(1)_R$ symmetry that is broken down to $Z_{2N}$ which prevents the gaugino masses. Indeed, once we gauge a subgroup of the flavour group and identify it as the Standard Model gauge group (or a GUT one like the $SU(5)$) the corresponding messengers are not all coupled with the meson field, which is the Goldstino superfield. Hence, the direct gauge mediation ISS model belongs to the type-I models of section 4.2. In the model of Murayama and Nomura \cite{Murayama:2006yf} the coupling to messengers breaks it down to $Z_2$, so that the model does not have any $R$-symmetry beyond $R$-parity. Note that the model (\ref{mur-3}) does not respect the $R$-symmetry and cannot classified according to the section 4.2. Nonetheless, it can be said that it is a type-II model supplemented with messengers mass and in the \cite{Murayama:2007fe} version of the model, stabilized due to K\"ahler corrections.

The magnetic theory has a local mimimum at $M^{ij}=0$ and $q=\bar{q}=(\mu,0,...,0)^{T}$ where $\mu^2 \sim m\Lambda$ where a $m$ a stands for, collectively, the mass of the electric quarks and $\Lambda$ the dynamical scale\footnote{The dynamical scale $\Lambda$ appearing in the superpotential of these SQCD theories should not be confused with the cut-off scale $\Lambda$ in the K\"ahler potential.}. The resulting low energy superpotential for the messengers takes the form
\begin{equation} \label{Mmag-3}
W_{mag}=-m\Lambda M^{ij}+ \frac{\lambda_{ij}\Lambda}{M_P} M^{ij} \phi \bar{\phi}+M\phi\bar{\phi} \, \equiv \, FX+\lambda X\phi\bar{\phi}+M \phi\bar{\phi}
\end{equation}
The meson $M^{ij} \equiv X$ field is not stabilized at the origin due to the $R$-violating messenger mass. The shift should be smaller than specific values for the ISS analysis to be valid and to avoid tachyonic messengers. Introducing a non minimal K\"ahler of the form (\ref{nmK-3}) the meson field can be stabilized at the origin, dominating over $R$-violating one-loop correction from the interaction with the messengers.

\subsubsection{A constant at the Superpotential}

Until now, we study the supersymmetry breaking in the global limit neglecting the gravity corrections. This approximation is good as long as the effects that stabilize the fields of the theory dominate over $M_P$ suppressed operators. However, firstly, in the global supersymmetry the supersymmetry breaking vacua have a positive cosmological constant with value $F^2$  which is huge compared to the tiny (or zero, if the dark energy has a dynamical origin) observed value. A constant term has to be added at the superpotential in order to tune the potential to zero. Secondly, once a constant is included the $R$-symmetry is explicitly broken and this breaking is visible to gravity. Hence, gravity effects will shift the minimum at the $X$-direction to $M_P$ vevs and with $F^2/M^2_P$ curvature. This is in contrast with the Coleman-Weinberg corrections due to messengers that tend to stabilize the $X$ at (or close to) the origin. One-loop corrections from the integrated out heavy fields, i.e. non-canonical K\"ahler, contribute further to the stabilization of the spurion about the origin. It is generally possible that the total effect of gravity and one-loop contributions is the stabilization of the $X$ at an "intermediate" value. Hence, once the minimal superpotential (\ref{min-3}) is supplemented with a constant $c$
\begin{equation} \label{const-3}
W=FX+\lambda X\phi \bar{\phi} + c
\end{equation}
the K\"ahler generated $X$-mnimum can be shifted outside the tachyonic region $|X|<\sqrt{F/\lambda}$. Generally, in the gauge mediation schemes the gravity can be neglected. Here this is not the case: gravity is responsible for the stabilization of the spurion and we will refer to this model as \itshape "Gravitational Gauge Mediation" \normalfont \cite{Kitano:2006wz}.

The origin of the constant $c$ is of particular importance. At the chapter 7 it will be interpreted as the superpotential of the string moduli field that are stabilized at a vacuum of negative energy. A different interpretation related to dynamically generated scales (connected to the supersymmetry breaking) is also possible; we will comment on it at the 8th chapter.  

\subsection{ A Sixth Order Correction at the K\"ahler}

It is possible to have the "very minimal" superpotential
\begin{equation} \label{min-3}
W=FX+\lambda X\phi \bar{\phi}.
\end{equation}
In the case of canonical K\"ahler there are only supersymmetric vacua. However, this "minimalism" may indicate a further structure at higher energies that generate the scales, as we already discussed above and in the following sections. Hence, corrections at the K\"ahler are expected. The Coleman-Weinberg potential usually results in minimum at the origin that is not stable at the messenger direction. However, in the case that the (\ref{nmK-3}) has an opposite sign (plus instead of minus) then the sixth order correction $\delta K=|X|^6/\Lambda^4$ should be taken into account. Then a balance between the 4th and 6th order correction can result in a supersymmetry and $R$-symmetry breaking metastable vacuum \cite{Lalak:2008bc, Dalianis:2010yk}. The K\"ahler reads
\begin{equation} \label{KSBR-3}
K=|X|^2+\epsilon_4 \frac{|X|^4}{\Lambda^2}+\epsilon_6\frac{|X|^6}{\Lambda^4}
\end{equation}
with $\epsilon_4<0$ and $\epsilon_6>0$. The $X$-direction minimum is found to be at 
\begin{equation}
|X|^2\sim \frac{|\epsilon_4|}{\epsilon_6}\Lambda^2
\end{equation}
and breaks the $U(1)_R$ symmetry spontaneously. The reason of a K\"ahler of this form (\ref{KSBR-3}) can be a massive integrated out sector that breaks the $R$-symmtery spontaneously. An example is the O'Raifeartaigh-type model proposed recently by Shih with superpotential 
\begin{equation} \label{Shih-3}
W=FX+kX\varphi_1\varphi_2+m_1\varphi_1\varphi_3+\frac12 m_2\varphi^2_2.
\end{equation}
The particular feature of (\ref{Shih-3}) is that it is the simplest O'Raifeartaigh-type model containing a field with $R \neq 0,\,2$ and as shown at \cite{Shih:2007av} the mass squared of the spurion can be negative at the origin and stabilized at $X \neq 0$. The extrema of the potential consists of a pseudo-moduli space 
\begin{equation}
\varphi_{1,2,3}=0, \quad X \,\,\, \text{arbitrary}
\end{equation}
There is a runaway behaviour as $\varphi_3 \rightarrow \infty$
\begin{equation}
X=\left(\frac{m^2_1m_2 \varphi^2_3}{k^2F} \right)^{1/3}, \quad \varphi_1=\left(\frac{F^2m_2}{k^2m_1\varphi_3} \right)^{1/3}, \quad \varphi_2=\left( \frac{Fm_1\varphi_3}{km_2} \right)^{1/3}, \quad \varphi_3 \rightarrow \infty
\end{equation}
The runaway behaviour at large fields implies that the pseudo-moduli space is not an absolute minimum of the potential. It is a local minimum as long as
\begin{equation} \label{bound-3}
|X|< \frac{m_1}{k}\frac{1-y^2}{2y}
\end{equation}
where
\begin{equation}
y=\frac{kF}{m_1 m_2}\, .
\end{equation}
We consider the theory (\ref{Shih-3}) as the UV completion of (\ref{min-3}) thus the masses $m_1$ $m_2$ are relatively heavy and integrated-out from the low energy theory.  For example, if $m_1 \sim m_2 \sim 10^{15}$ GeV, i.e. the GUT scale, $k= {\cal O}(1)$ and $F= {\cal O}(10^{18})$ GeV$^2$ then 
\begin{equation}
y \ll 1 \quad \text{and} \quad \frac{m_1}{k}\frac{1-y^2}{2y} \gg {\cal O}(10^{15}) \text{GeV}
\end{equation}
and the $|X|$ is safely below the bound (\ref{bound-3}). If not, then a linear combination of the $\varphi_{1,2,3}$ fields becomes tachyonic and the system can roll down classically into a runaway direction.

At the (\ref{Shih-3}) we add the ordinary messengers $\lambda X \phi \bar{\phi}$ and integrating out the $\varphi_1$, $\varphi_2$ and $\varphi_3$ introduce a correction to the K\"ahler that can be expanded in powers of $X$ as \cite{Lalak:2008bc}:
\begin{equation}
\delta K =-\frac{n_\varphi \bar{m}}{(4\pi)^2} \sum^\infty_{l=0} f_{2l} \left(\frac{k |X|}{\bar{m}} \right)^{2l}
\end{equation}
where $\bar{m}$ a representative mass of $m_1$, $m_2$, $n_\varphi$ the dimensional conjugate representation of the gauge group the $\varphi_i$ belong and $f_{2l}$ dimensionless coefficients. When 
\begin{equation}
r \equiv \frac{m_2}{m_1} >2.11
\end{equation}
then $X$ is stabilized at $|X| \neq 0$ breaking the $U(1)_R$ symmetry spontaneously. We can then write the (\ref{KSBR-3}) as
\begin{equation}
K=|X|^2 - \frac{|X|^4}{\Lambda^2_1} +\frac{|X|^4}{\Lambda^4_2}
\end{equation}
where
\begin{equation}
\Lambda^2_1 = \frac{(4\pi)^2 \bar{m}^2}{|f_4| k^4 n_\varphi}, \quad \Lambda^4_2 = \frac{(4\pi)^2 \bar{m}^4}{f_6 \, k^6 n_\varphi}.
\end{equation}
The parameters $f_4$ and $f_6$ read \cite{Shih:2007av}, \cite{Lalak:2008bc}
\begin{equation}
f_4 = - \frac{1+2r^2 -3r^4 +r^2(r^2+3)\ln r^2}{(r^2-1)^3} 
\end{equation}
\begin{equation}
f_6 =  \frac{1+27r^2 -9r^4-19r^6 6r^2(r^4+5r^2+2)\ln r^2}{3(r^2-1)^5} 
\end{equation}
and $f_4$ is negative for $r>2.11$ and the $f_6$ positive for $r>0.5$. The minimum for the spurion is then found to be at 
\begin{equation}
|X|^2= \frac{8|f_4|\bar{m}^2}{36 f_6 k^2} =\frac29 \frac{\Lambda_2^4}{\Lambda_1^2}.
\end{equation}

In the limit that the 4th order K\"ahler correction becomes negligible then the $X$-direction minimum is at the origin which, however, is unstable due to the messengers. 

This interesting behaviour, attributed to \itshape generalized K\"ahler potentials \normalfont of polynomial type, will be examined in detail at the section 4.9 including also the always present gravity effects.

\section{Dynamical Generation of Scales}

Following a top-bottom approach, as Witten states at \cite{Witten:1981nf}, if supersymmetry is broken at tree level, the breaking will have a strength of the same order as the natural mass scale of the theory. For supersymmetry to be broken only at, say $10^3$ GeV which is $10^{-16}$ times the $M_{P}$ we require a theory where supersymmetry is unbroken at the tree level and is broken only by extremely small corrrections. The quantum corrections are presumably non-perturbative since perturbative corrections cannot break supersymmetry. In this way a hierarchy can be generated. 

If supersymmety is unbroken at tree level, it is typically unbroken to all orders of perturbation theory. This follows from a set of non-renormalization theorems. As it was demonstrated at \cite{Affleck:1983rr} via instanton effects these theorems do not extend beyond the perturbation theory, opening the possibility of generating a large hierarchy, of order $e^{-c/g^2}$.
\\
\\
\itshape Non-Renormalization Theorems \normalfont
\\
A way to understand the non-renormalization theorems was suggested by Seiberg \cite{Seiberg:1993vc}. The couplings in the superpotential and the gauge couplings can be considered as expectation values of chiral fields. These fields must appear holomorphically in the superpotential and gauge couplings functions and this greatly restricts the coupling dependence of these quantities. For example in the simple Wess-Zumino model
\begin{equation}
W=\frac{1}{2} m \phi^2 +\frac{1}{3}\lambda \phi^3.
\end{equation}
For $\lambda=0$ this model possesses an $R$-symmetry, under which $\phi$ has charge $1$. Thus, we can think the coupling $\lambda$ as chiral field with $R=-1$. Since the potential is holomorphic the only allowed terms,  polynomial in the $\phi$'s have the form 
\begin{equation}
\Delta W= \sum_n \lambda^n \phi^{n+3}.
\end{equation}
The tree level diagrams with $n+3$ external legs have precisely this dependence on the $\lambda$. This heuristic arguments shows that there are no loop corrections to the superpotential. There is no corresponding argument for the K\"ahler potential which is already renormalized at one loop. Therefore, the physical masses and couplings are corrected in this model. 

Turning to the gauge theories the coupling can again be represented as a complex field
\begin{equation}
{\cal L}=- \frac{1}{4} \int d^2 \theta \tau W^2_\alpha
\end{equation}
where $\tau=g^{-2}+i\theta/(8\pi^2)$ and $\theta$ is the usual CP-violating parameter of the gauge theory.  Perturbation theory is insensitive to $\theta$ and there is the shift symmetry
\begin{equation}
\tau \rightarrow \tau + i \epsilon\, .
\end{equation}
The only combination of $\tau$ and $W^2_\alpha$ which is invariant under the shift symmetry, apart from the Lagrangian, is the $\int d^2\theta W^2_\alpha$ which is the structure of the one loop corrections. These arguments lead to the claim that the there is at most a one-loop correction to other gauge coupling and there are no loop corrections to the superpotential\footnote{The fact that there is a two loop correction to the beta-function in supersymmetric gauge theories can be understood by realizing that the cut-off of the theory is not, in general, a holomorphic function of the coupling. The resolution of this issue can be found at  \cite{Shifman:1991dz}.}.
\\
\\
\itshape Non-Perturbative Effects - Generation of Scales \normalfont
\\
However, the shift symmetry of perturbation theory is anomalous and is broken beyond the perturbation theory. For example, in an $SU(N)$ gauge theory instantons generate an expectation value for 
\begin{equation}
\left\langle (\lambda \lambda)^N \right\rangle \propto e^{\frac{8\pi^2}{g^2}\, +i\theta}= e^{8\pi^2 \tau} \,.
\end{equation}
A discrete $Z_N$ symmetry is left by the expectation value of the instantons. Moreover, in this theory gluinos are expected to condense 
\begin{equation}
\left\langle W^2_\alpha \right\rangle = \left\langle \lambda \lambda \right\rangle =\Lambda_c^3 e^{i\theta/N} \propto e^{-3\tau/b_0}.
\end{equation}
One can think of this as a constant superpotential, so it represents a breakdown of the non-renormalization theorems. However, in global supersymmetry physics is not sensitive to a constant $W_0$ and it is of interest only in the local case. But, an important result is obtained when we couple the gauge theory to a singlet, $X$, with no other couplings:
\begin{equation}
{\cal L}=\left(\tau+\frac{X}{M_*}  \right)W^2_\alpha. 
\end{equation}
Then an effective superpotential is generated
\begin{equation}
W_{\text{eff}} (X) \propto e^{-\frac{1}{3b_0}\left(\tau+\frac{X}{M_*}\right)} 
\end{equation} 
and a classical moduli space of $X$ has been lifted. This corresponds to a breakdown of the non-renormalization theorems and dynamical supersymmetry breaking through non-perturbative effects. The potential for $X$ falls to zero at large vaules without a stationary point at the calculable region. A minimum found in the region where the effective coupling is large  cannot be reliable since one cannot calculate. However, it is possible to have a perturbative theory that results in a minimum at $X=0$. Introducing singlet fields $A$, $B$ the superpotential and gauge couplings
\begin{equation}
{\cal L} =\frac{X}{M_*}W^2_\alpha + X A^2 + M_{AB} AB.
\end{equation}
At energies below the scale of the strong gauge group, $\Lambda_c$, the gauge interaction can be inegrated out leaving the effective superpotential
\begin{equation}
W= \Lambda_c^3 e^{-\frac{1}{N}\frac{X}{M_*}}+ XA^2 +M_{AB} AB.
\end{equation}
Near $X=0$ this is like the O'Raifeartaigh models and one loop correction can generate a local minimum of the potential. At large $X$ the potential falls away to zero. 

One important aspect of this paradigm is that the $F$ scale, $\mu^2$, is generated dyanamically. Expanding the potential in powers of $X$ the linear term reproduces the orginal O'Raifeartaigh model with $\mu^2$
\begin{equation}
-\mu^2 = \frac{\Lambda_c^3}{N M_*} \, .
\end{equation}
The $M_*$ is a high scale corresponding to new physics such as grand unified or Kaluza-Klein scales. The constant term of the expansion , $\bar{\Lambda}^3$, can be of particular importance once gravity is taken into account.  We will return to this model later in supergravity theories and in the chapters 7 and 8.

\section{SQCD Theories of Dynamical Supersymmetry Breaking}

Let us consider a supersymmetric QCD (SQCD) with $N_f$ flavours and gauge group $SU(N)$. The $N_f$ quarks $Q_f$ are in the $N$ representation and $N_f$ $\bar{Q}_{\bar{f}}$ in the $\bar{N}$ representation. A theory with massless quarks has a global symmetry, free from anomalies, $SU(N_f)\times SU(N_f) \times U(1)_B \times U(1)_R$. The $Q$ and $\bar{Q}$ transform as 
\begin{equation}
Q: \,\, \left(N_f, 1, \frac{1}{N}, \frac{N_f-N}{N_f}  \right)
\end{equation} 
\begin{equation}
Q: \,\, \left(N_f, 1, -\frac{1}{N}, \frac{N_f-N}{N_f}  \right)
\end{equation} 
It can be shown that in this theory the anomalies cancel.  The characteristic structure of the massles theory is the large moduli space of susy vacua. A potential can arise from the $D^2$ terms of the gauge fields. These vanish up to gauge and flavour transformations for $N_f<N$ and $Q=\bar{Q}$, where $Q$ are $N_f\times N$ matrices. In these directions the gauge symmetry is broken to $SU(N-N_f)$. There are $N^2-(N-N_f)^2= 2N N_f -N^2_f$ broken generators. Each broken generator "eats" one chiral field. There are also a set of broken flavour symmetries. In the simple example that the non-zero $N_f$ vevs are all equal the unbroken flavour symmetry is $U(1)\times SU(N_f)$ so there are $N^2_f-1+1$ Goldstone fields that also arise from the chiral fields. One can construct $N^2_f$ gauge invariant meson fields:
\begin{equation}
M_{f,\bar{f}} =\bar{Q}_{\bar{f}} Q_f \,.
\end{equation} 
Perturbatively these directions remain flat. Non-perturbatively there is a unique superpotential consistent with the symmetries 
\begin{equation}
W_\text{eff}=  \frac{{\Lambda_c}^{-{\frac{3N-N_f}{N-N_f}}}}{\det{M_{f,\bar{f}}}}.
\end{equation}
In the case $N_f=N-1$ the superpotential can be computed in a straightforward semiclassical analysis. 

Modifying the SQCD theories supersymmetry breaking can be realized. It can be either stable, in the sense that at the renormalizable level there are only supersymmetry-breaking ground states, or metastable, where there are additional supersymmetric states seperated by a large barrier or sizable distance in the field space.

\subsubsection{Supersymmetry Breaking at the Ground State}

Models with stable supersymmetry breaking are rare. They are generally characterized by two features: First, their potential have no flat directions classicaly i.e. there is no a moduli space of vacua. If this is not the case there are typically regions in the moduli space  where the potential tends to zero corresponding to asymptotic restoration of supersymmetry. Second, they exhibit global symmetries which are spontaneously broken in the ground state.  These two features  constitute a sufficient condition for supersymmetry breaking: A spontaneously broken global symmetry implies the existence of a Goldstone boson and unbroken supersymmetry leads to an additional massless scalar (non-compact flat direction) to complete the supermultiplet. This extra massless mode corresponds to a non-compact flat direction, in contradiction with the first feature. Whence, superesymmetry is broken. Even though this argument applies to any global symmetry, in all known models satisfying these criteria, the spontaneously broken symmetry is actually an $R$-symmetry.

The simplest example of such a theory in which it is possible to do systematic calculations is known as the $3-2$ model because the gauge group is $SU(3)\times SU(2)$. Its particle content is like that of a single generation of the Standard Model minus the singlet:
\begin{equation}
Q:(3,2)\,\,\, \bar{U}: (\bar{3},1)\,\,\,  \bar{D}: (\bar{3},1)\,\,\, L: (1,2)
\end{equation}
There is a unique superpotential allowed by the symmetries, up to field redefinitions
\begin{equation} \label{3-2-3}
W= \lambda Q L \bar{U} \, .
\end{equation}

Let us consider the first case where the $SU(2)$ coupling is assumed to be much smaller than the $SU(3)$ coupling. Without the superpotential this is supersymmetric QCD with $N=3$, $N_f=2$. The theory possesses a non-anomalous $R$-symmetry. Without the continuous $R$-symmetry, e.g. by adding higher dimension opperators like $(QL\bar{U})^2$ suppressed by a large mass scale, there will be supersymmetric minima. The theory has a set of flat directions and generates a non perturbative superpotential. However, the classical superpotential (\ref{3-2-3}) already lifts all of the flat directions. 

For small coupling $\lambda$ the effective potential is 
\begin{equation}
W_\text{eff} = \frac{\Lambda_c^6}{QQ\bar{U}\bar{D}} +\lambda QL\bar{U} \, 
\end{equation}
and the resulting potential exhibits a supersymmetry-breaking minimum.

The second case is the limit where the $SU(2)$ coupling is much greater than the $SU(3)$ coupling so that the $SU(2)$ becomes first strong. The theory looks now like a QCD with $N=2$, $N_f=2$. In this theory there is no non-perturbative superpotential: there exists an exact moduli space that only is modified by quantum corrections.

We note that in the case of a model with a large flavour group that can include e.g. the $SU(5)$ then an interesting scenario is possible: a subgroup of the flavour group can be gauged and identified with the Standard Model. However, in that case there is the difficulty that the SM $QCD$ and other gauge couplings are violently non-asymptotically free. Unification is lost and typically the models constructed to avoid the problems are rather complicated. 

\subsection{ The ISS Model}

In the case that $N_f < N$ there is a potential generated non perturbatively on the classical moduli space. On the other hand for $N_f \geq N$ there is always an exact moduli space. The dyanmics on this moduli space and in particular in the region of the strong coupling is rather complicated and have a strong dependence on the values of $N$ and $N_f$. The range $N+1<N_f<3/2 N$ is especially interesting \cite{Intriligator:2006dd}. Here the theory is dual to a theory with gauge group $SU(N_f-N)$ with $N_f$ flavours of quarks, $q_f$ and $\bar{q}_f$ and a set of mesons, $M_{f, \bar{f}}$. The effective superpotential of this theory is $W\sim q M \bar{q}$. The duality is not meant as an exact equivalence, but rather as a statement about the infrared behaviours of the two theories.  

A massive QCD has $N$-supersymmetric vacua as Witten has proven \cite{Witten:1982df} a fact that made these theories not attractive for supersymmetry breaking model bulding. Intriligator, Seiberg and Shih (ISS) made the remarkable observation about these theories in this range of $N_f$. They considered adding a small mass term for the quarks $mQ\bar{Q}$ in the ultraviolet theory. Then in the infrared, "magnetic" theory the superpotential is $W_\text{mag} =qM\bar{q} + \text{Tr} m \Lambda M$. This superpotential does not have supersymmetric minima i.e. the supersymmetry is broken contrary, at first sight, with the Witten's argument. At the classical level the magnetic Lagrangian gives rise to moduli space and one of these is the meson direction $M_{f,\bar{f}}$ which is not fixed. Another remarkable fact is that potential for the meson $M$ and the quarks $q$, $\bar{q}$ can be computed although the theory there is not weakly coupled. The outcome is that the minimum of the potential is at the origin for the $M$ field and does not break the $R$-symmetry. 

However, since the original electic-theory is massive with $mQ\bar{Q}$ there should be $N$ supersymmetric vacua in the dual magnetic description as well. At large values of the messon $M$ the dual quarks are massive and the theory is asymptotically free. There, a gaugino condensation leads to an additional term for the $M$ in the superpotential which yield $N$-supersymmetric vacua.

Let us be more precise and present the ISS model in some detail. The paradigm is a
supersymmetric $SU(N)$ QCD with $N_f$ flavours.  If one lies
in the free magnetic range, $N<N_f<\frac32 N$, then the low energy
theory is strongly coupled, but admits a dual interpretation in terms
of IR-free, magnetic variables. The tree-level superpotential in the magnetic theory is given by:
\begin{align} \label{ISS}
  W_{\mbox{tree}}=h\mbox{Tr}\bigl(q M \bar{q}\bigr) - h \mu^2
  \mbox{Tr}(M)
\end{align}
where $\mu^2=\Lambda m$. The $M$ transforms as $N_f \times \overline{N_f}$, $q$: $(\overline{N_f},N)$, $\bar{q}$: $(N_f,\overline{N})$,
$N_m=N_f - N$ is the number of squark colours in the magnetic theory and we
denote the parts of $M$ that  obtain expectation values
as follows: $M=\left(\begin{array}{cc}
    M_1 & 0 \\
    0 & M_0\end{array} \right)$.   The K\"ahler potential is canonical.

Considering the tree-level superpotential in isolation one finds that
the lowest energy state is a moduli space parameterized by

\begin{equation}
  M=\left( \begin{array}{cc}
      0 & 0 \\
      0 & M_0\end{array} \right), \quad \quad q= \left(\begin{array}{c} q_0 \\ 0 \end{array} \right), \quad \quad \bar{q}^{T}= \left(\begin{array}{c} \bar{q}_0 \\ 0 \end{array} \right), \quad \quad q_0 \bar{q}_0 = \mu^2 \, \mathbb{I}_{N_c \times N_c}. 
  \end{equation}
Supersymmetry is broken by the rank condition since $N<N_f$. When the one-loop effects
are included the moduli space is lifted and, aside from flat
directions identified with Goldstone bosons, a unique minimum is found
at:
\begin{equation} 
  M=0, \quad \quad q_0=\bar{q}_0=\mu \,
  \mathbb{I}_{N \times N}. 
\end{equation}
In addition one must include the non-perturbative, R-symmetry
violating contribution:
\begin{align} \label{nnpert}
  W = Nh^{N_f/N}\bigl(\Lambda_m^{-(N_f-3N)}\det(M)\bigr)^{1/N}.
\end{align}
Notice that the exponent of $\Lambda_m$,
$-\left(N_f-3N_m\right)=-(3N-2N_f)$, is always negative in the free
magnetic range.  Hence the coefficient of the determinant grows as the
cut-off shrinks. Since the non-perturbative piece is R-symmetry violating a susy preserving
minimum does exist. The $\Lambda_m$ is the dynamically generated scale of the infrared theory, i.e. the scale of the Landau pole. The supersymmetric vacuum lies at 
\begin{equation}
M_0 =  \frac{\mu/h}{\epsilon^{(N_f-3N)/(N_f-N)}}
\end{equation}
with $\epsilon \equiv \mu/\Lambda_m$.

\section{Gravitationally Stabilized Metastable SUSY breaking} \normalsize

A simple model proposed by Kitano \cite{Kitano:2006wz} achieves metastable susy breaking due to gravitational effects. The K\"ahler potential and the superpotential are
\begin{equation} \label{KitanoK}
K=X^{\dagger}X-\frac{(X^\dagger X)^2}{{\Lambda}^2}+\phi^\dagger \phi+\bar{\phi}^\dagger \bar{\phi}
\end{equation} 
\begin{equation} \label{KitanoW}
W=F X-\lambda X\phi\bar{\phi}+c.
\end{equation} 
The chiral superfield $X$ is a gauge singlet, while $\phi$ and $\bar{\phi}$ are the messenger fields which carry standard model quantum numbers and $\lambda$ is a coupling constant. The constant term $c$ does not have any effect if we neglect gravity interactions, but it is important for the cancellation of the cosmological constant. If we neglect the constant  $c$, the Lagrangian has an $R$-symmetry with charge assignments R($X$)=2, R($\phi$)=R($\bar{\phi}$)=0.

It is also necessary to estimate the perturbative quantum corrections to the K\"ahler potential coming from the interaction term $\lambda X \phi \bar{\phi}$ which may be more important than the gravity effect. At one loop level the correction is \cite{Kitano:2006wz, Intriligator:2006dd}
\begin{equation} \label{KitanoK-loop}
K_\text{1-loop}=-\frac{\lambda^2 N_\phi}{(4\pi)^2}X^\dagger X \log \frac{X^\dagger X}{Q ^2}
\end{equation} 
where $N_\phi$ the number of components in $\phi$ and $\bar{\phi}$ and $Q$ is the UV cut off scale. For example, $N_\phi=5$ if $\phi$ and $\bar{\phi}$ transform as 
$\bold{5}$ and $\bold{\bar{5}}$ under $SU(5)_{GUT}$. At low energies messengeres are integrated out and their effects are incorporated at (\ref{KitanoK-loop}). The cut off scale $\Lambda$ also originates from microscopic scale physics. For example, it may account for massive fields of the UV completion of the susy breaking sector which have been integrated out at the energy scales $E<\Lambda$ that we are considering here.

The scalar potential of the supergravity Lagrangian is given by 
\begin{equation} \label{}
V=e^G\left(G_X G_{X^\dagger}G^{X X^\dagger}+G_\phi G_{\phi^\dagger} G^{\phi \phi^\dagger}+G_{\bar{\phi}} G_{\bar{\phi}^\dagger} G^{\bar{\phi} \bar{\phi}^\dagger}-3\right)+\frac{1}{2}D^2,
\end{equation} 
where $G = K+\log(|W|^2/M^6_P)$ and $D^2/2$ represents the $D$-term term contributions that we neglect. 
The supersymmetric minimum is \footnote{We note that in the global susy limit we have the susy preserving \itshape flat direction \normalfont : $\phi\bar{\phi}=F/\lambda, \, X=0$.}
\begin{equation} 
\phi=\bar{\phi}=\sqrt{\frac{F}{\lambda}}-{\cal O}\left( \frac{c}{\lambda M^{2}_P}\right), \   \  \  X={\cal O}\left(\frac{c}{\lambda M^2_P}\right)
\end{equation}

Along the $X$ direction the potential simplifies to 
\begin{equation} \label{V-zero}
V(X)\simeq F^2-3\frac{c^2}{M^2_P}-2\frac{c}{M^2_P}F (X+X^\dagger)+4F^2 \frac{|X|^2}{{\Lambda}^2} + \frac{\lambda^2 N_\phi}{(4\pi)^2} \, F^2 \log \frac{X^\dagger X}{Q^2},
\end{equation} 
which for $\lambda^2 N_q/(4\pi)^2 <(\Lambda/M_P)^2$ gives the non-supersymmetric minimum
\begin{equation}
  \left\langle X \right\rangle=\frac{c {\Lambda}^2}{2F M^2_P}.
\end{equation}
The parameters $c$ and $F$ are connected via the cancellation of the cosmological constant
$F^2 \simeq 3c^2/M^2_P\, .$
With the help of this condition the minimum can be written as
\begin{equation} \label{S-min}
\left\langle X \right\rangle\simeq \frac{\sqrt{3}\Lambda^2}{6M_P} \, .
\end{equation}
Supersymmetry is broken with $F_X = F$. One can see that in the global susy limit  $M_{P}\rightarrow \infty$ the minimum moves to $X\rightarrow 0$ and the metastable vacuum disappears. It is the presence of gravity that reveals the non-supersymmetric vacuum. 
The dominant terms in the potential, the tree level plus the one loop correction (\ref{KitanoK-loop}), up to 4th order in fields reads:
\begin{eqnarray}  \label{Kitano-tree} \nonumber
&V \simeq F^2-3\frac{c^2}{M^2_P}-2\frac{c}{M^2_P} F (X+X^\dagger)+4F^2 \frac{|X|^2}{{\Lambda}^2}
- \lambda F (\phi \bar{\phi}+\phi^\dagger\bar{\phi}^\dagger)+2 \frac{c}{M^2_P} F \frac{|X|^2}{\Lambda^2} (X+X^\dagger)& \\ \nonumber 
&
+\lambda^2 |X|^2 (|\phi|^2+|\bar{\phi}|^2)+ \lambda^2 |\phi|^2 |\bar{\phi}|^2 + 
\frac{\lambda^2 N_\phi}{(4\pi)^2} \, F^2 \log \frac{X^\dagger X}{Q^2} & 
\end{eqnarray}
where $F^2 \approx 3c^2/M^2_P$.

The mass matrices for  $X$ and $\phi$ are:
\begin{equation}
m^2_X \simeq \left(\begin{array}{cl}      
4\frac{F^2}{\Lambda^2} &  -\frac{\lambda^2 N_\phi}{(4\pi)^2}\frac{F^2}{X^{\dagger \,2}}  \\     
-\frac{\lambda^2N_\phi}{(4\pi)^2}\frac{F^2}{X^2} & 4\frac{F^2}{\Lambda^2}   \\     
\end{array}\right),
\,\,\,\,\,\, m^2_\phi \simeq
\left(\begin{array}{cl}      
\lambda^2 |X|^2 & -\lambda F  \\     
-\lambda F & \lambda^2 |X|^2   \\     
\end{array}\right)
\end{equation}
and should be evaluated at the susy braking vacuum: $\left\langle  X \right\rangle \sim \Lambda^2/M_P$ and $\phi=\bar{\phi}=0$.
The susy breaking minimum is stable in the $\phi$, $\bar{\phi}$ directions when the determinant of the $\phi$-$\bar{\phi}$ mass matrix is positive which yields
\begin{equation} \label{vac-stab}
\lambda^2 \left\langle X \right\rangle^2 >  \lambda F_X \; \; {\rm i.e.} \;\; \Lambda^4/M^2_P> F/\lambda. 
\end{equation}
Thus, the susy breaking metastable minimum is further away from the origin than the susy preserving one. It is stable in the $X$-direction when $\lambda^2 N_\phi/(4\pi)^2 <{\Lambda}^2/M^2_P$ or roughly 
\begin{equation} \label{vac-stab-qua}
\lambda < \Lambda/M_P .
\end{equation}
This condition renders the perturbative quantum corrections harmless for the vacuum meta-stability. It says that the closer to the origin is the metastable vacuum, the smaller has to be (the square of) the coupling $\lambda$. 

Moreover, there is a phenomenological requirement that the gaugino masses should be of the order of $m_{\text{gaugino}}={\cal O}(100\ \text{GeV}-1\ \text{TeV})$. This fixes the relation between the parameters $F$ and $\Lambda$ as follows:
\begin{equation} \label{muLa}
F = \left(\frac{\alpha}{4\pi} \right)^{-1} m_{\text{gaugino}} \left\langle X \right\rangle \simeq 10^{-14} \Lambda^2.
\end{equation}
With fixed gaugino masses we have two parameters: $\Lambda$ and $\lambda$. 
The two conditions (\ref{vac-stab}) and (\ref{vac-stab-qua}), necessary for  gravitational stabilization, can both be fulfilled for 
\begin{equation} \label{Lambda}
\Lambda>10^{-14/3}M_P .
\end{equation}
This lower bound on $\Lambda$ is high  enough to keep the metastable vacuum far away from the supersymmetric one and  to suppress sufficiently  the tunneling rate. The shape of the zero temperature potential around the minima is depicted at the Figure 5.1, the constraints on the parameters from the susy breaking vacuum stability are presented in the Table 4.1 and illustrated at the Figure 4.3.

\section{Metastable Supersymmetry Breaking Vacua with Generalized K\"ahler Potential} \normalsize

We assume that at tree level the fields have canonical K\"ahler potential and that their interactions are described by the most general superpotential consistent with the $U(1)$ $R$-symmetry:
\begin{equation} \label{W-gen}
W=F X + \tilde{\varphi_i}(m_{ij}+\lambda_{ij}X)\varphi_j.
\end{equation}
Interactions of $X$ with $\tilde{\varphi_i}$ and $\varphi_j$ induce perturbative quantum corrections in the classic theory ($\ref{W-gen}$). The leading contribution of these corrections is the one-loop Coleman-Weinberg potential for $X$ \cite{Intriligator:2006dd}, 
\begin{equation}
V_{CW}=\frac{1}{64\pi^2}Tr(-1)^F {\cal M}^4 \text{log}\frac{{\cal M}^2}{Q^2}
\end{equation}
where ${\cal M}_{ij} = m_{ij}+\lambda_{ij} S $ and $Q$ the UV cut-off of the theory.
This can be approximately accounted for by introducing a correction to the K\"ahler potential
\begin{equation} \label{K-eff}
\delta K=-\frac{1}{16\pi^2}Tr\left[{\cal M}^\dagger{\cal M}\log \left( \frac{{\cal M}^\dagger {\cal M}}{Q^2} \right)\right]
\end{equation}
where ${\cal M}_{ij} = m_{ij}+\lambda_{ij} X $ and $Q$ is the UV cut-off of the theory. The $R$-symmetry present in ($\ref{W-gen}$) implies that the $R$ charge assignment for $X$ is $R(X)=2$, and guarantees that to leading order around $X=0$ , the effective potential takes the form $V_{CW}=V_0\pm m^2_X|X|^2+ {\cal O}(|X|^4)$ \cite{Shih:2007av}. Therefore, the correction to the K\"ahler potential ($\ref{K-eff}$) is a function only of $|X|^2$ and can be expanded in powers of $|X|^2$ 
with  the dimensionful parameters $g_{2l}$:
\begin{equation} \label{K-eff3}
\delta K=\sum_{l\geq2}g_{2l}|X|^{2l}=-g_4|X|^4-g_6|X|^6-...
\end{equation}
The $g_2$ term simply rescales the canonical term in the K\"ahler potential. Hence, starting from the the superpotential ($\ref{W-gen}$) of a generalized O'R model we can integrate out the heavy chiral superfields $\varphi_i$ ending up with an effective low energy superpotential $\delta W_{low}= F X$
and a K\"ahler potential with the correction ($\ref{K-eff3}$). Actually, we integrate the supersymmetric rafertons out  for simplicity - alternatively we could keep them in the low energy Lagrangian together with the 
1-loop correction they generate. It turns out that under rather general conditions  the vacuum found in the full theory coincides to a good accuracy with the 
vacuum found in the simpler model with decoupled rafertons. Again, we neglect the gauge bosons, assuming that the nonstandard ones are very heavy, 
and noticing that the SM gauge boson masses do not depend explicitly on $\left\langle X \right\rangle$.  This gives in the global limit the effective potential 
\begin{equation}
 V_\text{eff}=(K_{\text{eff} XX^{\dagger}})^{-1}|F_X|^{2}.
\end{equation}
Including gravity and messengers (not integrated out) we take 
\begin{equation} \label{W-low}
\delta W_\text{low}=F X+\lambda X \phi \bar{\phi}+c.
\end{equation}
and  $K_{eff}=|X|^2-g_4|X|^4-g_6|X|^6$.

Setting for economical reasons $\phi=\bar{\phi}$ one obtains the supergravity effective potential
\begin{equation} \nonumber 
V_0=F^2-3\frac{c^2}{M^2_P}-2\frac{c}{M^2_P} F (X+X^\dagger)+ 4 g_4 F^2 |X|^2 - 2\lambda F |\phi|^2 +  4 \frac{c}{M^2_P} F g_4 |X|^2(X+X^\dagger)
\end{equation} 
\begin{equation} \label{Vg}
  \;\;\;   + 2 \lambda^2 |X|^2 |\phi|^2 +4 F^2 \frac{g_4}{M^2_P} |X|^4 + 9 F^2 g_6 |X|^4 + \lambda^2 |\phi|^4.
\end{equation}
Again, we are going to cancel the cosmological constant by assuming $c\approx F M_P / \sqrt{3}$. 
The dimensionful parameters $g_4$ and $g_6$ are of the form $g_4\sim \epsilon_4 \Lambda^{-2}$ and $g_6\sim \epsilon_6 \Lambda^{-4}$ where $\Lambda$ is the mass scale of the particles which are integrated out and $\epsilon_4, \epsilon_6$ are coefficients of an unspecified sign. We absorb the scale into the coefficients and write them as $|g_4|\equiv 1/\Lambda^2_1$ and $|g_6| \equiv1/\Lambda^4_2$. The model discussed  earlier in this paper corresponds to $g_4=1/\Lambda^2_1= 1/ {\Lambda^2}$ and $\Lambda_2\rightarrow\infty$. The interesting observation lies in the fact that the parameters $g_4$ and $g_6$ can be positive or negative. When the $g_4$ is positive the $g_6$ correction is negligible and we obtain the minimum already known from the previous section
\begin{equation} \label{min1}
\left\langle  X \right\rangle=\frac{\sqrt{3} \Lambda^2_1}{6 M_P}, \;  \; \; g_4>0.
\end{equation}
However, $g_4$ can be negative. We are going to keep $g_6$ positive in this case to ensure the existence of a minimum. The higher order corrections are of course considered to be $g_{2l}|X|^{2l}<|X|^2$. In terms of $\Lambda_{1,2}$ this means that our theory is valid in the regime $|X|<\Lambda_1$ and $|X|<\Lambda_2$. A simple realization of a model that leads to such  a situation has been given in \cite{Lalak:2008bc}. 

For negative $g_4$, $g_4=-|g_4|=-1/\Lambda^2_1$ the K\"ahler potential reads
\begin{equation}
K_{eff}=|X|^2+\frac{|X|^4}{\Lambda^2_1}-\frac{|X|^6}{\Lambda^4_2}
\end{equation}
and we can select the following two cases:
\begin{itemize}
	\item $X |g_4|>1/M_P \Rightarrow X>\Lambda^2_1/M_P$,
\end{itemize}
which means that the ${\cal O}(X)$ is subdominant in ($\ref{Vg}$) and leads to the minimum
\begin{equation} \label{min2}
|\left\langle X \right\rangle|^2 \simeq \frac29 \frac{|g_4|}{g_6} = \frac29 \frac{\Lambda^4_2}{\Lambda^2_1}. 
\end{equation}
The condition $g_{2l}|X|^{2l}<|X|^2$ in this minimum implies that $\Lambda_2<\Lambda_1$. This minimum doesn't have any dependence on the $M_{P}$ which means that it survives  in the global susy limit. This can also be seen that omitting the ${\cal O}(X)$ in ($\ref{Vg}$), which contains the dimensionful constant $c$, the potential is to leading order the global susy one. The condition $X|g_4|>1/M_P$ in this minimum
translates into $\Lambda^3_1/M_P<\Lambda^2_2$ and one arrives at
\begin{equation}
\frac{\Lambda^{3/2}_1}{M^{1/2}_P}<\Lambda_2<\Lambda_1.
\end{equation}
The corresponding parameter space is illustrated in the Figure 4.1. The stability of the above vacuum gives us more constraints on the parameters. Asking for stability in the $\phi$-direction, i.e positive determinant of the $\phi$ mass matrix, one finds the condition 
\begin{equation} \label{vac-stab-2}
|X|^2>\frac{F}{\lambda}\Rightarrow |X|^2\sim \frac{\Lambda^4_2}{\Lambda^2_1} >\frac{F}{\lambda}.
\end{equation}
Loop corrections coming from the messengers are irrelevant if $\lambda^2 N_{\phi}/(4\pi)^2<(\Lambda_2/\Lambda_1)^4$. Thus, approximatelly, we take 
\begin{equation} \label{vac-stab-qua2}
\lambda<\left(\frac{\Lambda_2}{\Lambda_1}\right)^2.
\end{equation}
We note that contrary to (\ref{vac-stab-qua}) the above bound on the coupling $\lambda$ is not $M_{P}$ suppressed, and this is  as it should be for susy breaking minima that survive in the global susy limit. Gaugino massses of the order of 100 GeV - $1\, {\rm TeV}$ relate the value of $\langle X \rangle$  to the $F$ as follows
\begin{equation}
F = \left(\frac{\alpha}{4\pi} \right)^{-1} m_{\text{gaugino}} \left\langle X \right\rangle \simeq 10^{-14} \frac{\Lambda^2_2}{\Lambda_1}M_P.
\end{equation}
Combining the above constraints on $\lambda$ we find that the vacuum is metastable for  $\Lambda_2>10^{-7} \left (\frac{\Lambda_{1}^{3/2}M^{1/2}_P}{\Lambda_2} \right )$.
\begin{figure} \label{SBRp}
\centering

\includegraphics [scale=1.2, angle=0]{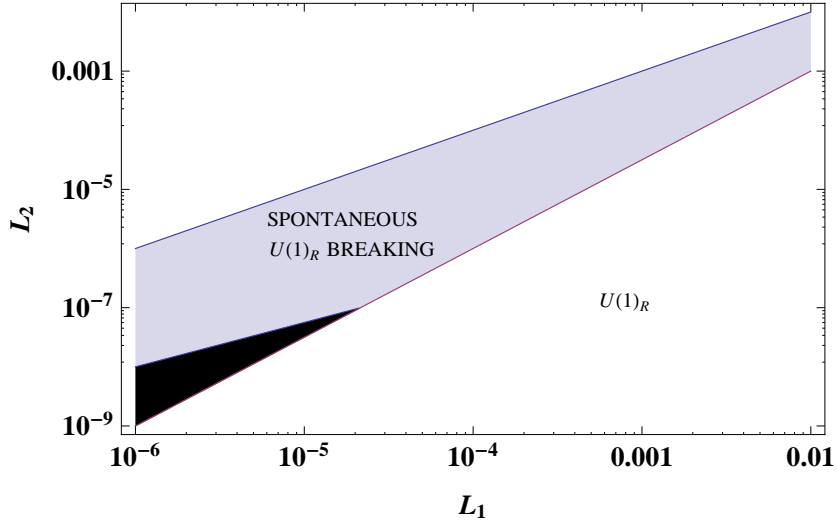} 
\caption{($L_1 \equiv \Lambda_1, L_2 \equiv\Lambda_2$) The parameter space (light blue region) that realize the spontaneous breaking of the $U(1)_R$ symmetry. The region above, $\Lambda_2>\Lambda_1$, corresponds to non-positive definite K\"ahler metric and the region below to $U(1)_R$ symmetric solutions -in the global susy limit. In the black region there is no metastable vacuum due to the 1-loop effects from the interaction of the $X$-field with the messengers $\phi$, $\bar{\phi}$.  } 
\end{figure}
\begin{itemize}
	\item The second case is $ X |g_4|<1/M_P \Rightarrow X<\Lambda^2_1/M_P$,
\end{itemize}
which leads to the minimum 
\begin{equation} \label{min3}
\left\langle X \right\rangle ^3 \sim \frac{1}{9}\frac{c}{F M^2_P}\frac{1}{g_6}  \simeq  \frac{\Lambda^4_2}{M_P}.
\end{equation}
The condition $X |g_4|<1/M_P$ is fulfiled for 
\begin{equation}
\Lambda_1 \left( \frac{\Lambda_1}{M_P} \right)^{1/2} > \Lambda_2 
\end{equation}
The special limit $g_4\approx 0$ belongs to this domain. Stability of this vacuum in the $\phi$-direction implies
\begin{equation}
|X|^2>\frac{F}{\lambda}\Rightarrow X^2\sim \frac{\Lambda^{8/3}_2}{M^{2/3}_P} >\frac{F}{\lambda}
\end{equation}
and the loop correction given by messengers is irrelevant  if
\begin{equation} \label{vac-stab-qua3}
\lambda<\left(\frac{\Lambda}{M_P}\right)^{2/3}
\end{equation} 
for $\lambda^2 N_\phi/(4\pi)^2 \sim \lambda^2$, in accordance with (\ref{vac-stab-qua}). ${\cal O}$(TeV) gaugino masses give us in this vacuum the relation
\begin{equation}
F  \simeq 10^{-14} \Lambda^{4/3}_2 M^{2/3}_P.
\end{equation}

\begin{table} \label{tabb1}
\small \small
\begin{center}
$$
\begin{array}{|r|r|r|r|r|}
\hline
\bold{K=|X|^2  \mp \frac{|X|^4}{\Lambda^2_1}-\frac{|X|^6}{\Lambda^4_2}} & \rm \bold{Metastable} \; \bold{VEV}  & \bold{\Lambda_1} \,\,\,\,\,& \bold{\Lambda_2} \,\,\,\,\,\,\,\,\,\,\,\,\,\,\,& \bold{\lambda} \,\,\,\,\,\,\,\,\,\,\,\,\,\,\,\,\,\,\,\, \\
\hline 
1.\bold{\, ~(-),\, \; \, \Lambda_2 = \Lambda_1 \equiv  \Lambda} 
& \left\langle X \right\rangle \sim \Lambda^2 \,\,\,\,\,\,\,\,\,\,\,\,\,\,\,\,\,& \Lambda > 10^{-14/3} & -\,\,\,\,\,\,\,\,\,\,\,\,\,\,\,\,\,\,\,\, & \frac{10^{-14}}{\left\langle X\right\rangle}<\lambda <\Lambda \,\,\,\,\, \\
\hline
\bold{2. ~~~~~  (+), \, \; \, \Lambda^{3/2}_1<\Lambda_2}
& |\left\langle X \right\rangle| \sim \frac{\Lambda^2_2}{\Lambda_1}\,\,\,\,\,\,\,\,\,\,\,\,\,\,\, & \Lambda_1>\Lambda_2 & \Lambda_2> 10^{-7} \left(\frac{\Lambda^{3/2}_1}{\Lambda_2}\right)& \frac{10^{-14}}{\left\langle X \right\rangle}< \lambda < \left(\frac{\Lambda_2}{\Lambda_1}\right)^2 \\
\hline
\bold{3. ~~~~~  (+), \, \; \, \Lambda^{3/2}_1>\Lambda_2}
& \left\langle X \right\rangle \sim \Lambda^{4/3}_2\,\,\,\,\,\,\,\,\,\,\,\,\,\,\, & \Lambda_1>\Lambda_2 & \Lambda_2> 10^{-7} & \frac{10^{-14}}{\left\langle X\right\rangle}< \lambda  < \Lambda_2^{2/3} \\
\hline
\end{array}
$$
\end{center}
\caption{\small Zero temperature vacuum stability constraints ($M_P=1$) in the three cases of the generalized K\"ahler potential. With gaugino masses $m_{\text{gaugino}} \propto F/ \left\langle X\right\rangle$ fixed at ${\cal O}(100\ \text{GeV}-1\ \text{TeV})$ the free parametrs left are the cut-off scale and the coupling $\lambda$. The first case corresponds to $\Lambda_1=\Lambda_2=\Lambda$. Since the 4th order correction has the same sign as the 6th order one, the 6th order term is negligible. In the other two cases the 6th order correction is necessary for the stabilization of the metastable vacuum. The hierarchy between $\Lambda_1$ and $\Lambda_2$ i.e. $\Lambda_1>\Lambda_2$ keeps the corrections to the K\"ahler potential under control. The lower bound on the cut-off scales $\Lambda$, $\Lambda_2$ originates from the vacuum meta-stability in the messenger sector. The upper bound on the coupling k renders the one-loop correction to the $X$ potential irrelevant for the stability of the vacuum and the lower bound prevents  messengers from becoming tachyonic.}
\end{table}
Hence, here a gravitational stabilization of the vacuum is possible for $\Lambda_2>10^{-7}M_P$ which is a less stringent bound on the cut off scale compared to (\ref{Lambda}). The minimum ($\ref{min3}$), as the ($\ref{min1}$), disapears if we neglect gravity. As $M_{P}\rightarrow \infty$ the susy breaking minima ($\ref{min1}$), ($\ref{min3}$) enter the domain $|X|<\sqrt{\frac{F}{\lambda}}$ and they become tachyonic in the $\phi$-direction. In the absence of messengers these $M_{P}$ suppressed minima could exist even in the global susy limit where the minimum would be at $X=0$ preserving also the $R$-symmetry. However, in the presence of messengers, the gravity allows the existence of these metastable vacua where both susy and $U(1)_R$ break down. The zero temperature constraints of all the cases are assembled and presented in the Table 4.1.

\section{Constraints on the SUSY Breaking}

Here we are going to review the constraints coming from the vacuum stability at zero temperature which have been already considered and combine them with constraints coming from gauge mediation  domination over gravity. 
The zero temperature vacuum stability constraints which have already been discussed have been summarized in the table 4.1.

Asking for gauge mediation domination over gravity mediation we can further constrain the cut-off parameters $\Lambda, \, \Lambda_1$ and $\Lambda_2$.  If gravity mediation contribution to squark mass squared is suppressed to ${\cal O}(1\%)$ then FCNC are sufficiently suppressed \cite{Feng:2007ke}. Hence, in the metastable vacuum we ask for 
\begin{equation} \label{grav/gau}
\frac{m_{3/2}}{m_{\text{gaugino}}}= \frac{4\pi}{\alpha\sqrt{3}}\frac{\left\langle X \right\rangle}{M_{P}} \lesssim {\cal O}(1\%).
\end{equation}
Obviously this gives us an upper bound on the value of the spurion $X$ field. For $\alpha=0.04$ and 
it yields
\begin{equation} \label{Smax}
\left\langle X \right\rangle \leq{\cal O}(10^{-4}-10^{-3})M_P
\end{equation}
and for the rest of the paper we take the bound $\left\langle X \right\rangle \leq 10^{-4}M_P$. 

For the first case where the 6th order correction to K\"ahler potential is negligible the above constraint translates into $\Lambda\lesssim10^{-2} M_P$. For the second and third cases where the 6th order correction to K\"ahler potential is necessary we respectively have $\Lambda_2\lesssim \sqrt{10^{-4}\Lambda_1 M_P}$ and $\Lambda_2\lesssim 10^{-3} M_P$. In the second case, the $\Lambda^{3/2}_1 /M^{1/2}_P<\Lambda_2$ condition gives us a numerical upper bound on $\Lambda_2$: $\Lambda_2\lesssim \sqrt{10^{-4}\Lambda_1 M_P}<10^{-3} M_P$. We see that the stringent upper bounds on the cut-off scales apply for the case of K\"ahler potential corrected up to 6th order and especially for the second case in which susy breaking vacua survive in the global limit. 

\begin{figure} \label{SBRp2}
\centering

\includegraphics [scale=1.2, angle=0]{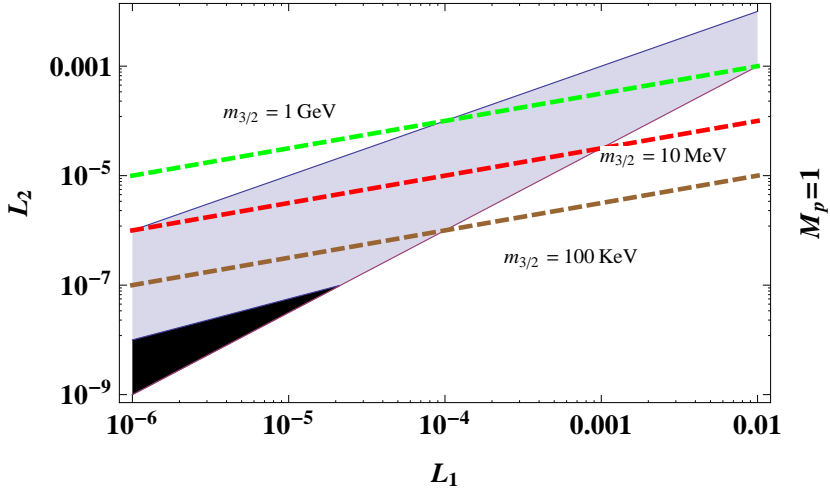} 
\caption{($L_1 \equiv \Lambda_1, L_2 \equiv\Lambda_2$) The parameter space (light blue region) that realize the spontaneous breaking of the $U(1)_R$ symmetry. The dashed lines corrrespond to parameters $\Lambda_1$ and $\Lambda_2$ of the coloured region that give gravitino of mass 1 GeV, 10 MeV and 100 KeV respectively from top to bottom.  } 
\end{figure}

\begin{figure} 
\centering
{(a)} \includegraphics [scale=1.2, angle=0]{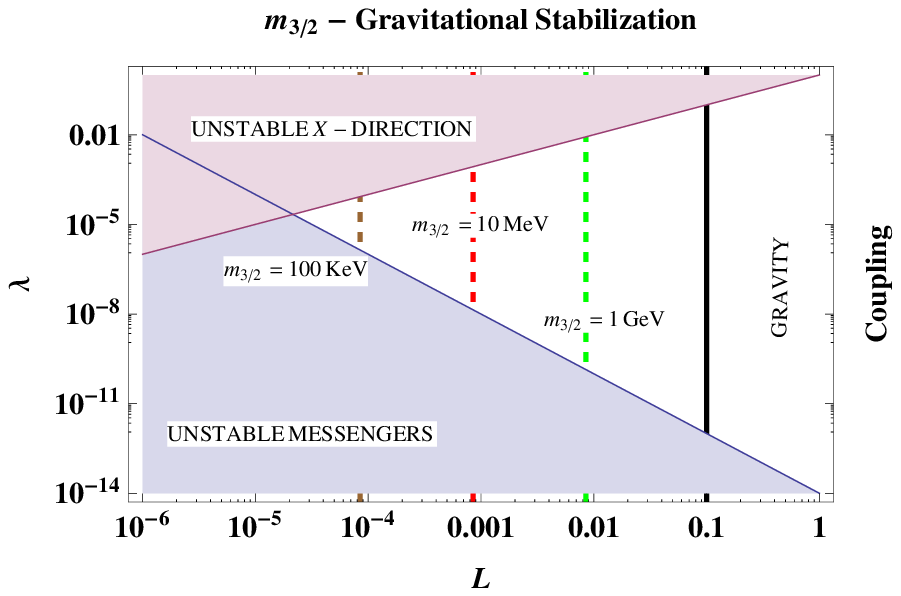} \\
{(b)} \includegraphics [scale=1.2, angle=0]{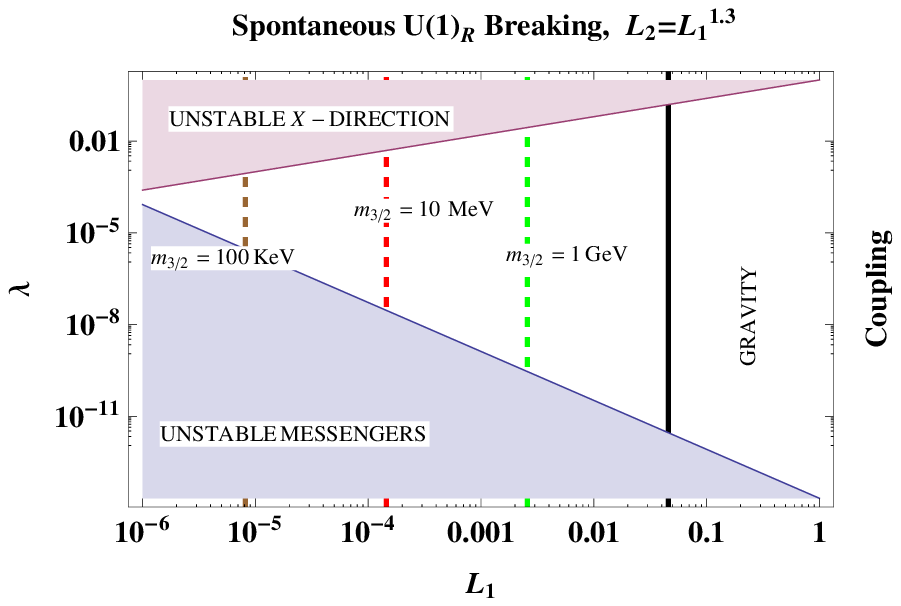}  \\
\caption{\small{At the upper panel, the region of the parameter space (white part) where the supersymmetry broken minimum is metastable for the case of gravitational stabilization is depicted. At the lower panel the parameter space (white part) where supersymmetry and $U(1)_R$ break spontaneously at a metastable minimum is depicted for $\Lambda_2=\Lambda^{1.3}_1$ in Planck units. The black continuous vertical line corresponds to gravitino mass of 100 GeV. The dashed lines, from right to left, correspond to 1 GeV, 10 MeV and 100 KeV gravitino mass respectively.  At the figure, $L$ stands for the $\Lambda$. }}
\end{figure}

\chapter{The Supersymmetry Breaking Sector at Finite Temperature}

The metastability of the supesrymmetry breaking vacuum raises the issue of the vacuum selection in the thermalized early universe. Here, the basics of finite temperature field theory are presented. Then, they are applied to the case of $F$-term supersymmetry breaking sectors. Considering only the supersymmetry breaking sector it is shown that the global sypersymmetric minima are thermally selected \cite{Dalianis:2010yk}. On the other hand the ISS paradigm, without ordinary messengers, is thermally favourable.
 
\section{ Field Theory at Finite Temperature}

The formalism used in conventional quantum field theory is suitable to describe observables, e.g. the cross-section, in empty spacetime as particle interactions in an accelerator. Nevertheless, in the early stages of the universe, at high temperature, the enviroment had a non-negligible matter and radiation density, making the hypothesis of conventional field theories impracticable.  Under those circumstances the methods of conventional field theories are no longer in use and should be replaced by others closer to thermodynamics, where the background state is a thermal bath. This field has been called field theory at finite temperature \cite{Dolan:1973qd, Weinberg:1974hy}
 and it is very usuful to study all phenomena which happened in the early universe: phase transitions, inflationary cosmology, baryogenesis, etc.  A textbook devoted to the subject is the \cite{Kapusta:1989tk}  and some textbooks that we follow here are the \cite{Kolb:1990vq, Bailin:2004zd, Bailin:1986wt, Mukhanov:2005sc} together with the review article \cite{Quiros:1999jp}.

One of the fundamental objects in the statistical thermodynamics of a finite temperature system is the partition function $Z$ defined by 
\begin{equation} \label{Tr-4}
Z=\text{Tr} e^{-\beta \hat{H}}
\end{equation}
where the $\hat{H}$ is the Hamiltonian operator and 
\begin{equation}
\beta=(k_B T)^{-1}= T^{-1}
\end{equation}
in units where the Boltzmann constant $k_B$ is set equal to 1. The trace in (\ref{Tr-4}) means that we are to sum over the the diagonal matrix elements of $e^{-\beta \hat{H}}$ for all independent states of the system. Once the partition function has been evaluated the Helmholtz \itshape free energy \normalfont is given by
\begin{equation}
Z=e^{-\beta F}
\end{equation}
where according to the thermodynamics $F$ is related to the internal energy $E$ and the entropy $S$ by 
\begin{equation}
F=E-TS.
\end{equation}
The pressure $p$ and the entropy are obtained from the free energy as 
\begin{equation}
\left. P=-\frac{\partial F}{\partial V}\right|_{T=\text{const}} 
\end{equation}
and
\begin{equation}
\left. S=-\frac{\partial F}{\partial T}\right|_{V=\text{const}}.
\end{equation}
The energy density is given by
\begin{equation}
\varepsilon= {\cal F}+ Ts
\end{equation}
where ${\cal F}$ and $s$ are the free energy and entropy densities, with
\begin{equation}
E=\int d^3 x \, \varepsilon \,\,\,\,\,\,\,\,\,\,\,\, \text{and} \,\,\,\,\,\,\,\,\,\,\,\, s=\int d^3 x S.
\end{equation}
Therefore, the calculation of the partition function provides us with a determination of the energy density.

The partition function in a gauge field theory is most efficiently calculated using path integrals methods. The simplest contribution to the partition function comes from the free neutral real scalar fields. The Lagrangian for such a field $\varphi$ having mass $m$ is given by 
\begin{equation}
{\cal L }(\varphi, \partial_\mu \varphi)= \frac{1}{2}\left( \frac{\partial \varphi}{\partial t} \right)^2 - \frac{1}{2} \left(\nabla \varphi \right)^2 -\frac{1}{2}m^2 \varphi^2.
\end{equation}
In field theory at finite temperature, scalar fields $\varphi(t, \textbf{x})$ are replaced by fields $\varphi(\tau, \textbf{x})$ periodic in $\tau$ with period $\beta$,
\begin{equation}
\varphi(\tau = 0, \textbf{x})=\varphi(\tau=\beta, \textbf{x})
\end{equation}
where $\tau=i t$. The usual convention of referring to non-zero temperature as \itshape "finite temperature"\normalfont. The partition function is formulated in terms of these periodic fields as
\begin{equation}
Z= \tilde{N}(\beta) \int_\text{periodic} {\cal D} \varphi \, e^{ \int^\beta_0 d\tau \int d^3 x {\cal L}(\varphi, \bar{\partial}_\mu \varphi) }
\end{equation}
where
\begin{equation}
\bar{\partial}_\mu \varphi \equiv \left(i \frac{\partial \varphi}{\partial \tau}, \nabla \varphi \right)
\end{equation}
and $\tilde{N}(\beta)$ is a temperature dependent renormalization. The ${\cal D} \varphi$ is a path integral. Evaluation of the path integral gives for the contribution of a real scalar field to the free energy
\begin{equation} \label{F-4}
-\beta F = \ln Z = - \int d^3 x \int \frac{d^3 p}{(2\pi)^3} \left( \frac{\beta}{2}\sqrt{\bold{p}^2 + m^2} + \ln \left[1-e^{-\beta\sqrt{\bold{p}^2+m^2 }} \right]. \right)
\end{equation}
Writing explicitly the temperature, $\beta=1/T$, the above equation reads
\begin{equation}
 -T\ln Z= F \equiv \int d^3 x\left( {\cal F}_0+{\cal F}_T \right)
\end{equation}
where 
\begin{equation}
{\cal F}_0=\frac{1}{4\pi^2} \int^\infty_0 dp\, p^2 \sqrt{p^2+m^2} 
\end{equation}
is the temperature independent one-loop zero point (for a quantum theory) energy and
\begin{equation}
{\cal F}_T=\frac{1}{2\pi^2} T\int^\infty_0 dp\, p^2 \ln \left(1-e^{-\sqrt{(p^2+m^2)/T^2}} \right) = \frac{T^4}{2\pi^2} \int^\infty_0 dy \, y^2 \ln \left(1- e^{-\sqrt{y^2+m^2/T^2}}  \right)
\end{equation}
is the temperature dependent  part. When the mass of the scalar field is negligible compared with the temperature the free energy simplifies to 
\begin{equation}
{\cal F}_\text{scalar} \cong -\frac{\pi^2 T^4}{90}.
\end{equation}

For the case of gauge bosons there are subtleties, because for a typical choice of gauge, the Lagrangian involves all four degrees of freedom of the gauge field $A^\mu(x)$ and also involves the Fadeev-Popov ghost fields which occur in the construction of a consistent renormalizable theory but are not physical particles. A massles vector field has only two degrees of freedom, the two polarization states. The extra degrees of freedom are not physical and cannot be in equilibrium with a heat bath nor can the Fadeev-Popov ghosts. This technical problem is addressed once one works in a gauge that the gauge field has only two degrees of freedom and there are no Fadeev-Popov ghosts. In such a gauge the partition function can be directly related to the Lagrangian densiy. Then, the contribution to the free energy density from a massless vector gauge field is found to be
\begin{equation}
{\cal F}_\text{vector} = -2\frac{\pi^2 T^4}{90}.
\end{equation}
In the case of Dirac fields $\psi$ the corresponding development of the finite temperature field theory involves fields $\psi(\tau, \bold{x})$ that are antiperiodic in $\tau$ in the interval $(0, \beta)$
\begin{equation}
\psi(\tau=0, \bold{x})=-\psi(\tau=\beta, \bold{x})
\end{equation}
and the contribution to the free energy density when $T \gg m$
\begin{equation}
{\cal F}_\text{fermion} \cong -\frac{7}{8}4\frac{\pi^2 T^4}{90}.
\end{equation}
For massless fermions, with one helicity state of the particle i.e. Weyl, or Majorana fermions the free energy is the half. 

Summing up, the free energy of an ideal ultra relativistic gas, $T \gg m$, is given by
\begin{equation}
{\cal F} \cong -\left(N_B +\frac{7}{8}N_F \right)\frac{\pi^2 T^4}{90}
\end{equation}
where $N_B$ and $N_F$ are the numbers of bosonic and fermionic degrees of freedom respectively. 

\section{Effective Potential at Finite Temperature}

The effective potential, of e.g. a scalar field $\varphi$, at finite temperature is the free energy density associated with the $\varphi$ field ${\cal F} \equiv \bar{V}^T(\varphi)=\rho_\varphi-T s_\varphi$. The derivation of the $\bar{V}^T$ is to account for the effect of the ambient gas in the calculation of higher order quantum correction to the classical potential. Hence, it is a finite temperature renormalization of the theory.

\subsection{ Path Integral Approach}

At zero temperature the expectation value $\varphi_c$ of the scalar field is determined by minimizing the effective potential, $V(\varphi_c)$, that may contain quantum corrections from various loop orders. The one loop quantum correction is calculated by considering excitations $\tilde{\varphi}$ about the expectation value (shift) and isolating the ${\cal L}_\text{quad}(\varphi_c, \tilde{\varphi})$ in the Lagrangian which are quadratic to the excitations  $\tilde{\varphi}$.  We can write
\begin{equation}\label{Vtotloop-4}
V(\varphi_c)= V_0(\varphi_c)+V_1(\varphi_c)
\end{equation}
where $V_0$ is the tree level contribution and $V_1$ is the one-loop quantum correction. Then it holds
\begin{equation} \label{exp-4}
e^{-i\int d^4x V_1(\varphi_c)}= \int {\cal D} \tilde{\varphi} \, e^{i\int d^4 x {\cal L}_\text{quad}(\varphi_c,\tilde{\varphi})}
\end{equation}
where the $\int {\cal D} \tilde{\varphi}$ denotes a path integral. Similarly, the finite temperature effective temperature 
\begin{equation} \label{VT-4}
\bar{V}^T(\varphi_c)= \bar{V}_0(\varphi_c) + \bar{V}_1(\varphi_c)=\bar{V}_0(\varphi_c) + \bar{V}^0_1(\varphi_c) + \bar{V}^T_1(\varphi_c)
\end{equation}
where the expectation value $\varphi_c$ is now a thermal average. The $\bar{V}_0$ and $\bar{V}_1$ are the finite temperature renormalized tree level and one-loop terms. At the right hand side of (\ref{VT-4}) the one-loop terms were seperated to the temperature independent part $\bar{V}^0_1$ and the temperature dependent $\bar{V}^T_1$.  The (\ref{exp-4}) is modified to 
\begin{equation} \label{exp2-4}
e^{\int^\beta_0 d\tau \int d^3x \bar{V}_1(\varphi_c)} = \int_\text{periodic}  {\cal D} \tilde{\varphi} \, e^{\int^\beta_0 d\tau \int d^3 x {\cal L}_\text{quad}(\varphi_c,\tilde{\varphi})}.
\end{equation}
When gauge fields $A^\mu_a$ and fermions $\psi_r$ are included then the (\ref{exp2-4}) also contains, in an arbitrary gauge,  path integrals over the gauge fields and their associated Fadeev-Popov ghosts and path integrals over antiperiodic fermion fields. Then the terms in the Lagrangian of quadratic order in fields are of the form
\begin{equation} \nonumber
{\cal L}_\text{quad} (\varphi_c, \tilde{\varphi}) =\frac{1}{2} \bar{\partial}_\mu\tilde{\varphi}_i\bar{\partial}^\mu \tilde{\varphi}_i -\frac{1}{4}(\bar{\partial}^\mu A^\nu_a-\bar{\partial}^\nu A^\mu_a) (\bar{\partial}_\mu A_{\nu a}-\bar{\partial}_\nu A^{\mu a}) -\frac{1}{2\xi} (\bar{\partial}_\mu A^\mu_a)^2 +\bar{\partial}_\mu \eta^*_a\bar{\partial}^\mu \eta_a
\end{equation}
\begin{equation} \label{quad-4}
\,\,\,\,\,\,\,\,\,\,\,\,\,\,\,\, -\frac{1}{2} \left[\hat{M}^2_S(\varphi_c)\right]_{ij} \tilde{\varphi_i}\tilde{\varphi}_j+ \frac{1}{2} \left[\hat{M}^2_V(\varphi_c)\right]_{ab} A^\mu_a A_{b\mu} - \left[\hat{M}^2_F(\varphi_c)\right]_{rs} \bar{\psi}_r\psi_s.
\end{equation}
The $\varphi_c$ denotes the complete set of expectation values of the scalar fields $\tilde{\varphi}_i$ and the $\psi_r$ are Dirac fermions. For the fermions and the vectors there is no expectation value in order to avoid breaking Lorentz invariance. The $\eta_a$ are the Fadeev-Popov ghost fields introduced in the construction of a consistent renormalizable theory of gauge fields but do not correspond to physical particles and $\xi$ is the gauge fixing parameter. The gauge adoped is the Landau gauge where there is no coupling of scalar fields to Fadeev-Popov ghosts.

The $\hat{M}^2_S$, $\hat{M}^2_V$ and $\hat{M}^2_F$ are the mass squared matrices; their eigenvalues are denoted by $(M^2_S)_i$, $(M^2_V)_a$ and $(M^2_F)_r$ respectively. Then the one-loop term in the effective potential $V^T_1$ takes the form 
\begin{equation} \nonumber
\bar{V}^T_1=\frac{T^4}{2\pi^2}\int^\infty_0 \left\{ \sum_i\ln \left[ 1-e^{-\sqrt{y^2+(M^2_S)_i /T^2}} \right]  + \sum_a \left( 3\ln \left[ 1-e^{-\sqrt{y^2+(M^2_V)_a /T^2}} \right]  -\ln(1-e^y) \right) \right.
\end{equation} 
\begin{equation} \label{VT-4}
\left. -4 \sum_r \ln \left[ 1+ e^{-\sqrt{y^2+(M^2_F)_i /T^2}} \right] \right\}.
\end{equation}
There are two limits in which $\bar{V}^T_1$ is particularly simple. First in the limit where all mass-squared eigenvalues are very much greater than $T^2$ all terms in $\bar{V}^T_1$ approach zero exponentially and $\bar{V}^T_1$ become negligible. Second, in the high temperature limit where the $T^2$ is very much greater than the mass-squared eigenvalues the integrals (\ref{VT-4}) can be analytically calculated and the one-loop finite temperature potential is given by the expression
\begin{equation} \nonumber 
\bar{V}^T_1(\varphi_c)\simeq -\frac{\pi^2 T^4}{90} \left(N_B+\frac{7}{8}N_F\right)+\frac{T^2}{24}\left[\sum_{i}(M^2_S)_i+3\sum_{a}(M^2_V)_a+2\sum_{r}(M_F)^2_r\right]
\end{equation} 
\begin{equation} \nonumber 
-\frac{T}{12\pi}\left[\sum_{i}(M^3_S)_i+3\sum_{a}(M^3_V)_a\right] 
\end{equation}
\begin{equation}\label{VT1-4}
- \sum_i \frac{(M^2_S)_i^2}{64\pi^2}\,\ln\left[ \frac{(M^2_S)_i}{a_bT^2} \right]- \sum_a \frac{(M^2_V)^2_a}{64\pi^2}\,\ln \left[\frac{(M^2_V)_i}{a_bT^2}\right]- \sum_r \frac{(M^2_F)^2_r}{64\pi^2}\,\ln \left[\frac{(M^2_F)_r}{a_f T^2}\right]...
\end{equation}
where the constants $\ln a_b\simeq 5.41$ and $\ln a_f =2.63$ numbers related to the Euler constant appearing in the integrals. Omiting the negligible logarithmic terms the effective potential  reads in terms of the mass matrices squared 
\begin{equation} \nonumber
\bar{V}^T_1(\varphi_c)\simeq -\frac{\pi^2 T^4}{90} \left(N_B+\frac{7}{8}N_F\right)+\frac{T^2}{24}\left[\text{tr} M^2_S(\varphi_c)+3\text{tr}M^2_V(\varphi_c)+2\text{tr} M^2_F(\varphi_c)\right]
\end{equation} 
\begin{equation} \label{VT2-4}
-\frac{T}{12\pi}\left[\text{tr} \{M^2_S(\varphi_c)\}^{3/2}+3\text{tr}\{M^2_V(\varphi_c)\}^{3/2}\right]+...
\end{equation}
We note that for Weyl spinor fields there should be no factor of $2$ in front of the $\hat{M}^2_F$ and the eigenvalue squared $(M_F)^2_r$. The $N_B$ and $N_F$ count the relativistic bosonic and fermionic degrees of freedom and the traces over the mass matrices should be evaluated only for light fields since the heavy fields do not contribute. 

\subsection{ Canonical Quantization Approach}

Another way to derive the effective potential at finite temperature (\ref{VT-4}) is through the cannonical quantization of the fields. Considering a self interacting real scalar field with Largangian
\begin{equation}
{\cal L} =\frac{1}{2}(\partial \varphi)^2 -V(\varphi).
\end{equation}
The field $\varphi$ can be decomposed into the the homogeneous and inhomogeneous components\footnote{The correspondance $\varphi_c \rightarrow \varphi_c$ and $\delta\varphi \rightarrow \tilde{\varphi}$ with the path integral approach is apparent.} 
\begin{equation}
\varphi(t, \bold{x})=\varphi_c(t) + \delta\varphi(t,\bold{x}).
\end{equation}
so that the spatial average of $\delta\varphi(t, \bold(x,t))$ is equal to zero. Then the Lagrangian after Taylor expanding around $\varphi_c$ and averaging over spacee reads
\begin{equation}
{\cal L}=\frac{1}{2}(\partial_t \varphi_c)^2+\left\langle \frac{1}{2}(\partial_\mu \delta\varphi)^2\right\rangle -V(\varphi_c)-\frac{1}{2}V''(\varphi_c)\left\langle \delta\varphi^2\right\rangle +...
\end{equation}
where the $'$ is derivative with respect to the field $\phi$.  We note that the $V'(\varphi_c)\left\langle \delta\varphi_c\right\rangle$ is zero by definition as well as the  $\left\langle \partial_\mu\delta\varphi\right\rangle=0$ appearing at the cross terms of kinetic energy. This last can be understood e.g. by assuming that the fluctuation behaves like $\delta\varphi \sim A e^{k^\mu x_\mu}$. By applying the Euler-Lagrange variational procedure
\begin{equation} \label{EL-4}
\partial_\mu \frac{\delta {\cal L}} {\delta (\partial_\mu \chi_i)} -\frac{\delta {\cal L}}{\delta \chi_i}=0
\end{equation}
with $\chi_i= \varphi_c,\, \delta\varphi$ we find the equation of motion for the space average and the fluctuations. For the homogeneous mode $\varphi_c$ we take
\begin{equation}
\partial^2_t\varphi_c+V'(\varphi_c)+\frac{1}{2}V'''(\varphi_c)\left\langle \delta\varphi^2 \right\rangle = 0
\end{equation}
where the higher order terms $\sim \left\langle \delta\varphi^3 \right\rangle$ have been neglected. In quantum field theory this corresponds to the so called one-loop approximation and the one-loop effective potential
\begin{equation} \label{ef-4}
V_\text{eff}=V(\varphi_c)+\frac{1}{2}V''(\varphi_c)\left\langle \delta\varphi^2\right\rangle
\end{equation}
can be defined. 

In the lowest, linear order the inhomogeneous modes $\delta\varphi(t, \bold{x})$ obey the equation of motion (\ref{EL-4})
\begin{equation} \label{fl-4}
\partial_\mu\partial^\mu \delta\varphi + V''(\varphi_c) \delta \varphi =0.
\end{equation}
Assuming that the mass 
\begin{equation}
m^2_{\delta \varphi}(\phi_c) \equiv V''(\varphi_c)\geq 0 
\end{equation}
does not depend on time\footnote{In different case we may have exponential growing modes, see Appendix Preheating section.} 
the solution of (\ref{fl-4}) is 
\begin{equation}
\delta \varphi(t, \bold{x}) = \frac{1}{(2\pi)^{3/2}}\int \frac{1}{\sqrt{2\omega_k}}\left(a^-_\bold{k} e^{-i\omega_kt+i\bold{k}\bold{x}} +a^+_\bold{k} e^{-i\omega_kt+i\bold{k}\bold{x}} \right)
\end{equation}
where $a_\bold{k} \equiv a(\bold{k})$ and  $a^+_\bold{k}= (a^-_\bold{k})^\dagger$. Also, 
\begin{equation}
\omega_k =\sqrt{k^2+V''(\varphi_c)}=\sqrt{k^2+m^2_{\delta\varphi}}
\end{equation}
with $k\equiv=|\bold{k}|$. At the equation of motion we did not considered the background geometry, however neglecting the expansion of the space is a good approximation for our purposes since an interacting field means that the interaction rate outrun the expansion rate i.e. $\Gamma> H$. 

The determination of the effective potential (\ref{ef-4} ) means calculation  of the $\left\langle  \delta \varphi^2 \right\rangle$ considering both quantum and thermal contributions. Starting from the quantum contributions we recall that in a quantum theory the $\delta\varphi(t, \bold{x})$ is treated as a position operator $\hat{\delta\varphi}_x(t)$ that should satisfy the Heisenberg commutation relations with the conjugated momenta $\partial {\cal L}/\partial \delta\varphi_t$. This translates to commutation relations for the operators $\hat{a}^+_\bold{k}$, $\hat{a}^-_\bold{k}$ that behave like creation and annihilation operators of harmonic oscillators acting in the Hilbert space. The vacuum state is defined via
\begin{equation}
\hat{a}^-_\bold{k} \left|0 \right\rangle=0
\end{equation}
for all $k$ and corresponds to the minimal energy state. The vectors 
\begin{equation}
\left|n_\bold{k} \right\rangle= \frac{(\hat{a}^+_\bold{k})^n}{\sqrt{n!}}\left|0 \right\rangle 
\end{equation}
are interpreted as describing $n_\bold{k}$ particles per single quantum state characterized by the wave vector $k$. Hence, calculating the square of $\delta\varphi$ and averaging over space and taking into account that the only nonzero combination is the $\left\langle \hat{a}^+_\bold{k} \hat{a}_{\bold{k}'}\right\rangle = n_\bold{k} \delta( \bold{k}-\bold{k}')$ one finds
\begin{equation} \label{QT-4}
\left\langle \delta\varphi^2 (\bold{x})\right\rangle = \frac{1}{2\pi^2} \int \frac{k^2}{\sqrt{k^2+m^2_{\delta\varphi}(\varphi_c)}}\left(\frac{1}{2}+n_\bold{k} \right) dk.
\end{equation}
The zero temperature vacuum contribution correspponds to $n_\bold{k}=0$. The integral (\ref{QT-4}) is divergent as $k\rightarrow\infty$. Introducing a cutoff scale $M$ the correction to the correction to the potential of the homogeneous mode $\varphi_c$ reads
\begin{equation}
V^0_1 (\varphi_c) \equiv \frac{1}{2}V''(\varphi_c) \left\langle \delta\varphi^2 (\bold{x})\right\rangle  = \frac{1}{4\pi^2}\int^M_0 \sqrt{k^2+m^2_{\delta\varphi}(\varphi_c)}\, k^2 dk.
\end{equation} 
This is the energy density of the vacuum fluctuations that is characterized by quatric, quadratic and logarithmic to the cut-off scale $M$ divergences. For a renormalizable theory like the $V=\lambda\varphi^4$ the divergencies can be absorbed by the introduction of counterterms and the constant appearing at the lagrangian at first place (bare constants) are replaced by the physical constants (renormalized) and the effective potential (\ref{ef-4}) takes the form
\begin{equation}
V_\text{eff}= V(\varphi_c)+\frac{m^4_{\delta\varphi}(\varphi_c)}{64\pi^2}\, \ln \frac{m^2_{\delta \varphi}(\varphi_c)}{\mu^2}
\end{equation}
where the $\mu$ is an energy scale introduced to render the effective potential, and in particular the logarithm, free from the arbitrary cut-off scale $M$. 

In the early universe a thermal equilibrium was established of temperatures at least of order the MeV scale. Therefore, at such an enviroment the occupation numbers are expected to be nonzero. A thermalized plasma of scalar particles is characterized by occupation numbers given by the Bose-Einstein formula
\begin{equation}
n_\epsilon=\frac{1}{e^{(\epsilon-\mu)/T}-1}
\end{equation}
where $\mu$, here, is the chemical potential that can be neglected and $n_\epsilon$ the occupation numbers for a single kind of particle per microstate of energy $\epsilon=\omega_k$. Applying this to (\ref{QT-4}) the fluctuations average value obtains a dependence on the temperature according to the expression
\begin{equation}
\left\langle \delta\varphi^2 (\bold{x})\right\rangle_T = \frac{1}{2\pi^2} \int^\infty_0 dk \frac{k^2}{\omega_k(e^{\omega_k/T}-1)}=\frac{T^2}{4\pi^2} J^{(1)}_- \left(\frac{m_{\delta\varphi}(\varphi_c)}{T},0 \right)
\end{equation}
where the integration variable have been changed to $k\rightarrow \omega_k/T$. Note that the zero temperature energy from the vacuum fluctuations (the $n_\bold{k}=0$ contribution) was not inluded. The function $J^{(1)}_-$ is an integral given by
\begin{equation} \label{J-4}
J^{(1)}_-\left(\frac{m}{T},0 \right)  \equiv \int^\infty_{m/T} \frac{\sqrt{x^2-m^2/T^2}}{e^{x}-1}dx
\end{equation}
and the thermal correction reads
\begin{equation}
V^T_1(\varphi_c)\equiv \frac{1}{2}V''(\varphi_c) \left\langle \delta\varphi^2 (\bold{x})\right\rangle_T= \frac{T^4}{4\pi^2} \int^{\frac{m_{\delta\varphi}}{T}}_0 a J^{(1)}_-(a,0) da
\end{equation}
and after computing the integral via (\ref{J-4}) at the high temperature limit $T \gg m_{\delta \varphi}$ it is expressed by the expansion
\begin{equation} \label{Muk-4}
V^T_1(\varphi_c) \simeq \frac{m^2_{\delta \varphi}}{24}T^2- \frac{m^3_{\delta \varphi}}{12\pi}T- \frac{m^4_{\delta \varphi}}{64\pi^2}\, \ln \frac{m^2_{\delta \varphi}}{a_bT^2}+...
\end{equation}
where $\ln a_b \simeq 5.4$. 

We mention that this result is identical with the $\bar{V}^T_1$ of (\ref{VT1-4}) apart from the absence of ${\cal O}(T^4)$ terms. The reason is that the (\ref{VT1-4}) is the high temperature expansion of the integral (see (\ref{F-4}))
\begin{equation}
V^T_1(\varphi_c) = T \int \frac{d^3 k}{(2\pi^2)}  \,\ln(1-e^{\omega_k/T})
\end{equation} 
that includes the contribution from the kinetic energy of the thermal fluctuations as can be seen from the expressions (\ref{exp2-4}) and (\ref{quad-4}) whereas, the (\ref{Muk-4}) does not account for the $(\partial\delta\varphi)^2$ term. Hence, one has to add to the expression (\ref{Muk-4}) the ideal gas contribution to the free energy $(N_B+7/8 N_F)\pi^2 T^4/30$ in a theory that includes bosonic and fermionic degrees of freedom since the expectation values of the fields are where the free energy minimizes and the entropy maximizes.

\section{Gravitational Gauge Mediation at Finite Temperature}

In the Kitano model the messengers $\phi$ and $\bar{\phi}$ carry standard model quantum numbers and we expect them to be in thermal (and chemical) equilibrium in the early universe. The chiral superfield $X$ of the secluded sector is coupled with a coupling $\lambda$ to the messenger sector and, in principle,  it can also achieve thermal equilibrium.  The messengers interact directly with the electroweak gauge bosons, which are light, hence their averaged interaction rate with the Standard Model fields is of the order of $\alpha_{SM}^{2} T$, which is larger than the expansion rate for temperatures smaller than $10^{14} - 10^{15}$ GeV. However, mesengers have a mass of the order of $m_{\phi} = \lambda \langle X \rangle$, which is typically  large with respect to the Fermi scale. According to the standard lore, see \cite{Mukhanov:2005sc}, for the temperatures smaller than $T_\phi = m_\phi/ 20$ the messengers decouple from the expansion and become irrelevant as the source of thermal corrections to the potential for $X$. However, as long as messengers are in equilibrium, they contribute thermal corrections to the effective potential, as they couple directly to the spurion field via the term $\lambda^2 X^2 \phi^2$. 
Even if the spurion $X$ is not in thermal equilibrium with the heat bath, these corrections are there, as the excited messengers with high energies also interact with $X$. In fact, $X$ interacts with the SM particles as well, with the strength which is suppressed by their coupling to the messengers and by the messenger propagators. Finally, the spurions may communicate with the Standard Model via the additional gauge boson, such as the "anomalous" $U(1)$ gauge boson which appears in string-derived models. Although this additional gauge boson is typically only one or two orders of magnitude lighter than the string scale,
it may be sufficient to bring the hidden sector into equilibrium for a period of time. In any case, we shall assume for now that the equilibrium at least for the messengers holds down to the low temperatures $T_\phi \sim m_\phi/20$. Let's assume\footnote{In subsequent section we shall examine whether this assumption holds.}, the sake of completence that the whole system to be in thermal equilibrium at temperature $T$; the formuli for a decoupled spurion can then be readily derived.

The interactions induce  a thermally corrected potential with a shape  different from the zero temperature one.
The minimum will be different, and we call it thermal average  $(X_c, \phi_c,\bar{\phi}_c)$. At a temperature $T$ we consider excitations (shifts) about the thermal average values. We can find the finite temperature mass matrix of the fields which depends on the thermal average values $X_c,\phi_c,\bar{\phi}_c$ and of course on the temperature $T$. From the equations (\ref{VT1-4}), (\ref{VT2-4}) we can calculate the finite temperature potential for the Kitano model, with field variables the shifts $\tilde{X},\tilde{\phi},\tilde{\bar{\phi}}$, i.e. ${\cal L}(\tilde{X},\tilde{\phi},\tilde{\bar{\phi}})$. As the temperature decreases the thermal average values of the fields move and the evolution of the ${\cal L}(X_c,\phi_c,\bar{\phi}_c)$ can be found.

The first step is to write the zero temperature tree level potential (\ref{KitanoW}) substituting in the place of the zero temperature excitations $X, \phi, \bar{\phi}$ the finite temperature excitations about the thermal average, i.e. $X,\phi,\bar{\phi}\rightarrow X_c+\tilde{X},\phi_c+\tilde{\phi},\bar{\phi}_c+\tilde{\bar{\phi}}$. To simplify the notation we set $X \equiv\tilde{X},\phi\equiv\tilde{\phi}, \bar{\phi}\equiv\tilde{\bar{\phi}}$. In the case of a supersymmetric theory with non-canonical K\"ahler potential (and including gravity) the traces of the mass matrices are given by the following formulae \cite{Binetruy:1984yx}
\begin{equation} \label{sca-mass}
\text{tr}M^2_S=2  K^{i\bar{j}}\frac{\partial^2 V_0}{\partial\phi^i\partial\ \bar{\phi}^{\bar{j}}}
\end{equation}
and
\begin{equation} \label{ferm-mass}
\text{tr}M^2_F=  e^G\left[K^{i\bar{j}}K^{k\bar{l}}(\nabla_iG_k+G_iG_k)(\nabla_{\bar{j}}G_{\bar{l}}+G_{\bar{j}}G_{\bar{l}})-2\right] .
\end{equation}
The term $-2$ in (\ref{ferm-mass}) takes into account the mixed goldstino-gravitino contribution. We adopt the notation $\nabla_i G_j=G_{ij}-\Gamma^k_{ij}G_k$ with the connection 
\begin{equation}
\Gamma^{k}_{ij}=K^{k\bar{l}}\partial_i K_{j\bar{l}}.
\end{equation}
Differentiating the $\bar{V}_0$ with respect to the $X,\phi,\bar{\phi}$ we find the matrix of second derivatives about the thermal average minimum:
\\
\\
\begin{math}
\bar{V}^{\prime\prime}_0= \left(\begin{array}{clllrrr}      
V_{X^\dagger X} & V_{X^\dagger X^\dagger} &V_{X^\dagger \phi} & V_{X^\dagger \phi^{\dagger}} & V_{X^\dagger \bar{\phi}}& V_{X^\dagger \bar{\phi}^\dagger}  \\     
V_{XX} & V_{XX^\dagger} & V_{X\phi} & V_{X\phi^\dagger } & V_{X \bar{\phi}} & V_{X\bar{\phi^{\dagger}}} \\
V_{\phi^\dagger X} & V_{\phi^\dagger X^\dagger} &V_{\phi^\dagger \phi} & V_{\phi^\dagger \phi^\dagger }&V_{\phi^\dagger \bar{\phi}} & V_{{\phi}^\dagger \bar{\phi}^\dagger} \\
V_{\phi X} & V_{\phi X^\dagger} & V_{\phi\phi} &V_{\phi\phi^\dagger }&V_{\phi\bar{\phi}}& V_{\phi\bar{\phi}^\dagger} \\
V_{\bar{\phi}^\dagger X} & V_{\bar{\phi}^\dagger X^\dagger} &V_{\bar{\phi}^\dagger \phi} & V_{\bar{\phi}^\dagger \phi^\dagger }&  V_{\bar{\phi}^\dagger \bar{\phi}}& V_{\bar{\phi}^\dagger \bar{\phi}^\dagger} \\
V_{\bar{\phi} X} & V_{\bar{\phi} X^\dagger} &V_{\bar{\phi} \phi} &V_{\bar{\phi} \phi^\dagger }& V_{\bar{\phi} \bar{\phi}}& V_{\bar{\phi} \bar{\phi}^\dagger} 
\end{array}\right) \text{at} 
\end{math} 
\begin{math}
\begin{array}{c}      
X=X^\dagger=0 \\
\phi=\phi^\dagger=0 \\
\bar{\phi}=\bar{\phi}^\dagger=0 
\end{array}
\end{math}
\\
\\
\\
In order to find the scalar mass matrix  $M_{X \phi \bar{\phi}}$ we have also to compute the inverse K\"ahler metric $K^{i\bar{j}}= (K^{-1})_{i\bar{j}}$:
\begin{equation}
K_{i\bar{j}}\equiv\frac{\partial^2 K}{\partial \varphi^i \partial \bar{\varphi}^{\bar{j}}}= 
\left(\begin{array}{clll}      
K_{XX^\dagger} & K_{X\phi^\dagger} & K_{X\bar{\phi}^\dagger} \\     
K_{\phi X^\dagger} & K_{\phi \phi^\dagger} & K_{\phi\bar{\phi}^\dagger}   \\     
K_{\bar{\phi}X^\dagger} & K_{\bar{\phi}X^\dagger} & K_{\bar{\phi}X^\dagger} 
\end{array}\right)
\simeq
\left(\begin{array}{clll}      
1-\frac{(X^\dagger X)^2}{\Lambda^2} & 0 & 0 \\     
0 & 1 & 0   \\     
0 & 0 & 1 
\end{array}\right)
\end{equation}
\begin{equation}
K^{i\bar{j}}\equiv(K^{-1})_{i\bar{j}}= 
\left(\begin{array}{clll}      
1+4\frac{(X^\dagger X)^2}{\Lambda^2} & 0 & 0 \\     
0 & 1 & 0   \\     
0 & 0 & 1 
\end{array}\right)
\end{equation}
where $\varphi=X,\phi, \bar{\phi}$. The approximation $(1-|X|^2/\Lambda^2)^{-1} \simeq 1+|X|^2/\Lambda^2$ is justified for $X \ll \Lambda$ i.e. where the low energy effective theory is valid.  From (\ref{VT2-4}) the eigenvalues of the scalar mass matrix at the thermal average read 
\begin{equation} \label{trace}
\text{tr}M^2_{X \phi \bar{\phi}}=2K^{i\bar{j}}\frac{\partial^2 \bar{V}_0}{\partial\varphi^i\partial \bar{\varphi}^{\bar{j}}}=K^{XX^\dagger}V_{XX^\dagger}+K^{\phi \phi^\dagger}V_{\phi \phi^\dagger}+K^{\bar{\phi}\bar{\phi}^\dagger}V_{\bar{\phi}\bar{\phi}^\dagger} 
\end{equation}
\begin{equation}
=2\left[\left(1+4\frac{X^\dagger X}{\Lambda^2}\right)V_{XX^\dagger} +V_{\phi \phi^\dagger}+V_{\bar{\phi} \bar{\phi}^\dagger}\right]
\end{equation}
Considering for simplicity  real values for the fields the trace (\ref{trace}) reads
\begin{equation}
\text{tr}M^2_{X\phi\bar{\phi}} \simeq 2\left[4\frac{F^2}{\Lambda^2}+X_c \left( -8F c\frac{1}{\Lambda^2}\right)+X^2_c(2\lambda^2)+2\lambda^2(\phi^2_c+\bar{\phi}^2_c)+\phi_c\bar{\phi}_c\left(-8 Fc\frac{1}{\Lambda^2}\right)\right]
\end{equation}
The scalars $X,\phi, \bar{\phi}$ are the scalar part of chiral superfields. Therefore, each of them has a fermionic superpartner, a Weyl spinor which has to be taken into account. 
According to  (\ref{ferm-mass}) we have
\begin{equation} \nonumber
\text{tr} M^2_F=\left\langle e^G\left[K^{A\bar{B}}K^{C\bar{D}}(\nabla_AG_C+G_AG_C)(\nabla_{\bar{B}}G_{\bar{D}}+G_{\bar{B}}G_{\bar{D}})-2\right] \right\rangle
\end{equation}
\begin{equation} \nonumber
=\left\langle  e^G\left[2K^{XX^\dagger}K^{\phi\phi^\dagger}\left|\nabla_X G_\phi+G_X G_\phi \right|^2+2K^{XX^\dagger}K^{\bar{\phi}\bar{\phi}^\dagger}\left|\nabla_XG_{\bar{\phi}}+G_XG_{\bar{\phi}}\right|^2 \right.\right.
\end{equation}
\begin{equation} \nonumber
+2K^{\phi \phi^\dagger}K^{\bar{\phi}\bar{\phi}^\dagger}\left|\nabla_\phi G_{\bar{\phi}}+G_\phi G_{\bar{\phi}}\right|^2+K^{XX^\dagger}K^{XX^\dagger}\left|\nabla_XG_X+G_XG_X\right|^2
\end{equation}
\begin{equation}
\left. \left. 
K^{\phi {\phi}^\dagger} K^{\phi {\phi}^\dagger} \left| \nabla_{\phi} G_\phi + G_\phi G_\phi \right| ^2 + 
K^{\bar{\phi}{\bar{\phi}}^\dagger}K^{\bar{\phi}{\bar{\phi}}^\dagger}\left|\nabla_{\bar{\phi}}G_{\bar{\phi}}+G_{\bar{\phi}}G_{\bar{\phi}}\right|^2-2 
\right] 
\right\rangle
\end{equation}
The most important terms are found to be the first three:
\begin{equation} \nonumber
\text{tr}M^2_F \simeq \left\langle e^K\left[2\lambda^2\bar{\phi}^\dagger \bar{\phi}+2\lambda^2 \phi^\dagger \phi +2\lambda^2 X^\dagger X+...  \right] \right\rangle\Rightarrow
\end{equation}
\begin{equation} 
\text{tr}M^2_F \simeq 2\lambda^2 \bar{\phi}_c^\dagger \bar{\phi}_c+2\lambda^2 \phi^\dagger_c \phi_c +2\lambda^2 X^\dagger_c X_c
\end{equation}
result which also coincides with that of the global supersymmetric limit. According to (\ref{Vtotloop-4}) we can now write the tree level Kitano potential plus the quatric and quadratic to temperature thermal corrections:
\begin{equation} 
\bar{V}=V_0+\bar{V}^T_1=V_0-\frac{\pi^2T^4}{90}N+\frac{T^2}{24}\left[\text{tr}M^2_S+ \text{tr}M^2_F \right] +{\cal O}(T).
\end{equation}
In terms of the fields it reads 
\begin{equation} \nonumber
\bar{V}= V_0- \frac{\pi^2 T^4}{90} N + \frac{T^2} {12} \left[4\frac{F^2}{\Lambda^2}+(X_c+X^\dagger_c)(4 F c\frac{1}{\Lambda^2})+|X_c|^2 (2\lambda^2)+2\lambda^2(|\phi|^2+|\bar{\phi}|^2)+ \right.
\end{equation}
\begin{equation} \nonumber
 \left. + (\phi_c \bar{\phi}_c+\phi^\dagger_c \bar{\phi}_c^\dagger) (-4 F c\frac{1}{\Lambda^2}) +\lambda^2 |\bar{\phi}_c|^2 +\lambda^2 |\phi_c|^2 +\lambda^2 |X_c|^2 \right]+{\cal O}(T) 
\end{equation}
For precise description of the evolution of the fields we need the ${\cal O}(T)$ contribution, although in the high temperature limit the ${\cal O}(T^2)$ dominates. The linear in $T$  part of the finite temperature corrected potential is
\begin{equation} \nonumber
V \supset 
-\frac{T}{12\pi} \text{tr}\{M^2_S\}^{3/2}.
\end{equation}
For sake of simplicity we take here $\phi=\bar{\phi}$, actually the symmetry between $\phi$ and $\bar{\phi}$ in the potential allows this simplification. The mass matrix has two positive eigenvalues $\lambda_{1,2}=(M^2_S)_{1,2}$ (the negative eigenvalues give rise to the imaginary part of the potential which we neglect). 
Dropping out the index '$c$' from the notation of the thermal average values $X_c, \phi_c, \bar{\phi}_c$ the full thermally corrected potential reads:
\begin{equation} \nonumber
V \simeq \frac{c^2}{M^4_P}{\Lambda}^2-2\frac{c}{M^2_P}F (X+X^\dagger)+4 F^2 \frac{|X|^2}{\Lambda^2}- 2\lambda F |\phi|^2-4 F \frac{c}{M^4_P} (X+X^\dagger) |\phi|^2 +2\lambda^2 |X|^2 |\phi|^2+ {\cal O}(X^3)+\lambda^2 |\phi|^4
\end{equation} 
\begin{equation} \nonumber
-\frac{\pi^2 T^4}{90}N+ \frac{T^2}{12}\left[4\frac{F^2}{{\Lambda}^2}+(X+X^\dagger)(4F^2\frac{c}{M^2_P}\frac{1}{{\Lambda}^2})+|X|^2(3\lambda^2)+|\phi|^2(6\lambda^2)\right]
\end{equation}
\begin{equation} \label{fullT-Kitano}
-\frac{T}{12\pi}\left[\left(4\frac{F^2}{\Lambda^2}-2\frac{Fc}{\Lambda^2 M^2_P}(X+X^\dagger)+2\lambda^2|\phi|^2\right)^{3/2}+\left( 2\lambda F+2\lambda^2|X|^2+2\lambda^2|\phi|^2 \right)^{3/2}\right].
\end{equation}

\subsection{The Minima and the Evolution of the Finite Temperature Scalar Potential} \normalsize

In what follows we have in mind thermal expectation values of real fields unless stated otherwise. 
To start with we shall examine the shape of the finite temperature potential in various directions in the field space.
\\
\\ 
\bfseries \itshape The $\phi$-direction.  \normalfont
\\
Firstly, in the $\phi$-direction i.e. taking $X=0$, the (\ref{fullT-Kitano}) reads
\\
\begin{equation} 
V^{\phi} = \text{const}(T)-2\lambda F \phi^2 +\lambda^2 \phi^4 +\frac{T^2}{2}\lambda^2\phi^2-\frac{T}{12\pi} \left[\left(4\frac{F^2}{\Lambda^2}+2\lambda^2 \phi^2\right)^{3/2}+\left( 2\lambda F+2\lambda^2 \phi^2 \right)^{3/2}\right].
\end{equation}
We can write it in a simpler form
\begin{equation} \label{q-dir}
V^{\phi} \simeq  \text{const}(T)+\frac{1}{2} m^2_{\phi} (T)\phi^2-{\cal O}(10^{-1})\lambda^3T  \phi^3+\lambda^2 \phi^4.
\end{equation}
where we defined the effective $\phi$-mass as $m^2_{\phi}(T)\equiv\lambda^2T^2-4\lambda F $. The expectation value of the $\phi$ scalar field is obtained by minimizing the potential. For sufficiently high temperatures there is only one solution of $\partial V/\partial \phi=0$, namely 
\begin{equation}
\phi=0,
\end{equation}
and this is a minimum so long as $m^2_{\phi}(T)$ is positive. The effective mass changes sign from positive to negative (becomes tachyonic) at temperature $T_0$ 
\begin{equation} \label{T-0}
T_0=2\sqrt{\frac{F}{\lambda}}
\end{equation}
and we may write it as 
\begin{equation}
m^2_{\phi}(T)=-4\lambda F (1-T^2/T^2_0)
\end{equation}
We can see that the $\partial V/ \partial \phi=0$ of (\ref{q-dir}) has two more solutions: a maximum and a second (local) minimum when $m^2_{\phi}(T)\leq (9/16){\cal O} (10^{-2}) \lambda^4 T^2$ and this occurs when temperature drops below $T_1$ where
\begin{equation} \label{T-1}
T^2_1=\frac{T^2_0}{1-(9/16){\cal O}(10^{-2})\lambda^2}>T^2_0.
\end{equation}
The second minimum and the maximum are at
\begin{equation} \label{q-sol}
\phi_{\pm}(T)=\sqrt{\frac{F}{\lambda}} \left[\frac{{\cal O}(10^{-1})\lambda^{3/2}}{\mu}T\pm\left(1-\frac{T^2}{T^2_1}\right)^{1/2}\right].
\end{equation}
At $T=T_1$ the $\phi_{\pm}(T_1)$ is an inflection point. Below $T_1$ we have the formation of the second local minimum. At $T=T_{cr}$ this minimum becomes degenerate with the global minimum $\phi=0$. This happens when $m^{2}_\phi (T) \leq (1/2){\cal O}(10^{-2}) \lambda^4 T^2$. This relation gives a critical temperature slightly lower than $T_1$:
\begin{equation} \label{T-cri}
T^2_{cr}=\frac{T^2_0}{1-(1/2){\cal O}(10^{-2})\lambda^2}<T^2_1.
\end{equation}
We see that $T_0<T_{cr}<T_1$. Below the critical temperature the global minimum changes discontinuously from $\phi=0$ to $\phi_{+}(T_{cr})$. The origin $\phi=0$ becomes metastable until $T_0$ when the barrier that keeps the origin locally stable disappears and then $\phi=0$ becomes a local maximum. However, comparing the ($\ref{T-0}$), ($\ref{T-1}$) and ($\ref{T-cri}$) we see that they are only slightly different. Therefore, we can safely say 
\begin{equation}
T_1 \cong T_{cr} \cong T_0.
\end{equation}
In other words, the origin becomes tachyonic simultaneously with the appearence of a second asymmetric minimum, to a good approximation. This is equivalent to neglecting the term linear in temperature in equation ($\ref{q-dir}$). This phase transition is practically of second order \cite{Quiros:1999jp} and takes place at the critical temperature
\begin{equation} \label{T-cr}
T_{cr} \cong 2 \sqrt{\frac{F}{\lambda}}.
\end{equation}
The solution ($\ref{q-sol}$) gives the late-time minima in the $\phi$-direction: $\phi_{\pm}(T\rightarrow 0)=\pm\, F/\sqrt{\lambda}$, which are the supersymmetric minima of the tree level potential, while $\phi = 0 $ ends up as a local maximum.
\\
\\ 
\bfseries \itshape The $X$-direction.  \normalfont
\\
Secondly, in the $X$-direction, i.e. for $\phi =\bar{\phi}=0$ the potential (\ref{fullT-Kitano}) reads
\\
\begin{equation} \nonumber
V^X\simeq \text{const}(T)-4\frac{c}{M^2_P}F X+4 F^2 \frac{X^2}{{\Lambda}^2} +{\cal O}(X^3)
-\frac{\pi^2T^4}{90}N+ 
\end{equation}
\begin{equation}
+\frac{T^2}{12}\left[X(8 F\frac{c}{M^2_P}\frac{1}{{\Lambda}^2})+X^2(3\lambda^2)
\right]-\frac{T}{12\pi}\sqrt{8}\lambda^3 X^3.
\end{equation}
\\
The minimum along the $X$-direction is given by 
\begin{equation} \label{S-min0}
X_{min}(T)=\frac{4\frac{c}{M^2_P} F -\frac{2Fc}{3\Lambda^2M^2_P}T^2}{8\frac{F^2}{\Lambda^2}+\frac{1}{2}\lambda^2T^2}.
\end{equation}
\\
We note the position of the minimum at different temperatures: \\
$\alpha$) for $T>{\Lambda} \  \Rightarrow \  X_{min} \simeq -\frac{4}{3}\frac{F c}{\lambda^2{\Lambda}^2M^2_P}$ , \\
$\beta$) for $T \sim T_{cr}=2 \sqrt{\frac{F}{\lambda}} \  \Rightarrow \  X_{min}\sim \frac{c}{\lambda M^2_P}$ ,\\
$\gamma$) for $T \ll T_{cr}=2\sqrt{\frac{F}{\lambda}} \  \Rightarrow \  X_{min}\simeq \frac{4c F/M^2_P}{8\frac{F^2}{{\Lambda}^2}+\frac{1}{2}\lambda^2 T^2}\rightarrow  \frac{\Lambda^2}{M_P} \ \text{as} \ T\rightarrow 0 $, \\
$\delta$) for $T=0  \Rightarrow \  X_{min}= \frac{\sqrt{3}}{6}\frac{\Lambda^2}{M_P}$. \\
\begin{figure} 
\textbf{\,\,\,\,\,\,\,\,\,\,\,\,\,\,\, T=0 \,\,\,\,\,\,\,\,\,\,\,\,\,\,\,\,\,\,\,\;\;\;\;\;\;\;\;\;\;\;\;\;\;\;\;\;\;\;\;\;\,\,\,\,\,\,\,\,\,\,\,\,\,\;\;\;\;\;\;\;\;\;\;\;  T$>$T$_{cr}$}
\centering
\begin{tabular}{ccc}

{(a)} \includegraphics [scale=0.5, angle=0]{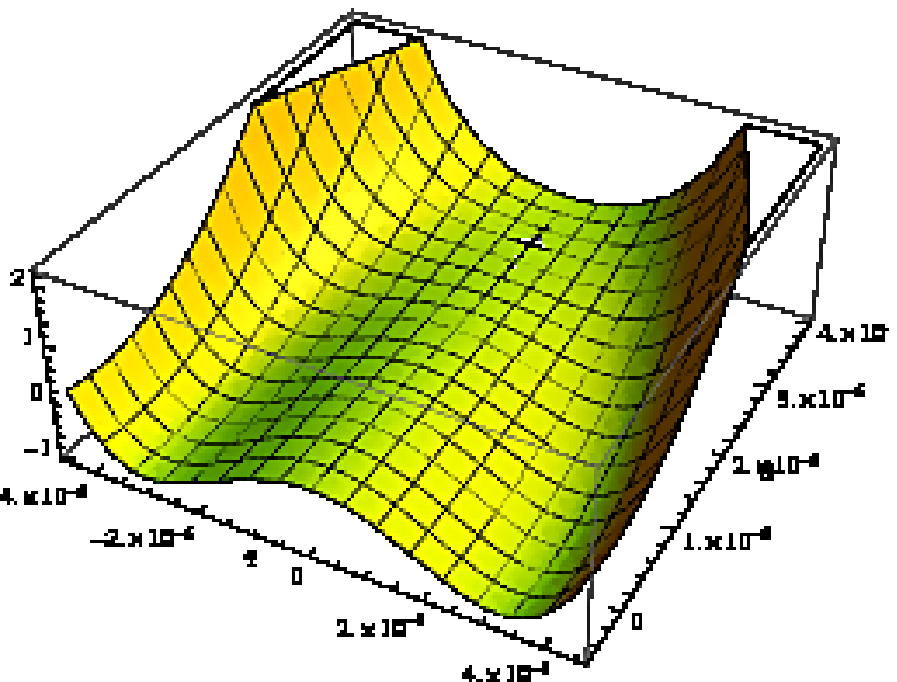} &
{(b)} \includegraphics [scale=0.5, angle=0]{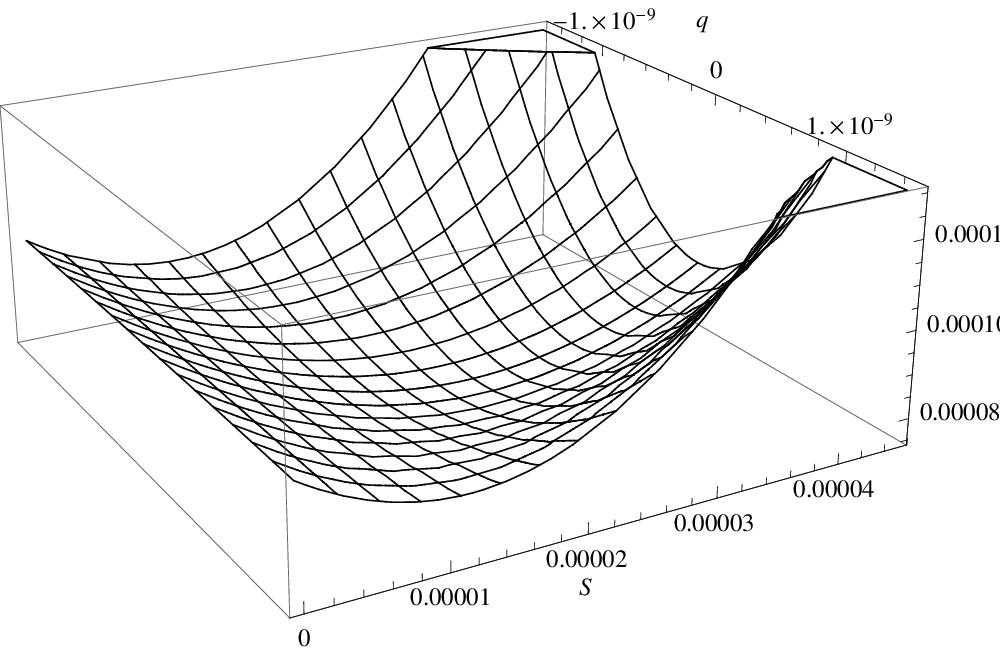}  &
{(c)} \includegraphics [scale=0.5, angle=0] {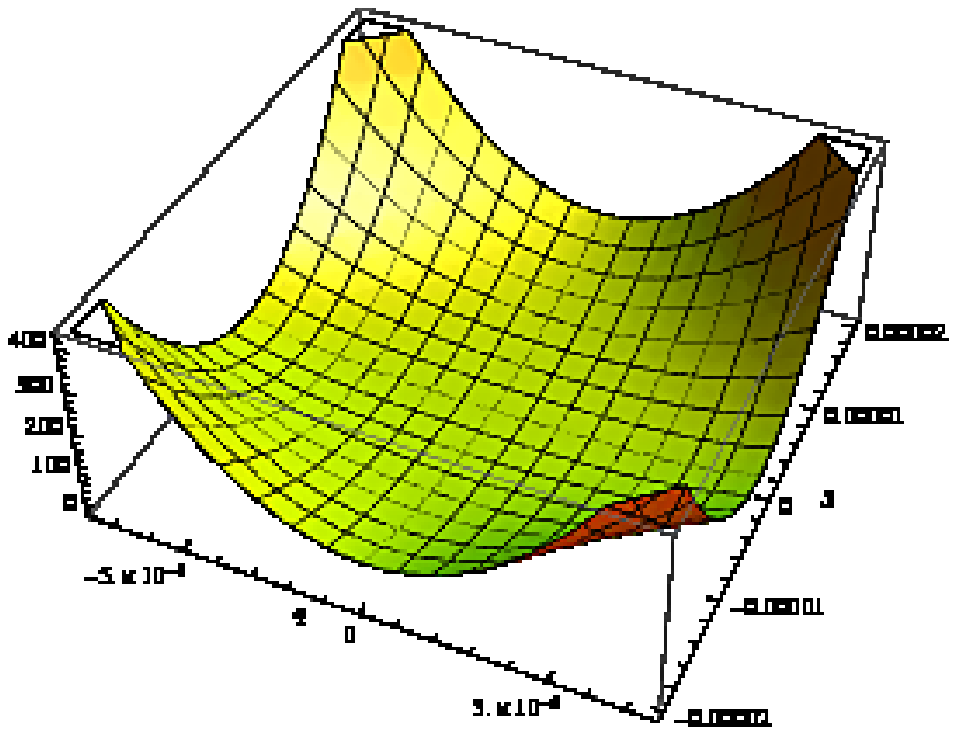}  \\
\end{tabular}
\caption{The zero temperature and finite temperature vacuum structure of the theory (\ref{KitanoK})-(\ref{KitanoW}) for $\Lambda=10^{-2}, \sqrt{F}=10^{-9}$ and $\lambda=10^{-7}$. Real values for the fields are assumed. The plot in the left panel (a) depicts the supersymmetry preserving global minima at $\phi^2=\sqrt{\frac{F}{\lambda}}=(10^{-5.5})^2$. The shallower supersymmetry breaking metastable minimum lies further away along the $X$-direction at $X =\sqrt{3}/6\Lambda^2 \simeq 0.3 \times 10^{-4}$ and can be seen in the middle panel (b). The right panel, (c), depicts the shape of the finite temperature potential for $T>T_{cr}$ without taking into account the shift due to the ${\cal O}(T^4)$ term. The $X$, $\phi$ axes are scaled in Planck units while the $V(X,\phi)$ is given in units of $F^2$.} 
\end{figure}
\\
\bfseries \itshape The evolution of the $X-\phi$ system.  \normalfont
\\
Until now we have examined independently the evolution of the $X$ and $\phi$ directions. However, what we deal with  is a coupled $X$-$\phi$ system that could evolve in a different way. To see what actually happens we write the scalar potential truncated to the most relevant terms:
\begin{equation} \nonumber
V \simeq \text{const}-4\frac{c}{M^2_P}\mu^2 X\left(1-\frac{T^2}{6\Lambda^2}\right)+X^2\left(\frac{4 F^2}{\Lambda^2}+\frac{\lambda^2 T^2}{4}\right)
+\phi^2\left(-2\lambda F + 2\lambda^2 X^2 + \frac{T^2}{2}\lambda^2 \right)+\lambda^2 \phi^4.
\end{equation}
The first remark is that the complete effective mass squared of the $\phi$ field is here $\partial^2 V(X,\phi,T)/\partial \phi^2\equiv m^2_{\phi,eff} \equiv \lambda^2 T^2 - 4 \lambda F +4 \lambda^2 X^2$. For high enough temperatures  the condition $\partial V/\partial \phi = 0$ is satisfied for $\phi=0$. On the other hand, due to the linear term, the high temperature minimum of the $X$ field is $X_{min}=-4 F c/(3 \lambda^2 \Lambda^2 M^2_P)$. Thus, the effective mass of $\phi$ changes at high temperatures  and it has a non-zero contribution from the $X$ field. However, we see from  the zero temperature vacuum meta-stability condition ($\ref{vac-stab}$) that the $\lambda^2 X^2_{min}$ term is negligible compared to $-\lambda F$ and the critical temperature ($\ref{T-cr}$) is practicaly unaffected. At this temperature the $\phi$ field becomes tachyonic and the global minimum of the potential moves from $\phi=0$ to $\phi= \sqrt{\frac{F}{\lambda}}$ which is the supersymmetry preserving minimum.

A second important remark is that the coupled $X$-$\phi$ system makes the high temperature minimum $S_{min}$ unstable (saddle point) for temperatures lower than the critical one. However, we saw that at $T=0$ there is a metastable minimum at the $X$-direction (\ref{S-min}). The temperature at which the unstable $X$-direction ($\phi$=0) minimum becomes a metastable one is given by the condition:
\begin{equation}
\frac{\partial^2 V(X,\phi)}{\partial \phi^2} \geq 0 \ \  \text{at} \ \ \phi=0, \  X=X_{min}(T) \  \text{and} \ T<T_{cr}.
\end{equation}      
First of all, we can see from ($\ref{fullT-Kitano}$) that 
\begin{equation}
\frac{\partial V(X,\phi)}{\partial \phi} = 0 \ \  \text{at} \ \ \phi=0.
\end{equation}      
Therefore the $V(X=X_{min},\phi=0)$ is an extremum of the potential. When $\partial^2 V(X=X_{min},\phi=0)/ \partial \phi^2<0$ it is a saddle point. When $\partial^2 V(X = X_{min},\phi=0)/ \partial \phi^2>0$, $X_{min}$ becomes a stable local minimum. For $T<T_{cr}$ and $\lambda> F M^2_P/\Lambda^4$ we obtain from (\ref{fullT-Kitano})
\begin{equation} \label{V''}
\frac{\partial^2 V(X =X_{min}(T),\phi=0)}{\partial \phi^2}\simeq -4\lambda F+ 4\lambda^2 X^2_{min}(T)
-\frac{T}{12\pi}\left[6\lambda^2\left(2\lambda F +2\lambda^2X^2_{min}(T)\right)^{1/2}\right]. 
\end{equation}      
It is easy to check that the linear  temperature correction can be safely neglected. Therefore
\begin{equation} 
\frac{\partial^2 V(X_{min}(T), \phi = 0)}{\partial \phi^2}\geq 0 \Rightarrow \lambda F \leq \lambda^2 X^2_{min}(T)
\end{equation}      
and 
\begin{equation} \label{SminS}
X^2_{min}(T_X)\simeq\frac{F}{\lambda},
\end{equation}      
where $T_X$ the temperature at which the $X$ minimum becomes locally stable.
Equating the last relation and the equation (\ref{S-min0}) we can find the temperatue $T_X$:
\begin{equation} \label{S-minT}
X_{min}(T)=\frac{4\frac{c}{M^2_P} F -\frac{2F c}{3\Lambda^2 M^2_P}T^2}{8\frac{F^2}{\Lambda^2}+\frac{1}{2}\lambda^2 T^2}\simeq \sqrt{\frac{F}{\lambda}} 
\end{equation}
which gives
\begin{equation} \label{TS-1}
T^2_X \sim \frac{c \sqrt{F}}{M^2_P \sqrt{\lambda^3}}.
\end{equation}
This is a temperature typically a few orders of magnitude higher than  the tree level mass of $X$, $m_X=F/ \Lambda$, hence the high temperature expansion is valid. Below this temperature  the metastable susy breaking vacuum forms and the temperature dependent terms start becoming negligible.
Let's note that $T_X$  vanishes in the limit $M_{P}\rightarrow \infty$ which makes sense since the susy breaking vacuum disappears in this limit.
\begin{figure} 
\textbf{\,\,\, Varying the coupling $\lambda$ \,\,\,\,\,\,\,\,\,\,\,\,\,\,\,\,\,\,\,\,\,\;\;\;\;\;\;\;\;\;\; Varying the scale $\Lambda$}
\centering
\begin{tabular}{cc}
\\
{(a)} \includegraphics [scale=.85, angle=0]{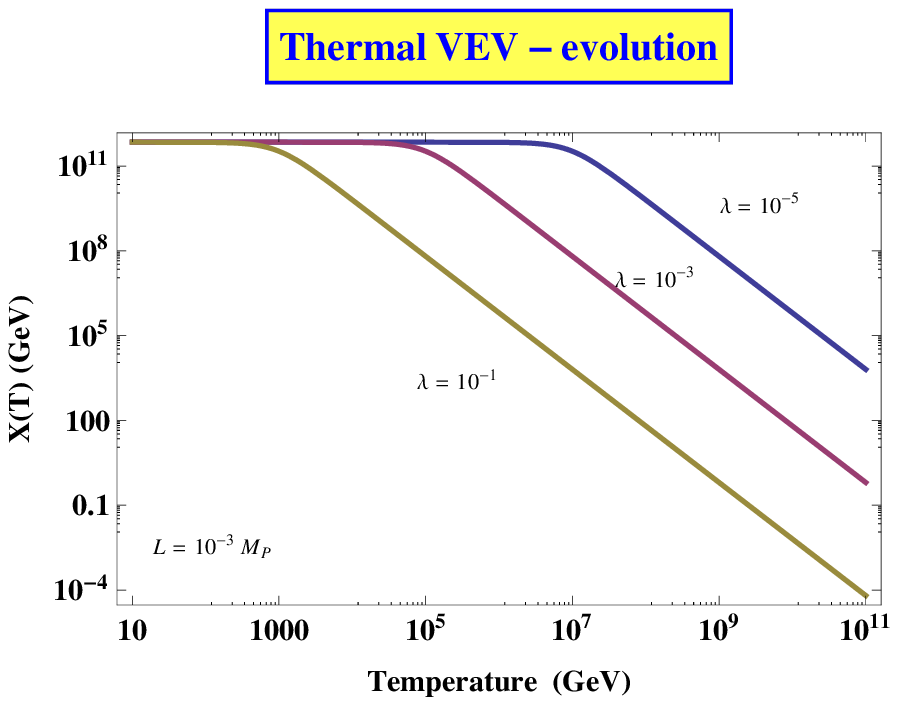} &
{(b)} \includegraphics [scale=.85, angle=0]{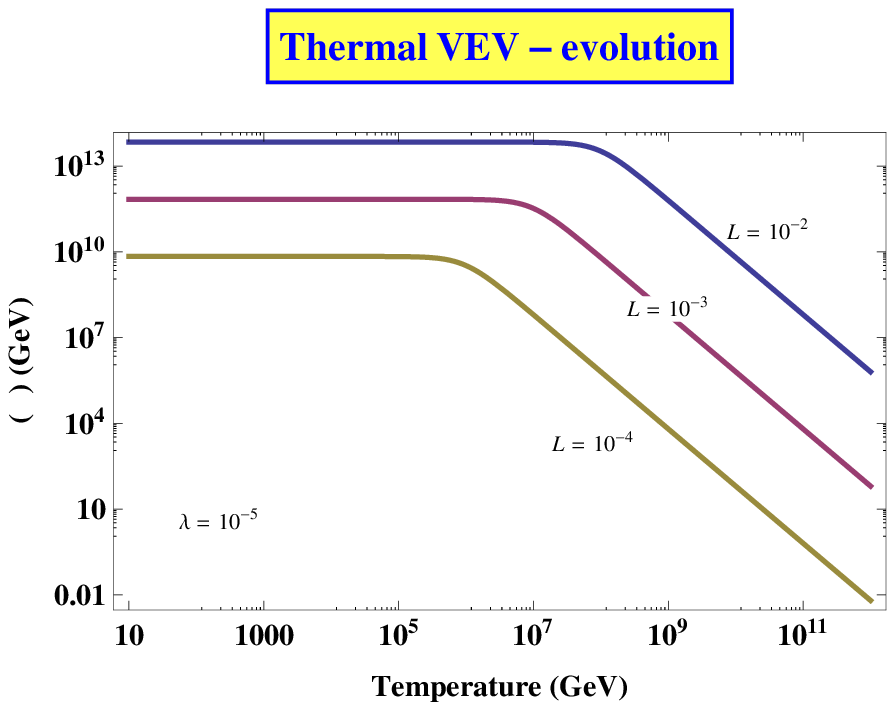}  \\
\end{tabular}
\caption{\small{The figures show the thermal evolution of the VEV, $X(T)$, of the spurion for the case of gravitational stabilization. The essential feature is that as the temperature increases the spurion has a thermal average value close to the origin. At  the left hand side panel the three curves correspond (from bottom to the top) to couplings $\lambda=10^{-1}, \, 10^{-3}, 10^{-5}$ respectively and cut-off scale $\Lambda=10^{-3}M_P$. We see that the smaller the coupling the faster the thermal VEV moves to the zero temperature one $X(T=0)=\sqrt{3}\Lambda^2/(6M_P)$. At the right hand side panel the coupling is constant and equal to $\lambda=10^{-5}$ and the parameter we vary is the cut-off $\Lambda$: from bottom to the top $\Lambda=10^{-4},\, 10^{-3}, \, 10^{-2}$. Therefore, the curves tend to different VEVs as the temperature decreases. At the figure, $L$ stands for the $\Lambda$. }}
\end{figure}
\begin{figure} 
\centering
\begin{tabular}{cc}
\\
{(a)} \includegraphics [scale=.85, angle=0]{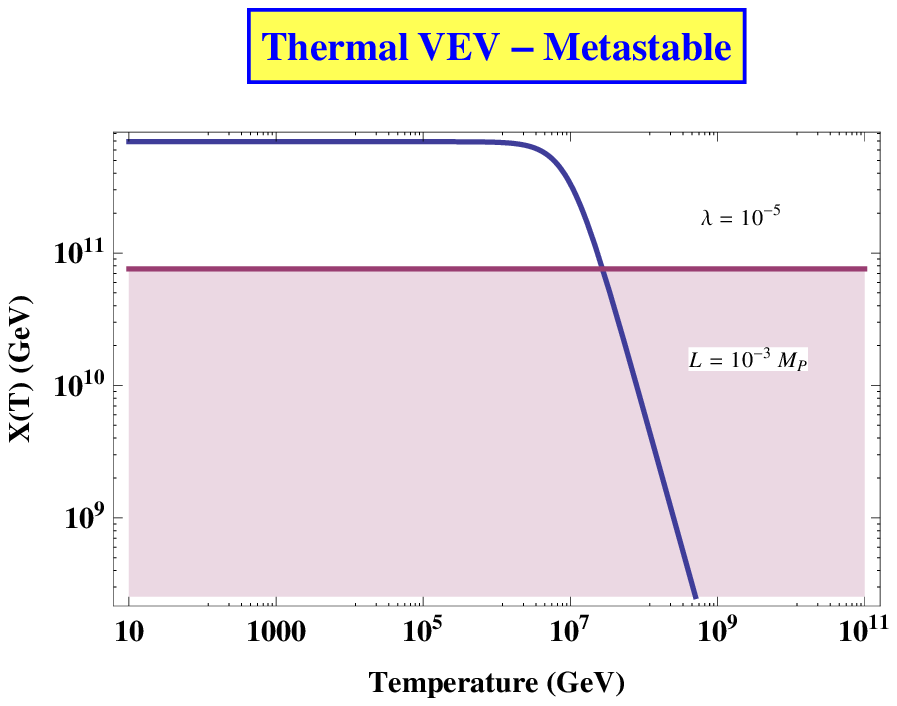} &
{(b)} \includegraphics [scale=.85, angle=0]{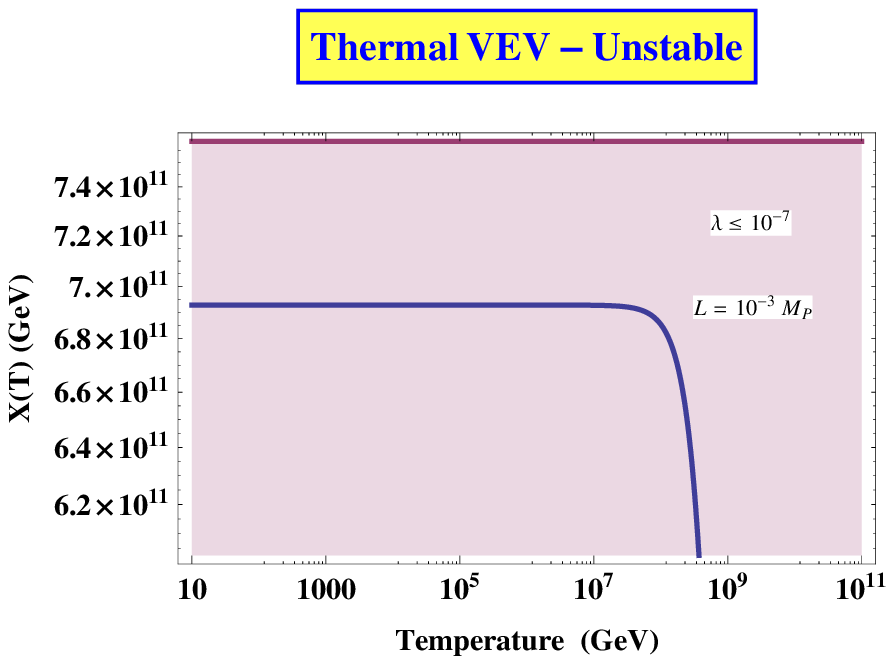}  \\
\end{tabular}
\caption{\small{The figures demonstrate the thermal evolution of the VEV, $X(T)$, of the spurion for the case of gravitational stabilization. The left hand side panel corresponds to the case of metastable zero temperature vacuum and the right hand side to an unstable one, i.e. the vacuum does not exist. However, at high temperatures the $X$-direction is stabilized. The painted region corresponds to zero temperature tachyonic region. The temperature the spurion VEV, $X(T)$, exits the painted region is the definition of the $T_X$. The parameters for the metastable case (left panel) are $\Lambda=10^{-3}$ and $\lambda=10^{-5}$ and for the unstable (right panel) $\Lambda=10^{-3}$ and $\lambda \lesssim 10^{-7}$.}}
\end{figure}
\\
\\ 
To summarize, the study of the evolution of the thermal averages of the fields (minima of the potential) from a phase of high temperature thermal equilibrium towards the zero temperature potential seems to disfavour the simple model of gravitational gauge mediation (see also \cite{Craig:2006kx}):  the susy breaking metastable vacuum is not reached if i) the univese has experienced at  high temperature  a hot thermal phase in which the hidden/messenger sector  fields ($X,
\phi,\bar{\phi}$) were part of the interacting plasma  and ii) that phase sets the thermal initial conditions for the evolution of these fields, which means that they aren't displaced from the symmetric thermal minimum of the potential.

\section{Temperature Corrections in Models  with Generalized K\"ahler Potential}

If we retain only the term ${\cal O}(T^2)$ and with the tree level potential $V_0$ given by ($\ref{Vg}$), the effective potential at finite temperature takes the form
\begin{equation}
V= V_0 
+\frac{T^2}{12}\left(4g_4 F^2 +4 \frac{c}{M^2_P} F g_4(X+X^\dagger)+3\lambda^2|X|^2+36 F^2 g_6|X|^2 + 6 \lambda^2 |\phi|^2 \right) + {\cal O}(T^4).
\end{equation}
We see that the effective potential for the $\phi$ field coincides with the one studied previously. There is a critical temperature $T_{cr}=2\sqrt{\frac{F}{\lambda}}$ at which a nearly second order phase transition takes place from the high temperature minimum $\phi=0$ to the tree level one $\phi^2= F /\lambda$. On the other hand, the evolution for the $S$ field changes and we study it in the following.
For $g_4<0$ we substitute $g_4=- 1/\Lambda^2_1$ and $g_6=1/\Lambda^4_2$. Near the origin, for $\phi=0$, the potential along the $X$-direction reads
\begin{equation} \label{V-S-T-2}
V^X=-2 \frac{c}{M^2_P} F (X+X^\dagger)-4\frac{F^2}{\Lambda^2_1}|X|^2+9\frac{F^2}{\Lambda^4_2}|X|^4 +\frac{T^2}{12}\left(-4\frac{c F}{\Lambda^2_1 M^2_P}(X+X^\dagger)+3\lambda^2|X|^2+36 \frac{F^2}{\Lambda^4_2}|X|^2 \right),
\end{equation}
up to terms which do not depend on $X$ and are quartic in $T$. The $\lambda^2 X^2$ is the ${\cal O}(X^2)$ term that dominates in the parenthesis above since the tree level stability condition of the metastable vacuum (\ref{vac-stab-2}) implies
$\lambda^2>F^2/\Lambda^4_2$. We also took into account $\Lambda_1>\Lambda_2$ condition necessary for the K\"ahler potential to be well defined.
At high temperatures $T>\Lambda_1$ that the tree level terms are completely negligible the minimum is close to the origin at $X \sim c F/\lambda^2\Lambda^2_1 M^2_P$. For $T<\Lambda_1$ the ${\cal O}(X T^2)$ can be omitted in favour of the ${\cal O}(X)$ term. As the temperature decreases the $X$-minimum moves away from the origin we, again,  distinguish two cases:
\begin{itemize}
	\item $\Lambda_2> \Lambda^{3/2}_1/M^{1/2}_P \Rightarrow X(T)>\Lambda^2_1/M_P$
\end{itemize}
The temperature $T_X$ at which the minimum exits the unstable tachyonic region $|X|<\sqrt{\frac{F}{\lambda}}$ and becomes metastable is found to be 
\begin{equation} \label{T-S-2}
T_X \sim \frac{F}{\lambda}\frac{1}{\Lambda_1} .
\end{equation}
At $T_X$ the minimum is of the order of the tree level one $X^2_{min}\sim \Lambda^4_2/\Lambda^2_1$. A fast way one to find $T_X$ is to note that the above result ($\ref{T-S-2}$) would be exact in the case of absence of the gravity terms ${\cal O}(X)$ and ${\cal O}(T^2 X)$ from ($\ref{V-S-T-2}$). Then the effective mass squared for $X$ would be $m^2_{X,eff}=-8F^2/\Lambda^2_1+T^2\lambda^2$/2 and at that temperature the $X$-minimum would move by a second order phase transition to non-zero values. This would correspond to a spontaneous $U(1)$-$R$ symmetry breaking simultaneously with the supersymmetry breaking; for another example of spontaneous $U(1)$-$R$ symmetry breaking of O'Raifeartaigh models \cite{Shih:2007av} at finite temperature, see \cite{Moreno:2009nk}. It is easy to check that the $T_X$ ($\ref{T-S-2}$) is less than the $T_{cr}=2 \sqrt{\frac{F}{\lambda}}$ at which a second order phase transition to the $q$ direction takes place. A way to understand this qualitatively is that the direction that opens first, i.e. becomes tachyonic, is the one towards the minima closest to the origin. The susy breaking minima are shallower than the susy preserving ones and can be locally stable only if they are further than the susy minima (condition (\ref{vac-stab-2})).
\begin{itemize}
	\item $\Lambda_2< \Lambda^{3/2}_1/M^{1/2}_P \Rightarrow X(T)<\Lambda^2_1/M_P$
\end{itemize}
Here, the temperature at which the minimum exits the tachyonic region $|X|<\mu/\sqrt{\lambda}$ and becomes metastable is  
\begin{equation} \label{T-S-3}
T^2_X \sim \frac{c\sqrt{F}}{M^2_P\sqrt{\lambda^3}}
\end{equation}
This can be seen from the fact that for $X(T)<\Lambda^2_1/M_P$ the ${\cal O}(X)$ term dominates over the ${\cal O}(X^2)$ in ($\ref{T-S-2}$). At this temperature the minimum is to a good approximation the zero temperature (tree level) one $X^3_{min} \sim \Lambda^4_2/M_P$. We see that ($\ref{T-S-3}$) coincides with ($\ref{TS-1}$) as actually expected from the similarities of the two models. We again remark that ($\ref{T-S-3}$) is $M_{P}$ suppressed which means that it disappears on the global susy limit together with the susy breaking minimum.

To summarize, we have discussed models that possess susy preserving vacua close to the origin and metastable susy breaking vacuua at VEVs defined by a power of an intermediate scale $\Lambda$ (or $\Lambda_1,\Lambda_2$) which is the cut-off for  these theories. We saw that at high temperatures the field is trapped near the origin. As the universe cools down, at the temperature $T_{cr}=2 \sqrt{\frac{F}{\lambda}} > T_X$, there is a second-order phase transition. The origin becomes unstable, since the $\phi$-direction becomes tachyonic, and the fields land in the supersymmetric global minimum. The small non-zero expectation value along the $X$-direction cannot block the transition to the supersymmetric vacuum. The conclusion seems to be that the susy-breaking metastable vacuum is not realized in the early universe.

\section{Including a Bare Messenger Mass}

We generalize the superpotential ($\ref{KitanoW}$) including an explicit mass term $M$ for the messengers:
\begin{equation} \label{MNW}
W=F X-\lambda X \phi \bar{\phi}\pm M \phi \bar{\phi}+c.
\end{equation} 
keeping the same structure ($\ref{KitanoK}$) for the K\"ahler potential 
\begin{equation} \label{MNK}
K=X^{\dagger}X-\frac{(X^\dagger X)^2}{\Lambda^2}+ {\cal O} \left(\frac{(X^\dagger X)^3}{\Lambda^4}\right)+\phi^\dagger \phi+\bar{\phi}^\dagger \bar{\phi}.
\end{equation} 
The mass terms violates the $U(1)$ $R$-symmetry down to a $Z_2$ one. We are assuming that the  messenger mass is fixed in the fundamental theory.
This model with $\delta W= -M \phi \bar{\phi}$ and the K\"ahler potential (\ref{MNK}) was discussed  in \cite{Murayama:2007fe} in the global susy framework. Here, we will couple it to gravity and comment on the thermal behaviour of such a model. 

In the global susy the theory the tree level potential reads
\begin{equation}
V_0=|F-\lambda \phi \bar{\phi}|^2\left(1+\frac{|\phi|^2}{\Lambda^2}+ {\cal O}(|\phi|^4/\Lambda^4)\right)+|\lambda \phi \bar{\phi}\pm M\bar{\phi}|^2+|\lambda X \phi \pm M \phi|^2
\end{equation}
and it has a susy minimum at 
\begin{equation} 
X =\mp \frac{M}{\lambda}, \; \; \phi \bar{\phi}=\frac{F}{\lambda},
\end{equation}
and a susy  breaking minimum at 
\begin{equation} \label{MNnS}
X=\phi=\bar{\phi}=0. 
\end{equation}
The $X$ gets a mass and is stabilized at the origin due to the non-canonical terms in the K\"ahler potential. Loops interactions with messengers, which do not respect the $U(1)_R$ because of the mass term, generate the following Coleman-Weinberg effective potential for $X$ \cite{Murayama:2007fe}:
\begin{equation}
V_{CW} \simeq \frac{5 F^2}{(4\pi)^2}\left(\frac{\lambda^3}{M}(X+X^\dagger)-\frac{\lambda^4}{2M^2}(X^2+X^{\dagger\, 2})+...\right)
\end{equation}
The meta-stability of the ($\ref{MNnS}$) susy breaking vacuum can be checked by looking at the mass matrices of the $X$, $\phi$ and $\bar{\phi}$:
\begin{equation}
m^2_X \simeq \left(\begin{array}{cl}      
4\frac{F^2}{\Lambda^2} &  -\frac{5 F^2 \lambda^4}{(4\pi)^2M^2}  \\     
-\frac{5 F^2 \lambda^4}{(4\pi)^2M^2} & 4\frac{F^2}{\Lambda^2}   \\     
\end{array}\right),
\,\,\,\,\,\,\,\,\,\,\,\,\,\,\,\,\,\,\,\,\,\,\,\,\,\,\, m^2_\phi=
\left(\begin{array}{cl}      
|\lambda X \pm M|^2 & -\lambda F  \\     
-\lambda F & |\lambda X \pm M|^2   \\     
\end{array}\right)
\end{equation}
\\ 
at the origin. Radiative corrections due to the coupling of the spurion to messengers can render the mass of $X$ tachyonic unless $M> \lambda^2\Lambda/(4\pi)$. In fact, the local minimum is shifted due to the $R$-violationg interaction with the messengers to a value $|\left\langle X\right\rangle| \simeq \lambda^2\Lambda^2/(16\pi^2M)$ neglecting the constant $c$. This vev is generally much smaller than $M/\lambda$. The $\phi$, $\bar{\phi}$ directions are stable as long as $M^2>\lambda F$ is satisfied. Otherwise one of the messengers becomes tachyonica and the susy breaking vacuum disappears. The exchange of messengers gives rise to the gaugino masses of the order of \cite{Dine:1994vc}
\begin{equation}
m_{1/2}\equiv m_\lambda \simeq \frac{\alpha}{4\pi}\frac{\lambda F}{M}
\end{equation}
where $\alpha$ represents a generic standard model gauge coupling. The fact that this model lacks a $U(1)_R$ symmetry is the reason why it is claimed to give viable phenomenology while having a metastable susy breaking vacuum at the origin contrary to models that respect $U(1)_R$ \cite{Cheung:2007es}. 

We shall demonstrate  that the inclusion of gravity changes drastically the vacuum structure. The dominant terms in the  scalar potential read
\begin{eqnarray} \nonumber
&V_0\simeq -2\frac{c}{M^2_P} F (X+X^\dagger)+4F^2 \frac{|X|^2}{\Lambda^2}- \lambda F (\phi\bar{\phi}+\phi^\dagger\bar{\phi}^\dagger)-2 F \frac{c}{M^2_P} (X+X^\dagger) (|\phi|^2+|\bar{\phi}|^2)&  \\ \nonumber 
&+|\lambda X \pm M|^2 (|\phi|^2+|\bar{\phi}|^2)+\lambda^2 |\phi|^2|\bar{\phi}|^2.&  \nonumber
\end{eqnarray}
It has been taken into account that the dimensionful constant $c/M_P$ must be of the order of the susy breaking scale $F$ for the vanishing cosmological constant. A few remarks are in order here. Firstly, the susy breaking minimum is shifted from the origin $X=0$ to the non zero value $X \simeq c\Lambda^2/(2M^2_P F)$. Secondly, since $\Lambda \geq M$,  the $X \sim \Lambda^2/M_P$ can be close to the susy preserving vacuum, although it is easy to arrange $\Lambda^2/M_P \ll M/\lambda$. 
Hence, this model has the interesting feature that the susy breaking minima are closer to the origin compared to the susy preserving ones. This fact raises the question whether the susy breaking minima are thermally preferred. In order to check this we write the finite temperature potential assuming a temperature higher than the messenger scale $M$
\begin{equation} \nonumber
V= V_0+V^T_1=V_0+\frac{T^2}{12}\left[4\frac{F^2}{\Lambda^2}+4 F \frac{c}{M^2_P}\frac{X+X^\dagger}{\Lambda^2}+ 3|\lambda X\pm M|^2+3\lambda^2(|\phi|^2+|\bar{\phi}|^2)\right]+ {\cal O}(T^4).
\end{equation}
At temperatures $T>\Lambda$ the thermal average field values are $\phi=\bar{\phi}=0$ and $X=\mp M/\lambda + {\cal O}( F c/(M^2_P\lambda^2\Lambda^2))$. As the temperature decreases the mass squared of messengers at this minimum, taking $\phi=\bar{\phi}$, is 
\begin{equation}
m^2_{\phi}\simeq -2\lambda F +2|\lambda X \pm M|^2 +\frac{1}{2}T^2\lambda^2 \simeq -2\lambda F +\frac{1}{2}T^2\lambda^2.
\end{equation}
Therefore, the situation is similar to those presented in the previous chapters. If the susy breaking sector and messenger are in thermal equilibrium with the thermal bath then the system at the critical temperature $T_{cr} \simeq 2  \sqrt{\frac{F}{\lambda}}$ will evolve to the phenomenologically unacceptable susy preserving vacuum. The situation could change if we assume further couplings of $X$ of the form $\delta W=\lambda' X \varphi\bar{\varphi}$ where $\varphi, \; \bar{\varphi}$ are fields uncharged under the SM gauge group \cite{Katz:2009gh}. This could shift the high temperature minimum of $X$ closer to the origin and change the thermal history. However, the  evolution of the system becomes  then highly model dependent. We note that the idea of adding to the superpotential a term $\delta W=\lambda' X \varphi\bar{\varphi}$ was first presented in the paper by Ellis, Llewellyn Smith and Ross \cite{Ellis:1982vi} where their mezzo-O'Raifeartaigh model was modified in this manner to push the thermal minimum towards the susy breaking metastable vacuum. 

As explained in the previous section, an inflationary phase is expected to displace the fields towards the region  of relatively large vevs. The results of that section can be also applied in the present  case with the important difference that here the susy breaking and susy preserving minima may exchange their roles in the arguments. For instance,
if the vev  of the spurion after inflation gets shifted into the vicinity of the susy preserving minimum, then the system can find itself to be trapped there. 
\begin{figure} 
\centering
\begin{tabular}{cc}
\\
{(a)} \includegraphics [scale=.85, angle=0]{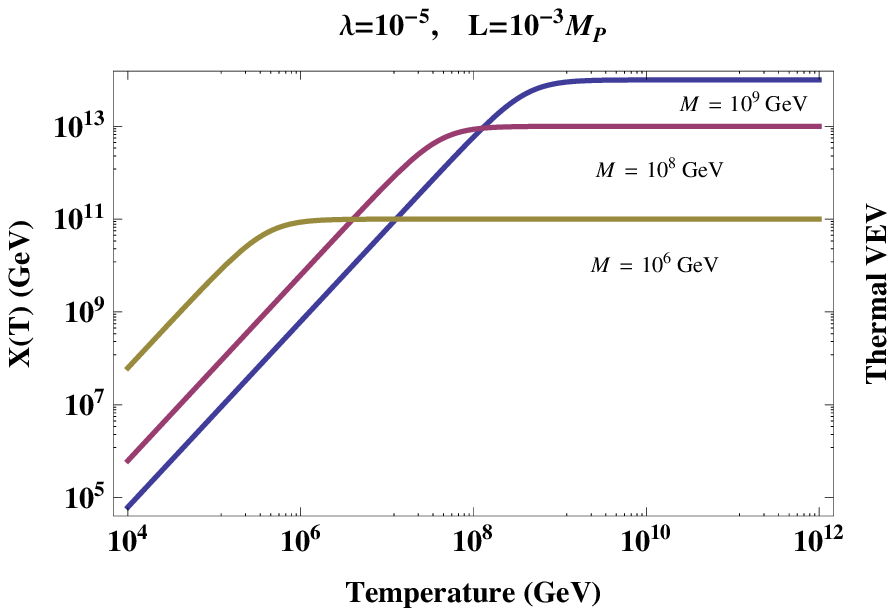} &
{(b)} \includegraphics [scale=.85, angle=0]{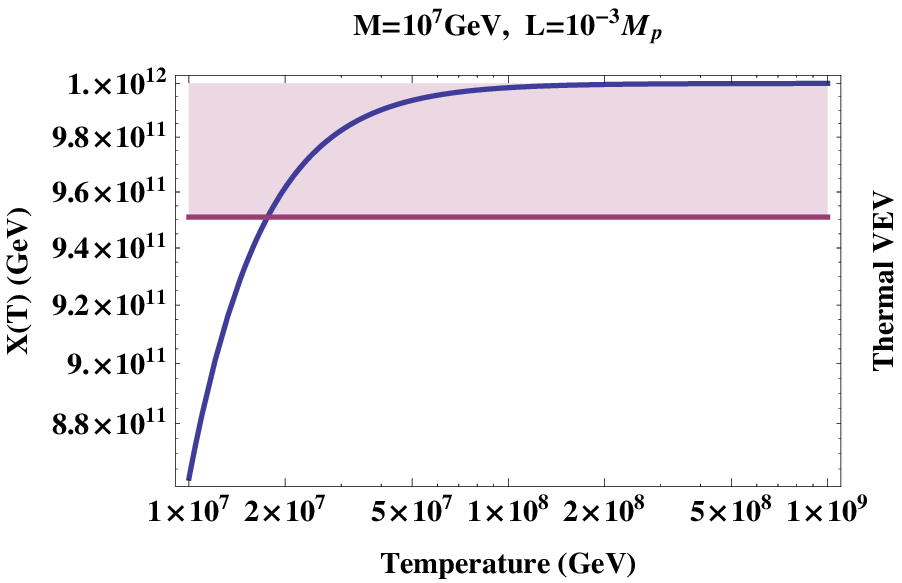}  \\
\end{tabular}
\caption{\small{In this figure the thermal evolution of the VEV, $X(T)$, of the spurion for the case of Messenger Mass is demonstrated. The left hand side panel corresponds to different parameters of Messnger Mass $M=10^{6}, \, 10^8$ and $10^9$ GeV for $\Lambda=2.4\times 10^{15}$ GeV and $\lambda=10^{-5}$. At the right hand side panel $M=10^{7}$ GeV, $\Lambda=2.4\times 10^{15}$ GeV and $\lambda=10^{-5}$; the painted area is the zero temperature tachyonic region; the $T_X$ is defined as the temperature that the $X(T)$ exits the painted region. The figure shows the zero temperature minimum $X=0$ and the high temperature $X \simeq M/\lambda$.}}
\end{figure}

\section{Thermalization}

At the end of inflation the inflaton field $I$ oscillates about the minimum $I_0$ of the inflationary potential. At some point it decays completely and the universe becomes reheated. After reheating $I=I_0$ and there are no Hubble induced terms in the potential of $X$, $\phi$ and $\bar{\phi}$. The value of the spurion at that time is denoted as $X_{rh}$. If the interaction rate of these fields with the thermal plasma is larger than the expansion rate $H$, the fields will  thermalize and the potential will be corrected by temperature dependent terms. Otherwise, their potential will be the zero temperature one, i.e. the tree level potential enhanced by  Coleman-Weinberg loop corrections. 

The messengers $\phi$ and $\bar{\phi}$ are coupled to the MSSM plasma via the SM gauge forces and they can achieve thermal equilibrium. If the reheating temperature is higher than $m_{\phi}$ where $m_{\phi} \approx \lambda \left\langle X \right\rangle$ then they get thermalized and stabilized at the origin. If not, one can still expect a thermal mass for messengers  
due to thermalized MSSM gauge bosons. The messengers SM gauge numbers couple to the gauge bosons as follows: 
\begin{equation} \label{thmass}
V \subset g^2(|\phi|^2+|\bar{\phi}^2|)\left\langle A_\mu A^\mu\right\rangle_T. 
\end{equation}
This induces a thermal mass of the order of $g\,T$,  large enough to push  the vevs of the messengers towards  $\left\langle \phi \right\rangle=\left\langle \bar{\phi}\right\rangle=0$  even for $T_\text{rh}<m_{\phi}$.

On the other hand, the spurion  $X$ is coupled to MSSM degrees of freedom via loop diagrams with the messengers as the heavy fields in the loops.  For $T>m_{\phi}$ the  thermally averaged cross section for 2-2 processes  with two spurions is of the order of 
\begin{equation} 
\Gamma_{int}=\left\langle \sigma v n\right\rangle_{T}\sim \frac{\lambda^4 \alpha^2 }{16 \pi^2} T
\end{equation}
($\alpha$ corresponds to the SM fine structure constant) and the equilibrium in a radiation dominated universe, $\Gamma\geq H=\sqrt{g_*} \, T^2/M_P$, may be achieved below the temperature 
\begin{equation} \label{Teq}
T_{eq} \sim \frac{\lambda^4 \alpha^2 }{16 \pi^2 \sqrt{g_*}}M_P={\cal O}(10^{-3})\lambda^4 \alpha^2 M_P.
\end{equation}
When messengers become  non-relativistic, i.e. for  $T< m_{\phi}$, the thermalization could also be achieved and the relevant  averaged cross section becomes $ \left\langle \sigma v n \right\rangle_{T}\sim \alpha^2 \lambda^4T^5/ m^4_{\phi}$ .  The requirement that this interaction rate is larger than the expansion rate gives a lower bound on the temperatures at which  $X$ can be  thermalized. If the coupling $\lambda$ is small enough then the window of temperatures where the spurion  thermalizes can be closed. Actually, the zero temperature constraints on the coupling $\lambda$ (\ref{vac-stab-qua}), (\ref{vac-stab-qua2}) and (\ref{vac-stab-qua3}) don't allow the $X$ field to thermalize. Even when $T>m_{\phi}$, i.e. when the messengers running in the loops are light compared to the temperature, the spurion $X$ is out of equilibrium.

However, when $T_\text{rh}>m_{\phi}$, the messengers get thermalized and they can contribute thermal corrections to the potential of the spurion. 
One can see this if one takes into account that thermal averaging of the  term $\sim \lambda^2 |X|^2 |\phi|^2$ in the tree level scalar potential leads to 
\begin{equation} \label{S-thermal}
\lambda^2 |X|^2 \langle |\phi|^2 \rangle_T \sim \lambda^2 |X|^2 T^2,
\end{equation}
for a thermal distribution of messengers. Once $T<m_{\phi}$ the evolution of the spurion is governed by the zero temperature potential.

It is interesting to note that thermalized MSSM degrees of freedom may alter the value of  the critical temperature. The exact modification of the critical temperature depends on the Lagrangian that describes the interactions of the messengers with the observable sector and it is generally model dependent. The study of the coupled system of hidden, messenger and observable sectors will be performed in the next chapter. 

In conclusion, the messengers being charged under the Standard Model gauge group obtain thermal masses which help to localize them at the origin of the field space. When the temperature is higher than their tree level mass $m_{\phi} \simeq \lambda \left\langle X \right\rangle$ they induce a thermal mass for the spurion $X$ according to (\ref{S-thermal}). This thermal mass may also drive the $X$ field to the origin. On the contrary, when the temperature is lower than $m_{\phi}$ the thermal excitations of messengers are Boltzmann suppressed hence the thermal induced mass of the spurion is negligible. 

\section{Thermal Evolution of the ISS}

Lets assume that the ISS hidden sector is thermalized. At high enough temperatures the origin is the minimum of the finite temperature effective potential. In the meson direction far away from the origin a second minimum (which becomes the susy preserving one at $T=0$) forms, but it is always seperated by a barrier from the minimum at the origin. At a temperature $T^{\text{ssb}}_c \sim \mu$ the curvature of the potential at the origin becomes negative in the quark direction
\begin{equation}
(T^\text{ssb}_c)^2= \frac{12 \mu^2}{3N_e+2N_m +g^2(1+5(N^2_m-1))/h^2N_m}.
\end{equation}
but stays positive in the meson direction:
\begin{equation}
\left. \frac{\partial^2 V}{\partial q^2}\right|_{M=0, q=0} <0\quad \quad \text{and} \quad \quad \left. \frac{\partial^2 V}{\partial M^2}\right|_{M=0, q=0} >0
\end{equation} 
A new minimum forms in the quark direction, a phase transition occurs and the fields move to the newly formed minimum. At a temperature $T^{\text{susy}}_c \sim (h\mu^2)^{1/2}$ the isolated minimum in the meson direction becomes degenerate with the one at the origin:
\begin{equation}
V(M=0, T^{\text{susy}}_c)= V(M^T_0, T^{\text{susy}}_c)
\end{equation}
The $T^{\text{susy}}_c$ when the minima have degerate potential energies and a tunneling through the barrier can start is
\begin{equation}
(T^M_c)^2 =\left[ \frac{24 N_f}{\pi^2\left(N_f+N_m)^2 -1  \right)} \right]^{1/2} h \mu^2 + {\cal O}(h).
\end{equation}
The potential barrier between them implies that the  transition could be accomplished through quantum tunneling between the two vacua which  is much more strongly suppressed than the classical  transition in the quark direction. As the temperature decreases the minimum in the meson direction becomes the global one and the other minimum, close to the origin in the quark direction, becomes metastable. The minima are always seperated by a potential barrier and the phase transition into the supersymmetric phase is suppressed by tunneling at all stages. The conclusion is that the transition to the non-susy vacuum is thermally favoured. Also, it was shown in \cite{Abel:2006cr,Abel:2006my} that even if the fields start in the supersymmetric minimum, e.g. due to non-adiabatic initial conditions, high enough temperatures will thermally drive them to the susy breaking minimum. In particular, if the reheating temperature is $T_{rh}> {\cal O} (1) \, \mu $ the universe ends up in the non-susy vacuum.

The ISS evolution is exactly the opposite of what has been  discussed in the previous chapters, where susy vacua were thermally preferred. Large thermal masses stabilize the fields at the origin of field space. At such small vevs of the fields the non-perturbative piece which creates the supersymmetry preserving vacuum is irrelevant\footnote{The overall power of the components of $M$ in the nonperturbative piece is $N_f /(N_f - N)$, which is larger than 2 given the validity range of the model.}. It is the tree level superpotential which determines the behaviour near the origin, enhanced by thermal corrections. The basic difference between the models considered earlier and the  ISS is  due to the multi-field structure of ISS and to the rank condition breaking, which relegates the supersymmetric vacuum from the vicinity of the origin.  As a result, the number of light degrees of freedom is larger near the origin, i.e. near the nonsupersymmetric vacuum. Hence, as the temperature drops, the closest, that  is supersymmetry breaking, minimum is naturally selected in the case of ISS. In O'Raifeartaigh models studied here the situation is different - at high temperatures the corrections which are responsible for the stabilization of the spurion at the supersymmetry breaking minimum are irrelevant, and the supersymmetric minimum gets naturally selected.

However, it should be noted, that the models studied in this work belong to the class known as ordinary gauge mediation with explicit messengers and with K\"ahler potential stabilization, whereas in the original ISS analysis explicit messengers have not been considered. In the literature there are several deformations of the ISS model with ordinary or direct gauge mediation. An example of explicit messenger sector  added to the ISS (\ref{ISS}) is $W_\text{mess}=-\lambda \text{Tr}(M)\phi\bar{\phi}+M_B \phi \bar{\phi}$ \cite{Murayama:2006yf,Abel:2009ze}. In principle one could imagine a large number of messengers (at least of the order of $N\times N_f$ sets of messengers) which become light far from the origin, for instance due to the presence of an explicit mass term. Although the squarks of the ISS sector are light at $M = 0$,  the actual physical masses of messengers  could vanish at $M \sim M_B/\lambda$.  Then the thermal minimum, preferred by the large number of light states, could be at $M \neq 0$, contrary to the previous conclusions concerning  the pure ISS sector.  Taking also into account that the presence of messengers increases the number of susy preserving vacua in the field space, this could result in thermal selection of a phenomenologically wrong vacuum. However, such a setup  is non-generic. 

\subsection{ A Note on the Hidden Light Degrees of Freedom}

Assuming that a Goldstino superfied $X$ is coupled apart from the messengers to extra degrees of freedom, as the quarks of the ISS model, then the number of the degrees of freedom coupled to the $X$ field becomes crucial. The minimum of the thermal potential is the minimum of the free energy and, it is there that the entropy maximizes. A heuristic way to see where the minimum lies is to count the light degrees of freedom in the field space i.e. to determine the $\delta V^T=-N \pi^2 T^4/90$ part of the free energy. For a supepotential of the form $W\sim hXq\bar{q}+(\lambda X+M_B)\phi\bar{\phi}$ a parametrization of the number of light degrees of freedom $N$ coupled to the $X$ field as
\begin{equation} \label{entr-4}
N(X) \sim N_q\left(1- \frac{h^2 X^2}{T^2}  \right) +N_{\phi} \left(1-\frac{(\lambda X+ M_B)^2}{T^2} \right)
\end{equation}
can give a qualitative picture of the position of the thermal minima in the direction of the $X$-field. At (\ref{entr-4}) the $N_q$ is the numbers of hidden sector quarks and $N_\phi$ the number of messengers with bare mass $M_B$. Of course this heuristic formula holds only when the temperature is larger than the masses of the fields (the number $N$ does not account for the non-relativistic degrees of freedom). Moreover, in the case the hidden sector, consisting of the $X$ and $q$, $\bar{q}$ degrees of freedom, is not thermalized, e.g. when the inflaton decays dominantly to MSSM degrees of freedom and does not reheat the dynamical supersymmetry breaking sector, one should not count the $q$, $\bar{q}$ degrees of freedom. Even in the case that the hidden sector quarks are light they may be decoupled from the thermal bath. For instance, the Standard Model QCD axion particles though light are non-relativistic and they consist cold dark matter candidates. Indeed, as we have shown the spurion $X$ is decoupled. In such a case the $N_q$ of (\ref{entr-4}) is effectively zero and the hidden sector quarks do not drag the thermal minimum towards the origin. There is also the interesting possibility that the hidden sector is thermalized but it is not in a complete thermal equilibrium with the MSSM. Then a different temperature should be considered for the two systems -much similar to the temperature of the cosmic background neutrinos which is different than the CMB photon's temperature.

\chapter{Thermally Favourable Gauge Mediation Schemes}

In the chapter 5 we studied the thermal evolution of the system of fields $X, \phi, \bar{\phi}$. The $X$ field is by definition Standard Model gauge singlet and the $\phi$, $\bar{\phi}$, as messengers, are charged under the $SU(3)_c\times SU(2)_L\times U(1)_Y$. However, the coupling of the messengers to the Standard Model degrees of freedom was not taken into account -apart from the consideration of the thermalization. In this chapter, the Standard Model degrees of freedom will be included and we will demonstrate that the metastable vacuum can be thermally selected under general condition.
Nonetheless, before we consider the Standard Model effects we consider Standard Model gauge singlets $\varphi$, $\bar{\varphi}$ coupled to the spurion and examine first whether this isolated system $X, \varphi, \bar{\varphi}$ can select the false vacuum. This study is of special interest since generically the $X$ can be coupled to Standard Model singlet fields. We can readily extend the hidden sector. In the lines of (\ref{eogm2-3}) we consider 
\begin{equation} \label{gen-5}
W=FX+ \sum^N_i\sum^N_j \lambda_{ij}X \phi_i\bar{\phi}_j \equiv FX+ X \left(\sum^{N_\text{HS}}_i \sum^{N_\text{HS}}_jk_{ij}\varphi_i\bar{\varphi}_j + \sum^{N_\text{mess}}_i \sum^{N_\text{mess}}_j\lambda_{ij}\phi_i\bar{\phi}_j \right)  
\end{equation}
i.e. we split the supersymmetry breaking sector into the Standard Model singlet fields $\varphi$, $\bar{\varphi}$ and the Standard Model charged $\phi$, $\bar{\phi}$. The indices $i, j$ run from $1$ to $N$ where $N=N_\text{HS}+N_{\text{mess}}$ and $N_{HS}$ the copies of $\varphi$, $\bar{\varphi}$ and $N_\text{mess}$ the copies of the $\phi$, $\bar{\phi}$. The thermal evolution of the ($\ref{gen-5}$) with some interesting extensions will be studied. In particular the superpotenial
\begin{equation}
W=FX+k X \varphi^2+ \lambda X \phi\bar{\phi}+M \phi\bar{\phi}+ c
\end{equation}
where the $\phi$, $\bar{\phi}$ are, as usual, the messenger fields and the $\varphi$ carry \itshape no \normalfont Standard Model numbers. 

We will show that extra degrees of freedom that restore supersymmetry and carry no Standard Model charges make the selection of the metastable supersymmetry breaking vacuum a not-generic phenomenon \cite{Dalianis:2010yk}.  On the other hand, messengers although restore supersymmetry as well they can make the metastable vacuum favourable \cite{Dalianis:2010pq}.

The essential result of this chapter is based on the paper \cite{Dalianis:2010pq}  that demonstrated the generality of the thermal selection of the metastable supersymmetry breaking vacuum in ordinary gauge mediation models.

\section{Conditions for selection of the susy breaking vacuum in systems without Messengers}

We start the study of the condition that implement the selection of the metastable supersymmetry breaking vacuum by considering, as a first step, the system
\begin{equation} \label{hs-5}
W_{HS}= FX+kX \varphi^2+c
\end{equation}
and the corrected for the spurion K\"ahler
\begin{equation}
K= X^\dagger X-\frac{(X^\dagger X)^2}{\Lambda^2}+\varphi^\dagger \varphi
\end{equation}
where the $\varphi$ are \itshape not \normalfont charged under the $SU(3)\times SU(2)\times U(1)$ Standard Model gauge group. Although they look like ordinary messengers they are actually not, hence we can refer to them as \itshape  "hidden messenger-like" \normalfont fields. Since the Standard Model degrees of freedom were not directly included in the chapter 5 the results derived there can be readily applied in the example (\ref{hs-5}). The purposes of this study are twofold. Firstly, we generalize the supersymmetry breaking sector; actually the spurion is expected to be coupled to other hidden degrees of freedom: Direct mediation modeles are typical examples. Secondly, this study will exhibit the crucial r\^ole of the MSSM gauge bosons, that are absent at (\ref{hs-5}), in the selection of the phenomenologically correct metastable vacuum.

According to the section 5.4 it is straightforward to conclude that the condition which controls the selection of a susy breaking vacuum of the system (\ref{hs-5}) is 
\begin{equation} \label{condition1}
\sqrt{F/k}< |X|<{\Lambda}. 
\end{equation}	
This is the prospective basin of attraction of the susy-breaking minimum. It is bounded from above by the condition that the quantum corrected kinetic energy stays positive definite. The lower bound is the tachyonic region about the origin. This  condition can be broken into two: i) $X$ at the end of inflation should have a vev in  the regime of our effective theory, i.e. $|X_\text{INF}|<\Lambda$. 
ii) $X$ during reheating, or during a non-thermal phase, must obey $|X|>\sqrt{F/k}$. In the case of a generalized K\"ahler potential 
\begin{equation}
K=X^\dagger X + \frac{(X^\dagger X)^2}{\Lambda^2_1}-\frac{(X^\dagger X)^3}{\Lambda^2_2} + \varphi^\dagger\varphi.
\end{equation}
the cut-off scale is $\Lambda_2$ and the condition reads: $\sqrt{F/k} < |X|<{\Lambda_2}$.
\\
\\
Unless the above condition is fulfilled the system lands either in a susy preserving vacuum or in the region of  large field values where our IR-effective theory is not valid. For the selection of the susy breaking vacuum the hidden messenger-like fields must have a vev $\left\langle \varphi \right\rangle=0$. Otherwise, the spurion has a tree level mass contribution $k \left\langle \varphi \right\rangle$ and can be attracted to the origin. The hidden fields may have a vev $ \left\langle \varphi \right\rangle=0$ thanks to the thermal mass that possibly  receive from thermalized hidden sector gauge bosons even in the case that the $\varphi$ itself is not thermalized. 

However, in the case the $\varphi$ fields are thermalized, e.g. due to their self interaction, the spurion also receives a thermal mass. This mass drives the spurion to the origin. A simple  way out would be to impose the condition  $T_\text{rh} <m_\varphi$. 
In fact, the more general condition which makes it likely  that the metastable vacuum becomes actually selected is the following: 
\begin{equation} \label{Trh-suf} 
		T_\text{rh}<\text{max}\{m_\varphi, T_X\}.
\end{equation}
We note that the temperature $T_X$ can be larger than the mass of hidden messengers and hence they can be thermalized.  
If there are further contribution to the thermal mass of $\varphi$ the system of fields is expected to be aligned along the $X$-direction (in the complex $X$-plane) and the tree level $\varphi$-contribution to the mass of the $X$ field, $k \left\langle \varphi \right\rangle$, vanishes. 
The upper bound on the reheating temperature (\ref{Trh-suf}) guarantees that the susy breaking vacuum in the $X$ direction has formed: it is locally stable. Thus, the field can land in the susy breaking vacuum. In particular, below the temperature $T=F/(k \Lambda)<T_X$ the relevant potential is approximately the zero temperature one (\ref{V-zero})
\begin{equation}
V(X)\simeq F^2-\frac{3c^2}{M^2_P}-2\frac{c}{M^2_P} F (X+X^\dagger)+4F^2 \frac{|X|^2}{{\Lambda}^2} + \frac{k^2 N_\varphi}{(4\pi)^2} F^2 \log \frac{X^\dagger X}{Q^2} +{\cal O}(T).
\end{equation}
The logarithmic term, originating from the interaction of $X$ with the $\varphi$, is important near $X=0$ but its effects are negligible for small values of coupling $k \ll \Lambda$. As shown in \cite{Ibe:2006rc}, where a non-thermal evolution of the system of fields was considered, the $X$ field feels at most points of the complex $X$-plane a much stronger force towards the supersymmetry breaking vacuum than towards the supersymmetric one. It has been shown numerically that for initial conditions Re$(X)$=Im$(X)=\Lambda$ the $X$ field settles into the supersymmetry breaking minimum. If we additionally want the energy stored in the oscillations of the $X$ field not to dominate the energy density of the universe then the spurion at the time of reheating should be localized around the metastable minimum. This could be realized via a possible tracing of the minimum after inflation. Otherwise, it is possible that the late decay of the spurion will cause a late entropy production diluting the dark matter abundance and the baryon asymmetry.

The condition (\ref{Trh-suf}) is not a strict constraint on the reheating temperature. Actually, the smaller the coupling $k$  the larger the reheating temperature can be. Also, decreasing the $k$ opens the window $m_\varphi<T_\text{rh}<T_X$, see Table 6.2.

An interesting observation here is that for couplings $k$ so small that $k<T_X/M_P$ the upper bound on the reheating temperature at (\ref{Trh-suf}) is not necessary. The reheating temperature can be arbitrary high. The reason is that due to the very small coupling $k \ll 1$ the thermal mass $k T$ is smaller  than the Hubble scale $H \simeq \sqrt{g_*}\,T^2/M_P$ and the field $X$ starts rolling down only for temperatures $T<T_X$. This can be seen by examining the dynamics of the spurion. In an FRW universe the homogeneous spurion field obeys the equation of motion
\begin{equation}
\ddot{X}+3H\dot{X}+ dV/dX=0.
\end{equation}
In a radiation dominated phase $H=1/(2t)\simeq \sqrt{g_*}\,T^2(t)/M_P$ and   $dV/dX \sim F^2/\Lambda^2 X+k^2 T^2 X+k^2 \varphi^2 X$. Taking into account that the thermalized hidden messengers are driven fast to the origin, one can see that  for $T\geq T_X$ the thermally  induced spurion mass dominates the potential and the equation of motion reads
\begin{equation}
\ddot{X}+3\sqrt{g_*}\,T^2(t)\dot{X}+ k^2T^2(t)X \approx 0.
\end{equation}
The $X$ starts rolling down only after $H\sim m_X(T)$ i.e. when $k \sim T/M_P$. Hence, the condition $k<T_X/M_P$ means that the thermal corrections to the spurion will not drive it to the origin before the metastable susy breaking vacuum has appeared, independently  of how large the reheating temperature actually is. However, there is a price to pay: the smallest the coupling $k$ the largest is the tachyonic region $|X|<\sqrt{F/k}$ about the origin and therefore, the area of the initial values for the field $X$ which realize the metastable vacuum gets reduced.

\subsection{Cosmological Constraints on the Supersymmetry Breaking Sector}

Apart from the zero temperature constraints presented in the section 4.10 we can additionally apply the cosmological constraints of this section. The necessary condition for selection of the metastable susy breaking vacuum is that the initial value of the spurion $X$ is smaller than the cut-off scale $\Lambda$, $\Lambda_2$ ($\Lambda_1>\Lambda_2$).
We have set this initial value right after inflation to be $ X_\text{INF}$.
We shall  assume  $|X_\text{INF}| \ll M_P$. 
 Hence, we ask for $\Lambda>|X_\text{INF}|$ and $\Lambda_2>|X_\text{INF}|$. The larger the $|X_\text{INF}|$ is the larger the cut-off scale has to be. 
 For the three types of models these constraints are presented in the Table 6.1.

As discussed, there exists  an additional condition which favours the selection of the susy breaking vacuum for the theory (\ref{hs-5}). Namely: $T_\text{rh}<\text{max}\{m_\varphi, T_X\}$. The temperature, at which the metastable vacuum appears, $T_X$, depends inversely on the coupling $k$, e.g.
\begin{equation}
T^2_X \sim \frac{c\sqrt{F}}{M^2_P \sqrt{k^3}}
\end{equation}
for the (\ref{hs-5}) model.  Hence, the smaller the coupling is the higher the $T_X$. In other words, small coupling $k$ means a small coupling to the thermal bath i.e. the tree level potential dominates over the 1-loop temperature dependent corrections even for high temperatures. On the other hand, the larger the coupling $k$ the heavier the hidden messenger-like fields are. Of course, the value of $k$ is model dependent. But, it cannot be arbitrary large or arbitrary small. For example, considering the first case of gravitational stabilization, where the order 6 correction is negligible, the coupling lies in the range $10^{-14}(\Lambda/M_P)^{-2}<k<\Lambda/M_P$, see Table 6.1. Otherwise, the vacuum is unstable either due to tachyonic direction or due to quantum correction coming from the coupling $k  X \varphi \bar{\varphi}$. 
\begin{table} \label{tabb2}
\begin{center}
$$
\begin{array}{|r|r|r|r|}\hline
\bold{K=|X|^2\mp \frac{|X|^4}{\Lambda^2_1}-\frac{|X|^6}{\Lambda^4_2}} & \bold{\Lambda_1} \,\,\,\,\,\,\,\,\,\, & \bold{\Lambda_2} \,\,\,\,\,\,\,\,\,\,\,\,\,\,\,\,\,\,\,\, & \bold{k} \,\,\,\,\,\,\,\,\,\,\,\,\,\,\, \\
\hline 
\bold{1.\, ~(-), \; \, \Lambda_2=\Lambda_1\equiv \Lambda}   & |X_\text{INF}|<\Lambda\lesssim 10^{-2} & - \,\,\,\,\,\,\,\,\,\,\,\,\,\,\,\,\,\,\,\,& \frac{10^{-14}}{\Lambda^{2}}<k< \Lambda \,\,\,\,\, \\
\hline
\bold{2.~~~~~ (+), \, \; \, \Lambda^{3/2}_1<\Lambda_2}  & \Lambda_1>\Lambda_2 \,\,\,\,\,\,\,\,\,\,& |X_\text{INF}| <\Lambda_2\lesssim \left(10^{-4} \Lambda_1 \right)^{1/2}& \frac{10^{-14}}{(\Lambda^2_2/\Lambda_1)}<k< \left(\frac{\Lambda_2}{\Lambda_1}\right)^2 \\
\hline
\bold{3. ~~~~~ (+), \, \; \, \Lambda^{3/2}_1>\Lambda_2}   & \Lambda_1>\Lambda_2 \,\,\,\,\,\,\,\,\,\,& |X_\text{INF}| <\Lambda_2\lesssim 10^{-3} & \frac{10^{-14}}{\Lambda_2^{4/3}}< k < \Lambda_2^{2/3} \\
\hline\end{array}
$$
\end{center}
\caption{\small Combined constraints ($M_P=1$) coming from zero temperature vacuum stability, cosmological considerations necessary for susy breaking vacuum selection  and gauge mediation domination over gravity. The $X_\text{INF}$ is the initial value of the spurion field right after the end of the inflationary phase. It constrains the cut off scales from below and the requirement for gauge mediation domination over gravity constrains them from above. The larger the $X_\text{INF}$ is the more the gravity contributes to the susy breakdown mediation. The constraints on the coupling $k$ are the zero temperature ones.}
\end{table}
\\
\begin{table} \label{tabb3}
\begin{center}\begin{tabular}{|r|r|r|r|r|}
\hline
$\bold{\Lambda}$ & $ \bold{k} $ & $ kX_{min} <\bold{ m_\varphi}<kX_{max} \, $ & $ \bold {T_{X}}\;\;$ & $ \bold{T_\bold{\text{rh}}} \,\,\,\,\,\,\,\,\,\,\,\,\,\,\,\,\,\;\;$ \\
\hline
$10^{-2}$ & $10^{-3}$ & $ 10^{-7}<m_\varphi < 4\times 10^{-5} $ & $10^{-11.25}$ &  $T_\text{rh}<m_\varphi$\,\,\,\,\,\,\,\,\,\,\\
\hline 
$10^{-2}$ & $10^{-5}$ & $ 10^{-9}<m_\varphi < 10^{-7} $ & $10^{-9.75}$ &  $T_\text{rh}<m_\varphi$\,\,\,\,\,\,\,\,\,\,\,\, \\
\hline
$10^{-2}$ & $10^{-7}$ & $ 10^{-11}<m_\varphi <10^{-9} $ & $10^{-8.25}$ &  $T_\text{rh} <m_\varphi $ \,\,\,\,\,\,\,\,\,\, \\
\hline
$10^{-2}$ & $10^{-8}$ & $ 10^{-12}<m_\varphi <10^{-10} $ & $10^{-7.5}$ & $  \text{unbounded} $ \,\,\,\,\,\,\,\,\,\, \\
\hline
$10^{-2}$ & $10^{-9}$ & $ 10^{-13}<m_\varphi <10^{-11}$ & $10^{-6.25}$  & $ \text{unbounded} $ \,\,\,\,\,\,\,\,\,\, \\
\hline
$10^{-3}$ & $10^{-4}$ & $10^{-10}<m_\varphi <10^{-7} $ & $10^{-12}$ & $T_\text{rh}<m_\varphi$ \,\,\,\,\,\,\,\,\,\, \\
\hline
$10^{-3}$ & $10^{-7}$ & $10^{-13}<m_\varphi <10^{-10} $ & $10^{-9.75}$ & $T_\text{rh}<T_X$ \,\,\,\,\,\,\,\,\,\, \\
\hline
$10^{-4}$ & $10^{-5}$ & $10^{-13}<m_\varphi <10^{-9} $ & $10^{-12.75}$ & $T_\text{rh}<\text{max}\{m_\varphi, T_X\}$\\
\hline
\end{tabular}\end{center}
\caption{\small Some bounds on the reheating temperature that favour, for specific initial values for the spurion, the selection of the susy breaking vacuum by the system of fields (\ref{hs-5}). The value of the spurion field $X$ is the one at the time of reheating; it must be $\Lambda^2 \lesssim X_{rh}<\Lambda$. The $\Lambda$ cannot exceed $\sim 10^{-2}$ which is the maximum value allowed by the requirement of gauge mediation domination. The reheating temperature is \itshape not \normalfont bounded for couplings $k<T_X$. If not, the upper bound is either $T_X$ or $m_\varphi$ if $m_\varphi >T_X$. Approximately, the maximum value of $m_\varphi$ is $ k \Lambda$ and the minimum $ k \Lambda^2$. We used $M_P=1$ and these results are for the first case i.e.  the K\"ahler potential for the spurion is $K=|X|^2-|X|^4/\Lambda^2$.}
\end{table}

As explained the hidden messenger-like fields can be thermalized without ruling out the selection of the metastable vacuum. This is achieved when $m_\varphi <T_\text{rh}<T_X$. This translates in a constraint on the coupling which for large $T_X=(10^{-42}\Lambda^6/(k^3 M^2_P))^{1/4}$ has to be small.  Furthermore if the coupling $k$ is smaller than $T_X$ the selection of the metastable vacuum does not imply any bound on the reheating temperature. The reason is that if $k<T_X/M_P$ i.e. $k<10^{-6}(\Lambda/M_P)^{6/7}$ the spurion will not roll unless the temperature drops below $T_X$.

For the first case of negligible 6th order K\"ahler correction,  Table 6.2 demonstrates the range of values of the reheating temperature that can stabilize the spurion at the susy breaking vacuum for different values of the parameter $k$ given that the spurion vev at the time of reheating is $\Lambda^2 \lesssim X_{rh}<\Lambda$.

\subsection{Conclusions}

The \itshape conclusion \normalfont of this section is that when the spurion is coupled to Standard Model gauge singlet degrees of freedom via a superpotential $\delta W = k X \varphi^2$ then the metastable vacuum (it is metastable due to this interaction) can be selected only under very constrained initial conditions: \\
1) the spurion $X$ has to be close to the zero temperature metastable vacuum at the time of reheating and, \\
2) the reheating temperature has to be smaller than the $max\{m_\varphi, T_X \}$ and, \\
3) the coupling $k$ has to be rather small (for extremely small coupling $k$ the second condition in not necessary.)
\\
\\
Once the $\varphi$-fields get charged under the Standard Model gauge numbers (i.e. $\varphi \rightarrow \phi$, $k\rightarrow \lambda$) then the thermal behaviour radically changes. The MSSM degrees of freedom decrease the effective temperature that the supersymmetric vacuum forms i.e. the $T_{susy}$ becomes smaller. Then a system described by the superpotential 
\begin{equation}
W=FX+\lambda X \phi \bar{\phi}
\end{equation}
as we will show, is free from all the above constraints: 
\begin{itemize}
	\item the spurion initial vev can be generic,
\end{itemize}
\begin{itemize}
	\item the reheating temperature is not constrained from above and
\end{itemize}
\begin{itemize}
	\item the coupling, although there is an an upper bound, takes natural values.
\end{itemize}
The thermal behaviour of this system, i.e. a spurion field coupled to ordinary messengers, will be presented in the next sections and is the main topic of this chapter.

\section{Gauge Mediation with Ordinary Messengers}

Gauge mediation is an attracive way of generating soft susy breaking in the Supersymmetric Standard Model. There exist viable models of gauge mediation, complete with detailed hidden sectors where susy is broken dynamically through strong dynamics; for a recent review see \cite{Kitano:2010fa}. Since the details of the hidden sector are often phenomenologically irrelevant, the hidden sector is parameterized by a singlet field $X$ which is a spurion of susy breaking and messengers $\phi$, $\bar{\phi}$ that through gauge interactions communicate susy breaking from the hidden sector to the Supersymmetric Standard Model fields. The most general renormalizable, gauge invariant and $R$-symmetric superpotential  is
\begin{equation} \label{W-OGM}
W=FX + (\lambda_{ij} X + m_{ij} )\phi_i \bar{\phi}_j.
\end{equation}
We consider that all the fields $\phi$, $\bar{\phi}$ are messengers, i.e. $k_{ij}=0$, for the (\ref{gen-5}).  

We assume that the messenger quarks and leptons are vector like under the Standard Model gauge group. One might imagine constructing a model in which the messenger fields are chiral rather than vector-like under the Standard Model gauge group. The general problem of this type is that radiatively gaugino masses are too small. The one-loop diagram responsible for generating a gaugino mass  necessarily involves chirality flips on the fermion and scalar lines. These chirality flips are proportional to the messenger fermion masses $\tilde{m}_\phi$ and therefore the resulting masses of the gauginos are of order $(\alpha/4\pi)\tilde{m}^2_\phi/M_{SUSY}$. Since $\tilde{m}_\phi$ is of the order of weak scale the gauginos in this scenario are unacceptably light.

Hereafter we consider that the messenger quarks and leptons are vector-like under the Standard-Model gauge group. For example, assuming $SU(5)$ GUT unification then the gauge structure of the theory is $SU(5)\times G_{DSB}$, where the $G_{DSB}$ stands for any nonstandard gauge groups of the hidden sector that may be necessary for communicating susy breaking to the messenger sector. SU(5) gauge invariance implies that the messenger quark and leptons form complete $SU(5)$ representations. In the minimal case of a $\bold{5}+\bold{\bar{5}}$ messenger sector the messenger superpotential at the GUT scale has the form
\begin{equation}
W_{mediation}=\lambda X \bold{5}\bold{\bar{5}}.
\end{equation}
Below the GUT scale, $SU(5)$ is broken and the superpotential takes the form (\ref{ql-6}), see below. The assumption of unification makes the computation of $\lambda_q$ and $\lambda_\ell$ in terms of $\lambda$, by running these couplings down to the messenger scale and threshold corrections are calculable. This is also true for representations larger than $\bold{5}+\bold{\bar{5}}$. 
Introducing additional $SU(5)$ multiplets preserves gauge unification however, the gauge coupling at the GUT scale $\alpha_5(M_{GUT})$ increases as we add multiplets. If we require that the $\alpha_5(M_{GUT})$ remains perturbative then we may add only 1, 2, 3 or 4 $\bold{5}+\bold{\bar{5}}$ pairs or a single $\bold{10}+\bold{\bar{10}}$ pair or ($\bold{5}+\bold{\bar{5}}$)+($\bold{10}+\bold{\bar{10}}$) to the particle content of the minimal $SU(5)$ GUT. Additional $\bold{5}$s or $\bold{10}$s, or larger $SU(5)$ representations will render $\alpha_5(M_{GUT})$ nonperturbative \cite{Carone:1995kp}.

However, increasing the messenger scale the upper bound on the messenger multiplets increases. In particular, defining \cite{Giudice:1998bp}
\begin{equation}
N_\phi=\sum^{N_f}_{i=1} n_i
\end{equation}
where $N_f$ the flavours of the chiral superfields $\phi_i$ and $\bar{\phi}$, $(i=1,.., N_f)$ transforming as the representations $\bold{r}+\bold{\bar{r}}$ under the gauge group; $n_i$ is the Dunkin index of the gauge representation with flavour index $i$; for an $SU(5)$ $\bold{5}$ $n=1$ and for a $\bold{10}$ $n=3$. The $N_\phi$ is referred as the messenger index. The perturbativity of gauge interactions up to the scale $M_{GUT}$ implies 
\begin{equation} \label{Nindex}
N_\phi\lesssim \frac{150}{\ln \left( \frac{M_{GUT}}{M_{mess}}\right)}
\end{equation}
where $M_{mess}$ the messenger scale. If $M_{mess}$ is as low as $100$ TeV then $N_\phi$ is, as explained above, less than five. As the messenger scale increases the upper bound on $N_\phi$ is relaxed. For instance, for $M_{mess}=10^{10}$ GeV, the eq. (\ref{Nindex}) shows that $N_\phi$ as large as $10$ is allowed.

\subsubsection{Gaugino Masses and Vacuum Structure}

The ordinary gauge mediation theory (\ref{W-OGM}) can give either vanishing or non-vanishing gaugino masses at the leading order and a classification of the different cases can be found in \cite{Cheung:2007es}. It was shown in \cite{Komargodski:2009jf} that the gaugino masses are closely related to the vacuum structure of the theory. The formula for the gaugino masses at leading order in susy breaking is
\begin{equation} 
m_{\tilde{g}} \sim F^\dagger \frac{\partial}{\partial X} \log \det (\lambda_{ij} X +m_{ij})
\end{equation} 
and one can see that they vanish when $\det(\lambda X +m)=\det m$. In this case the origin $X=0$ is locally stable because there the scalar  messengers have positive squared masses. On the other hand, when $\det(\lambda X +m)$ depends on $X$, the gaugino masses are nonzero at the leading order. But there is a price to pay: there are no bare masses to protect all the messengers from becoming tachyonic for $|X|<X_{\text{min}}$, i.e. at the origin of field space. This implies the necessity the spurion field $X$ to be stabilized at a nonzero vacuum expectation value (vev) far from the origin. The superpotential (\ref{W-OGM}) doesn't determine the vev of $X$ which is, at tree level, a pseudo-modulus. It can get a potential from perturbative quantum corrections in the effective theory which lift the degeneracy. The Coleman-Weinberg potential usually stabilizes the pseudo-modulus at $X=0$ which implies that the potential runs off to infinity or to a susy vacuum at $X=0$, $\phi$, $\bar{\phi}\neq0$. 

Therefore, one has either to turn to models with locally stable origin and a mass hierarchy between sfermions and gauginos (ISS \cite{Intriligator:2006dd} and other direct mediation models fall to this category however, deformations of the ISS can evade this problem, \cite{Kitano:2006xg} is a first example) or to look for ways to stabilize the spurion at an $X\neq 0$ minimum. The former direction conflicts with a light Higgs necessary for the generation of the electroweak scale, except if one is ready to accept a more severe fine tuning in the Higgs sector. The later direction, fortunately, is not a blind siding. Gravitational effects and the need to cancel the cosmological constant in the phenomenologically acceptable vacuum can shift the susy breaking minimum at $X\neq0$ outside the tachyonic region \cite{Kitano:2006wz}. Also, it has been shown \cite{Shih:2007av} that when there are fields with $R$-charges $R\neq0,2$ the 1-loop corrections can create an $R$-symmetry breaking minimum at $X\neq0$.
Adding an explicit $R$-symmetry breaking mass term for messengers can stabilize the susy breaking minimum as well \cite{Murayama:2006yf}.

Despite the above positive results the theories (\ref{W-OGM}) with metastable vacua that give non-vanishing gaugino masses are cosmologically questioned. The thermal evolution of the hidden sector-messenger fields disfavours the selection of these susy breaking minima \cite{Craig:2006kx, Katz:2009gh, Dalianis:2010yk}. The free energy density minimizes at the origin of the field space and as the temperature decreases a phase transition towards the susy preserving vacuum takes place. On the contrary, vacua that give vanishing leading order gaugino masses are generally thermally preferred, see \cite{Abel:2006cr, Craig:2006kx, Fischler:2006xh} for the ISS model. 

The cosmological selection of these phenomenologically viable theories can be accomplished assuming a non-thermal evolution as in \cite{Ibe:2006rc} or even with thermalized messengers \cite{Dalianis:2010yk}. 
Nevertheless, whatever the proposed solution was, the spurion $X$ had to be in a particular way displaced at the time of reheating and obviously, the exact value of displacement is highly model dependent. 

In this chapter we show that the Supersymmetric Standard Model (SSM) degrees of freedom can change the thermal history of the gauge mediation models in the limit of small coupling between the susy breaking and the messenger sector. We continue the discussion of \cite{Dalianis:2010yk} taking into account the  SSM fields explicitly. We show that when $\lambda \ll 1$ the metastable susy breaking vacuum can be thermally selected. Thermal selection means that the messengers are thermalized, i.e. the reheating temperature is large enough. The metastable vacuum is realized without the domination of the energy density of the universe by the spurion and hence, without a late entropy production.

\section{Metastable Gauge Mediated Susy Breaking} \normalsize

The minimal model of ordinary gauge mediation (OGM) is 
\begin{equation} \label{mm} 
W=FX-\lambda X \phi\bar{\phi}
\end{equation}
where $X$ is a standard model gauge singlet field and $\phi$, $\bar{\phi}$ messenger fields carrying Standard Model quantum numbers.
The scalar potential in the global limit reads 
\begin{equation}  
\, \,\,\,\,\,\,\,\,\,\,\,\,\,\,\,\,\,\,\,\,\,\,\,\,\,\,\,\,\,\,\,\,\,\,\,\,\,\,\,\,\,\,\, V_F = |F|^2 + |\lambda|^2 |X|^2 \left(|\phi|^2+|\bar{\phi}|^2\right) - (F^\dagger \lambda \phi\bar{\phi}+ \text{h.c.} ) + |\lambda|^2 |\phi|^2 |\bar{\phi}|^2 \,\,\,\,\,\,\,\,\,\,\,\,\,\,\,\,\,\,\,\,\,\,\,\,\,\,\,\,\,\,\,\,\,\,\,\,\,\,\,\,\,\,\,\,
\end{equation}
for canonical K\"ahler. The  $\lambda$ and $F$ can be considered real after a phase rotation. The $X=0,\,\phi\bar{\phi}=F/ \lambda$ is a supersymmetric flat direction. The $\phi=\bar{\phi}=0, X$ is the susy breaking flat direction with $X$ not determined at tree level.
An $R$-symmetric extension of the minimal model (\ref{mm}) is to include an extra set of messengers plus a mass parameter:
\begin{equation} \label{mm2}
W=FX+\lambda X \phi_1 \bar{\phi}_1+\lambda X \phi_2 \bar{\phi}_2+m \phi_1 \bar{\phi}_2.
\end{equation}
The directions $\bar{\phi}_1$ and $\phi_2$ in the field space are not protected by the mass term. The area about the origin $|X|^2<F/\lambda$ is tachyonic for both models (\ref{mm}) and (\ref{mm2}).

The degeneracy along the $X$-direction can be lifted. The interaction term $\lambda X \phi\bar{\phi}$ induces at one-loop level a correction to the K\"ahler potential $\delta K  \simeq -(\lambda^2/16\pi^2) |X|^2\log(|X|^2/M^2)$ which attracts $X$ to the origin. In addition, the initial K\"ahler for the spurion may be non-canonical and take the form
\begin{equation} \label{K}
K=X^\dagger X -\frac{(X^\dagger X)^2}{\Lambda^2}
\end{equation}
with a cut-off scale $\Lambda$. For $|X|<\Lambda$ the potential scales like $V \sim |X|^2m^4/\Lambda^2$. Above that scale another (microscopic) theory takes over. A simple example is an O'Raifeartaigh type superpotential $W=m_{o}\chi_1\chi_2+k_o X\chi_1^2+FX$. For $\sqrt{F}\ll m_o$ the O'Raifeartaigh fields $\chi_1$ and $\chi_2$ are integrated out in and the effective superpotential is $W_{\text{low}}=FX+messengers$. The presence of the raifeartons is encoded in the K\"ahler potential which includes the one-loop contribution from $\chi$ fields 
and at low energies is effectively described by ($\ref{K}$) with $\Lambda^2 \sim m^2_o/k^4_o$. Another possibility is that $X$ is a composite particle 
which forms a bound state below the scale $\Lambda$.

The next question concerns the expectation value of the pseudo-modulus spurion. Obviously it has to be stabilized at $|X|>X_\text{min}$. Generally, this can happen thanks to gravity. Adding to the superpotential a dimensionful constant $c$ in order to cancel the cosmological constant at the susy breaking vacuum and for the K\"ahler (\ref{K}) the minimum is at \cite{Kitano:2006wz}
\begin{equation}
\left\langle X\right\rangle\simeq \frac{\sqrt{3}\Lambda^2}{6M_P}.
\end{equation}
There is also a way to give a vev to $X$ for the superpotential ($\ref{mm}$) even in the global limit. Assuming a K\"ahler potential 
\begin{equation} \label{6}
K=X^\dagger X + \frac{(X^\dagger X)^2}{\Lambda^2_1}-\frac{(X^\dagger X)^3}{\Lambda^4_2} 
\end{equation}
for $\Lambda^{3/2}_1/M^{1/2}_P<\Lambda_2<\Lambda_1$, the spurion is stabilized at $\left\langle X\right\rangle= \Lambda^2_2/\Lambda_1$, where $U(1)_R$ is spontaneously broken. 

For canonical K\"ahler the Coleman-Weinberg correction can give a $\left\langle X\right\rangle\neq0$, breaking also spontaneously the $R$-symmetry, if there are exotic messenger $R$-charges  \cite{Shih:2007av}. In particular for the superpotential ($\ref{mm2}$) there is a minimum at $\left\langle X \right\rangle \simeq 0.3 m/\lambda$. We note that the minimal model (\ref{mm}) cannot exhibit such a behaviour because there is no field with charge $R\neq 0, 2$.

Another simple solution to the problem of the spurion stabilization is to add to the superpotential (\ref{mm}) an explicit $U(1)_R$ violating mass term $M\phi \bar{\phi}$ for the messengers \cite{Murayama:2006yf, Murayama:2007fe}
\begin{equation}
W=FX-\lambda X \phi \bar{\phi}-M\phi \bar{\phi}.
\end{equation}
This relegates the susy vacua to $X\neq 0$ and a K\"ahler of the form (\ref{K}) can stabilize safely the spurion at the origin. This model has similarities with the gravitational stabilization. Here, instead of the constant $c$ it is the mass $M$ that violates the $R$-symmetry. After the transformation $X \rightarrow \tilde{X}=X+M/\lambda$ the superpotential and the K\"ahler metric read respectively $W=F\tilde{X}-\lambda \tilde{X}\phi \bar{\phi}- FM/\lambda$ and $K_{\tilde{X}^\dagger\tilde{X}}=1-4\left(|\tilde{X}|^2-(\tilde{X}+\tilde{X}^\dagger)M/\lambda+(M/\lambda)^2\right)/\Lambda^2$. This will result in a term linear in $\tilde{X}$ that shifts the minimum of the susy breaking vacuum to $\left\langle \tilde{X} \right\rangle=M/\lambda$. The susy preserving is at $\tilde{X}=0,\, \phi\bar{\phi}=F/\lambda$.
\\
\\
The fact that the susy breaking vacuum is a local minimum in the field space, with an unstable origin, makes these theories cosmologically doubtful.
The messengers carrying $SU(3)\times SU(2)\times U(1)$ quantum numbers get thermalized and, also, induce thermal masses on the spurion $X$ \cite{Dalianis:2010yk}. The unstable origin becomes the minimum of the finite temperature effective potential since the thermal masses compensate the tachyonic ones. For coupling $\lambda$ of order ${\cal O} (1)$ there is a second order phase transition towards the susy preserving vacua.

However, as we will demonstrate in the next section,  in the limit $\lambda \ll 1$ the thermal evolution radically changes and the selection of the metastable vacuum can take place naturally. The small coupling is necessary in order the thermal mass of the spurion to stay small and hence, the metastable vacuum to emerge from the thermal corrections at high temperatures. On the other hand, while $\lambda$ decreases, the messenger thermal masses cannot become arbitrary small thanks to the SSM degrees of freedom.  Thus, the messenger tachyonic masses are 'covered' by the thermal ones until lower temperatures.  Asking for a particularly small coupling between messengers and the spurion prompts us to check whether other interactions could alter this picture. Actually, only if the exact interactions of messengers and spurion fields are known one can trace the thermal evolution of the system. Below, we will briefly summarize some extensions of the minimal interactions (\ref{mm}) of $X$, $\phi$ and $\bar{\phi}$ with SSM fields.
 
Firstly, one can assume that there is a mixing of the messenger fields with ordinary fields. This could enhance further the thermal effects.
The messenger superfields $\bar{\phi}$ have the same quantum numbers as the ordinary, visible $\bar{d}$ superfields. The difference is that the former have couplings of the form $X \phi\bar{\phi}$ whereas, the later have Yukawa couplings, $QH_D \bar{d}$.  Thus, one can consider a simple modification that takes place in the Yukawa sector. In particular a messenger-matter mixing \cite{Dine:1996xk}
\begin{equation} \label{mixing}
H_D L_iY^l_{ij}\bar{e}_j+H_D Q_iY^d_{ij}\bar{d}_j
\end{equation}
with each of $L_i$ and $\bar{d}_i$ refers to the four objects with the same quantum numbers. The convention is that the $L_4$ and $\bar{d}_4$ are a linear combination of fields which couple to the spurion $X$. $Y^l$ is a $4\times 3$ matrix while $Y^{d}$ is a $3\times 4$ matrix, and the $Y^l_{4i}$ and $Y^d_{i4}$ are the "exotic" Yukawa couplings. The above messenger-matter mixing, if present can also contribute to the thermal mass squared of messengers with an additional $(|Y^{l}_{4i}|^2+|Y^{d}_{i4}|^2)T^2$ term in the effective potential. However, this mixing results in non-universal contributions to scalar masses and FCNC constraints 
the exotic Yukawa couplings to be weaker than the ordinary Yukawa couplings. 

Another possibility is that the messengers couple to the Higgs superfields in the superpotential  
\begin{equation} \label{Hq}
W=k' H_U\phi_1\phi_2+\bar{k}' H_D\bar{\phi}_1\bar{\phi}_2.
\end{equation}
This coupling was proposed in \cite{Dvali:1996cu} in order to generate a $\mu$-term at one-loop level. For $k'={\cal O}(1)$ this coupling can induce a significant thermal mass on messengers. 

On the other hand the gauge singlet $X$ may have direct couplings to SSM Higgs superfields  
\begin{equation} \label{HX}
W \supset \epsilon X H_U H_D  
\end{equation}
with a small coefficient $\epsilon$. This interaction was introduced in order to generate at tree level a $B_\mu$-term for the Higgs sector \cite{Dine:1995ag}. For low energy phenomenological reasons it has to be negligible small. If $\epsilon< \lambda $ then it is negligible in the finite temperature effective potential as well.

To sum up, the couplings in the case of (\ref{mixing}) are negligible, whereas the (\ref{Hq}) and (\ref{HX}) may be important and could modify the critical temperature of the phase transitions. In the next section we will present the thermal evolution of the spurion and messengers taking into account only the gauge vector fields which by definition are present and probably account for the most significant thermal contributions. We will comment on the possible effects of (\ref{Hq}) and (\ref{HX}) combined with the cosmological constraints in the conclusions.

\section{Thermal Evolution of OGM} 

If one neglects the SSM degrees of freedom, the thermalized system of fields evolves towards the susy preserving vacua \cite{Dalianis:2010yk}\footnote{Except if the system of fields, for $\lambda \ll 1$, is trapped close to the metastable vacuum after inflation \cite{Dalianis:2010yk}.}.
This makes perfect sense. Having only one coupling $\lambda$ to the thermal plasma messengers and spurion are equally influenced by the thermal equilibrium. The messengers, having tree level masses, are heavier than the spurion which receives a mass of quantum origin (either due to non-minimal K\"ahler or to Coleman-Weinberg corrections from the interaction with messengers). Therefore, higher temperatures are required to overwhelm the messenger tree level masses rather than the small 'quantum' mass of $X$.

Including the Standard Model gauge bosons introduces an extra contribution $gT$ to the messenger thermal masses but \itshape not \normalfont to the thermal mass of the spurion. Decreasing the coupling $\lambda$ the spurion thermal mass, $\lambda T$, is suppressed while the messengers' remains $gT$. Therefore, there is a threshold value of the coupling, $\lambda_\text{max}$, that below this value the phase transitions get inversed: the transition to the metastable susy breaking vacuum precedes the transition to the supersymmetric one. 

At finite temperature the fields that interact with the thermal plasma are no longer in their vacuum state. The occupation numbers $n_\bold{k}$ are given by the Bose-Einstein formula. The temperature dependent 1-loop effective potential is of the form \cite{Mukhanov:2005sc, Quiros:1999jp}  
\begin{equation}
V^T_1 \sim T^4 \int dx x^2 \ln \left(1 \pm \text{exp}\left(-\sqrt{x^2+M^2_i/T^2}\right)\right)
\end{equation}
where $M^2_i$ is an eigenvalue of the mass squared matrices. In the high temperature limit where $T$ is much greater than the mass eigenvalues the scalar potential reads 
\begin{equation} \label{hT}
\bar{V}^T_1(\phi_c)\simeq -\frac{\pi^2 T^4}{90} \left(N_B+\frac{7}{8}N_F\right)+\frac{T^2}{24}\left[\sum_{i}(M^2_S)_i+3\sum_{a}(M^2_V)_a+ \sum_{r}(M_F)^2_r\right]
\end{equation} 
where we have omitted the negligible terms linear in temperature. We are interested in how the thermal effects change the shape of the potential and in particular in the position of the high temperature minima and how they evolve as the temperature decreases. 
\begin{figure} 
\textbf{\,\,\,\,\,\,\,\,\,\,\,\,\,\, GRAVITATIONAL STABILIZATION \,\,\,\,\;\;\;\;\;\;\;\;\;}
\centering
\begin{tabular}{c}

\includegraphics [scale=1.2, angle=0]{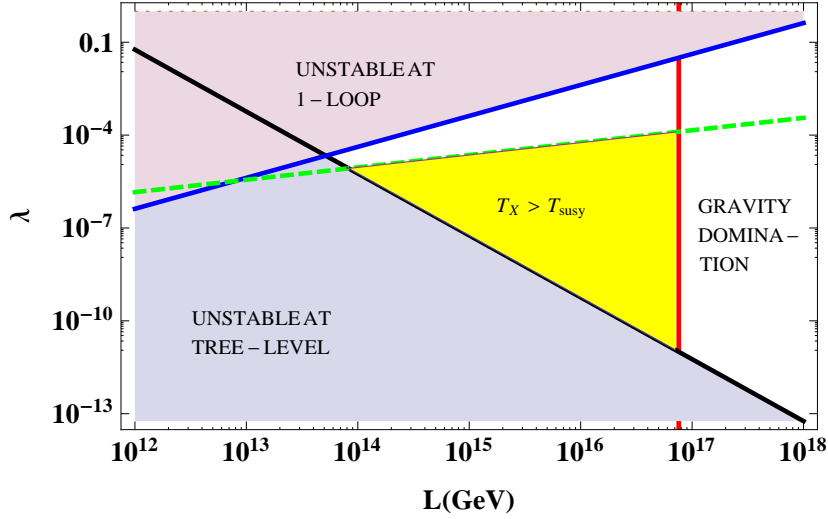} 
\end{tabular}
\caption{\small{The figure show the region of parameter space where the supersymmetry breaking minimum is metastable (white and yellow region) for the case of gravitational stabilization. In the yellow region, below the green dashed line, the thermal selection of the metastable vacuum is realized. The red line separates gauge mediation, $m_{3/2}<0.1 m_{\tilde{g}}$, from gravity mediation.}}
\end{figure}
\begin{figure} 
\textbf{\,\,\,\,\,\,\,\,\,\,\,\,\,\,\,\,\,\,\,\,\,\,\,\,\,\,\,\,\,\,\,\,\,\,\,\,\,\,\,\,\,\,\,\,\,\,\, GENERALIZED K\"AHLER \,\,\,\,\;\;\;\;\;\;\;\;\;}
\\
\\
\centering
\begin{tabular}{c}

\includegraphics [scale=1.2, angle=0]{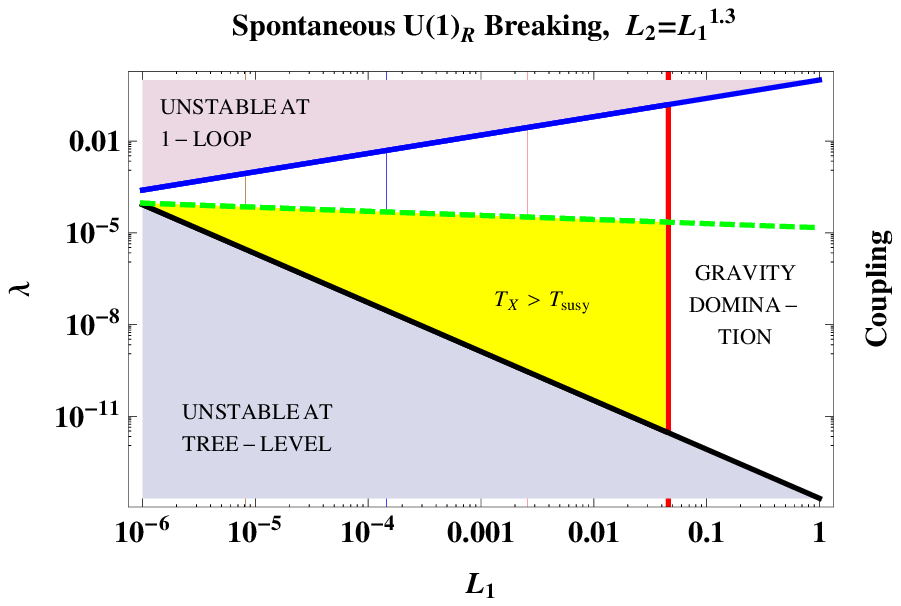} 
\end{tabular}
\caption{\small{The figure show the region of parameter space where the supersymmetry breaking minimum is metastable (white and yellow region) for the case of spontaneous $U(1)_R$ breaking due to 6th order correction to the K\"ahler. Here we have considered the case where $\Lambda_2=\Lambda_1^{1.3}$ in Planck units. In the yellow region, below the green dashed line, the thermal selection of the meta-stable vacuum is realized. The red line separates gauge mediation, $m_{3/2}<0.1 m_{\tilde{g}}$, from gravity mediation. The verical thin lines in the white region correspond to gravitino masses $1$, $10^{-2}$ and $10^{-4}$ GeV, from right to left. The $L$ stands for $\Lambda$.}}
\end{figure}
\begin{figure} 
\textbf{\,\,\,\,\,\,\,\,\,\,\,\,\;\;\;\;\;\;\;\;\;\; MESSENGER MASS}
\centering
\begin{tabular}{cc}

{(a)} \includegraphics [scale=.85, angle=0]{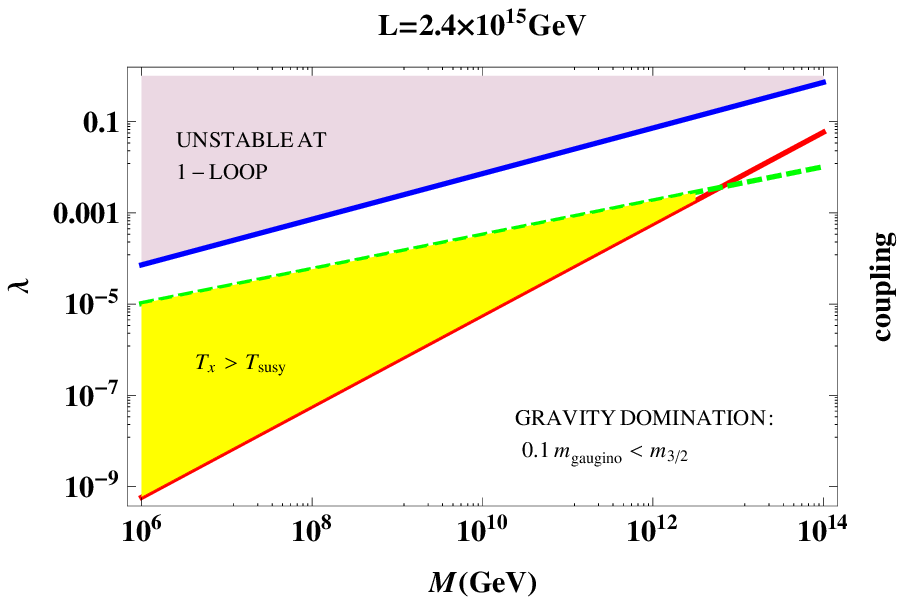} &
{(b)} \includegraphics [scale=.85, angle=0]{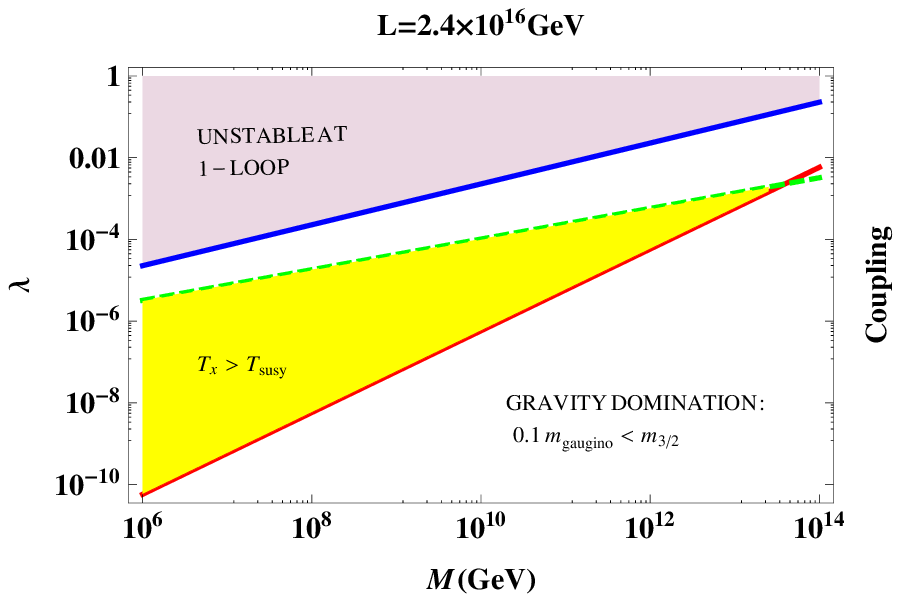}  \\
\end{tabular}
\caption{\small{The figures show the region of parameter space where the supersymmetry breaking minimum is metastable (white and yellow region) for the case of  messenger mass $M\phi\bar{\phi}$. In the yellow region, below the green dashed line, the thermal selection of the meta-stable vacuum is realized. The red line separates gauge mediation, $m_{3/2}<0.1 m_{\tilde{g}}$, from gravity mediation. In the panel (a) the K\"ahler correction scale is fixed at $\Lambda=2.4\times 10^{15}$ GeV and in the (b) $\Lambda=2.4\times 10^{16}$ GeV}}
\end{figure}
In principle we should include the $D$-terms in the scalar potential i.e. $V=V_F+V_D$.
The $D$-term contribution to the sfermion masses can vanish if one imposes the "messenger parity" proposed in \cite{Dimopoulos:1996ig} -but not for the superpotential (\ref{mm2}).
Although the $D$-terms may be problematic for the low energy phenomenology they do not change essentially the thermal evolution of the fields. 

The messenger superfields $\phi+\bar{\phi}$ can be decomposed into colour triplets $q+\bar{q}$ and weak doublets $\ell+\bar{\ell}$ which couple to the spurion like 
\begin{equation} \label{ql-6}
\lambda_q X q \bar{q}+\lambda_\ell X \ell \bar{\ell}\, ,
\end{equation}
where we considered the general case of different $\lambda_\ell$ and $\lambda_q$. In the case of unification, e.g. when the messengers transform in the $\bold{5}+\bold{\bar{5}}$ representations of $SU(5)$ one has $\lambda_\ell=\lambda_q=\lambda$. We explicitly write it in this way instead of the compact form (\ref{mm}) because the doublets $\ell +\bar{\ell}$ are coupled to the thermal plasma weaker than the colour triplets $q+\bar{q}$. 

At high temperatures the thermal masses squared compensate the negative ones and the effective minimum lies at the origin\footnote {Except if there are $U(1)_R$ violating terms like the constant $c$ or messenger mass terms of the form $M\phi\bar{\phi}$ in the superpotential. Then, the thermal minimum is at $X\neq 0$.} of field space. 
Apart from the spurion\footnote{For weak coupling $\lambda_{\ell,\,q}$ the spurion is actually out of equilibrium \cite{Dalianis:2010yk}.} and the self-coupling of the messengers, the gauge bosons induce thermal masses for $\ell +\bar{\ell}$ and $q+\bar{q}$. The interaction between the observable gauge bosons and the messengers is identified in the kinetic terms. For the scalar messenger fields the covariant derivatives read
\begin{equation}
D_\mu \ell= \partial_\mu \ell - i g_2 W^a_\mu \frac{\tau ^a}{2} \ell - i y_\ell \frac{g_1}{2}B_\mu \ell
\end{equation}
and
\begin{equation}
D_\mu q= \partial_\mu q - i g_3 G^a_\mu \frac{\tilde{\lambda}_{a}}{2}q - i y_q \frac{g_1}{2}B_\mu q
\end{equation} 
with $y_{\ell,q} =-1,-2/3$ and $y_{\bar{\ell},\bar{q}}=1, 2/3$. The triplets $q +\bar{q}$ couple to the thermal plasma mainly via the strong gauge coupling $g_3$. 
The doublets $\ell+\bar{\ell}$ couple with $g_2$ and $g_1/2$ of $SU(2)$ and $U_Y(1)$ respectively.  At high energy $g_3$ runs weaker and $g_2$, $g_1$ stronger. Using the one loop $\beta$-function the solution to the renormalization group equation of gauge coupling strengths is given by
\begin{equation} \label{run-6}
\frac{1}{g^2_i(T)} \simeq \frac{1}{g^2_i(m_Z)}- \frac{b_i}{8\pi^2} \ln \left(\frac{T}{m_Z} \right),
\end{equation}
with $b_1=11$, $b_2=1$ and $b_3=-3$. 
For the temperatures discussed here we consider the approximate values $g^2_3 \sim 4\pi/17$, $g^2_2\sim 4\pi/28$ and $g^2_1\sim 4\pi/43$.

In the vicinity of the origin i.e., $q=\bar{q}=X=0$, the relevant terms in the temperature corrected scalar potential for the doublets $\ell+\bar{\ell}$ are
\begin{equation}
V \supset -\lambda_\ell F\left(\ell \bar{\ell}+\text{h.c.}\right) + \lambda^2_\ell |\ell \bar{\ell}|^2+ \frac{T^2}{24}\left[\left(6\lambda^2_\ell+\frac{9}{2}g^2_2+\frac{3}{2}g^2_1\right)\left(|\ell|^2+|\bar{\ell}|^2\right)\right].
\end{equation}
In the ${\cal O}(T^2)$ part of the potential the contribution of the fermions and the spurion to the thermal masses of $\ell$,$\bar{\ell}$ has been taken into account. A decoupled spurion would decrease the coefficient in front of the $\lambda_\ell$ by a factor of 3. We recall that in the above expression the $\ell$,$\bar{\ell}$ refer to the thermal average values since the potential is corrected by the temperature dependent part. 
The critical temperature, i.e. the temperature when the mass squared at the origin turns from positive to negative can be seen easier after a diagonalization of the mass matrix. We rotate the fields to $L_1= (\bar{\ell}^\dagger+\ell)/\sqrt{2}$ and $L_2=(\bar{\ell}-\ell^\dagger)/\sqrt{2}$ and the mass terms in the potential are transformed to
\begin{equation}
V \supset -\lambda_\ell F \left(|L_1|^2-|L_2|^2\right) + \frac{T^2}{24}\left(6\lambda^2_\ell+\frac{9}{2}g^2_2+\frac{3}{2}g^2_1\right) \left(|L_1|^2+|L_1|^2\right). 
\end{equation}
The direction $L_1$ becomes tachyonic at temperature 
\begin{equation} \label {Tcr1}
T^\ell_{\text{susy}} = 4\sqrt{\frac{\lambda_\ell F}{3g^2_2+g^2_1+4\lambda^2_\ell}} \, .
\end{equation}
At this temperature, to a good approximation, a second order phase transition towards the supersymmetric vacuum takes place. In the case that $\lambda_\ell \sim 1$ so, $\lambda_\ell >g_2$ the critical temperature will be 
\begin{equation}
T^\ell_\text{susy} \simeq 2\sqrt{\frac{F}{\lambda_\ell}}, \,\,\,\,\,\,\,\,\,\,\,\,\, \text{for $\lambda_\ell \sim 1$}.
\end{equation}
We see that a small coupling, $\lambda_\ell \ll g_1$, between the spurion and the messenger fields decreases the critical temperature $ \sqrt{(3g^2_2+g^2_1)/4\lambda^2_\ell}$ times compared to the case of negligible gauge bosons contribution.  The values of gauge couplings, as mentioned above, are $(3g^2_2+g^2_1) \simeq 1.64$ hence, for $\lambda_\ell \ll 1$ the decrease can be significant.

In the case that the $D$-terms don't vanish the relevant potential reads
\begin{equation}
V=V_F+\frac{1}{2}g^2_2\left(\ell^\dagger\frac{\vec{\tau}}{2}\ell+\bar{\ell}^\dagger\frac{\vec{\tau}}{2}\bar{\ell}\right)^2+\frac{1}{2}\left(\frac{g^2_1}{2}\right)^2\left(\ell^\dagger \ell- \bar{\ell}^\dagger \bar{\ell} \right)^2.
\end{equation}
The effective (thermal) masses of $\ell$, $\bar{\ell}$ will obtain an extra contribution, but it is of the same order of magnitude as the previous one and the critical temperature is not essentially changed.

Following the same steps for the triplets $q+\bar{q}$ it is straightforward one to see that the critical temperature in this case is
\begin{equation}
T^q_\text{susy} = 4\sqrt{\frac{\lambda_q F}{8g^2_3+(4/9)g^2_1+4\lambda^2_q}}.
\end{equation}
For weak couplings, $\lambda_q$,$\lambda_\ell \ll 1$, the critical temperature for the triplets $q+\bar{q}$ is lower than the critical temperature ($\ref{Tcr1}$) for the doublets $\ell+\bar{\ell}$ provided that $\lambda_q/\lambda_\ell<(8g^2_3+(4/9)g^2_1)/(3g^2_2+g^2_1)$. Plugging in the values of the running gauge couplings for temperatures of the order $T\sim 10^{9} $ GeV the previous condition reads $\lambda_q/\lambda_\ell < 5$. Hence, considering $\lambda_q\sim\lambda_\ell$, the first (larger) critical temperature for the transition to the susy vacua is the one for the doublets. Hereafter we will assume the (\ref{Tcr1}) as the critical temperature for the system of fields. 

The metastable susy breaking vacua appear at temperature $T_X$. The exact value depends on the way the spurion $X$ is stabilized. We consider separately the cases of stabilization with and without gravity.

\subsection{Gravitational Stabilization}

\subsubsection{Gravitational Gauge Mediation}

A minimal model is the one described in \cite{Kitano:2006wz} with $W=FX-\lambda X\phi \bar{\phi}+c$ and $K=|X|^2-|X|^4/\Lambda^2$. At zero temperature the origin is unstable in the direction of messengers for $|X|<\sqrt{F/\lambda_{\ell,\, q}}$ ; at temperatures $T>T_\text{susy}$ messengers thermal masses overtake the tachyonic ones. The spurion receives a thermal mass of order $\lambda_\ell T$ and $ \lambda_q T$ from doublets and triplets that stabilize $X$ close to zero. As the temperature decreases the thermal effects weaken and the minimum in the $X$-direction shifts towards the zero temperature value. The moment that it exits the (would-be at $T_\text{susy}$) tachyonic region, i.e. $X>\sqrt{F/\lambda_{\ell,\,q}}$, 
the metastable vacuum forms \cite{Dalianis:2010yk}. This takes place at temperature squared\footnote{For temperatures $T > T_\text{susy}$ there is single global minimum of the finite temperature effective potential. There are no tachyonic directions. In the case that $T_X > T_\text{susy}$ at the temperature $T_X$ the "would-be metastable" minimum forms; hence, initially the minimum in the $X$-direction is global and at $T_\text{susy}$ it becomes local i.e. metastable, but it never becomes unstable. It would become unstable only if $T_X<T_\text{susy}$.}
\begin{equation}
{T^\ell_X}^2 \simeq 8  \frac{c \sqrt{F}}{(2\lambda^2_\ell+3\lambda^2_q) M^2_P}\sqrt{\lambda_\ell}\,\, , \,\,\,\,\,\,\,\,\,\,\,\,\,\,\,\, {T^q_X}^2 \simeq  8\frac{c \sqrt{F}}{(2\lambda^2_\ell+3\lambda^2_q) M^2_P}\sqrt{\lambda_q}
\end{equation}
for doublets and triplets respectively. Considering $\lambda_\ell \simeq \lambda_q =\lambda$ the $T_X$ temperature reads
\begin{equation} \label{T_X}
T^2_X \simeq \frac{8}{5} \frac{c}{\lambda M^2_P}\sqrt{\frac{F}{\lambda}}
\end{equation}
with $c=FM_P/\sqrt{3}= m_{3/2}M^2_P$ for vanishing cosmological constant in the metastable vacuum.  
The $T_X$ can be larger than $T_\text{susy}$ for small coupling $\lambda$, namely
\begin{equation}
\lambda< \left(\frac{3g^2_2+g^2_1}{10}\frac{\sqrt{F}}{\sqrt{3}M_P}\right)^{2/5} \simeq \left(0.16\frac{\sqrt{F}}{\sqrt{3}M_P}\right)^{2/5}.
\end{equation} 
To present an example for $\sqrt{F} =2.4\times 10^9$ GeV  ($m_{3/2} \simeq 1$ GeV) the coupling has to be lower than $\lambda \lesssim 1.0 \times 10^{-4}$ and for $\sqrt{F} =2.4 \times 10^8$ GeV  ($m_{3/2}\simeq 10^{-2}$ GeV), $\lambda \lesssim 3.9 \times 10^{-5}$. The scaling of temperatures $T_\text{susy}$ and $T_X$ is demonstrated in the figure 6.4. We recall that gaugino mass of the order of ${\cal O}(100)$ GeV relates the parameters $F$ and $\Lambda$ according to $F\simeq  10^{-14}\left\langle X \right\rangle M_P$. We also remind the reader that the coupling $\lambda$ cannot become arbitrary small or large  because of the constraints from the zero temperature stability conditions on the susy breaking vacuum: $10^{-14}(\Lambda/M_P)^{-2}<\lambda<\Lambda /M_P$ which are illustrated in figure 6.1.

In the case of gravitational stabilization, at $T_X$ there is no  phase transition; only a smooth shift of the vacuum to larger values. Hence, the system of fields lands at the metastable vacuum if the effective mass of the spurion $X$ is sufficiently larger than the Hubble scale. Following \cite{Linde:1996cx} we assume that when $M_X>30H$ the $X$ field tags along the position of the temperature dependent minimum and its oscillations are efficiently damped. In a radiation dominated phase $H=0.33 g^{1/2}_*T^2/M_P$ and $M_X \simeq \lambda T/\sqrt{2}$ as one finds from the finite temperature potential. This gives a lower limit on the ratio $\lambda/T>30\times 0.33\sqrt{2}g^{1/2}_* M^{-1}_P ={\cal O}(100) M^{-1}_P$. Otherwise, the spurion either stays frozen (when $M_X<H$) to its postinflationary value until lower temperatures or its oscillations about the metastable vacuum are not efficiently damped (when $H<M_X<30H$). For the values of coupling considered here, i.e. $ 10^{-8} \lesssim \lambda \lesssim 10^{-4}$, the limit on the ratio $\lambda/T$ is not problematic.

\subsection{Global Limit}

\subsubsection{Higher Order K\"ahler Corrections}
For the case of 6th order corrected K\"ahler function (\ref{6})
the metastable vacuum survives in the global limit $M_P \rightarrow \infty$. We consider again $\lambda_\ell \simeq \lambda_q =\lambda$. The metastable vacua appear at \cite{Dalianis:2010yk}
\begin{equation} \label{gl-1}
T_X= \frac{4}{\sqrt{5}} \frac{1}{\lambda} \frac{F}{\Lambda_1}.
\end{equation}
neglecting gravity. Also here, we see that decreasing $\lambda$ increases the temperature $T_X$. For 
\begin{equation}
\lambda< \left( \frac{3g^2_2+g^2_1}{5} \frac{F}{\Lambda^2_1}\right)^{1/3} \simeq \left(0.33\frac{F}{\Lambda^2_1}\right)^{1/3}
\end{equation}
$T_X$ is larger than $T_\text{susy}$ and there is a second order phase transition to the metastable vacuum. Supersymmetry and $U(1)_R$ break spontaneously. For $\sqrt{F} =2.4 \times 10^8$ GeV and $\Lambda_1=2.4\times 10^{14}$ GeV the coupling has to be smaller than  $\lambda \lesssim 6.9 \times 10^{-5}$ and for $\sqrt{F} =2.4 \times 10^{7.5}$ GeV and $\Lambda_1=2.4\times 10^{13}$ GeV we take $\lambda \lesssim 1.5 \times 10^{-4}$.

\subsubsection{Messenger Mass} \normalfont
In the case that there is an extra messenger mass term $M\phi\bar{\phi}$ in (\ref{mm}) with K\"ahler $K=|X|^2-|X|^4/\Lambda^2$ susy breaks down at $\phi=\bar{\phi}=X=0$ while the susy preserving minimum lies at $X=-M/\lambda, \phi\bar{\phi}=F/\lambda$, for  $\lambda_\ell \simeq \lambda_q =\lambda$. With a field transformation $X\rightarrow \tilde{X}=X+M/\lambda$ the vacua switch positions along the $X$-axis. The potential, then, has a form similar to the potential of the gravitational stabilization. Following the same steps, we find that the temperature at which the metastable vacuum exits the tachyonic region $|\tilde{X}|<\sqrt{F/\lambda}$ is 
\begin{equation} \label{TM}
T^2_{{X}} \simeq \frac{16}{5}\frac{FM}{\lambda^2\Lambda^2}\sqrt{\frac{F}{\lambda}}
\end{equation}
which is of course the same for $X$ and $\tilde{X}$. The temperature (\ref{TM}) is the analogue of (\ref{T_X}) with the correspondence $c/M^2_P \rightarrow 2 FM/(\lambda \Lambda^2)$. Fixing the gaugino mass to be of the order ${\cal O}(100)$ GeV, gives $F\simeq 10^{-14} M_P \left\langle \tilde{X} \right\rangle =10^{-14}M_PM/\lambda$. The main difference, is that here one has three parameters ($M,\Lambda, \lambda$) instead of two ($\Lambda, \lambda$) of the gravitational stabilization. The fact that the messengers have explicit mass $M$ that doesn't depend on the coupling $\lambda$ changes the behaviour of the critical temperature of the transition towards the susy vacua. Namely, from the last relation we take that $\lambda F \simeq 2.4 \times 10^{4}M$ GeV and the critical temperature (\ref{Tcr1}) reads
\begin{equation}
T^\ell_\text{susy} \simeq  10^{7} \, \text{GeV}\,\left(\frac{M}{2.4\times 10^{8}\,\text{GeV}}\right)^{1/2}\left(\frac{1}{3g^2_2+g^2_1}\right)^{1/2}.
\end{equation}
For a given mass $M$ it has a fixed value. The $T_X$ is larger than $T^\ell_\text{susy}$ for 
\begin{equation} \label{messM-5}
\lambda< \left(0.41 \times 10^{-2} \frac{M^2}{\Lambda^2}\right)^{1/4}\left(\frac{2.4 \times 10^{8}\,\text{GeV}}{M}\right)^{1/8}.
\end{equation}
Hence, for messenger mass $M=2.4\times 10^{8}$ GeV and cut-off scale $\Lambda=2.4\times 10^{15}$ GeV the transition to the susy breaking vacuum takes place first if $\lambda \lesssim 8.0 \times 10^{-5}$; for $M=2.4\times 10^{6}$ GeV, $\Lambda=2.4\times 10^{15}$ GeV if $\lambda \lesssim 1.4 \times 10^{-5}$, see figure 6.4. The coupling $\lambda$ cannot become arbitrary small because gravity contributions to the soft masses start to dominate, see figure 6.3. Note that, here, the vev of the spurion is $\left\langle \tilde{X}\right\rangle=M/\lambda$ and the coupling $\lambda$ is a free parameter. Hence, for fixed gaugino masses the gravitino mass will scale like 
\begin{equation}
m_{3/2} \simeq \left(\frac{M}{2.4\times 10^{8}\,\text{GeV}}\right)\left(\frac{10^{-6}}{\lambda}\right)\,\, \text{GeV}.
\end{equation} 

We recall here the zero temperature constraints that render the susy breaking vacuum metastable: $\lambda F < M^2$ and $\lambda^2<4\pi M/\Lambda$, see figure 6.3.

\subsubsection{Canonical K\"ahler}

Finally, for the case of canonical K\"ahler, if there is e.g. a double set of messengers with $\delta W =m \phi_1 \bar {\phi}_2$ (\ref{mm2}) which have exotic $R$-charges both $U(1)_R$ and supersymmetry can break down spontaneously via a second order phase transition \cite{Cheung:2007es, Shih:2007av}. Although there are similarities with the case of K\"ahler corrected up to 6th order, here the stabilization of the spurion is basically different. It is due to the perturbative quantum correction coming from the interaction of the messengers with the $X$ field. The Coleman-Weinberg potential has a higher order dependence on the coupling i.e. $\lambda^2F^2$ to leading order in $F^2$. It is not straighforward to see from the effective potential which is not of a polynomial type the critical temperature analytically. The mass squared of the spurion scales like $\lambda^4 F^2/m^2$ and hence we expect that the $T_X$ takes the approximate form 
\begin{equation}
T_X \sim \frac{\lambda F}{m}.
\end{equation}
\begin{figure} 
\textbf{\,\,\,\,\,\,\,\;\;\;\;\;\;\;\;\;\;\;\,\,\,\,\,\,\,\,\,\, GRAVITATIONAL STABILIZATION \,\,\,\,\,\,\,\,\,}
\\
\\
\centering
\begin{tabular}{cc}
{(1)} \includegraphics [scale=.85, angle=0]{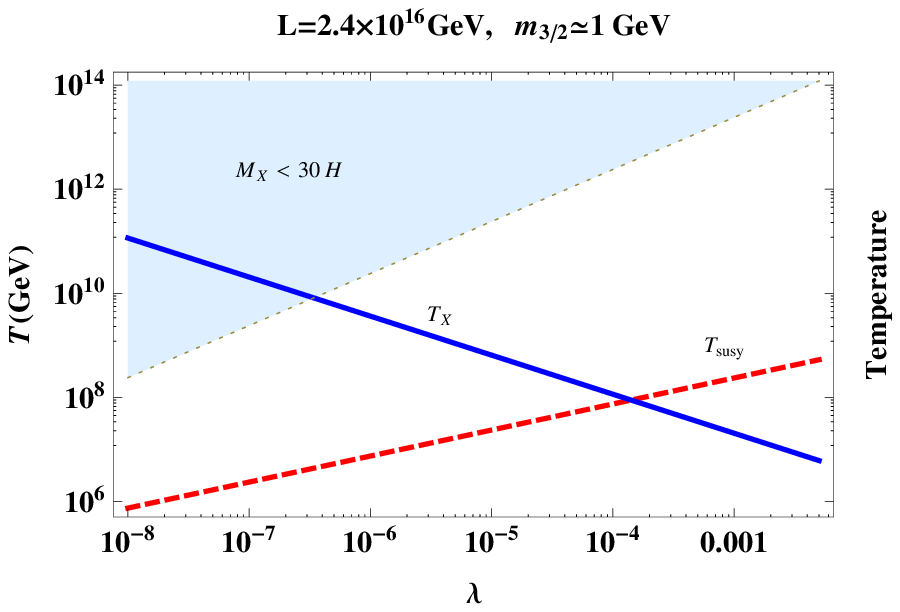} &
{(2)} \includegraphics [scale=.85, angle=0]{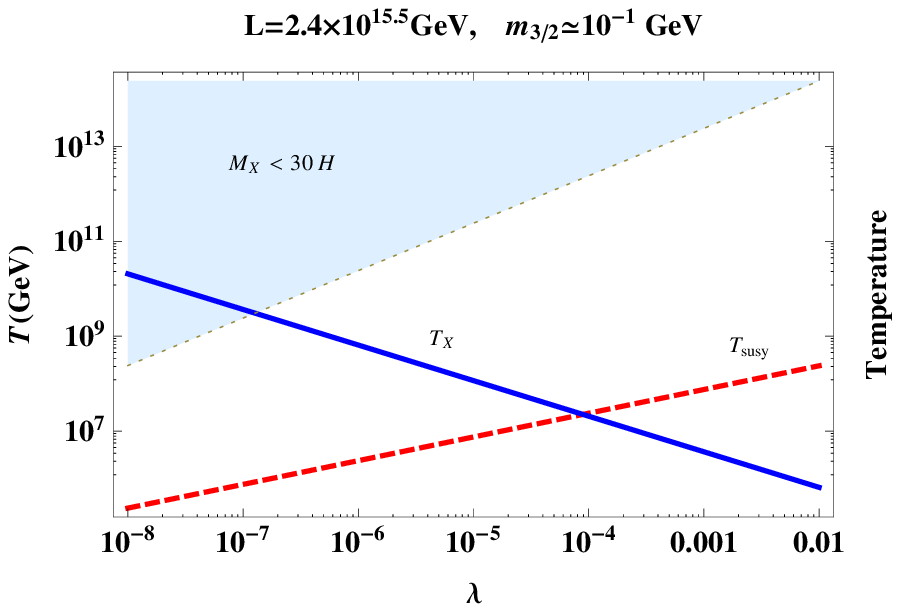}  \\
\end{tabular}

\begin{tabular}{cc}
{(3)} \includegraphics [scale=.85, angle=0]{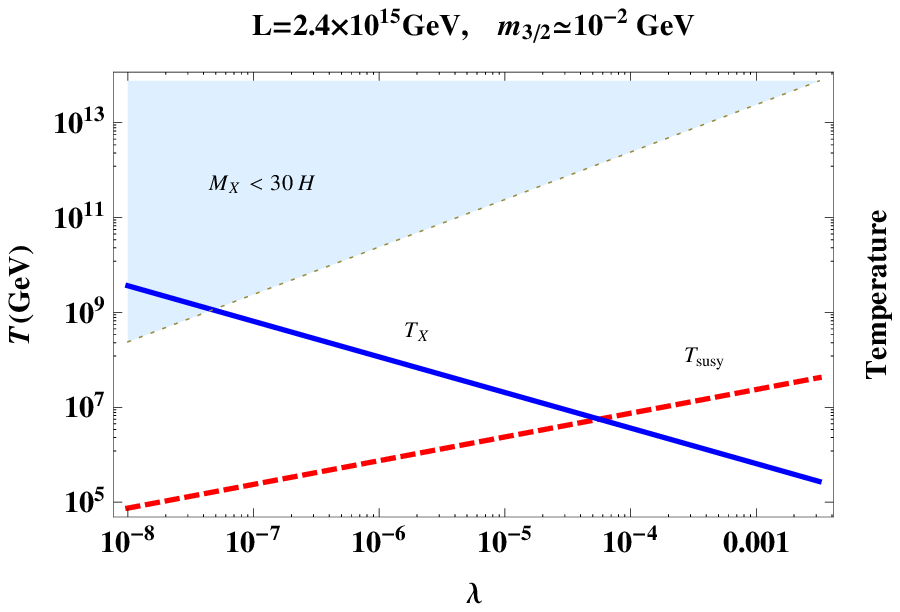} &
{(4)} \includegraphics [scale=.85, angle=0]{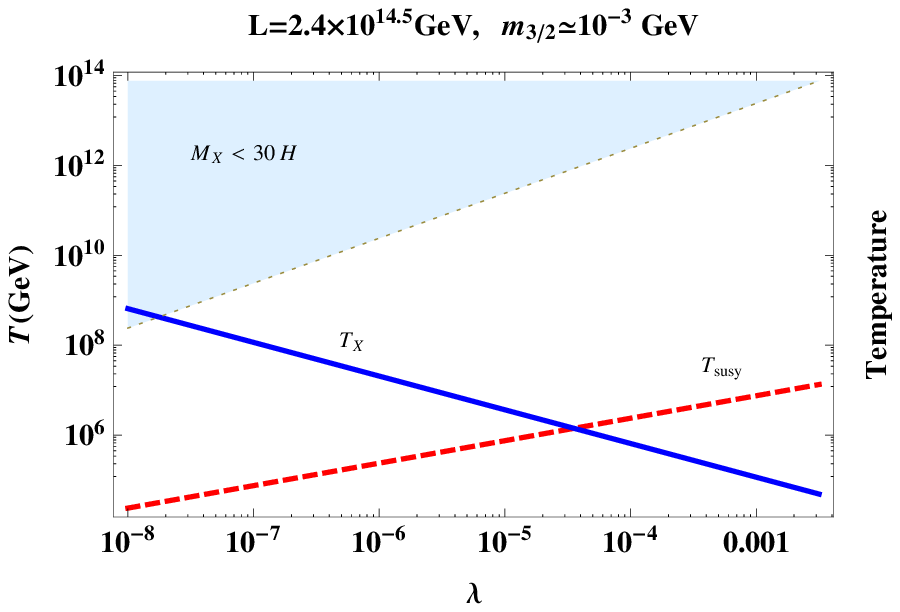}  \\
\end{tabular}
\caption{\small The plots show how the critical temperature $T_\text{susy}$ (red dashed line) of the transition to the susy vacua and the temperature $T_X$ (blue line) of the transition to the metastable vacuum scale with the coupling $\lambda$, for the case of gravitational stabilization. It demonstrates that as the coupling decreases the $T_X$ becomes larger than the $T_\text{susy}$. In the white region there is an efficient damping of the spurion oscillations thanks to a large enough thermal mass i.e. $M_X>30H$. We consider $\lambda=\lambda_\ell=\lambda_q$.  ($L\equiv\Lambda$).} 
\end{figure}

\begin{figure} 
\textbf{\,\,\,\,\, \,\,\,\,\,\,\,\,\,\,\,\,\,\,\,\,\,\,\,\;\;\;\;\;\;\;\;\;\;\; MESSENGER MASS}
\\
\\
\centering
\begin{tabular}{cc}
{(a1)} \includegraphics [scale=.85, angle=0]{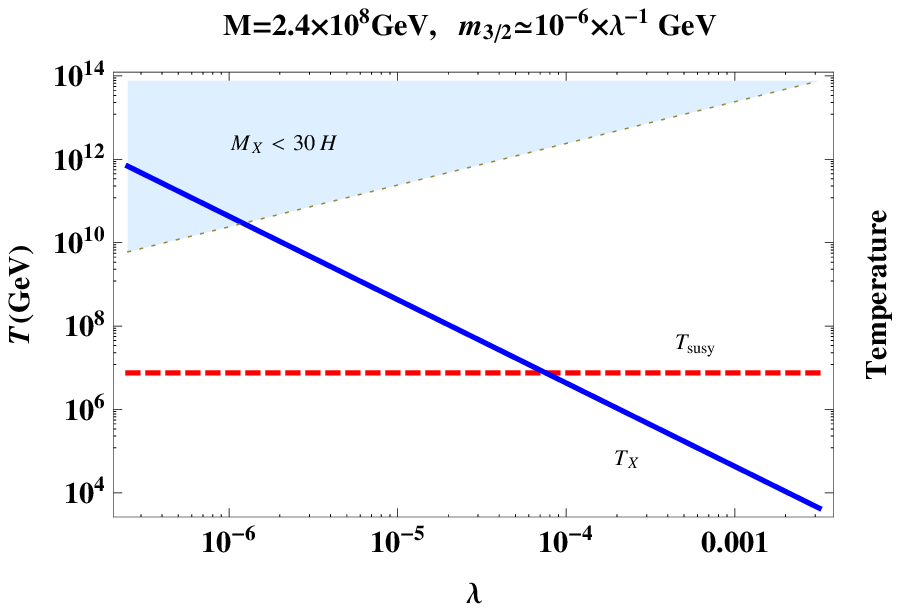} &
{(b1)} \includegraphics [scale=.85, angle=0]{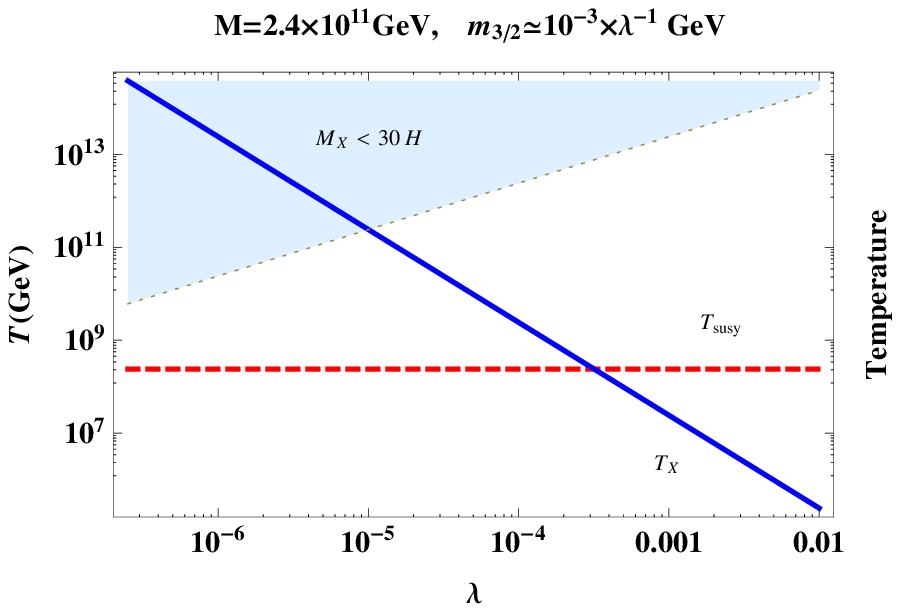}  \\
\end{tabular}
\begin{tabular}{cc}

{(a2)} \includegraphics [scale=.85, angle=0]{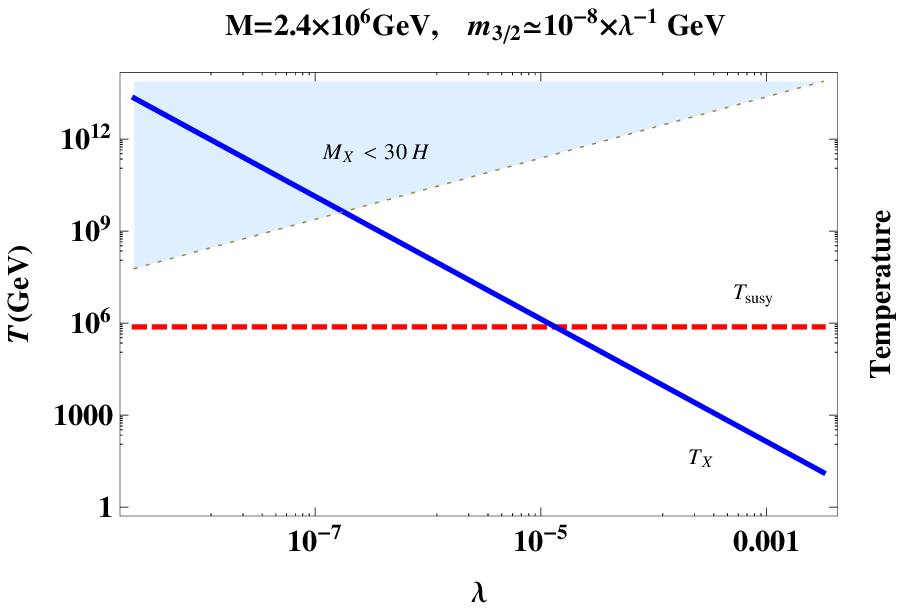} &
{(b2)} \includegraphics [scale=.85, angle=0]{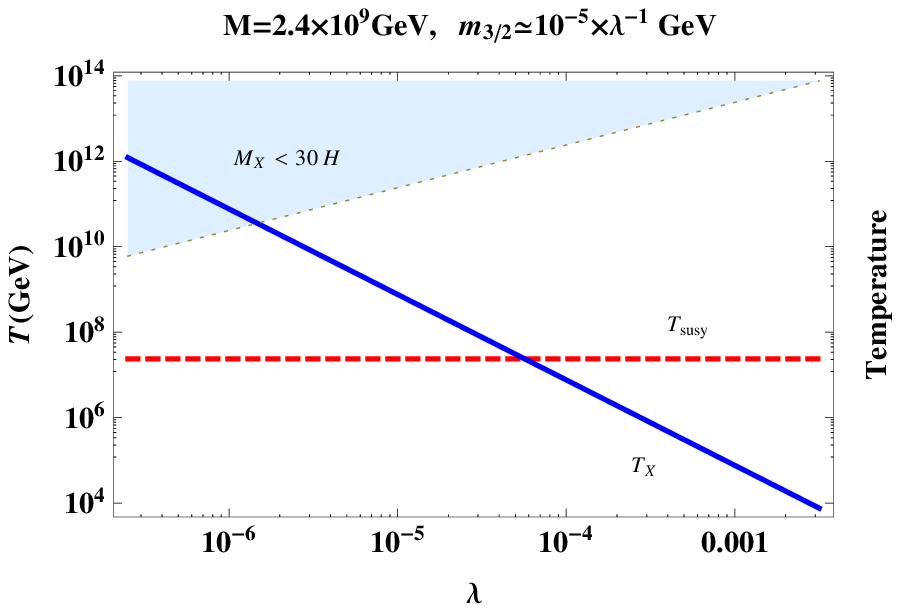}  \\
\end{tabular}
\caption{\small The plots show how the critical temperature $T_\text{susy}$ (red dashed line) of the transition to the susy vacua and the temperature $T_X$ (blue line) of the transition to the metastable vacuum scale with the coupling $\lambda$, for the case messenger mass $M\phi\bar{\phi}$. It demonstrates that as the coupling decreases the $T_X$ becomes larger than the $T_\text{susy}$. In the white region there is an efficient damping of the spurion oscillations thanks to a large enough thermal mass i.e. $M_X>30H$. We consider $\lambda=\lambda_\ell=\lambda_q$ and $\Lambda=2.4\times 10^{15}$ GeV for the (a) panels and $\Lambda=2.4\times 10^{16}$ GeV for the (b) panels.} 
\end{figure}

Decreasing the coupling $\lambda$ also decreases the $T_X$ and generally it cannot get larger that $T_\text{susy}$. Weak $\lambda$ means even weaker stabilization of the spurion i.e. smaller mass. Therefore, models of ordinary gauge mediation with canonical K\"ahler where the susy vacuum is stabilized due to the interactions with the messengers (minimal UV completion) cannot become thermally favourable. 

In the first case of the global limit the spurion is stabilized by the corrections in the K\"ahler function. These corrections have also a perturbative quantum origin but they come from the interaction of the spurion with the integrated out degrees of freedom, e.g. with the heavy raifeartons with different coupling $k_o$. So in that case, decreasing the $\lambda$ doesn't change the mass of the spurion.

\section{Features of Thermally Favourable Gauge Mediation} \normalsize

We have shown that metastable susy breaking vacua of ordinary gauge mediation with non-vanishing gaugino masses can be thermally selected.
The thermal selection favours a small coupling between the messengers and the spurion and it can be realized for generic initial vevs of the fields. 

The stronger a field is coupled to the thermal plasma the larger is the 'thermal screening' effect on the tree level parameters. Decreasing the coupling makes the zero temperature potential to dominate quickly over the finite temperature corrections. In the case of messengers, SM gauge bosons don't let the thermal mass to drop below $gT$. On the other hand, the spurion being coupled with the coupling $\lambda$ feels only slightly the thermal effects  if $\lambda$ is small enough. Hence, the spurion zero temperature potential can emerge at higher temperatures than the tree level potential of messengers (which is responsible for the tachyonic origin). The conclusion is that the temperature $T_X$ at which the metastable susy breaking vacuum appears can be larger than the critical temperature $T_\text{susy}$ of the transition towards the susy vacuum. This happens in models where the spurion is stabilized due to K\"ahler corrections. Therefore, the spurion zero temperature mass is unaffected by decreasing $\lambda$  because it originates from the interaction with integrated out heavy fields and not from the interaction with messenger fields. A coupling $10^{-8} \lesssim \lambda  \lesssim 10^{-4}$ can make the metastable vacuum thermally favourable for gravitino with ${\cal O}(10^{-3}-1)$ GeV mass. 

We note that such small values of the coupling $\lambda$ are reasonable if the $X$ field is a composite operator above the scale $\Lambda$ as is often the case in dynamical supersymmetry breaking scenarios. Then $\lambda $ is suppressed by a factor of $(\Lambda/M_P)^{d(X)-1}$ where $d(X)$ the dimension of the operator $X$ above the scale $\Lambda$ \cite{Ibe:2007km}; see also the next subsection. Small values of $\lambda$ imply stabilization of the spurion at relatively large vevs, hence, the gravitino mass lying in the GeV range.

Let us note that if the messengers have a Yukawa coupling $k$ to SSM fields, e.g. to Higgses like in (\ref{Hq}), and $k'>g_2$ the thermal mass of messengers is further enhanced. This can relax the upper bound on the coupling $\lambda$ for the thermal selection of the metastable vacuum. On the the other hand, we ask for a weakly interacting spurion, which implies that $X$ should not directly couple to any observable field, like in (\ref{HX}), with a coupling $\epsilon>\lambda$.

The selection of the metastable susy breaking vacuum takes place thermally. The oscillations of the spurion are efficiently damped and thus there is no late entropy production. The reheating temperature has to be high enough in order that the messengers to get thermalized and also, the system of fields to get localized in the origin. For small coupling $\lambda$  and temperatures higher than about ${\cal O}(10^7-10^9)$ GeV the thermal selection of the metastable vacuum is realized. It is interesting to note that leptogenesis scenarios can be accommodated in these ordinary gauge mediation models.

\subsection{A UV completed Example}
Small values for the coupling $\lambda$ are natural in several models. Especially, it characterizes models where the spurion $X$ is a composed particle in microscopical level. For completeness, we present an example that a small coupling is natural consequence of the hidden sector supersymmetry breaking dynamics \cite{Murayama:2006yf}. A supersymmetric QCD $SU(N)$ with massive vector quarks $Q^i$ and $\bar{Q}^i$ ($i=1,..., N_f$) and massive messengers $\phi$, $\bar{\phi}$:
\begin{equation}
W_\text{tree}= m_{ij} Q_i\bar{Q}^j + \frac{\bar{\lambda}_{ij}}{M_P} Q^i \bar{Q}^j \phi\bar{\phi}+M \phi\bar{\phi}.
\end{equation}
When $ N+1 <N<\frac32 N$ the magnetic dual $SU(N_f-N)$ breaks supersymmetry on a metastable local minimum if the quark masses $m_{ij}$ are much smaller than the dynamical scale $\bar{\Lambda}$ \cite{Intriligator:2006dd}. At energies below the dynamical scale the superpotential is described by
\begin{equation}
W_{tree}= \frac{1}{\bar{\Lambda}^{2N_f-3}} \left(\bar{B}_i M^{ij}B_j -\det M^{ij} \right) +\left(\frac{\bar{\lambda}_{ij}}{M_P} M^{ij}+ M \right) \phi\bar{\phi}
\end{equation}
where $M^{ij}=Q^i \bar{Q}^j$ a meson field and $B_i= \epsilon_{ii_1...i_N}Q^{i_1}...Q^{i_N}/N!$, $\bar{B}_i= \epsilon_{ii_1...i_N}\bar{Q}^{i_1}...\bar{Q}^{i_N}/N!$ the baryon and antibaryon chiral superfields respectively. After redefinition of the meson, baryon and antibaryon fields to fields with canonical dimenions  $X^{ij}=M^{ij}/\bar{\Lambda}$, $b_i=B_i/\bar{\Lambda}^{N_f-2}$ and $\bar{b}_i=\bar{B}_i/\bar{\Lambda}^{N_f-2}$ the low energy superpotential can be written as
\begin{equation}
W_{low}= X^{ij}b_i\bar{b}_j-m_i\bar{\Lambda} X^{ii}-\frac{\det X^{ij}}{\bar{\Lambda}^{N_f-3}}+\frac{\bar{\lambda}_{ij}\bar{\Lambda}}{M_P}X^{ij}\phi\bar{\phi}+M\phi\bar{\phi}.
\end{equation}
For $N_f>3$ the superpotential term $\det X^{ij}$ is irrelevant and can be ignored to discuss physics around the origin $X^{ij}=0$. The above superpotential can be rewritten in accordance with (\ref{gen-5})
\begin{equation}
W=F_iX^{ii}+k X^{ij}\varphi_i\bar{\varphi}_j + \left( \lambda_{ij} X^{ij} +M \right) \phi \bar{\phi}
\end{equation}
where the coupling to messengers $\lambda_{ij}=\bar{\lambda}_{ij}\bar{\Lambda}/M_P$. We note here that if the ISS model \cite{Intriligator:2006dd} is the microscopic description of the sypersymmetry breaking sector then for $\varphi, \bar{\varphi} \neq 0$ the supersymmetry is broken due to the rank condition. However, $\phi$, $\bar{\phi}$ restore supersymmetry.  An example of natural values for the model are $\bar{\lambda}_{ij} \sim 1 $, $\bar{\Lambda} \sim 10^{11}$ GeV the effective low energy coupling is $\lambda \sim 10^{-7}$.  Therefore, the metastable vacuum according to (\ref{messM-5}) is thermally preferred.
\\
\\
We believe that this cosmological constraint on the coupling can be a guide for the hidden sector gauge mediated susy breaking model building.

\chapter{Stringy Moduli Stabilization And Dynamical SUSY Breaking}

In this chapter the basic topic is the stabilization schemes for the stringy moduli fields. An introduction to some notions of superstring theories relevant for the discussion are presented. We follow the KKLT flux compactification proposal where the volume modulus is left unfixed in the low energy effective theory. It can be stabilized on either anti de Sitter or Minkowski minimum. In the first case we employ an uplifting mechanism based on matter superpotentials that break spontaneously supersymmetry. A central issue is the gravitino mass since, in the KKLT scheme, it is tightly related to the the high of the barrier that prevents the overall volume modulus from the decompactification limit. Alternative racetrack-like proposals where the volume modulus is stabilized at Minkowski vacuum are also considered. Some of the results of this chapter will appear at \cite{D}.      

\section {Theories of Higher Dimensions} \normalsize

We live in a universe with one time and three spatial dimensions. It could be, however, that the world is really a space of higher dimensionality but that we are limited in our ability  to experience all its dimensions. Already in the 1920s T. Kaluza and O. Klein proposed to consider an extra dimension to unify the theories of electromagnetism and gravitation by identifying some of the extra components of the metric tensor with the gauge fields of the four-dimensional, physical space time. Half a century later theorists asked for the extra dimensions in order to construct consistent string theories.

Apparently, the first problem that such theories have to address is the invisibility of the hypothetical extra dimensions. The simplst method that has been proposed for making the extra dimensions unobservable is to suppose that they are compactified with a scale parameter that is smaller than we can resolve. Imagining the presence of one extra dimension $y$ this can be compactified by identifying the points $y$ and $y+2\pi R$. This is equivalent to saying that the $y$ direction is curled up into a circle of radious $R$. Considering a scalar field $\varphi(x,y)$ living in this extended space then we require
\begin{equation} \label{a-6}
\varphi(x,y)=\varphi(x, y+2\pi R).
\end{equation}
and expanding it in the Fourier series it reads
\begin{equation}
\varphi(x,y)= \sum^{\infty}_{n=-\infty} \varphi_n(x) e^{iny/R}.
\end{equation} \label{aa-6}
The quantization of the momentum in the compactified dimension yields values ${\cal O} (|n| \hbar /R)$. Hence, for a sufficiently small $R$ only the $n=0$ state will appear in the low energy physics, $E \ll \hbar c/ R$. The observed states will be independent of the extra dimension $y$ which will be invisible from the low energy point of view. Actually, the smaller the lenght scale of compactification the harder is to probe them experimentally. A common proposal is to take the scale of compactification to be of the order of the Planck scale 
\begin{equation}
R \sim l_P \equiv (\hbar G_N/ c^3)^{1/2} \simeq 1.6 \times 10^{-35} \text{m},
\end{equation}
hence, the mass of the excited states, $n \neq 0$, would be of the order of the Planck mass rendering their direct observation out of reach. 

Extra dimensions where invoked in physics by Kaluza aiming to unify the Maxwell theory with the recently, at that time,  formulated Einstein gravity.  The idea was that electromagnetism can be regarded as a consequence of general relativity in five dimensional space. Thus, the metric is dimensionally extended to $g_{MN}$ with $M,N=0,1,2,3,4$ which from the 4D point of view contains the following degrees of freedom: the standard metric of spacetime $g_{\mu,\nu}$, a four vector field $g_{\mu 4}=g_{4 \mu}$ and a scalar field $g_{44}$. The indices $\mu, \nu$ span the observed spacetime and $x^4\equiv y$ is the coordinate of the extra dimension. The phenomenological requirement is the invisibility for the extra dimension. Hence, according to (\ref{a-6}), we Fourier expand the $g_{MN}(x,y)$ metric
\begin{equation} \label{ab-6}
g_{MN}(x,y)= \sum_n g^{(n)}_{MN}e^{iny/R}
\end{equation}
and parametrize $g^{(0)}_{MN}$ as \cite{Collins:1989kn}
\begin{equation} \label{ac-6}
g^{(0)}_{MN}= \varphi^{-1/3}
\left(\begin{array}{cll}      
g_{\mu\nu} + \varphi A_\mu A_\nu & \varphi A_\mu     \\     
\varphi A_\nu & \varphi 
\end{array}\right).
\end{equation}
This parametrization is general, it is only that the notation was particularly chosen for future convinience.  Including the extra dimension $y$ the action of general relativity has the form 
\begin{equation} \label{b-6}
S= - \frac{1}{2\kappa^2_5} \int d^4 x \, dy \, e^{(5)}R^{(5)}
\end{equation}
where $\kappa_5$ is the Einstein gravitational constant in 5-dimensional space, $\kappa^2/8\pi \equiv G^{(5)}_N$ and $e^{(5)}= [-\det (g_{MN})]^{1/2}$. From (\ref{ac-6}) it entails that 
\begin{equation} \label{c-6}
e^{(5)}= \varphi^{-1/3}e
\end{equation}
where $e$ is defined in terms of the physical $4 \times 4$ metric. Following standard texts on gravity, e.g.  the \cite{Hartle:2003yu},
 $R^{(5)}$ can be calculated and the action (\ref{b-6}) simplifies  
\begin{equation} \label{ca-6}
S=-(2\pi R) \int d^4 x \frac{e}{2\kappa^2_5} \left(R^{(4)}+\frac{1}{4}\varphi F^{\mu\nu}F_{\mu\nu}+ \frac{1}{6\varphi^2} \partial^\mu \varphi \partial_\mu \varphi \right)
\end{equation}
where the $4$-dimensional scalar curvature $R^{(4)}$ is calculated from $g_{\mu\nu}$. The tensor $F_{\mu\nu}$ is defined as $F_{\mu\nu}= \partial_\mu A_\nu - \partial_\nu A_\mu$. Identifying it as the electromagnetic contribution to the action a new feuture appears: the electromagnetic field is coupled to a massless scalar field $\varphi$. Since $\varphi$ is the scale parameter of the fifth dimension it is often called the "dilaton/moduli" field. Apart from the electromagnetism there is also an apparent modification of gravity. 
The gravitational force between two masses $m_1$ and $m_2$ decreases now as $r^3$ and for a general $D$-dimensional space as $r^{(D-2)}$; hence $F(r)\sim G_D m_1m_2/r^{D-2}$. For distances $r< R$, where $R$ the compactification scale, the gravity looks much different than the observed. Returning to the $5$-dimensional example, at distances larger than the compactification radius the gravitational force is $F(r)\sim G_5 m_1 m_2/(r^2 2 \pi R )$. This can be also obtained by the action $(\ref{ca-6})$ which relates the $4$-dimensional constant $\kappa^2= 8\pi G$ with the fundamental $5$-dimensional one $\kappa^2_5= 8\pi G_5$ :
\begin{equation} \label{d-6}
\kappa^2=\frac{\kappa^2_5}{2\pi R}.
\end{equation}
Another implication of this geometrical derivation of the electromagnetic field is that it couples in a particular way to the matter. Introducing an additional scalar field $\chi$ in the theory (without attributing to it any geometrical origin),
\begin{equation} \label{da-6}
S_\chi= \int{d^4x \,dy\, e^{(5)}}[g^{(0)MN}\partial_M \chi \partial_N \chi]
\end{equation}
and after expanding it as in $(\ref{aa-6})$ the above action reads
\begin{equation} \label{db-6}
S_\chi = 2\pi R \sum _n \int{d^4 x \, e \left[ g^{\mu\nu} \left( \partial_\mu + \frac{in}{R}A_\mu \right)\chi_n \left( \partial_\nu + \frac{in}{R}A_\nu \right)\chi_n - \frac{n^2}{\varphi R^2} \chi^2_n \right]  }.
\end{equation}
The (\ref{db-6}) manifests that the scalars $\chi$ are coupled in a locally gauge invariant way to the gauge field $A_\mu$ and their mass term is related to the "dilaton" vev and the scale of the compactification. Furthermore, after normalizing the photon field $A_\mu$ the (\ref{db-6}) gives a quantization condition for the electromagnetic charge according to $q_n = n \kappa R^{-1}\sqrt{2/\varphi}$ and quantized mass for the scalar $\chi$, $m^{\chi}_n= |n| /(R \sqrt{\varphi})$.

The deeper reason of this fascinating geometical derivation of electromagnetism is that both the compactification scheme and the Maxwell theory share a basic symmetry. The electromagnetism is a $U(1)$ gauge theory and has the internal symmetry under the gauge transformations 
\begin{equation} 
A_\mu \rightarrow A_\mu + \partial_\mu \theta(x).
\end{equation}
On the other hand, the extra dimension $y$ is compactified in a circle (torus) hence, there is the symmetry of the low energy theory under rotations 
\begin{equation} \label{f-6}
y \rightarrow y+ \theta(x).
\end{equation}   
Hence, the gauge transformation leaves the action invariant thanks to the reparametrization invariance (\ref{f-6}) of the theory. This remarkable fact encourages the assumption that all the forces of nature have a geometrical origin and that all the internal symmetries might eventually be understood as invariances under additional coordinate transformations.

Despite the impressive results of the Kaluza-Klein theory the theory is not phenomenologically viable. It implies that the gravitational force between two particles is equal to the electrostatic force. Also, the charged $\chi$ particles are ultraheavy whereas the massless $\chi$ are uncharged, i.e. singlets under the gauge group. These facts should have been expected since Kaluza-Klein theory treats electromagnetism as part of a $5$-dimensional gravity theory with a universal basic coupling constant $\kappa_5$. However, the Kaluza-Klein theory is very appealing and generalizations were proposed that can avoid the initial problems of the theory.

\subsection{N-Extra Dimensions}
The first step towards a phenomenologically acceptable higher dimensional theory is to incorporate the Standard Model gauge bosons into the higher dimensional metric. Increasing the number of the extra dimensions provides additional vectorial degrees of freedom. Considering $N$ extra dimensions compactified in the manifold $K$ with $y^\alpha$ the extra variables that span this space, $\alpha=1,..., N$, results in a generalization of the metric (\ref{ac-6}), $g^{(D)}_{MN}$, where $D=4+N$. It should be expected that the metric $g^{(D)}_{MN}$ is a solution of the vacuum Einstein equation 
\begin{equation}
R_{MN}=0
\end{equation}
or of the corresponding equation with a cosmological constant $R_{MN}+(1/2)R g_{MN}+\Lambda g_{MN}=0$. However, the dynamic determination of the $g^{(D)}_{MN}$ is  highly non-trivial. A different approach is to postulate a specific form for the compact space $K$. Choosing $K$ to be a flat space with Cartesian coordinates $y^\alpha$ in which we identify points $y^\alpha$  and $y^\alpha + 2\pi R^\alpha$. The resulting manifold $K$ is a $N$-torus and can be thought as $N$ small circles with radii $R^\alpha$, which is the generalization of the circle compactification of a single extra dimension. The symmetry group in this case is $[U(1)]^N$ corresponding to invariance under seperate rotations around each of the circles. Such a model have $N$ different type of photons.

A more general symmetry that can be obtained is the so called "isometry" group of the manifold $K$. This is the group of the transformations of coordinates that leave the metric unchanged. These transformations can be defined in terms of the "Killing vectors" of the metric, which are the directions, at any given point in the manifold, in which it is possible to move from that point while keeping the form of the metric unchanged. Let us assume we make a local coordinate transformation of the form, which defines the isometry group of the metric, 
\begin{equation}
y^\alpha \rightarrow y^\alpha + \sum^{n_B}_{n=1} \epsilon ^n(x) k^\alpha_n(y)
\end{equation}
where $k_n$ are $n_B$ independent Killing vectors and $\epsilon^n(x)$ are a set of infinitesimal parameters. Then, because of the definition of the Killing vectors, there will be no change in the $g^{(N)}_{\alpha \beta}$ part of the metric $g^{(D)}_{MN}$. However, there will be a change in the other components which is compensated by the transformation
\begin{equation}
A^n_\mu \rightarrow A^n_\mu + \partial_\mu \epsilon ^n(x)
\end{equation}
which is exactly a gauge tranformation. The conclusion is that the coordinate invariance of the $4+N$-dimensional theory leads to a locally gauge invariant Yang-Mills theory, i.e. the gauge group is the isometry group of the compact manifold.  

The inclusion of matter fields can take place without apparent phenomenological contradictions as in the Kaluza-Klein model. A set of scalar fields 
\begin{equation}
\chi^\alpha_n=\tilde{\chi}_n(x) k^\alpha_n(y)
\end{equation}
lies in the adjoint representation of the gauge group. These fields are massless in a manifold of zero cosmological constant but they can have nonzero couplings to the gauge fields. Moreover, these couplings can be related. Hence, given a particular manifold $K$ that includes the Standard Model $SU(3)$, $SU(2)$ and $U(1)$ the coupling strengths of each group can be connected even in the absence of Grand-Unified group.

The crucial point here, is whether such a manifold $K$ that contains the Standard Model gauge group exists and what its properties is. Assuming that the manifold  has at least the $SU(3)\times SU(2) \times U(1)$ isometry group then, as Witten (1981) has shown, this requires the compact manifold to have at least seven dimensions. Thus, in order that all interactions of the Standard Model can be obtained geometrically through the Kaluza method, we must live in a world of eleven or more dimensions.  Another important remark is that although the bosons that may have a geometrical origin according to the spirit of the model, the fermions have to be put in "by hand".

However, the requirement that the 7-dimensional compact manifold has the standard model as the isometry group does not determine it uniquely. The possibilities can be infinite. How to choose the properties of the compact manifold is one of the major problems of all the theories that begin in higher dimensions.  The resultant physics is determined by the metrical and topological properties of the compact manifold. Even if we require that this manifold must be a solution of the higher dimensional equation of motion there is still so much freedom that all the predictive power is lost.
\\
\\
\itshape{Fermions}\normalfont
\\
Fermions are basic constituents of our world, by default present in a supersymmetric framework and hence, they should be aslo considered in the Kaluza-Klein type of theories. In an arbitrary number of space dimensions the Dirac equation reads
\begin{equation}
(i\Gamma^M \partial_M -m)\Psi=0
\end{equation}
where $M=0,1..,D-1$ spatial dimensions and $\Gamma^M$ are unitary matrices satisfying 
\begin{equation}
\{\Gamma^M, \Gamma^N \}= 2 \eta^{MN}I_{[2^{D/2}]}.
\end{equation}
The $\eta^{MN}$ is the higher dimensional generalization of the Minkowski metric with the convention $\eta^{00}=+1$. The $\Gamma$ matrices have $2^{[D/2]}$ rows and columns, where $[D/2]$ is the larger integer not greater than $D/2$. 

It can be shown that whenever $D$ is even there is an analogue of the $4$-dimensional $\gamma^5$ matrices hence, Weyl spinors can be formed with $2^{D/2-1}$ components. On the other hand, Majorana spinors only exist for dimensions $D=2,3,4,8,9$ modulo 8. The requirement that a spinor can be both Majorana and Weyl at the same time is even more restrictive. Majorana and Weyl conditions can be imposed simultaneously for $D=2$ modulo $8$, i.e. $2,10,18, 26,...$ dimensions. Coming back to the requirement of $7$-dimensional manifold, i.e. $D=11$, which can have an isometry group that includes Standard Model we see, however, that there is no analogue of "chirality" in eleven dimensions. The fermions in $D=11$ cannot not be Weyl. It follows that a $D=11$ dimension mass is not forbidden by chirality and since a $11$-dimensional mass implies a $4$-dimensional mass too, it entails that $4$-dimensional mass is always possible and the observed fermions can not be chiral.

Fermions are fundamentally present in supersymmeric theories. Supersymmetric versions of higher dimensional theories have been constructed and extensively studied. The supersymmetric version of $N=1$ in eleven dimensions has many attractive features. Under certain conditions the Lagrangian of $N=1$ supergravity in eleven dimensions is unique. In this model all the symmetries, the forces and all the particles have geometrical origin. They appear all in the only one possible supermultiplet that includes a symmetric tensor, a vector spinor and an antisymmetric tensor. However, the chirality problem and the absence of a generally acceptable compactification that preserves supersymmetry at the $D=4$ physical world appears to rule out the $D=11$ Kaluza model. Decreasing the number of the dimensions to $D=10$ then the chirality problem is removed. Nevertheless, Yang-Mills multiplets are generally included resulting in two sorts of gauge vector fields in four dimensions: those that arise from the metric tensor $g_{MN}$ through the Kaluza mechanism and the fields $A^\alpha_\mu$ from the extra Yang-Mills multiplet. 

Despite the problems the attractive features of the higher dimensional theories are strong. Furhermore, string theory, which was introduced in particle physics for completely different reasons demands a space of a higher dimensionality. A brief summary of basic notions of string theory is given in the following sections necessary for a phenomenological analysis of stringy inspired models.

\section {String Theories} \normalsize

String theory is widely considered a compelling fundamental theory, see eg. \cite{Kiritsis:2007zz}. It assumes that the fundamental building blocks are one-dimensional objects called strings. The cornerstones of its success are basically two: firstly, string theory as a theory of extended objects seems to be finite and secondly, in the spectrum of the closed string states occurs a massless spin-$2$ particle that can be identified as the graviton.

The first remarkable aspect of string theory, the expected absence of ultraviolet divergences, is due to the 1-dimensional nature of the string. Quantum field theory of point particles interactions are associated with vertices where worldlines meet at a well defined point leading to diverging physical quantities. On the other hand string interactions are associated with worldsheets and there is no particular point to be identified with a vertex. Two different observers will see the two interacting strings to merge at different points. Hence, there is no Lorentz-invariant way of specifying the space-time point at which the two strings join. This fact entails that the potenially ultraviolet divergent regions of integration are missing from string diagrams and the string theory is apparently a finite one.

The string, being an extended object, has internal degrees of freedom. The oscillation modes of the string may appear as the "fundamental particles". Each oscillation mode of the string is an eigenstate of the energy and can be interpreted as a particle. In the closed string case one encounters a spin-$2$ resonance. Its long wavelength interactions are found to be in agreement with general relativity and hence, it is identified as the graviton field. Such an interpretation means that the energy scale of gravity, namely the Planck mass, becomes a characteristic scale of the string. Hence, in the low energy limit, one can reach only the fundamental modes with the excited modes chracterized by the $M_P$.

We require the string to have a finite length, so it can be either an "open" string with two free ends or, a "closed" srtring with no free ends. The trajectories followed by such strings are two dimensional surfaces in spacetime, the worldsheet. Parametrizing the coordinate along the string as $\sigma$, with $ 0 \leq\sigma \leq \pi$, and the time along the worldline of any point of the string as $\tau$ then the worldsheet is described in $D$-dimensional spacetime by the coordinates $X^M(\sigma, \tau), \, M=0,1,...,D-1$. In order to describe the dynamics of the string and write down its equation of motion we should construct the string action. For a relativistic free point particle the action is proportional to the invariant length of the particle trajectory 
\begin{equation}
S=-\alpha \int{ds}
\end{equation}
where the constant of proportionality has dimensions of mass. Turning to the case of a string moving in spacetime we expect that the action of a string is proportional to the surface area of the worldsheet:
\begin{equation}
S=-T\int dA\, .
\end{equation}
Here the constant $T$ turns out to be the string tension. Introducing the worldsheet coordinates the action can be recast in the form
\begin{equation}
S=-T \int^{\tau_f}_{\tau_i } d\tau \int^l_0 d\sigma \sqrt{(\dot{X} \cdot X^{'})^2 - (\dot{X})^2 (X^{'})^2 }\, .
\end{equation}
This is the \itshape Nambu-Goto \normalfont action and describes the dynamics of the classical relativistic string. As the motion of a point particle in spacetime serves to minimize the length of the worldline, the motion of a classical string in space-time acts to minimize the surface area of the worldsheet. In order to write a quantum theory of strings we need to find the equations of motion for the string which can then later be quantized. Quantization using the Nambu-Goto action is not convenient due to the presence of the square root in the Lagrangian. It is possible to write down an equivalent action, equivalent in the sense that it leads to the same equation of motion that does not have the cumbersome square root. This action goes by the name of the \itshape Polyakov action \normalfont or by the more modern term the \itshape string sigma model action. \normalfont This is done by introducing an intrisic metric $h_{\alpha \beta}(\tau, \sigma)$ which acts like auxiliary field. Using the notation $h=\text{det} h_{\alpha \beta}$, the Polyakov action can be written as
\begin{equation}
S_P=-\frac{T}{2} \int d\tau d\sigma \sqrt{-h} \, h^{\alpha \beta} \partial_\alpha X^\mu \partial_\beta X^\nu \eta_{\mu\nu}.
\end{equation}
Extremizing the action one finds the equation of motion which can take the form of a simple wave equation
\begin{equation}
\left(\frac{\partial^2}{\partial \tau^2} - \frac{\partial^2}{\partial \sigma^2} \right) X^M (\sigma, \tau)=0.
\end{equation}
The solution of a wave equation can be written in terms of a superposition of waves moving to the left on the string and waves moving on the right on the string:
\begin{equation}
X^M= X^M_L(\tau+\sigma)+X^M_R(\tau-\sigma).
\end{equation}

\subsection{The Closed String}

The closed string is characterized by the periodicity conditions 
\begin{equation} \label{perC-6}
X^M(\tau, \pi)= X^M(\tau, 0), \,\,\,\,\,\,\,\,\, \partial_\sigma X^M (\tau, \pi)=\partial_\sigma X^M (\tau, 0).
\end{equation}
The solution of the wave equation can be written as an expansion of Fourier modes. We denote these modes a $\alpha^M_k$ and the right and left moving components read
\begin{equation} \label{strmotL-6}
X^M_L(\tau, \sigma)= \frac{x^M}{2}+\frac{\ell^2_s}{2}p^M(\tau+\sigma)+i \frac{\ell_s}{\sqrt{2}} \sum_{n \neq 0} \frac{a^M_n}{n} e^{-in(\tau+\sigma)}
\end{equation}
\begin{equation} \label{strmotR-6}
X^M_R(\tau, \sigma)= \frac{x^M}{2}+\frac{\ell^2_s}{2}\tilde{p}^M(\tau-\sigma)+i \frac{\ell_s}{\sqrt{2}} \sum_{n \neq 0} \frac{\tilde{a}^M_n}{n} e^{-in(\tau-\sigma)}.
\end{equation}
The $x^M$ is the the center of mass coordinate and the $p^M$ is the total momentum of the string. The second term at (\ref{strmotL-6}) and (\ref{strmotR-6}) corespond to the motion of the string as a single unit ($\alpha^M_0$ and $\tilde{\alpha}^M_0$ Fourier modes) and the modes $\alpha^M_n$ describe the vibrations of the string. The $\ell_s$ is the characterisric lenght of the string which is related to the Regge slope parameter $\alpha'$ and hence to the tension in the string via
\begin{equation} \label{tension-6}
T=\frac{1}{2\pi \alpha'} \,\,\,\,\,\,\,\,\,\,\,\, \text{and} \,\,\,\,\,\,\,\,\,\,\,\, \frac{1}{2}\ell^2_s=\alpha'.
\end{equation}
The periodicity conditions (\ref{perC-6}) restricts the solutions to those for which the wavenumber $n$ takes on integral values. In addition they enforce the condition $p^M=\tilde{p}^M$ and we will refer to its implications after the quantization of the string spectrum. The main approaches of quantization are called covariant, light-cone and BRST quantization. It translates in commutation relations for the Fourier modes of the closed string reminiscent of an infinite set of harmonic oscillators \cite{Zwiebach:2004tj}
\begin{equation}
[\alpha^M_m, \alpha^N_n]=m\eta^{MN}\delta_{m+n,0} \,\,\,\,\,\, \,\,\,\,\,\, [\tilde{\alpha}^M_m, \tilde{\alpha}^N_n]=m\eta^{MN}\delta_{m+n,0}  \,\,\,\,\,\,\,\,\,\,\,\, [\alpha^M_m, \tilde{\alpha}^N_n]=0.
\end{equation}
The crucial point is that the presence of the Minkowski metric $\eta_{\mu\nu}$ means that we can have negative commutators, since $\eta_{00}=-1$. This results in negative norm states. One can get rid the theory of the negative states by applying the so called Virasoro constraints $L_m=(1/2)\sum_n \alpha_{m-n}\alpha_n$ and $\tilde{L}_m=(1/2)\sum_n \tilde{\alpha}_{m-n}\tilde{\alpha}_n$ which are promoted to operators in the quantum theory. The Virasoro operators satisfy the commutation relations
\begin{equation}
[L_m, L_n]=(m-n)L_{m+n}+\frac{D-2}{12}(m^3-m)\delta_{m+n,0}
\end{equation}
which is called the \itshape Virasoro algebra \normalfont with central extension. The central charge is the space time dimension $D$. The Virasoro operators can be used to eliminate the unphysical negative norm states. This takes place at the critical dimension
\begin{equation}
D=26
\end{equation} 
i.e. in a 26-dimensional space. The mass spectrum of the mass operator $M^2=-p^\mu p_\mu$ is given for the case of the closed string
\begin{equation} \label{Mstr-6}
M^2= \frac{4}{\alpha'} \left(\sum^{\infty}_{n=1} :\alpha^I_{-n} \alpha^I_n: - \frac{D-2}{24}\right). 
\end{equation}
where the summation is over the transverse degrees of freedom and the $(D-2)/24 \equiv a_0$ is the zero point energy. The ground state $\left |0,k \right\rangle$ of the closed string has mass squared $M^2=-4a_0/\alpha'$ and it is a tachyon. The first excited states are derived from the ground state by
\begin{equation}
 \varepsilon_{IJ}\alpha^I_{-1} \tilde{\alpha}^J_{-1} \left |0,k \right\rangle
\end{equation} 
with a mass squared $4(1-a_0)/\alpha'$. The tensor $\varepsilon^{IJ}$ is a square matrix of size $D-2$ and it can be decomposed into a symmetric traceless tensor field
\begin{equation}
h^{IJ}=\frac{1}{2}\left(\alpha^I_{-1}\tilde{\alpha}^J_{-1}+\alpha^J_{-1}\tilde{\alpha}^I_{-1}- \frac{2}{D-2}\delta^{IJ}\sum_P \alpha^P_{-1}\tilde{\alpha}^P_{-1} \right) \left|0,k \right\rangle
\end{equation}
an antisymmetric tensor field
\begin{equation}
b^{IJ}=\frac{1}{2}\left(\alpha^I_{-1}\tilde{\alpha}^J_{-1}-\alpha^J_{-1}\tilde{\alpha}^I_{-1} \right) \left| 0,k \right\rangle
\end{equation}
and a scalar field known as the \itshape string dilaton \normalfont
\begin{equation}
e^\phi= \sum_P \alpha^P_{-1}\tilde{\alpha}^P_{-1}
\end{equation}
The string dilaton plays a central role since it determines the value of the string coupling $\lambda= \left\langle e^\phi \right\rangle$. The indices $I$, $J$ correspond to the transverse spatial degrees of freedom. In a 26-dimensional space the above fields are massless and fall into representations of the transverse group $SO(D-2)$. The fact that there is a symmetric traceless tensor massless field in the spectrum is essential. It is interpreted as one particle \itshape graviton \normalfont states i.e. a fluctuation of the metric, $g_{IJ}=\eta_{IJ}+h_{IJ}$. This interpretation implies that the interaction of the strings relate the $1/\sqrt{\alpha'}$ with the Planck scale. The $b^{IJ}$ states correspond to the one particle states of the \itshape Kalb-Ramond \normalfont  field which is in many ways the tensor generalization of the Maxwell gauge field $A_\mu$. The Kalb-Ramond field couples to strings in a way that is analogous to the way that the Maxwell field couples to particles and thus, it is said that strings carry Kalb-Ramond charge. 

The ground state of the string has negative mass squared thus, is a tachyon. It turns out that the introduction of supersymmetry i.e. fermionic degrees of freedom gets rid of the tachyonic states and the massless excitations are the expected states of the string at low energies.

Turning back to the condition $p^M=\tilde{p}^M$ it implies, at first sight,  that no winding for the closed string is permitted. However, in the case that the ambient spacetime includes a compact extra dimension then this condition does not hold and the closed string can wind around the compactified dimension. Asuming that an extra dimension is compactified on a circle of radius $R$ then $X^M(\tau, \sigma+2\pi R)= X^M(\tau, \sigma)+ 2\pi R m$ where the $m$ is the winding number. If $X^{25}$ is the compactified dimension then the momentum of the solution (\ref{strmotL-6}) for the $X^{25}$ changes like $p^{25}\rightarrow p^{25}+m R /\alpha'$. In addition to the winding contribution the momentum the compactified $X^{25}$ dimension imposes a quantization of the momentum $p^{25}=n/ R$. Then the string mass spectrum (\ref{Mstr-6}) is modified to 
\begin{equation} \label{MKKW-6}
M^2=\frac{1}{\alpha'}\left[ 2(N+\tilde{N})-4+n^2\frac{\alpha'}{R^2}+m^2\frac{\rho^2}{\alpha'} \right]
\end{equation}
where $N-\tilde{N}=nm$ the difference between the number operators that are associated with the barred and un-barred operators \cite{Zwiebach:2004tj}. The $n^2/R^2$ term corresponds to the Kaluza-Klein modes and the $m^2 R^2$ comes from the winding sectors. The mass spectrum (\ref{MKKW-6}) is invariant under the transformation
\begin{equation} \label{T-6}
R \longleftrightarrow \alpha'/ R, \,\,\,\,\,\,\,\,\,\,\, m\longleftrightarrow n
\end{equation}
which the \slshape T-duality \normalfont for the closed string. This transformations exchanges the Kaluza-Klein modes with the winding modes and hence, the light modes of one theory become the heavy modes of the other theory. T-duality exchanges large and small radius compact manifolds. The special radius $R^*= \sqrt{\alpha'}$ is the unique radius that is mapped to itself under the transformation (\ref{T-6}). The duality implies that each radius smaller than $R^*$ is equivalent to some radius larger than $R^*$. In this sense $R^*$ represents the minimal radius that can be attained in toroidal compactification. The compactification radius $R$ is a parameter of a spacetime (allowed in string theory) which is often called \itshape moduli. \normalfont Hence, T-duality tells us that the moduli space of compactifications into a circle can be taken to include only radii larger than or equal to $R^*$.

\subsection{The Open String}

For an open string there are two possible boundary conditions 
\begin{equation}
\text{Dirichlet,} \,\,\,\,\, X^M=\text{constant}  \,\,\,\,\,\,\,\,\,\, \text{or \,\,\,\,\, Neumann,} \,\,\,\,\,  \frac{\partial X^M}{\partial \sigma}=0.
\end{equation}
For the case of Neumann boundary conditions one finds that $p^M=\bar{p}^M$, i.e. the open string cannot wind around itself, and $\alpha^M_k=\bar{\alpha}^M_k$, i.e. the modes are the same for left and right moving waves. Physically this means that for an open string with free endpoints the modes combine to form standing waves on the string. Thus the solution for this case reads
\begin{equation}
X^M(\tau,\sigma)=x^M+\ell^2_s p^M \tau + i \frac{\ell_s}{\sqrt{2}} \sum _{n \neq 0} \frac{\alpha^M_n}{n} \cos(n\sigma).
\end{equation}
For the case of open string with fixed endpoints, i.e. Dirichlet boundary conditions, it is straightforward to write down the solutions, however we omit them here. The reader can found them in standard string textbooks, Zwiebach's for example \cite{Zwiebach:2004tj}. 

For the open string in light-cone coordinates the mass shell conditions reads
\begin{equation}
M^2=\frac{1}{\alpha'}\left(N-\frac{D-2}{24} \right)
\end{equation}
and the ground state has mass squared
\begin {equation}
M^2\left|0,k \right\rangle=-\frac{1}{\alpha'}\frac{D-2}{24}\left|0,k \right\rangle.
\end{equation}
The first excited state 
\begin {equation}
M^2 \alpha^I_{-1} \left|0,k \right\rangle=-\frac{1}{\alpha'}\left(1-\frac{D-2}{24}\right) \alpha^I_{-1}\left|0,k \right\rangle
\end{equation}
where $I=1,...,D-2$ spatial index of the transverse degrees of freedom and the fields $\alpha^I_{-1} \left|0,k \right\rangle$ form a vector representation of the transverse group $SO(D-2)$. For the critical dimension $D=26$ the ground state is tachyonic and the vectorial first excitation is massless.

\subsection{String Phenomenology}

String theory/M-theory is believed to describe the fundamental elements of our universe in a unified fashion. Being a theory of the smallest possible scales it is currently impossible to be directly confronted with the experiment, to be "observed". However, string theory may have some low energy implications with interesting phenomenological aspects that have been pursued by string theorists. Also, phenomenology is the ruler that shapes every stringy model of particle physics. 

Firstly, our world is \itshape four-dimensional \normalfont a fact that does not automatically emerge from the basic conjecture of string/M-theory. Hence, one imposes the topological criterion that the ten-dimensional space is of the form $R^4 \times M_6$, where $M_6$ is some compact manifold or some abstract conformal field theory. All other solutions have to be excluded or relegated to the regime of the gedanken. Secondly, the \itshape supersymmetry \normalfont of our world, if present, it must be $N \leq 1$. This excludes some compactification schemes as the toroidal which yields 16 or 32 supersymmetries in four dimensions which cannot yield the chiral structure of the observed world. This entails that only internal manifolds such as \itshape Calabi-Yau \normalfont (CY) plus their orbifold limits have to be considered. A Calabi-Yau manifold is a complex K\"ahler manifold with a metric of $SU(3)$ holonomy i.e. manifold of complex dimensions 3. Such solutions are characterized by a number of parameters, the $h_{1,2}$ complex structure and the $h_{1,1}$ K\"ahler structures of the manifold. Moreover, supersymmetry must be broken with mass splitting of order $10^{-15}M_P$. Thirdly, string theory includes by construction the \itshape dilaton and moduli fields \normalfont that have to be massive given the observational and experimental fact that massless gravitationally interacting fields are excluded since they affect the Newtonian gravity at large scales. The moduli fields appear as massless four dimensional chiral superfields in the low energy four dimensional action that is supposed to describe the real world at scales below the string scale. Hence, a potential must be generated with the moduli and the dilaton fields. Fourthly, the \itshape cosmological constant \normalfont has to be small and, if not zero, positive of the order of $10^{-120}M_P$. The natural scale of string theory is $M \sim M_P$ and this implies a fine tuning which, although severe, seems to be possible in some spesific stringy set ups. Finally, the string theory has, as an ultimate goal, to yield three generations of chiral fermions and the \itshape Standard Model \normalfont gauge group.

\subsection{The Dilaton and the Moduli fields}

In the following, we will concentrate on the moduli fields, whose values determine the low energy physics and they are the most phenomenological relevant ingredients of string theories. The scalar degree of freedom of the closed string quantum states is the dilaton field. Its vacuum expectation value determines the string coupling 
\begin{equation}
g_s= \left\langle e^\phi \right\rangle
\end{equation}
which appears in all the stringy diagrams and hence, in the action. In the case of the weakly coupled heterotic string the 10-dimensional effective supergravity action in the string frame includes terms at the closed string tree level
\begin{equation} \label{s10-6}
{\cal S}_H = - \int{ \frac{d^{10} x }{(2\pi)^7} \sqrt{-g}\, e^{-2 \phi} \left[ \frac{1}{(\alpha')^4} \left( R^{(10)}+ 4 \partial^{\mu} \phi \partial_{\mu} \phi \right)+\frac{1}{(\alpha')^3} \frac{1}{4} \text{Tr}F^2+... \right]}
\end{equation}
where $\alpha'= M_H^{-2}$ the heterotic string scale. Actually, focusing on the gravity plus dilaton sector of the theory all the low energy effective actions for the different perturbative string theories reduce to the same form as (\ref{s10-6}) (from de alwis,Br, Novak). After the compactification on a 6-dimensional manifold of volume $V_6$ the action is recast \cite{Binetruy:2006ad}
\begin{equation} \label{s4.1-6}
{\cal S}_H = - \int{ \frac{d^{4} x }{(2\pi)^7} \sqrt{-g} \left( \frac{V_6}{(\alpha')^4} e^{-2 \phi}  R^{(4)} + \frac{V_6}{(\alpha')^3} e^{-2 \phi} \frac{1}{4} \text{Tr}F^2+... \right)}
\end{equation}
with $M_c \ \equiv V^{-1/6}_6$ the compactification scale. The above action shows that the string scale is of the order of the Planck mass.

Apart from the scalar dilaton field which fixes the string coupling the compactification of the extra dimensions entail the presence of further scalars, the scalar moduli fields that determine with their vevs the radii and shape of the compact manifold. The vacuum expectation values of the dilaton and the moduli fields are given in terms of the only fundamental scale, the string scale.

The ten dimensional low-energy effective action is reduced to a four dimensional action using the ansatz
\begin{equation}
ds^{10}= e^{a u (x^\mu)} g^{(4)}_{\mu\nu} dx^\mu dx^\nu +e^{b u (x^\mu)}g^{(6)}_{kl}dy^k dy^l 
\end{equation}  
where $m,n$ go over the dimensions of the internal space and $\mu,\nu$ over the physical space. The $u$ modulus field is the degree of freedom associated with the overall size of the compact manifold. This can be seen by writing the compact space part of the metric as 
\begin{equation}
g_{kl}(x,y)= e^{2 u (x)} g^{(0)}_{kl}, \,\, k,l=4,...,9
\end{equation}
and the $6$-dimensional compact manifold reads
\begin{equation}
V_6 = \int {d^6y \sqrt{-g}}= M^{-6}_s \left\langle e^{6 u} \right\rangle. 
\end{equation}
The compact manifold can be taken to be a Calabi-Yau 3-fold and we assume here, for simplicity, that has one K\"ahler modulus but may have an arbitrary number of complex structure moduli.  The $M_s$ is the string scale which in the string frame is the funadamental mass scale and the $\left\langle e^{2 u}\right\rangle$ measures $R^2$ of the compact space in string units. This last is the reason that the $e^u$ is called the breathing mode of the compact manifold. For the case of the weakly coupled heterotic string the 4-dimensional action (\ref{s4.1-6}) takes the form 
\begin{equation} \label{s4.2-6}
{\cal S}_H = \int{ \frac{d^{4} x }{(2\pi)^7} \sqrt{-g^{(4)}} \,  e^{-2 \phi+6 u} \left[ M^2_s \left(- R ^{(4)} + 12 D^\mu\partial_\mu u +42 \partial^\mu u \partial_\mu u -4\partial^\mu u \partial_\mu \phi \right) - \frac{1}{4} \text{Tr}F^{\mu\nu} F_{\mu\nu} \right]}.
\end{equation}
Defining the real scalar fields 
\begin{equation} \label{st-6}
s= \frac {1}{(2\pi)^7} e^{-2 \phi + 6 u}, \,\,\,\,\, t=\frac {1}{(2\pi)^7} e^{2 u}
\end{equation}
and after integration by parts the action (\ref{s4.2-6}) is recast in the more transparent form 
\begin{equation} \label{s4.3-6}
{\cal S}_H = \int{ d^{4} x  \sqrt{-g^{(4)}} \, s \left[ M^2_s \left(- R ^{(4)} + \frac{3}{2}\frac{\partial^\mu t \partial_\mu t}{t^2}- \frac{\partial^\mu s \partial_\mu s}{s^2} \right) - \frac{1}{4} \text{Tr}F^{\mu\nu}F_{\mu\nu} \right]}.
\end{equation}
From (\ref{s4.3-6}) we read the gauge coupling and the Planck mass respectively
\begin{equation}
\frac{1}{g^2} = \left\langle s \right\rangle \,\,\,\,\, \text{and} \,\,\,\,\, M^2_P= 2 \left\langle s \right\rangle M^2_s.
\end{equation}
Switching from the string frame to Einstein frame by performing a Weyl transformation on the four dimensional metric the kinetic term of the $s$ field has the standard positive sign
\begin{equation} \label{s5.3-6}
{\cal S}_H = \int{ d^{4} x  \sqrt{-g}  \left[ -\frac{1}{2}M^2_P \, R + \frac{3}{4}\frac{\partial^\mu t \partial_\mu t}{t^2} + \frac{1}{4} \frac{\partial^\mu s \partial_\mu s}{s^2} - \frac{1}{4} s \text{Tr}F^{\mu\nu}F_{\mu\nu} \right]}.
\end{equation}

In superstring theories these scalars have to be part of a supermultiplet. The string dilaton field $\phi$ is paired up with the antisymmetric tensor $b_{\mu\nu}$, where $\mu,\nu$ are four-dimensional indices. Together with the Majorana fermion, the dilatino, they form new kind of supermultiplet, the real \itshape linear supermultiplet \normalfont $L$. The radii moduli, i.e the K\"ahler moduli, are related to the antisymmetric tensor $b_{kl}$, where $k,l$ are six-dimensional compact indices. The fact that the imaginary part of the K\"ahler moduli has only derivative interactions like (it has a Peccei-Quinn symmetry just like the axion) and that the dilaton $\phi$ is a component of a real linear superfield implies that the superpotential cannot depend directly on the K\"ahler moduli and the dilaton fields.
Now, the fields $s$ and $t$ of (\ref{st-6}) appear to be the real parts of the complex scalar fields of the chiral supermultiplets $S$ and $T$ where
\begin{equation}
S= \frac{1}{(2\pi)^7}e^{-2\phi+6 u }+i\alpha, \,\,\,\,\,\, T=\frac{1}{(2\pi)^7}e^{2 u}-i\sqrt{2}\beta^{(h_1,1)}
\end{equation}
the complex scalars. From the classical action and the properties of the compactified manifold the $S$ and $T$ superfields have a K\"ahler potential
\begin{equation}
K(S,T)= -\ln (S+S^\dagger)-3 \ln (T+T^\dagger).
\end{equation}
 
\section {Stabilization Schemes for the Moduli fields} \normalsize 

Apparently, from (\ref{s5.3-6}) the dilaton and moduli fields appear to be massless at 4-dimensions. 
A moduli potential can be generated in many different ways. The earliest solution was to consider gaugino condensation in a gauge group \cite{Derendinger:1985kk, Dine:1985rz, Derendinger:1985cv}. Typically this yields a runaway potential for the moduli, hence the theory prefers to go to the zero coupling and decompactification limit \cite{Dine:1985he}. But, if there is a direct product of gauge groups  then one has the possibility of developing a critical point in the so called "race-track" models \cite{Krasnikov:1987jj, de Carlos:1992da}. Such a product of different gauge groups can be obtained by turning on discrete Wilson lines in the internal manifold. Similar effects can be obtained by considering brane instanton effects \cite{Witten:1996bn}. In addition contributions to the potential can be generated by turning on fluxes in internal compact directions \cite{Dine:1985rz}. However, the minima are supersymmetric with a string scale negative cosmological constant.

In supergravity theories it is possible to obtain a zero or tiny cosmological constant by fine tuning. In string theory there was no obvious mechanism that would allow for this fine tuning which is rather severe: 120 orders of magnitude smaller than the string scale. In the work of \cite{Bousso:2000xa} it was shown that such a fine tuning is possible in spite of the quantization of the parameters (fluxes) in terms of the sring scale. Another point is that the moduli have to be stabilized at weak or intermediate coupling region in order the stabilization result to be reliable. This was actually the main argument rejecting one of the first models \cite{Dine:1985rz} proposed for the dilaton stabilization based on the observation of the quantized fluxes.

The problem of finding minima with zero or positive cosmological constant is related to a "no-go" theorem which guarantees that such solutions cannot be obtained in string or $M$ theory by using only the lowest order terms in the 10D or 11D supergravity action \cite{Maldacena:2000mw}. However, corrections in the leading order Lagrangian in the $g_s$ or $\alpha'$ expansion or the inclusion of extended objects as branes can invalidate the "no-go" theorem for warped backgrounds \cite{Giddings:2001yu}. Moreover it was shown, at \cite{Giddings:2001yu}, in the context of type IIB string theory compactified on a Calabi-Yau manifold, that all the complex structure moduli and the dilaton can be stabilized by an appropriate choice of fluxes apart from the K\"ahler moduli of the compactification. The resulting 4D models are of the no scale type. In the work of Kachru, Kallosh, Linde and Trivedi (KKLT) \cite{Kachru:2003aw} a model where the K\"ahler modulus, i.e. the volume modulus, (which is assumed to be the only one) can be also stabilized with susy broken in a Minkowski or de Sitter vacuum  was proposed.

The KKLT starts with a classical $N=1$ supergravity potential which can be obtained by considering 10D low energy type IIB theory compactified on a Calabi-Yau orientifold with $D3$ and $D7$ branes. This model is essentially a limit of $F$-theory construction. Hence, the starting point is the effective action in the string frame
\begin{equation}
{\cal S}_\text{IIB}= \frac{1}{2 \kappa_{10}^2} \int d^{10}x \sqrt{-g_s} \left[e^{2\phi} \left( R_s+ 4 \partial^\mu \phi \partial_\mu \phi \right) - \frac{1}{2} F^2_{(1)} - \frac{1}{2 \cdot 3!} G_{(3)} \bar{G}_{(3)}- \frac{1}{4\cdot 5!} \tilde{F}^2_{(5)}\right]+...
\end{equation}
where $g_s$ the string metric and $R_s$ the string scalar curvature. The $G_{(3)}=F_{(3)}-\tau H_{(3)}$ is the combined three flux and $\tilde{F}_{(5)}=F_{(5)}-\frac{1}{2} C_{(2)} \wedge H_{(3)}+\frac{1}{2} B_{(2)} \wedge F_{(3)}$. The $\tau = C_{(0)}+ie^{-\phi}$ is the IIB complex axion dilaton field.
In this subsection we follow the notation of the original works \cite{Giddings:2001yu} and \cite{Kachru:2003aw} i.e. the dilaton field and the volume moduli that appear in the four-dimensional actions are denoted by $"\tau"$ instead of "$S$" and by "$\rho$" instead of "$T$". The compactification to be considered arises from $F$-theory so it is particular useful to reformulate the action in an SL(2, $\bold{Z}$) invariant form by defining the Einstein metric $g_{MN}=e^{-\phi/2}g_{sMN}$. Then the action reads
\begin{equation}
{\cal S}_\text{IIB}= \frac{1}{2 \kappa_{10}^2} \int d^{10}x \sqrt{-g} \left[ R -\frac{\partial_M \tau \partial^M \bar{\tau}}{2(\text{Im}\tau)^2}-\frac{G_{(3)}\cdot \bar{G}_{(3)}}{12\, \text{Im} \tau}- \frac{1}{4\cdot 5!} \tilde{F}^2_{(5)}\right]+...
\end{equation}
The warped metric maintaining four dimensional Poincar\'e symmetry is parametrized as 
\begin{equation}
ds_{10}^2= e^{2A(y)-6u(x)} \tilde{g}^{(4)}_{\mu\nu} dx^\mu dx^\nu +e^{-2A(y)+2u(x)}\tilde{g}^{(6)}_{kl}dy^k dy^l 
\end{equation}
in terms of four dimensional coordinates $x^\mu$ and coordinates $y^\mu$ on the compact manifold $M_6$. The $e^{A(y)}$ is a warp factor which effectively changes the scale of four dimensional physics at different poins in the internal manifold. 
In the approximation of large-radius limit, which implies an approximately constant warp factor $e^{A}\simeq 1$ and vanishing $\tilde{F}_5$, the effective four-dimensional action for the dilaton and the volume modulus reads
\begin{equation}
{\cal S} = \frac{1}{2\kappa^2_4} \int d^4x \sqrt{-g_4} \left(R_4- 2 \frac{\partial_\mu \tau \partial^\mu \bar{\tau}}{|\tau-\bar{\tau}|^2}- 6 \frac{\partial_\mu \rho \partial^\mu \bar{\rho}}{|\rho-\bar{\rho}|^2}  \right)
\end{equation}
where $\rho=ie^{4u-\phi}+b/\sqrt{2}$  a four dimensional superfield, the single volume modulus (the K\"ahler modulus).

\subsection {Stabilizing the Dilaton and the Complex Structure Moduli}

The kinetic terms for the $\tau$ and $\rho$ fields can be found from the K\"ahler potential, that results by the dimensional reducing of the 10D action. Including also the complex structure moduli $z^\alpha$ (coordinates on the complex structure moduli space) it reads
\begin{equation} \label{K1-6}
K(\rho, \tau, z^\alpha)=-3 \ln [-i(\rho-\bar{\rho})] - \ln[-i(\tau-\bar{\tau})]-\ln\left( -i \int_M \Omega \wedge \bar{\Omega}\right). 
\end{equation}
The last term gives the $K=K(z^\alpha,\bar{z}^\alpha)$ dependence and the $\Omega$ is the holomorphic three-form on the Calabi-Yau manifold $M$. 
Before turning on fluxes the K\"ahler and complex structure moduli, $\tau$ and  $z^\alpha$ respectively, correspond to massless scalar fields (for orientifold models, the dilaton field $\tau$ is massless, whereas in general $F$-theory models it is fixed in terms of the complex structure moduli). 
The fluxes generate a superpotential which takes the form \cite{Gukov:1999ya}
\begin{equation} \label{W1-6}
W=\int_M \Omega \wedge G_{3},
\end{equation}
which fixes the vevs of the dilaton and the complex structure moduli.
However, the (\ref{W1-6}) is independent of the volume modulus $\rho$. 

The $N=1$ supergravity potential reads
\begin{equation} \label{V1-6}
V_\text{sugra}= \frac{1}{2\kappa^2_{10}}e^K \left(G^{a\bar{b}}D_a W \overline{D_b W}-3|W|^2\right)
\end{equation}
where $D_a W = \partial_a W + W \partial_a K$ and $G_{a \bar{b}}=\partial_a \partial_{\bar{b}} K$. The $\rho$ field cancels out of the scalar potential. This is expected since the $\rho$ does not appear in (\ref{W1-6}) because the potential (\ref{V1-6}) is of no-scale type for the volume modulus. Hence, the $|D_{\rho} W|^2$ cancels the negative term at (\ref{V1-6}) leaving a non-negative potential:
\begin{equation} \label{V2-6}
V_\text{sugra}= \frac{1}{2\kappa^2_{10}}e^K \left(G^{i \bar{j}}D_i W \overline{D_j W}\right)
\end{equation}
with $i,j$ labeling indices excluding $\rho$.  When $D_aW=0$ the potential vanishes. This condition is independent of $\rho$. Generically at these solutions $W\neq 0$ so $D_\rho=-3W/(\rho-\bar{\rho})$ is nonzero and supersymmetry is broken. When $D_jW=0$ means that the dilaton and the complex structure moduli fields are stabilized at supersymmetric minima with a mass 
\begin{equation} \label{m-6}
m \sim \frac{\alpha'}{R^3}
\end{equation}
where $R$ is the radius of the manifold. The $R$ is undetermined since $\text{Im} \rho$ scales like $R^4$ ($(\text{Im}\rho)^3 \propto V^2_{CY} \sim R^{12}$). By tuning flux quanta, it is possible to fix $g_s$ at small values (by fixing the vev of the dilaton), though not arbitrary small.

\subsection{Stabilizing the Volume Modulus - the KKLT proposal}

The $\rho$ field is a flat direction i.e. not stabilized. The degeneracy in the $\rho$ flat-direction can be lifted once corrections to the no-scale models are included. One of the source that lifts the degeneracy comes from Euclidean D3-branes in type IIB compactifications \cite{Witten:1996bn}. These are instantons which at large volume yield a new term to the superpotential 
\begin{equation}
W_\text{ins}=T(z_\alpha) e^{2\pi i \rho}.
\end{equation}
$T(z_\alpha)$ is a complex structure dependent one-loop determinant. Since the $z_\alpha$ and the dilaton are fixed by the fluxes they can be integrated out at lower energies. The exponential dependence comes from the action of the Euclidean D3-brane wrapping a four cycle in the manifold $M$. Another source of corrections to the no-scale model (\ref{K1-6}), (\ref{W1-6}) comes from non-abelian gauge groups which may arise from stacks of D7-branes wrapping four-cycles in the manifold. Considering a stack of $N_c$ coincident branes results in Yang-Mills theory with a 4-dimensional coupling $8 \pi^2 /g^2_\text{YM}=2\pi R^4/g_s=2\pi\text{Im}\rho$. The low energy theory is a $N=1$ supersymmetric $SU(N_c)$ gauge theory and undergoes gluino condensation which results in a non-perturbative superpotential
\begin{equation}
W_\text{gauge}= \Lambda^3_{N_c}= A e^{2\pi i \rho/N_c}.
\end{equation}
The $\Lambda_{N_c}$ is the dynamically generated scale, where the gauge coupling explodes, and the coefficient $A$ determined by the energy scale below which the SQCD theory is valid. 

Therefore, given that the heavy (\ref{m-6}) complex structure moduli and the dilaton fields are intergated out the low energy theory includes the classically (without the corrections) massless volume-modulus with tree-level K\"ahler
\begin{equation} \label{K-KKLT-6}
K=-3 \ln \left[-i(\rho-\bar{\rho}) \right]
\end{equation}
and the corrected superpotential 
\begin{equation} \label{W-KKLT-6}
W=W_0 + A e^{ia\rho}.
\end{equation}
$W_0$ is a tree level contribution which arises from the fluxes and the exponential is the quantum correction to the no-scale potential for $\rho$. Corrections to the K\"ahler potential can be neglected in the case that the volume modulus is stabilized at values which are parametrically large compared to the string scale.

The potential of the above theory for the volume modulus is given by $V=e^K(G^{\rho {\bar{\rho}}}D_\rho W \overline{D_\rho W} -3 |W|^2)$. It has a supersymmetric vacuum at $D_\rho W=0$. At the \cite{Kachru:2003aw} they considered $\rho=i\sigma$ and took  the $A,a$ to be real and the $W_0$ to be real and negative . Then the supersymmetric minimum lies at $\sigma_c$ given by 
\begin{equation} \label{min-6}
W_0= -A e^{a \sigma_c}(1+\frac{2}{3} a \sigma_c). 
\end{equation}
The $\sigma_c \gg 1 $ and this is possible by tuning the fluxes and arranging by this way the $W_0 \ll 1$. Otherwise, generically, if fluxes break supersymmetry the expectation is $W_0 \sim {\cal O}(1)$ and the stabilization of the volume modulus at a calculable regime will not be possible. An illustrative example considered is one with parameters $W_0=-10^{-4}$, $A=1$ and $a=0.1$. This results in a minimum at $\sigma_c \simeq 113$. Writing the Planck mass explicitly the values tranlate to $W_0=-10^{-4} M^3_P$, $A=M^3_P$ and $a=0.1 M^{-1}_P$ and the minimum lies at $\sigma_c \simeq 113 M_P$.

The above paradigm, \cite{Kachru:2003aw}, exhibits a stabilization of the volume modulus using superpotential with one exponential utilizing the effect of a tunable constant $W_0$. Another possibility to get a minimum at large volume is to consider a racetrack model \cite{Krasnikov:1987jj}. In this case the fluxes preserve supersymmetry, the superpotential involves multiple exponential terms and the desired value of the volume modulus is obtained by tuning the ranks of the gauge groups. 

Going back at the supersymmetric $\sigma_c$ minimum (\ref{min-6}), it is easily seen that it has negative potential energy equal to
\begin{equation} \label{VAds-6}
V_\text{AdS}= (-3 e^K W^2)_\text{AdS}= - \frac{a^2 A^2 e^{-2 a\sigma_c}}{6\sigma_c}.
\end{equation}
In order to satisfy the observational constraint of zero or slightly positive cosmplogical constant, the KKLT proposed to add the contribution of a $\bar{D}_3$ brane to the four dimensional effective action. The anti-D brane gives a positive contribution to the potential 
\begin{equation} \label{Dup-6}
\delta V = \frac{D}{\sigma^3}
\end{equation}
where $D$ is positive and proportional to the $\bar{D}_3$ tension. The full potential then reads
\begin{equation} \label{KKLT-6}
V=\frac{aAe^{-a\sigma}}{2\sigma^2}\left(\frac{1}{3} \sigma a A e^{-a \sigma}+ W_0 + A e^{-a\sigma} \right) + \frac{D}{\sigma^3}.
\end{equation}
Tuning properly the last term results in an uplifting of the potential to a de-Sitter or even Minkowski vacuum. This vacuum is metastable with a barrier protecting it from the runaway towards the decompactification. However, the (\ref{Dup-6}) term \itshape breaks supersymmetry explicitly \normalfont  since the potential (\ref{KKLT-6}) is not derived by a four dimensional SUGRA theory \cite{Brustein:2004xn}.

\begin{figure} 
\textbf{ AdS KKLT \,\,\,\,\,\,\,\,\,\,\,\,\,\,\,\,\,\,\,\;\;\;\;\;\,\,\,\,\,\,\,\,\,\,\,\,\,\,\,\,\,\,\,\,\,\,\,\,\,\,\,\,\,\;\;\;\;\;\,\,\,\,\,\,\,\,\,\,\,\,\,\,\,\,\,\,\,\,\,\,\,\,\,\,\,\,\,\, dS KKLT}
\centering
\begin{tabular}{cc}

{(a)} \includegraphics [scale=.85, angle=0]{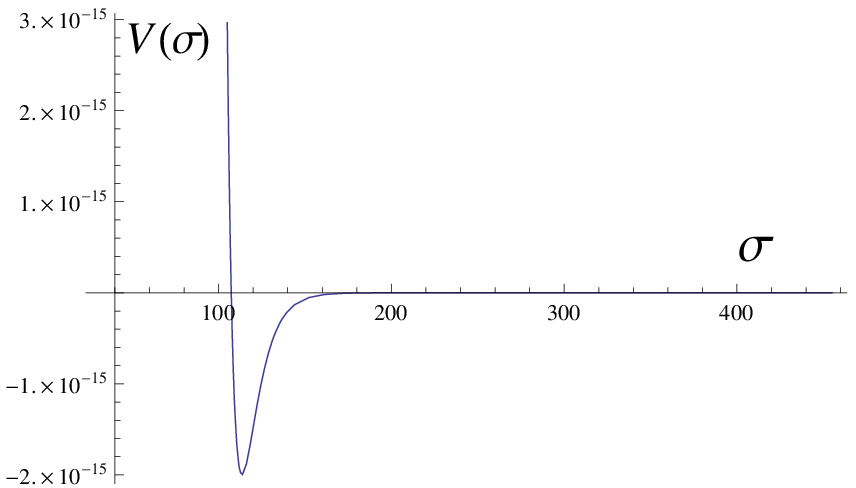} &
{(b)} \includegraphics [scale=.85, angle=0]{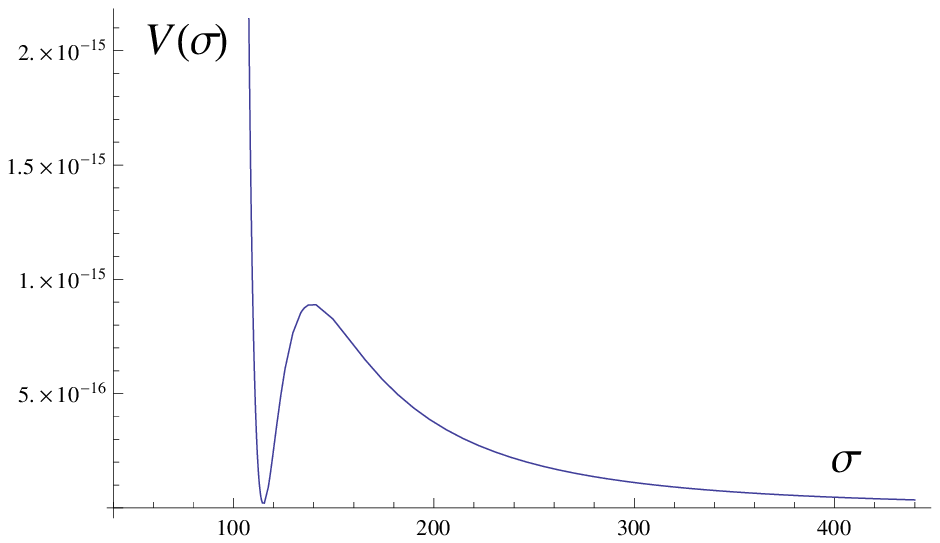}  \\
\end{tabular}
\caption{\small{ At the left panel is the KKLT potential for the overall volume modulus stabilized at the AdS vacuum for values of the parameters $W_0=-10^{-4}$, $A=1$, $a=0.1$. This results in a minimum at $\sigma_c \simeq 113$ which ensures that the calculation is under control.  At the right panel the vacuum is uplifted to a de-Sitter via the introduction of anti-D branes according to the KKLT proposal. The uplifting parameter has a value $D=3 \times 10^{-19}$, following \cite{Kachru:2003aw}}. }
\end{figure}

\section {Stabilization and Dynamical Supersymmetry Breaking } \normalsize

A minimal example that does not require an antibrane to source the uplifting of the volume modulus is given by \cite{Saltman:2004sn}. This is achieved by observing that non-zero local minima of the no-scale potential in the complex structure and the dilaton directions in moduli space can uplift the volume modulus in the same way as does the antibrane of the KKLT construction. This alternative uplifting retains the supersymmetric Lagrangian because it uses the $F$-terms of the fixed by fluxes fields.

The uplifting using a generic $F$-type supersymmetry breaking from a matter sector seems to be the most promising mechanism for the remaining volume modulus $\rho$. It was shown in \cite{Brustein:2004xn}, \cite{GomezReino:2006dk} that if the low energy theory contains a single modulus field then it is hard this field to break supersymmetry in a Minkowski vacuum unless more than one moduli fields remain unfixed (by fluxes) in the low energy. The K\"ahler potential and superpotential governing the dynamics of the moduli fields has the general structure 
\begin{equation}
K= - \sum^{n}_{a=1} n_a \ln(\Phi_a+ \Phi^\dagger_\alpha)+...
\end{equation}
\begin{equation}
W=W(\Phi_1,..,\Phi_n).
\end{equation}
Assuming that for these fields there is a stationary point then this point is stable when the eigenvalues of the mass matrix (Hessian) are all positive. This is possible only when 
\begin{equation} \label{nec-6}
\sum^{n}_{a=1} n_a>3.
\end{equation} 
The above condition is necessary for the stability of the vacuum. The result (\ref{nec-6}) explains why the racetrack models with one modulus that have been discussed in the literature from the middle '90s have failed to produce a model for stabilizing moduli with zero or positive cosmological constant. This is also why a supersymmety breaking dominated by the dilaton cannot be realized in a controlable way. 

In the case of the volume modulus $\rho$, that has $n_\rho=3$ (\ref{K1-6}), it violates only marginally the necessary condition when considered on its own. In this case subleading corrections to the K\"ahler potenials are crucial, since even a slight change in the curvature can stabilize it. Nevertheless, in the absence of corrections and an extra uplifting sector, the remaining volume modulus can be stabilized only in an anti de-Sitter minimum. If there are two light moduli then there are examples where stabilization can be achieved in regions where string perturbation theory is under control.

When the K\"ahler potential can be approximated by the canonical form then the stability condition that constrains the K\"ahler curvature is fulfilled. In principle it should be possible to realize the idea of uplifting assuming a supersymmetry breaking sector dominated by chiral superfields with approximatelly canonical K\"ahler with negligible mixing between the fields.  Hence, the effective theory that can stabilize the volume modulus at a Minkowski vacuum should be
\begin{equation} \label{uplifting-6}
K=K_\rho + K_\text{up}  \,\,\,\,\,\,\,\,\, \text{and} \,\,\,\,\,\,\,\,\, W=W_\rho + W_\text{up}.
\end{equation}
Due to gravitational effects the two sectors will unavoidable interact and influence its other. However, it is possible the uplifting sector to have a mild effect on the supersymmetric sector, thereby providing the uplifting without destabilizing the fields. The supersymetry breaking can be realized either by $D$-terms  \cite{Burgess:2003ic, Jockers:2004yj, Dudas:2005vv, Villadoro:2005yq, Achucarro:2006zf, Haack:2006cy}  or $F$-terms \cite{Saltman:2004sn, GomezReino:2006dk, Lebedev:2006qq, Dudas:2006gr, Kallosh:2006dv} in the matter sector. The former case generically leads to very heavy gravitino mass. The later can naturally produce the appropriate intermediate energy scale and uplift the volume modulus to a Minkowski vacuum with a TeV range gravitino mass. Also, the fact that the  $D$-term breaking cannot exist without an amount of $F$-term makes the $F$-term uplifting from a matter sector for the volume modulus preferable.

\section{ Stabilization of the Volume Modulus at a Minkowski Supersymmetric Vacuum}

Although it is not possible to stabilize the remaining volume modulus into a Minkowski supersymmetry breaking minimum, the stabilization can be realized when one asks for Minkowski vacuum that preserves supersymmetry. Keeping the K\"ahler potential  (\ref{K1-6}) of the KKLT model but assuming a \itshape racetrack  \normalfont superpotential 
\begin{equation}
W=W_0 + A e^{i a \rho} + B e^{ib\rho}
\end{equation}
a potential with two supersymmetric vacua, one Minkowski and one anti de Sitter can be found \cite{Kallosh:2004yh}. The $W_0$ is again a tree level contribution that arises from fluxes and the exponentials, that amount to corrections to the no-scale structure of the model,  arise either from Euclidean D3 branes or from gaugino condensation on D7 branes. A supersymmetric Minkowski vacuum lies at 
\begin{equation}
W(\sigma_c)=0, \,\,\,\,\,\,\,\,\,\,\,\,\,\, DW(\sigma_c)=0
\end{equation}
which has an acceptable lifetime. As in KKLT, the real part of the volume modulus $\rho$ (or the imaginary part of the $\varrho=i\rho$ ) is set to zero and $\rho=i\sigma$. In the supersymmetric Minkowski vacuum the gravitino mass vanishes. Hence, in this model the uplifting via e.g. $F$-terms is not required. This has important implications since the amount of the uplifting is tightly related to the scale of supersymmetry breaking, i.e. the $m_{3/2}$, as explained before. Here, the high of the barrier that seperates the Minkowski vacuum from the second supersymmetric AdS can be arbitrary high with respect to the $m_{3/2}$. 

Another advantage of this model is that it can accommodate high scale inflation. In stringy inspired models the inflaton potential induces an extra term at the volume modulus direction, an uplifting term  
\begin{equation} \label{inf-6}
V^\text{inf}_\text{tot} \approx V(\sigma) + \frac{V(I)}{\sigma^3}
\end{equation} 
where $I$ the inflaton field. Such a kind of inflation can be achieved by considering dynamics of branes in the compactified space. In the KKLT case, the uplifting due to the inflaton potential can be acceptable as long as the volume modulus is not destabilized and thus, the internal space not decompactified. This entails the constraint for the Hubble parameter
\begin{equation}
H^2 \approx V_B/3 \sim |V_\text{AdS}| /3 \sim m_{3/2}.
\end{equation}
The runaway $\sigma^{-n}$ dependence (\ref{inf-6}) of the energy densiy in string theory is quite generic and the $\sigma^{-3}$ appears explicitly in the $D$-term contribution to the vacuum energy like in D3/D7 inflation \cite{Dasgupta:2004dw}. In principle it might be possible to design inflationary models where the inflaton potential depends on $\sigma$ and $I$ in different way e.g. due to non-perturbative effects involving both fields. This could prevent vacuum destabilization at large enrgy density. However, no examples of such models are known.

\section{Gauge Mediated O'Raifeartaigh Supersymmetry Breaking as the Uplifting Sector for the Volume Modulus}

As explained above, a successful stabilization scheme for the remaining in the low energy theory volume modulus is the following: the volume modulus is stabilized via corrections to the no-scale structure at an anti de-Sitter supersymmetric vacuum and uplifted via $F$-terms originating from a sector that contains matter fields and breaks supersymmetry spontaneously. Looking also for a scheme that mediates the supersymmetry breaking through gauge interactions, hence a supersymmetry breaking scale $m_{3/2}< {\cal O}$(TeV), then the $F$-term uplifting is the profound choice contrary to the D-terms that generally give large $m_{3/2}$.

The first step is to combine the theory describing the "stringy relic" volume modulus and the one describing the matter sector. Following \cite{Lebedev:2006qq} we assume a K\"ahler potential of the form 
\begin{equation} \label{KX-6}
K=-3 \ln (\varrho+ \bar{\varrho})+ |X|^2
\end{equation}
where $\varrho=i\rho$ the volume modulus and $X$ the gauge singlet matter field that parameterize the supersymmetry breaking in the hidden matter sector. The "modular weight" of the field $X$ is considered to be zero. Systems of this type arise in type IIB and heterotic string theory, \cite{Witten:1985xb}. The effective superpotential is assumed to have the generic form
\begin{equation} \label{WKLN-6}
W=\sum_i \omega_i(X)e^{-a\varrho} + f(X)
\end{equation}
where the sums run over the gaugino condensates. The functions $\omega_i(X)$ and $f(X)$ can arise due to perurbative and non-perturbative interactions in the process of integrating out heavy fields. Once the supergravity potential (\ref{sugrascalar-1}) is computed the stationary points can be found and checked whether they are stable or not. The (\ref{KX-6}) suggests that stable stationary points can be expected.  We are interested in a gauge mediation scheme thus, a matter dominated supersymmetry breaking and for the $F$-terms holds $F_\varrho \ll F_X$ which corresponds to large $W_{\varrho\varrho}$.  The eigenvalues of the mass matrix are positive when roughly $(\varrho+\bar{\varrho})W_{\varrho\varrho} \gg (\varrho+\bar{\varrho}) W_{\varrho X}, W_X$ and $W_{\varrho\varrho} \gg |W_0|^2$, where $W_0$ the constant term of the (\ref{WKLN-6}). These conditions generally hold when the modulus is heavy compared to the gravitino which is the case for the gravity mediation and therefore for the gauge mediation as well. The limit $F_\varrho \ll F_X$ corresponds to large $W_{\varrho\varrho}$ and the stationary point is stable. This result agrees completely with the conclusions of the section 7.4: since the volume modulus is not accounted for the breakdown of supersymmetry it can be safely stabilized.  The stationary point $\partial V/ \partial \varrho$ implies 
\begin{equation}
W_{\varrho\varrho} F_\varrho + \text{smaller terms}=0
\end{equation}
hence, the volume modulus is stabilized close to the supersymmetric point. It can be said that the supersymmetric sector, including the moduli fields, is appropriately shielded from the matter supersymmetry breaking sector and therefore, the matter sector justifies its name as the uplifting one.

These results suggest that the Kitano model (\ref{KitanoK}), (\ref{KitanoW}) 
\begin{equation} \label{Kitano-6}
W=FX- \lambda X \phi \bar{\phi} + c,  \,\,\,\,\,\,\,\,\,\,\,\,\,\, K=\bar{X} X - \frac{(\bar{X}X)^2}{\Lambda ^2}
\end{equation}
can be safely used as the uplifting sector for the volume modulus potential. Here, the $\bar{X}$ denotes $X^\dagger$. We recall that the constant $c$ was introduced in order to tune the value of the potential at minimum, i.e. the cosmological constant, to zero. Since the positive contribution of the $F$-term of (\ref{Kitano-6}) is wanted for the uplifting we suspect that the constant $c$ is not necessary. From another point of view, the constant $c$ can be interpreted as the 'downlifting' contribution coming from the AdS vacuum of the volume modulus. Therefore, we consider the coupled system 
\begin{equation} \label{total-6}
W=W_0+ Ae^{-a\varrho}+FX  \,\,\,\,\,\,\,\,\,\,\,\,\,\,  K=-3 \ln(\varrho + \bar{\varrho})+\bar{X} X - \frac{(\bar{X}X)^2}{\Lambda ^2}
\end{equation}
that is of the form (\ref{uplifting-6}). At the (\ref{total-6}) the messenger dependent part of the superpotential has been ommited since it is irrelevant for the uplifting discussion. The two sectors do not mix and the stability conditions are satisfied. The no mixing assumption is justified since the Kitano model is expected to originate from an intermediate energy SQCD theory like the ISS model. Such theories can be effectively described by the prototype O'Raifeartaigh model which at energies below the scale $\Lambda$ reduces to the simple $W=FX$ form (see chapter 4). The total potential of the (\ref{total-6}) reads
\begin{equation} \nonumber
V_\text{K-KKLT}=\frac{e^{\bar{X} X- \frac{(\bar{X}X)^2}{\Lambda^2}}}{(\rho+\bar{\rho})^3} \left\{\left(1+4\frac{\bar{X}X}{\Lambda^2} \right)\left[F^2+\left( FXW_0  + F^2\bar{X}X + FX Ae^{-a\varrho}-2FW_0 \frac{\bar{X}X^2}{\Lambda^2} + \right. \right. \right.
\end{equation}
\begin{equation}\nonumber
\left. -2 F Ae^{-a\varrho}\frac{\bar{X}X^2}{\Lambda^2}- 2F^2\frac{(\bar{X}X)^2}{\Lambda^2} +\text{h.c.}\right) + 
\left. \left.   \left(\bar{X}X-4\frac{(\bar{X}X)^2}{\Lambda^2} +  4\frac{(\bar{X}X)^3}{\Lambda^4}\right) \left(W^2_0 +W_0(FX+ \right. \right. \right.
\end{equation}
\begin{equation} \nonumber
\left. \left. \left. Ae^{-a\varrho} +\text{h.c.}) +F^2\bar{X}X+ (FX Ae^{-a\varrho}+\text{h.c.}) +A^2e^{a(\varrho+\bar{\varrho})}\right) \right] +\frac{(\varrho+\bar{\varrho})a^2}{3}A^2 e^{a(\varrho+\bar{\varrho})}+    \right. 
\end{equation}
\begin{equation}
\left. + \left[(\varrho+\bar{\varrho})(W_0+Ae^{-a\varrho}+FX)Aae^{-a\varrho} +\text{h.c.} \right]\right\}.
\end{equation}
The above expression for the potential is not much illuminating. It can be decomposed and represented in a compact and more clear form as
\begin{equation} \label{pot-6}
V_\text{K-KKLT}=V_\text{KKLT}(\varrho, \bar{\varrho}) + \frac{V_\text{K}(X,\bar{X})}{(\varrho+\bar{\varrho})^3} +V_\text{mix}(X, \bar{X}, \varrho, \bar{\varrho}).
\end{equation}
The first term at (\ref{pot-6}) is the KKLT potential (\ref{KKLT-6}) without the unnecessary uplifitng explicit susy breaking term and the second term comes from the $V\supset e^K V_\text{K}$ which is of the standard uplifting form for the K\"ahler moduli. The $V_\text{K}=V_\text{K}(\bar{X},X)$ stands for the Kitano potential without messengers and the constant term $c$. It is of the form
\begin{equation} \nonumber
V_\text{K}= e^{\bar{X}X-\frac{(\bar{X}X)^2}{\Lambda^2}}\left(1+4\frac{\bar{X}X}{\Lambda^2} \right)\left\{ F^2 + \left(F^2\bar{X}X-2F^2\frac{(\bar{X}X)^2}{\Lambda^2} +\text{h.c.}  \right) +\right.
\end{equation} 
\begin{equation}
\left. + \left(\bar{X}X - 4\frac{(\bar{X}X)^2}{\Lambda^2} +  4\frac{(\bar{X}X)^3}{\Lambda^4}  \right) F^2\bar{X}X     \right\}
\end{equation}
and keeping only the dominant terms it can be approximated by
\begin{equation} \label{Kit-appr-6}
V_\text{K} \simeq F^2+ 4\frac{F^2}{\Lambda^2}\bar{X}X+ {\cal O}\left((\bar{X}X)^2 \right).
\end{equation}
Obviously the $V_\text{K}$ seperately yields a minimum at the origin $X=0$ and mass for the spurion $m_X=2F/\Lambda$. The last term at (\ref{pot-6}) contains the mixing terms. The main importance of the mixing terms is that provide a linear to $X$ term that shifts the minimum away from the the origin. They read 
\begin{equation}\nonumber
V_\text{mix} = e^K \left\{ K^{X\bar{X}} \left[ \left(FXW_0 + FXAe^{-a\varrho}+{\cal O} \right( \bar{X}X^2 \left)+\text{h.c} \right)+ \bar{X} X \left( W^2_0 + A^2 e^{-a(\bar{\varrho}+\varrho)}\right) \right] + \right.
\end{equation}
\begin{equation} \label{Xmin-6}
\left. + K^{\varrho \bar{\varrho}}\left(\frac{3}{\varrho+\bar{\varrho}} FX \, a Ae^{-a \varrho} +\text{h.c.} \right) \right\}
\end{equation}
where we kept for clarity the K\"ahler potential $K$ explicitly. The quadratic in $|X|$ terms are negligible with respect to the dominant quadratic term (\ref{Kit-appr-6}) and hence the $X$-mass is to a good approximation the $2F/\Lambda$. The field responsible for the breakdown of supersymmetry is the $X$ field:
\begin{equation}
D_\varrho W=A e^{-a\varrho}\left(\frac{\varrho+\bar{\varrho}}{3}a+1 \right)+W_0+FX
\end{equation}
\begin{equation}
D_X W=F+\left(\bar{X}-\frac{X^2\bar{X}}{\Lambda^2} \right) (W_0+e^{-a\varrho}+FX) 
\end{equation}
At the $(\varrho, X)=(\varrho_c, 0)$ the supersymmetry breaks due to the $X$ field $F$-term. The real values $\varrho_c=\sigma_c+ i\alpha_c=\sigma_c$ and $X=0$ are the minima when the KKLT and the Kitano models are calculated separately. The imaginary parts of the fields vanish even when the coupled system is considered. Once we assume the cancellation of the cosmological constant there will be a slight shift of the values. This can be demonstrated by approximating the potential at the volume modulus direction by the quadratic piece of the Taylor expansion and including the dominant uplifting contribution:
\begin{equation} \label{KKKLTa-6}
V_\text{K-KLLT} \simeq \frac{1}{2}V_{\sigma\sigma} (\sigma-\sigma_c)^2 +\frac{F^2}{(2\sigma)^3}
\end{equation}
This expression is more illuminating when it is written in terms of the modulus mass $m_\sigma$. Due to the non-canonical form of the K\"ahler potential the kinetic term of the $\varrho$ fields is canonically normalized after the rescaling $\varrho \rightarrow (\left\langle\varrho+\bar{\varrho} \right\rangle /\sqrt{3}) \varrho$; in the present case it reads $\sigma \rightarrow (2\sigma_\text{min}/\sqrt{3})\sigma$. The $\sigma_\text{min}$ is the actual minimum of the coupled system. Hence, at the scalar potential, there is aslo the rescaling $G_\varrho \rightarrow (\left\langle\varrho+\bar{\varrho} \right\rangle /\sqrt{3}) G_\varrho$ or $G_\sigma \rightarrow (2\sigma_\text{min}/\sqrt{3}) \sigma$. Then the (\ref{KKKLTa-6}) reads
\begin{equation} \label{KKKLTb-6}
V_\text{K-KKLT}\simeq\frac{1}{2} K_{\sigma \sigma} m^2_\sigma (\sigma-\sigma_c)^2 +\frac{F^2}{(2\sigma)^3}
\end{equation}
where the $K_{\sigma \sigma}=3/(2\sigma_\text{min})^2$ and 
\begin{equation}  \label{massm-6}
m_\sigma=m_\varrho=e^{G/2}G^{\varrho\bar{\varrho}}G_{\varrho\varrho}.
\end{equation}
From (\ref{KKKLTb-6}) we see that the $\sigma$-direction minimum will be shifted by $\delta\sigma\equiv (\sigma_\text{min}-\sigma_c) \sim F^2/(m^2_\sigma \sigma^2_\text{min})$. This leads to a new $F$-term for the volume modulus $F_\varrho \sim \delta\sigma  \partial_\varrho(D_\varrho W) \sim F^2/(m_\sigma \sigma^{5/2}_\text{min})$. However, the $D_\varrho W \ll D_X W$ at the actual minimum of the coupled system will hold. 

The stationary point at the $X$-direction, keeping only the leadind quadratic and linear to $X$ terms,  is given by 
\begin{equation} \nonumber
\frac{\partial V_\text{K-KKLT}}{\partial \bar{X}}=0 \,\,\,\,\,\,\, \Rightarrow \,\,\,\,\,\,\,\, X_\text{min} \simeq-\frac{\Lambda^2}{4F}\left(W_0+Ae^{-a\varrho}+(\varrho+\bar{\varrho})aAe^{-a\varrho} \right)
\end{equation}
\begin{equation}
\,\,\,\,\,\,\,\,\,\,\,\,\,\,\, \,\,\,\,\,\,\,\,\,\,\,\,\,\,\,\,\,\,\,\,\,\,\,\,\,\,\,\,\,\,\,\,\,\,\,\,\,\,\,\,\,\,\,\,\,\,\,\,\,\,\,\,\, \simeq -\frac{\Lambda^2}{4F}\left(W_0+Ae^{-a\sigma_c}+2\sigma_c a A e^{-a\sigma_c} \right)
\end{equation}
At the minimum of the potential the cosmological constant must vanish. The vacuum expectation value of the $X$-filed is  $X_\text{min} \ll 1$ and the mixing terms are negligible. The value at the minimum can be approximated by
\begin{equation} 
V_\text{K-KKLT}(\sigma_\text{min}, X_\text{min}) \simeq V_\text{KKLT}(\sigma_c) +\frac{V_\text{K}(X=0)}{(2\sigma_c)^3}+ \text{higher order terms}.
\end{equation}
From (\ref{VAds-6}) we take the relation between the AdS depth of the KKLT volume modulus and the magnitude of the $F$-term supersymmetry breaking:
\begin{equation} \label{CC-6}
-\frac{a^2A^2 e^{-2a\sigma_c}}{6\sigma_c}+\frac{F^2}{(2\sigma_c)^2}=0.
\end{equation}
The second condition is that $D_\varrho W \ll D_X W$ which for $FX_\text{min} \ll 1$ is given to a good approximation by (\ref{min-6}) i.e. the KKLT limit. Combining the (\ref{min-6}) with the (\ref{CC-6}) we take the relation
\begin{equation}
F=\sqrt{3}\left(\frac{2}{3}\sigma_c a A e^{-a\sigma_c} \right)=\sqrt{3}\left(-W_0-Ae^{-a\sigma_c} \right)= - \sqrt{3} \, W_\text{KKLT}(\sigma_c).
\end{equation}
Given the above result, the value of the $X$ minimum (\ref{Xmin-6}) takes the simple form
\begin{equation}
X_\text{min}= \frac{W_0+Ae^{-a\sigma_c}}{2F}\Lambda^2 =-\frac{\sqrt{3}}{6}\Lambda^2.
\end{equation}

This is exactly the minimum of the Kitano model apart from the minus sign that is transformed away when $FX \rightarrow -FX$ at the superpotential (\ref{total-6}).

This result has important implications for the Kitano-Polonyi model and, generally, for every O'Raifeartaigh-like model of $F$-term supersymmetry breaking that include a constant term $c$ in the superpotential in order that the cosmological constant to cancel at the vacuum. This constant $c$ may originate from a sector that yields an AdS vacuum as the KKLT volume modulus vacuum. This dyanamical interpretation of the constant term $c$ is rather attractive. In particular, here the constant $c$ of (\ref{KitanoW}) is the value of the KKLT superpotential at the minimum
\begin{equation}
c=-W_0-Ae^{-a\sigma_c}=-W_\text{KKLT}(\sigma_c).
\end{equation}

The lesson of this uplifting mechanism is that the sclale of supersymmetry breaking is connected with the expectation value of the superpotential and the high of the barrier that prevents the volume modulus from the runaway:
\begin{equation}
F \sim \left\langle W_\text{KKLT} \right\rangle \, ,\,\,\,\,\,\,\,\,\,\,\,\,\,\,\,\,\,\,  V_\text{barrier}  \sim \frac{F}{(2\sigma_c)^3}
\end{equation}
This has a twofold implication. Firstly, the gravitino mass is suppressed relatively to the $F$-term by the the expectaion value of the volume modulus cubed
\begin{equation}
-V_\text{AdS}= -3 e^K |W|^2=-3 \, m^2_{3/2} \simeq \frac{F^2}{(2\sigma_\text{min})^3}. 
\end{equation}
Hence, due to the logarithmic non-minimal K\"ahler potential the actual $F$-terms of the matter superpotentials are $(2\sigma_{min})^{3/2}$ larger. However, once the modulus is integrated out the effective $F$-term will be $F\rightarrow F/(2\sigma_{min})^{3/2}$ and the standard relation $m_{3/2}=F/\sqrt{3}$ holds.

Secondly, this imposes tight constraints on the inflationary model building rulling out all the high-scale models that apparently destabilize the volume modulus. Moreover, as we will present at the next chapter the high of the barrier also bounds the highest value of the reheating temperature experienced in the universe. The mass of the volume modulus is related to the high of the barrier and therefore to the $m_{3/2}$ mass at the class of uplifting models of this section. In particular at the minimum of the potential
\begin{equation}
m_\sigma=e^{G/2}G^{\varrho\bar{\varrho}}G_{\varrho\varrho} =  \left( \frac{V_{\sigma\sigma}}{K_{\sigma\sigma}} \right)^{1/2} \simeq \left(a^2 \sigma_c W^2_0 \right)^{1/2} \simeq \, 2a\sigma_c m_{3/2}
\end{equation} 
Thermal effects at high temperature are expected to wash out the $m_{3/2}$-related curvature of the modulus potential and hence, destabilize it. This is a reason to consider also the racetrack models like the KL proposal. Althought, the original KL model yields a Minkowski vacuum an appropriate AdS vacuum can be readily constructed and uplifted by matter superpotentials.

\begin{figure} 
\textbf{KKLT uplifted by Matter Superpotential}
\centering
\begin{tabular}{cc}

{(a)} \includegraphics [scale=.80, angle=0]{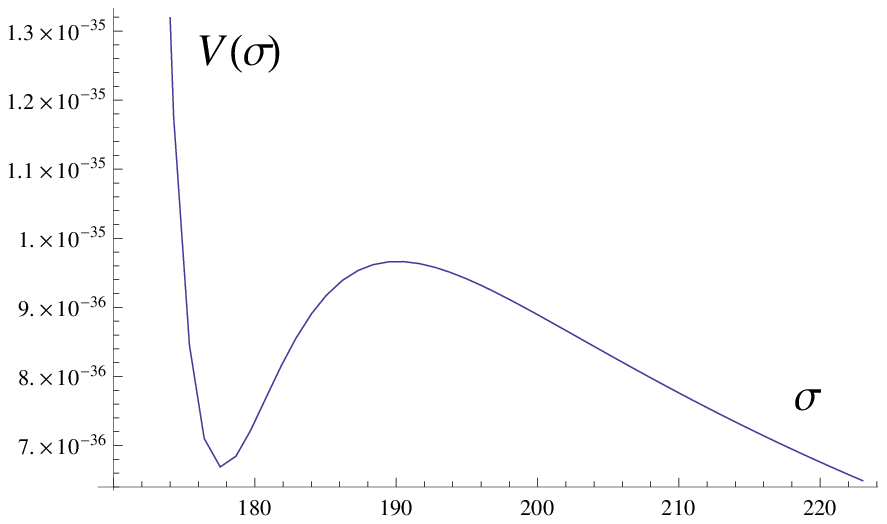} &
{(b)} \includegraphics [scale=.80, angle=0]{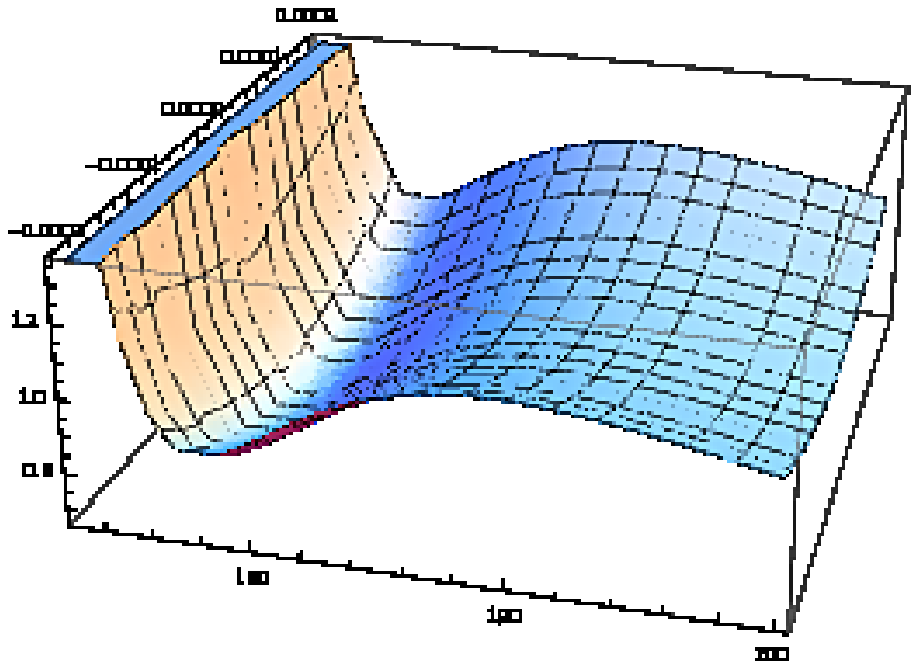}  \\
\end{tabular}
\caption{\small{ At the left panel is the KKLT potential for the overall volume modulus uplifted by the matter superpotential $W=FX$ with $K=|X|^2-|X|^4/\Lambda^2$ responsible for a low scale supersymmetry breaking.  The value of the KKLT parameters are $W_0=-10^{-14}$, $A=1$, $a=0.2$ and of the uplifting sector $F=1.2\times 10^{-14}$ and $\Lambda=10^{-3}$ tuned to yield a vanishing (of the order of ${\cal O}(F^2/(2\sigma_c)^3)$ and a light gravitino $m_{3/2}\simeq 10^{-18}$ i.e. of the order of GeV. The volume modulus minimum lies at  $\sigma_\text{min} \sim 178$ and the spurion at $X_\text{min} \simeq \sqrt{3}\Lambda^2/6$. We note that the uplifting sector shifts the value of $\sigma_{min}-\sigma_c \lesssim 1$. At the right panel the 2-dimensional $V(\sigma, X)$ potential  multiplied by $10^{36}$ is depicted}. }
\end{figure}

\subsection{Uplifting Racetrack Potentials}

The racetrack superpotential is characterized by two two exponents
\begin{equation} \label{rc-6}
W=W_0+ A e^{-a\varrho}+ Be^{-b\varrho}+FX
\end{equation}
and the same K\"ahler
\begin{equation}
K=-3\ln (\varrho+\bar{\varrho})+X\bar{X}- \frac{(X\bar{X})^2}{\Lambda^2}.
\end{equation}
The volume modulus is well possible to be stabilized in a Minkowski vacuum. Then, no uplifting is necessary. However, the positive contribution to the cosmological constant by the matter superfields that break supersymmetry has be cancelled. Hence, the construction of an Anti-de Sitter vacuum for the volume modulus is motivated in this case as well. But in the racetrack constructions there is an advantage: the high of the barrier uncorrelated with the scale of supersymmetry breaking. It can be arbitrary high. The price to pay is a second fine tuning in order to have a vanishing cosmological constant. The first tuning is the construction of the Minkowski vacuum. In other words, the tuning in this category of racetrack models is more severe because the natural scale generated here, measured by the high of the barrier, is much greater than gravitino mass $m_{3/2}\sim 1\text{GeV}-1\text{TeV}$. In the KKLT case, the scale of stabilization was comparable to the $m_{3/2}$. 
 
Here, we ask for a slight AdS minimum for the modulus to substract the $F^2$ value of the Kitano gravitational gauge mediation supersymmetry breaking minimum. According to the section 7.4 the difference in the scales implies that the uplifting sector will cause negligible shift in the vacuum expectation value for the $\varrho$. The potential for the volume modulus reads
\begin{equation}
V= \frac{e^{-2(a+b)\sigma}}{6\sigma^2} \left(bBe^{a\sigma}+aAe^{b\sigma} \right) \left[Be^{a\sigma}(3+b\sigma)+e^{b\sigma}\left(A(3+a\sigma) + 3e^{a\sigma}W_0 \right) \right]
\end{equation} 
and the slightly AdS minimum lies approximately at the value of the original Minkowski one:
\begin{equation}
\sigma_{c}= \frac{1}{a-b} \ln \left|\frac{aA}{bB} \right|.
\end{equation}
The departure from a precise Minkowski vacuum is achieved with a slight change in the parameters of the relation
\begin{equation}
-W_0= A\left|\frac{aA}{bB}\right|^{\frac{a}{b-a}} +B\left|\frac{aA}{bB} \right|^{\frac{b}{b-a}}.
\end{equation}
Hence, an extremely light gravitino but a large barrier seperating the AdS minimum from the next one and from the Minkowski decompactification vacuum at infinity is possible.

\begin{figure} 
\textbf{Racetrack}
\centering

\includegraphics [scale=.85, angle=0]{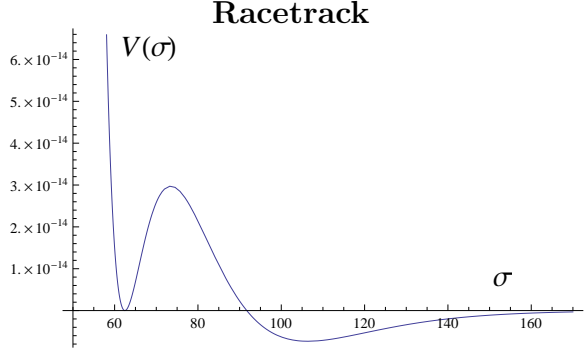} 
\caption{\small{A roughly Minkowski vacuum of the racetrack model. The precise value of minimum is $V\simeq - F^2/(2\sigma_c)^3\simeq 10^{-36}$ i.e. it is AdS and of course it is not visible since the high of the barrier is $10^{-14}$ in Planck units. The values of the parameters of the racetrack model are slightly different from those of the original KKLT for both exponentials: $A=1, a=2\pi/100, b=2\pi/99, B=1.03, W_0=-2\times 10^{-4}$.} }
\end{figure}

\chapter{Thermal Destabilization of the Volume Modulus}

In this chapter we briefly discuss cosmological issues when high reheating temperatures are assumed in the early universe. The first is the destabilization of the volume modulus due to thermal effects. We show that this potential  problem does not rule out high reheating temperatures and a standard cosmological evolution is possible. These results are going to be presented in a forthcoming publication \cite{D}\footnote{Some remarks have appeared at \cite{DL}}.

\section{ Thermal Effects on The Moduli Fields}

One important aspect of the string theory is the r\^ole of the moduli fields in the effective low energy theory. Their values determine the geometry of the compactified space as well as the gauge and Yukawa couplings. This last point, i.e. that the integrated out moduli fields are hidden in the couplings of the observable fields, is the essence of this chapter. The interaction of the light fields can affect the potential of the moduli and there is an energy threshold that destabilization takes place towards vanishing coupling and decompactification of the internal space. Highly energetic interactions $T\gg M_\text{EW}$ may have occured at the thermalized early universe. Hence, an upper bound on the highest temperatures realized in our universe has to be applied in order to prevent the moduli from the destabilization. We call this highest temperature as $T_{de}$ where the index "de" stands for destabilization.

\subsection{Thermal Average Values}

The four dimensional action for the zero modes of the real part of the volume modulus $\varrho$ contains a coupling to the supersymmetric gauge kinetic term $\int d^2\theta \, W^a W_a$. For instance
\begin{equation} \label{a-7}
S_4= \int d^4 x \left\{ \frac{g^2 \sigma}{M_P}\left(-\frac{1}{4} F^a_{\mu\nu}F^{a\mu\nu} - i\lambda^a\sigma^{\mu}(D_\mu \bar{\lambda})\right)-\frac12 m_{\lambda}(\lambda^a\lambda^a+\bar{\lambda}^a\bar{\lambda}^a) +...  \right\}
\end{equation} 
where $F^a$ the field strength, $\lambda^a$ the gaugino, $m_\lambda$ its soft mass and $\sigma$  the real part of the volume modulus $\varrho$. At finite temperature the gauge kinetic term acquires an expectation value which modifies the potential of the modulus. The energy-momentum tensor of the supersymmetric field strength $W^a$ is given by
\begin{equation}
{T^\mu}_\mu = -2 \frac{\beta(g)}{g} \left(-\frac14 F^a_{\mu\nu}F^{a\mu\nu} - i\lambda^a\sigma^{\mu}(D_\mu \bar{\lambda}) \right)
\end{equation} 
where $\beta(g)$ the $\beta$-function of the gauge coupling. The thermal average of the energy-momentum tensor
\begin{equation} \label{EM-7}
\left\langle {T^\mu}_\mu \right\rangle = \varepsilon -3 P
\end{equation}
where $\varepsilon = -P+Ts$ and according to the chapter 5 the free energy is ${\cal F}=-P$. The free energy has been calculated in perturbation theory for a gauge theory with fermions in the fundamental representation \cite{Kapusta:1989tk}. For a $SU(N)$ supersymmetric QCD theory with $N$ colours and $N_f$ matter multiplets in the fundamental representation (flavours) the free energy reads
\begin{equation} \label{free-7}
{\cal F }(g, T)= -\frac{\pi^2 T^4}{24} \left[\alpha_0 +\alpha_2 g^2 + {\cal O}(g^3) \right]
\end{equation}
with $g$ and $T$ being the gauge couplings and the temperature respectively. The zeroth order coefficient $\alpha_0= N^2 + 2N N_f-1$ counts the number of degrees of freedom and the one-loop coefficient is given by
\begin{equation} \label{a2-7}
\alpha_2 = -\frac{3}{8\pi^2}(N^2 - 1)(N+3 N_f).
\end{equation}
Considering that $N_f=0$ in the present case and for $s=\partial P / \partial T $ the thermal average of the energy momentum tensor the (\ref{EM-7}) is given by
\begin{equation}
\left\langle {T^\mu}_\mu \right\rangle = \left\langle  -\frac14 F^a_{\mu\nu}F^{a\mu\nu} - i\lambda^a\sigma^{\mu}(D_\mu \bar{\lambda})  \right\rangle = -\frac{\pi^2}{24}\, \alpha_2 g^2 T^4.
\end{equation}
This is a positive quantity and does not depend on the $\beta$-function. Because of the trace anomaly of the energy momentum tensor \cite{Grisaru:1983hc, Novikov:1984bt}  one no longer has $P=\varepsilon/3$. Hence, according to the action (\ref{a-7}) there is a thermally induced term for the modulus. Considering small fluctuations about the minimum of the potential the Lagrangian reads \cite{Buchmuller:2003is}
\begin{equation}
{\cal L} =\frac12 (\partial \sigma)^2- m^2_{\sigma}\, \sigma^2 -\frac{\pi^2}{24}\, \alpha_2 g^2 T^4 \frac{\sigma}{M_{P}}.
\end{equation}
where $\sigma$  the real part of $\varrho$. This linear term is a thermal effect on the volume modulus that modifies the shape of the potential. In principle, it can lead to destabilization of the volume modulus for small values the barrier, however the effect from the induced thermal masses is not the dominant one. The dominant destabilizing effect comes from the minimization of the free energy that drives the volume modulus to larger values.

\subsection{Thermal Effects on the Modulus Potential}

At finite temperature the energy of a system in thermal equilibrium is described in terms of the free energy. The free energy depends on the gauge couplings and subsequently the gauge couplings are determined by the vacuum expectation value of the moduli. The dynamical interpretaion of the coupling by string theory implies that the minimization of the free energy takes place when the couplings vanish i.e. at the decompactification limit. This is expressed by a linear negative term in the moduli effective potential at finite temperature. This result is, at first sight, unexpected since the moduli fields are gravitationally interacting fields and their interaction rate $\Gamma \sim  m^3/M^2_P$ always smaller than the expansion rate $H$. 

For a supersymmetric $SU(N)$ gauge theory with $N_f$ matter multiplets in the fundamental representation the free energy is given by (\ref{free-7}). 
The crucial fact is that the $\alpha_2$ is negative (\ref{a2-7}). Consequently, gauge interactions increase the free energy, at least in the weak coupling regime. This means that the gauge coupling, being a dynamical quantity, is pushed towards smaller values as temperature increases. It can be shown \cite{Buchmuller:2004xr} that even when the coupling $g={\cal O}(1)$ the perturbation theory results are quantitatively consistent with numerical lattice QCD results. 

For the case of KKLT stabilization the remaining single K\"ahler modulus $\varrho=\sigma+i\alpha$ has a superpotential and K\"ahler given by (\ref{K-KKLT-6}) and (\ref{W-KKLT-6}) with $\varrho=-i\rho$. The exponent in the superpotential may originate either from D-brane instantons or gaugino condensation. In the later case the exponent is given by the $\beta$-function of the corresponding gauge group: $4\pi \alpha =3/(2\beta)$. A gauge field in the 4D effective theory may arises from $D7$ branes wrapping a 4-dimensional cycle $\Sigma$ in a Calabi-Yau manifold. Then the low energy gauge coupling is determined by the volume of the cycle: $g^{-2} \propto \text{Vol} (\Sigma)$ which is the real part of the volume modules $\varrho$ in the IIB strings. Hence, an effective 4-dimensional Yang-Mills gauge coupling on the $D7$-branes is related to $\sigma$ as
\begin{equation}
\sigma=\frac{4\pi}{g^2}\,.
\end{equation}
We consider a KKLT model uplifted by $F$-terms from matter (\ref{total-6}). Hence,
\begin{equation}
W=FX+Ae^{-a\sigma}+W_0, \quad \quad K=-3\ln(2\sigma)+\bar{X}X-\frac{(\bar{X}X)^2}{\Lambda^2}
\end{equation}
At finite temperature the effective potential for the volume modulus reads
\begin{equation}
V^T(\sigma)= V_\text{K-KKLT}(\sigma) + {\cal F}(g(\sigma), T).
\end{equation}
As the temperature increases the local minimum of the uplifted KKLT becomes more and more shallow until a critical temperature $T_{de}$ where it disappears. At that temperature the maximium of the potential, i.e. the peak of the barrier, and the local minimum merge into a saddle point. The $T_{de}$ can be defined by the two equations \cite{Buchmuller:2004xr}
\begin{equation}
V'(\sigma_{de})+ {\cal F}'(g(\sigma_{de}), T_{de})=0
\end{equation}
\begin{equation}
V''(\sigma_{de})+ {\cal F}''(g(\sigma_{de}), T_{de})=0
\end{equation}
In the linear approximation the equations for the critical value of the volume modulus and the critical temperature  become $V''(\sigma_{de})=0$ and the critical temperature
\begin{equation}
T_{de}=\kappa \left[ V'(\sigma_{de}) \right]^{1/4}
\end{equation} 
where $\kappa= {\cal O}(1)$. The $\sigma_{de}$ lies between the zero temperature $V_\text{K-KKLT}$ local minimum and the barrier peak. Following \cite{Buchmuller:2004tz} \cite{Buchmuller:2004xr}, the critical temperature is given by
\begin{equation} \label{Tde-7}
T_{de} \sim \sqrt{m_{3/2}} \left(\frac{2}{B} \right)^{1/4} \left(\frac{3}{\beta}\right)^{1/4} \left(\frac{1}{g^2} \right)^{3/8}
\end{equation}
where $B=T^{-4} \partial {\cal F}/\partial g(\sigma_{de})$ and $\beta$ the beta function of the corresponding gauge group that lifts the no-scale structure of the overall volume modulus through non-perturbative effects. The destabilization temperature (\ref{Tde-7}) depends linearly on the square root of the gravitino mass. For $m_{3/2} \sim 100$ GeV and $\beta \sim 0.1$ the critical temperature of the destabilization is 
\begin{equation} \label{de-7}
T_{de} \sim 10^{11} \text{GeV}\, .
\end{equation}
Obviously, a low scale supersymmetry breaking as the gauge mediation scheme decreases the destabilization temperatures. This may be in tension with the results of chapter 6 where high reheating temperaure, of the order of $10^{9}$ GeV were necessary for a thermal selection of the metastable vacuum. For example a gravitino of mass $10$ MeV then temperatures at least of order $5\times 10^{7}$ GeV are necessary for the metastable vacuum selection whereas the destabilization of the volume modulus constrains the highest value of the reheating temperature to be $T_{rh}<T_{de}\sim 10^9$ GeV. Hence, a thermal implementation of supersymmetry breaking and the avoidance of decompactification is only marginally satisfied. In the following we claim that the destabilization temperature is found to be higher once the dynamics of the volume modulus are taken into account \cite{D}.

\begin{figure} 
\centering
\begin{tabular}{ccc}

{(a)} \includegraphics [scale=0.5, angle=0]{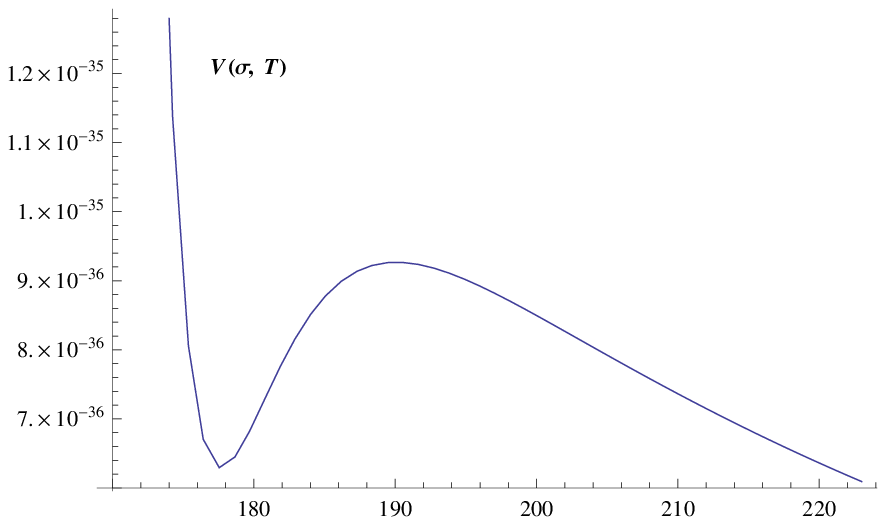} &
{(b)} \includegraphics [scale=0.5, angle=0]{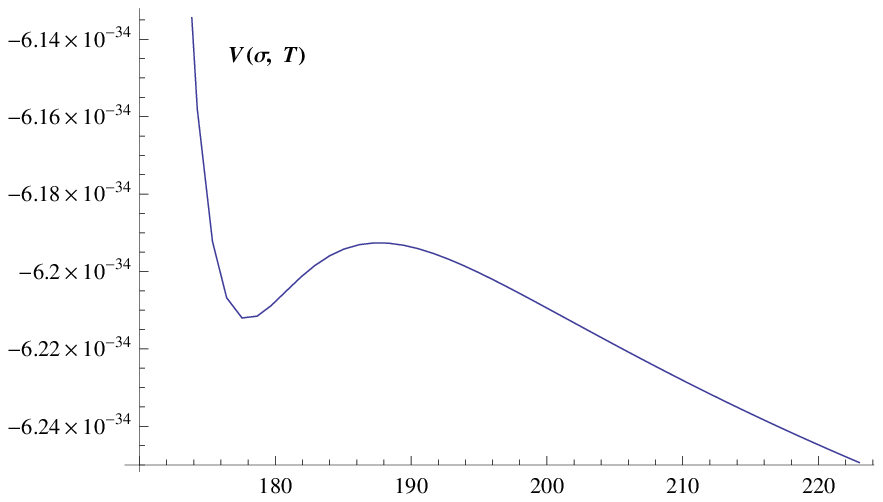}  &
{(c)} \includegraphics [scale=0.5, angle=0] {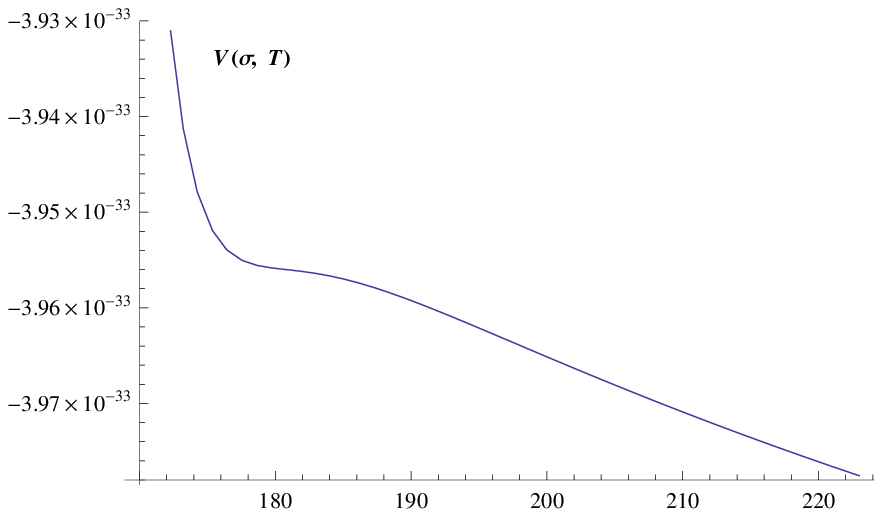}  \\
\end{tabular}
\caption{The effective potential of the overall volume modulus uplifted by matter superpotentials at finite temperature. From left to right the temperature increases.} 
\end{figure}

\section{Volume Modulus Kinematics}

The analysis presented above describes the change of the shape of the volume modulus potential and the $T_{de}$ is the critical temperature where the local minimum becomes an inflection point. However, at that moment the curvature of the potential and hence the volume modulus mass squared is zero. The system of the volume modulus $\varrho$ and the matter fields are parts of the energy density of a universe dominated by radiation. The Hubble parameter scales like $H\sim T^2/M_P$ and effectively acts like a friction term on the $\text{Re}(\varrho)=\sigma$. Assuming that initially the volume modulus is localized at the local minimum of the KKLT potential then, as the temperature increases the value of the minimum is shifted towards larger vevs. The volume modulus has no initial velocity and a "static" friction is acting on it.  Hence, at the temperature $T_{de}$ the modulus does not run-away. The run-away will take place at higher temperatures when the potential slope is steep enough to outweight the Hubble friction. The moment when 
\begin{equation} \label{mra-7}
m^2_\sigma(T) \sim  H^2(T_{ra})
\end{equation}
the modulus will run away and 
\begin{equation}
T_{ra}>T_{de}
\end{equation}  
where the index "ra" stands for the run-away. The $m^2_\sigma(T)$ is given by the effective potential written below (\ref{veff-7}). The relavant fields that rule the kinimatics of the modulus are described by the action 
\begin{equation}
S= -\int d^4x \sqrt{-g}\left(\frac12 M^2_P R - {\cal L}(\sigma, X, \gamma) + {\cal F}(g, T) \right)
\end{equation} 
where the $\gamma$ stands for the light fields part of the radiation plasma. The minimal approximation is to consider only the fields of the Yang-Mills $SU(N)$ gauge fields and the $N_f$ flavours of matter, although, in principle, a second component of relativistic particles that only interacts with the modulus gravitationally can also be present in the dynamics. The radiation energy density reads
\begin{equation}
\rho_r= -\frac{\pi^2}{24} T^4 \alpha_0(1+rg^2)T^4
\end{equation}
where $r=\alpha_2/\alpha_0$. This is actually the free energy. The equation of motion for the volume modulus is given by 
\begin{equation}
\ddot{\sigma}+3H\dot{\sigma} + K^{\sigma\sigma} V_\sigma = \frac{r\rho_r}{3(1+rg^2)}K^{\sigma\sigma}\partial_\sigma g^2(\sigma)\, 
\end{equation}
where the term $(K_{\sigma \sigma \sigma}/K_{\sigma\sigma})\dot{\sigma}^2$ has been neglected in the region of values of $\sigma$ considered here. An effective scalar potential for the $\sigma$ can be defined
\begin{equation} \label{veff-7}
V_{eff}(\sigma)=V(\sigma)-\frac{1}{3}\frac{r \rho_r g^2}{1+rg^2}.
\end{equation}
It is found that the modulus runs away for temperatures $T_{ra} \gtrsim{\cal O}(10) T_{de}$ \cite{D}.

The $T_{de}$ can become several orders of magnitude higher than the (\ref{de-7}) once racetrack models are employed. At these models the $m_{3/2}$ is replaced by the high of the barrier which is disconnected with the $m_{3/2}$. Then, the destabilization issue is basically addressed since it is claimed that the universe cannot thermalize above temperatures of the order of the GUT scale. The volume modulus then, cannot get destabilized by thermal effects. These stabilization  models for the volume modulus are rather attractive apart from the fact that the double  tuning is required. This problem can be addressed once the "downlifting" of the supersymmetry breaking vacuum, as those of O'Raifeartaigh models, is implemented by the supersymmetry breaking sector itself, see last section.

Another issue is that, for the KKLT case the thermal effects shift the volume modulus from its zero temperature value. This is another notorious cosmological problem related with the gravitational interacting fields. We present some of the details of this problem in the appendix. In the present case, there are two ways to encounter the problem. The first is to assume a relatively large gravitino mass. The volume modulus mass is $m_{\sigma} \sim \sigma m_{3/2}$ and for $m_{\sigma}>10$ TeV the decay of the modulus does not spoil the BBN. However, the energy stored in the oscillations of the $\sigma$ field cause a late entropy production that may dilute the baryon asymmetry and the dark matter abundance.  Moreover, the branching ratio of the moulus decay to gravitinos is too large and gravitinos are overproduced (see appendix, the moduli induced gravitino problem).
 The second way is to adopt a racetrack-type model where the modulus is rather heavy and decays at temperatures $T>10^8$ GeV.

\section{Heavy Moduli Decays}

Choosing a superpotential such that the volume modulus becomes parametrically much heavier than the gravitino a potential problem appears: the moduli decays may overproduce gravitinos (see also the appendix). The heavy moduli could either cascade through Standard Model superpartners ending in gravitinos and Standard Model particles or decay directly to a pair of gravitinos. The decay widths scale as 
\begin{equation}
\Gamma_\varrho= \frac{c}{8\pi} \frac{m^3_\varrho}{M^2_P}
\end{equation}
with a model dependent constant. For the case of modulus decays to the SSM particles the decay width is given by the coupling $\int d \theta^2 \varrho W^\alpha W_\alpha$ in the supersymmetric actions, the (\ref{a-7}) is an example. 
A straightforward calculation yields the partial widths 
\begin{equation} \label{decSM-7}
\Gamma(\varrho_{R,I} \rightarrow gg))=\Gamma(\varrho_{R,I} \rightarrow \tilde{g}\tilde{g})= \frac{n}{96\pi}\frac{m^3_\varrho}{M^2_P}
\end{equation}
where $n$ counts the degrees of freedom of the final states, e.g. $n=8$ for gluons and gluinos. 

The potential source of cosmological difficulties is the direct decay $\varrho \rightarrow 2 \psi_\mu$ i.e. to gravitinos \cite{Nakamura:2006uc, Endo:2006zj}. The coupling depends on the susy breaking $F$-term of the modulus field. If the $F_\varrho$ is nonzero then $G_\varrho \sim m_{3/2}/m_\varrho$ and the decay width is of the order $m^3_\varrho/M^2_P$ similar to the (\ref{decSM-7}). This causes cosmological problems since a gravitino of mass larger than KeV will not thermalize at low temperatures, and hence it will be overproduced and either overclosing the universe if stable or spoiling the BBN nucleosynthesis if unstable (see next chapter).

However, this problem does not always appear. It was pointed out at \cite{Dine:2006ii} that in the case of KKLT models the decay to gravitinos can be suppressed. The physical Goldstino is a linear combination of $\psi_X=x$ and $\psi_\varrho$ while the orthogonal combination is massive. Similarly, its massive scalar partner is also a linear combination of $X$ and $\varrho$:
\begin{equation}
\varphi =\epsilon X+\varrho.
\end{equation}
where the fields are canonically normalized. According to \cite{Dine:2006ii} the mixing is supersymmetric at ${\cal O}(m_{3/2}/m_\varrho)$ level and concequently the decay amplitude of the volume modulus, or more precisely of the massive scalar $\varphi$ starts at ${\cal O}((m_{3/2}/m_\varrho)^2)$. For the model studied in this and the previous chapter i.e. the (\ref{total-6}) the decay width is given by 
\begin{equation}
\Gamma(\varrho \rightarrow 2\psi_{3/2})=\frac{c_{3/2}}{288\pi}\frac{m^3_\varrho}{M^2_P},
\end{equation}
where 
\begin{equation}
c_{3/2}=\frac{27m^4_X}{(m^2_\varrho-m^2_X)^2}.
\end{equation}
Thereby, for the case of a heavy volume modulus $m_\varrho \gg m_X$ the partial decay width to gravitinos is significantly suppressed relatively to the (\ref{decSM-7}).  Hence, the moduli induced gravitino problem can be avoided. The gravitino can have the correct dark matter abundance only from scatterings off the thermalized particles in the plasma as we discuss at the next chapter.

Another important point is whether the modulus dominates the energy density of the universe, i.e. there is modulus matter-dominated era due to the modulus oscillations about its vacuum. As we saw in previous sections, thermal effects are expected to modify the potential for the volume modulus leading even to destabilization at the $T_{de}$ temperature. For temperatures lower than this value, the main modification to the potential is the shift of its minimum towards larger values. This entails an initial displacement for the volume modulus causing an oscillating phase. However, the displacement also depends on the high of the barrier and for a heavy modulus, where the barrier can be much larger than the $m^2_{3/2}M^2_P$, this displacement is much smaller. Therefore, the displacement can be small enough not to source a modulus matter-dominated era. In this case the decay of the modulus to gravitino may be irrelevant since its entropy production will be a subdominant effect, see (\ref{Trh-A}). In the opposite case that the volume modulus dominates the energy density of the universe, then according to (\ref{Trh-A}) the reheating temperature will be 
\begin{equation}
T_{rh} \sim 10^{-10}\, \text{GeV} \left(\frac{m_\varrho}{\text{GeV}} \right)^{3/2}
\end{equation}
For the parameters given at the \cite{Kachru:2003aw} the volume modulus can have a mass up to $10^{13}$ GeV and thus reheating the universe up to temperatures of the order of $10^9$ GeV. Such a scenario, requires a thermally favourable gauge mediation i.e. models presented at the chapter 6. 

It is possible that the volume modulus is light. In this case thermal effects or inflation (see appendix) may displace the modulus resulting to a late entropy production from its decay. Then, a low scale inflation or other sources of late entropy production may be necessary. For a recent work see \cite{Fan:2011ua}. This scenario, nevertheless, does not exclude the possibility that much higher reheating temperatures were realized in the early universe.
  
For the most interesting to us case of a very heavy modulus, stabilized at a Minkowski vacuum, a different mechanism for the cancellation of the cosmological constant has to be pursued. This is discussed in the next section.

\section{Avoiding The Uplifting Fine Tuning In Racetrack Models}

The constant $c$ appears in the supersymmetry breaking models with matter superpotentials in order to cancel the positive contribution of the non-zero $F$-terms at the potential i.e. the cosmological constant.  As we have demonstrated it can be interpreted as the expectation value of the superpotential of the overall volume modulus field, i.e. $\left\langle W(\varrho) \right\rangle$, stabilized in an anti-de Sitter vacuum. However, this interpretation is only economical for the KKLT scenario where the anti-de Sitter minimum is inevitable. In the case of the racetrack construction the volume modulus can be stabilized in a Minkowski vacuum and no uplifting is necessary. Asking for small negative values for the racetrack minimum is redundant and, looks like a rather contrived scenario. The minimal scheme of a modulus stabilized in a Minkowski vacuum at first place is more attractive.
 
On the other hand, positive energy contribution to the cosmological constant by the supersymmetry breaking matter fields have to be canceled and the arbitrary constant $c$ is finally "reintroduced"  by hand and to be fine-tuned at the value $c \simeq FM_P$. Since $\sqrt{F} \lesssim 10^{11}$ GeV (the upper bound corresponds to gravity mediation models) it has to be far smaller than $M^3_P$.  Thus, at first sight it cannot be directly related with microscopic dynamics like the string or other GUT scale theory. A simple assumtion is to identify the constant as an $\delta W =F(S+X)$ term where the $S$ an extra singlet chiral superfield that is stabilized at a value $S\sim M_P$ and $X$ the dominant Goldstino superfield coupled to messengers and stabilized at $X\ll M_P$. Nevertheless, the $S$-field has to be supplemented by additional dynamics to fix the Planck scale vev\footnote{An inverted hierarchy mechanism could be a reason if the $S$-field is gauged. However, dynamics become complicated, and the $M_P$ vev does not result automatically.} contrary to the dyanamics of low scale susy breaking that fix the vev of $X$. Moreover, it accounts for a Polonyi-kind field associated with the notorious cosmological problems.

A different direction is to attribute the smallness of the constant $\left\langle c \right\rangle$ to a symmetry. The $R$-symmetry is the only kind of symmetry that can suppress such a constant \cite{Dine:2009swa}. In the framework of retrofitted models an elegant interpretation for the origin of $c$ is possible. Considering the simplest of retrofitted models
\begin{equation}
W= \frac{k}{32\pi^2} \frac{X W^2_\alpha}{M_P} +XA^2 +M_{AB}AB
\end{equation}
A gaugino condensation can generate the scales, namely
\begin{equation}
\left\langle \lambda\lambda \right\rangle = N {\bar{\Lambda}}^3 e^{-kX/(N M_P)} \simeq N{\bar{\Lambda}}^3 - \frac{k\bar{\Lambda}^3}{M_P}
\end{equation}
and we can define
\begin{equation}
F \equiv \frac{k\bar{\Lambda}^3}{M_P}, \quad\quad W_0 \equiv N{\bar{\Lambda}}^3 .
\end{equation}
Hence, the low energy effective supperpotential, for $X \ll M_P$, is
\begin{equation}
W=W_0 + X(A^2-F) +M_{AB}AB.
\end{equation}
i.e a simple O'Raifeartaigh model with a constant $W_0$ of suitable order of magnitude to give a small cosmological constant. In these models this happens automatically; nonetheless tuning takes place also here.

\chapter{Gravitino Production In the Thermalized Universe}

In this final chapter we consider the thermal gravitino production in the context of hidden sector dynamical supersymmetry breaking models. We discuss the possible r\^ole of the $R$-symmetry that is generically present. We argue that due to the thermal restoration of the $R$-symmetry the Goldstino production rate is suppressed. Thereby, the bounds on the reheating temperature coming from the gravitino abundance are significantly relaxed \cite{arXiv:1110.2072}.

\section{Gravitino Cosmology}

In the gauge mediation mechanism of supersymmetry breaking, the gravitino is the Lightest Supersymmetric Particle (LSP) a fact that makes it a natural dark matter candidate. So far, almost nothing is known about the gravitino mass since it escapes the constraints applied on the Standard Model superpartners. Examples of considered mass ranges are \cite{Buchmuller:2009fm} $m_{3/2} < 1$ KeV, corresponding to hot dark matter, $1 \, \text{keV} <m_{3/2} < 15$ KeV corresponding to hot warm dark matter, $100 \, \text{KeV}<m_{3/2} < 1-10$ GeV, a range of cold dark matter and $100 \,\text{GeV}<m_{3/2} < 1$ TeV, a range of cold dark matter typical for gravity mediation. Cosmological problems \cite{Ellis:1984eq} and thereby cosmological constraints differ, depending on the mass range considered for the gravitino. If the gravitino is not the Lightest Supersymmetric Particle (LSP) or the $R$-parity is not a conserved number then it is an unstable particle, a fact that alters the cosmological bounds.

\section{Properties}
In an exact supersymmetric theory, the gravitino is a massless spin $3/2$ particle with two degrees of freedom. Once supersymmetry is broken the Goldstino fermion of supersymmetry breaking provides the longitudinal degrees of freedom and the gravitino becomes a massive spin $3/2$ particle with four degrees of freedom.  The resulting gravitino mass is $m_{3/2}=F/(\sqrt{3}M_P)$ i.e. depends linearly on the (squared) scale of the supersymmetry breaking $F$. Its properties are given by the Lagrangian
\begin{equation}
{\cal L} =- \frac12 \epsilon^{\mu\nu\rho\sigma}\bar{\psi}_\mu \gamma_5\gamma_\nu\partial_\rho\psi_\sigma -\frac14 m_{3/2}\bar{\psi}_\mu [\gamma^\mu, \gamma^\nu]\psi_\nu -\frac{1}{2M_P}\bar{\psi}_\mu S^\mu
\end{equation}
where $S_\mu$ is the supercurrect corresponding to supersymmetry transformations, see e.g. the appendix of \cite{Rychkov:2007uq}.
Hence, gravitinos $\psi_\mu$ couple Standard Model particles to their superpartners through gravitino-gaugino-gauge boson interactions and gravitino-sfermion-fermion interactions:
\begin{equation} \label{L-3/2}
{\cal L} = -\frac{i}{8 M_P} \bar{\psi}_\mu [\gamma^\nu, \gamma^\rho] \gamma^\mu \lambda^a F^a_{\nu\rho}
 -\frac{1}{\sqrt{2} M_P} D_\nu \tilde{f}^{*i} \bar{\psi}_\mu \gamma^\mu\gamma^\nu  f^i_R
 -\frac{1}{\sqrt{2} M_P} D_\nu \tilde{f}^{i} \bar{f}^i_L \gamma^\mu\gamma^\nu \psi_\mu.
\end{equation}
\\
\\
\itshape Light Gravitino \normalfont
\\
In spontaneously broken susy models the massless gravitino field $\psi_\mu$ acquires mass by absorbing Goldstino modes. Before becoming massive, the gravitino only possesses helicity $\pm \frac32$ modes and the Goldstino provides the helicity $\pm\frac12$ modes of the massive gravitino field. This suggests that the helicity $\pm \frac12$ mode of the gravitino field behaves like a Goldstino field. Actually, if the gravitino mass is much smaller than the mass splitting in the chiral and gauge supermultiplets the effective Lagrangian for the relativistic gravitino of helicity $\pm\frac12$ components is given in terms of the Goldstino field. For energies $\sqrt{s} \gg m_{3/2}$ the wave function of the gravitino of heliciy $\pm \frac12$ components is approximately proportional to $p_\mu/m_{3/2}$ where $p_{\mu}$ the momentum of the gravitino and the $\pm \frac12$ component. Hence, the gravitino field can be written as 
\begin{equation}
\psi_\mu \sim i\sqrt{\frac23} \frac{1}{m_{3/2}}\partial_\mu\psi.
\end{equation}
The $\psi$ represents the spin $\frac12$ fermionic field which can be interpreted as the Goldstino. 

Total amplitudes with the helicity $\frac12$ gravitino at external line should vanish unless susy is broken; thus it should be proportional to some susy breaking parameters. One can obtain an effective Lagrangian for the helicity $\frac12$ light gravitino field by replacing all the derivatives operated on $\tilde{f}$, $f$ and $\lambda$ by the masses $m_{\tilde{f}}$, $m_{f}$ and $m_\lambda$ of the sfermion, fermion and gaugino fields respectively. Hence, considering the Goldstino mode $\psi$ the gravitino Lagrangian density contains the terms \cite{Moroi:1995fs}
\begin{equation}\label{Gold-9}
{\cal L}_{eff,\,\text{Goldstino}} =- \frac{i m_\lambda}{m_{3/2}} \frac{1}{8\sqrt{6}M_P} [\gamma^\mu, \gamma^\nu] \bar{\psi} \lambda^a F^a_{\mu\nu} + \frac{m^2_{\tilde{f}}-m^2_f}{m_{3/2}}\frac{1}{\sqrt{3}M_P} \bar{\psi} f_R \tilde{f}^* +\text{h.c.} 
\end{equation}
where $\psi$ is the spin$\frac12$ Goldstino mode eaten by the gravitino, $(f, \tilde{f})$ a chiral multiplet and $(A^a_\mu, \lambda^a)$ a vector multiplet (see also section 3.6.3). In the supersymmetric limit i.e. $m_\lambda \rightarrow 0$ and $m^2_{\tilde{f}}-m^2_f \rightarrow 0$ the above effective Lagrangian vanishes and the helicity $\pm\frac12$ modes of the gravitino field decouple from the theory.

A light gravitino with $m_{3/2} \ll m_\lambda$ interacts effectively as the Goldstino field and it can be in thermal equilibrium, decouples when it is relativistic and accounts for hot dark matter. One sees from the Lagrangians (\ref{L-3/2}),(\ref{Gold-9}) that indeed the couplings are enhanced $m_\lambda/m_{3/2}$ for the case of interactions with particles members of a gauge supermultiplet and, $ (m^2_{\tilde{f}}-m^2_f)/(m_{3/2}p^\mu)$ for the interaction with chiral particles.

A light gravitino, given that the $R$-parity is conserved number, is a stable one. As the mass of the gravitino increases then the cosmological behaviour of the LSP gravitino changes. A ${\cal O}$(KeV) mass gravitino can have a thermal abundance that does not overclose the universe but as the mass increases it violates the overclosure bounds. A gravitino of mass at the ${\cal O}$(GeV) range can be thermalized only at ultra high temperatures that are not expected to have been realized in the early universe. Despite this fact, it can be produced by scatterings off the thermal plasma, by $2\rightarrow 2 $ reactions, and since the annihilation processes are negligible it can have a significant abundance, as we discuss in the next section.
\\
\\
\itshape Heavy Gravitino \normalfont
\\
If the mass of the gravitino is comparable to, or larger than, the mass splitting of the observable sector supermultiplets the interaction strenght of its helicity $\pm \frac12$ component is no longer enhanced compared to that of its helicity $\pm \frac32$ component. At high energies, the gravitino production cross section is dominated by the transverse spin $\frac32$ component. Hence, in this case its behaviour (as the leading high energy behaviour) is independent of the susy breaking parameters that are negligible in the total amplitude. The most important single gravitino production processes are again the $2 \rightarrow 2$ reactions where at least one of the three other external particles is a member of a gauge supermultiplet.

\section{Relic Density}

Gravitinos can be produced in the early universe via several different ways. The first is  to consider gravitino production as a result of freeze out from thermal equilibrium. The gravitino coupling is $E/M_P$ suppressed and only for hypothetical temperatures $T \sim M_P$ the gravitinos can be in a thermal equilibrium and a thermal abundance $n_{3/2} = n_{eq}$. After decoupling the gravitino thermal relic density would scale like $n_{3/2} \propto T^3$ and would have a similar to CMB photons number density. For stable gravitino, in order not to overclose the universe, i.e. $\Omega_{3/2} \lesssim 1$, their mass has to be $m_{3/2} \lesssim 1$ KeV, a bound similar to that on the relic neutrinos. In the case of unstable gravitino (either due to $R$-parity violation or when it is the LSP) there is no bound from overclosure. However, it can decay to the LSP and the gravitino lifetime is estimated to be \cite{Feng:2003zu}
\begin{equation}
\tau_{3/2} = \Gamma^{-1}_{3/2} \sim \frac{M^2_P}{m^3_{3/2}} \sim 0.1 \text{years} \left( \frac{100 \text{GeV}}{m_{3/2}} \right)^3
\end{equation}
Requiring decays to be completed before BBN commences, $t_\text{BBN}\sim 1$sec,  implies $m_{3/2} \gtrsim 10$ TeV. 

According to particle physics interactions and conventional cosmology the universe was not thermalized at temperatures close to $M_P$.  Despite the fact that the gravitino is not thermalized it can be produced by scatterings in the thermal plasma. After reheating, the universe is characterized by three seperate rates: the interaction rate of MSSM particles, $\sigma_\text{SSM} n$; the rate of interaction involving one gravitino, $\sigma_{3/2} n$; and the expansion rate $H$. The $n$ is the MSSM number density. For reheating temperatures larger than the MSSM masses, and for a heavy gravitino, the hierarchy of these three separated rates is
\begin{equation}
\sigma_\text{MSSM} n \sim T \gg H \sim \frac{T^2}{M_P} \gg \sigma_{\tilde{G}} n \sim \frac{T^3}{M_P}
\end{equation}
In the thermal bath the (M)SSM particles occasionally interact to produce a gravitino through the interactions like $A_{\rho} A_\sigma \rightarrow \lambda \psi_\mu$. The produced gravitinos then propagate through the universe essentially without interacting. If they are stable they contribute to the present dark matter density. 
The gravitino production cross sections, taking into account both the helicity $\pm \frac32$ and helicity $\pm \frac12$ contributions to the self energy, reads 
\begin{equation}
\sigma_{\tilde{G}}(p) \propto \sum^{3}_{N=1} \frac{g^2_{N}}{M^2_P} \left(1+\frac{m^2_{\lambda_N}}{3m^2_{3/2}} \right)
\end{equation}
and considering only processes involving gluons and gluinos that dominate the cross section with $g_3 \equiv g$ \cite{Bolz:2000fu} then
\begin{equation} \label{cross-9}
\sigma_{\tilde{G}}(p) \propto \frac{g^2}{M^2_P} \left(1+\frac{m^2_{\tilde{g}}}{3m^2_{3/2}} \right)
\end{equation}
for a gravitino with momentum $p$.
The gravitino abundance is determined by the Boltzmann equation
\begin{equation}
\frac{1}{a^3} \frac{d\left(n_{3/2}a^3 \right)}{dt}= \left\langle \sigma_{\tilde{G}} v\right\rangle (n^2_{rad}-n^2_{3/2})\,
\end{equation}
or equivalently
\begin{equation}
\frac{dn_{3/2}}{dt}+3Hn_{3/2}=\left\langle \sigma_{\tilde{G}} v\right\rangle (n^2_{rad}-n^2_{3/2}).
\end{equation}
Here, the source term $n^2_{rad}=n^2_{eq}$ arises from interactions such as $A_{\rho} A_\sigma \rightarrow \lambda \psi_\mu$; the sinking term originating from the inverse interaction, i.e. from gravitino annihilation processes, is negligible for gravitino with mass $m_{3/2} \gtrsim 10$ MeV. 
Taking into account that in a radiation dominated universe
\begin{equation}
H^2=\frac{g_*(T)\pi^2}{90M^2_P}T^4\equiv \frac{n_{rad}T}{3M^2_P}\, ,  \quad \quad t=\frac{1}{2H}= \frac{1}{2} \sqrt{\frac{90 M^2_P}{g_*(T)\pi^2}}\frac{1}{T^2}
\end{equation}
we can perform a variable change: $t \rightarrow T$ and $n\rightarrow Y\equiv n/n_{rad}$. Then the yield for the gravitino $Y_{3/2}$ is given by the equation
\begin{equation} \label{yield-9}
\frac{dY_{3/2}}{dT}= -\frac{\left\langle \sigma_{\tilde{G}} v\right\rangle n_{rad}}{HT}\, .
\end{equation}
For \itshape temperature independent \normalfont $\left\langle \sigma_{\tilde{G}} v\right\rangle$ the right hand side is independent of the temperature since $n_{rad}\propto T^3$ and $H\propto T^2$. The result is that the gravitino relic number density is linearly proportional to the reheating temperature $T_{rh}$. The constant of proportionality is the gravitino production cross section with the leading $2\rightarrow 2$ processes are the QCD interactions. The only temperature dependent effect as we integrate from $T_{rh}$ to $T$, with $T_{rh}\gg T$, is the dilution factor
\begin{equation}
\frac{g_*(T<1 \, \text{MeV})}{g_*(T_{rh})}=\frac{43/11}{915/4}
\end{equation}
for the MSSM.

For a \itshape light \normalfont gravitino a freezing out temperature can be defined at which the production of gravitino from the thermal plasma is effectively ceased
\begin{equation}
T^f_{3/2} \sim 2\times 10^{10} \text{GeV} \left( \frac{m_{3/2}}{10 \text{MeV}} \right)^2 \left(\frac{1 \text{TeV}}{m_{\tilde{g}}} \right)^2.
\end{equation}
Also, the relic abundance is given by
\begin{equation}
\Omega_{3/2}h^2 \simeq 0.2  \left(\frac{T_{min}}{10^{10}\,\text{GeV}} \right) \left(\frac{100\,\text{GeV}}{m_{3/2}} \right) \left(\frac{m_{\tilde{g}}(\mu)}{1\, \text{TeV}} \right)^2
\end{equation}
where $T_{min} \equiv \text{min}(T_{rh}, T^f_{3/2})$ i.e. the gravitino relic density is determined as a function of reheating temperature.  Therefore, for relatively heavy gravitino, e.g. of the order of GeV, decreasing the reheating temperature the gravitino abundance is truncated. Asking for $\Omega_{3/2}h^2 \lesssim 0.1$ a bound is applied on the reheating temperature 
\begin{equation} \label{Tlight-9}
T_{rh} \lesssim 10^6 \text{GeV} \left(\frac{m_{3/2}}{10 \text{MeV}} \right)
\end{equation}
when the gluino has mass $m_{\tilde{g}}\sim 1$ TeV. 
For gravitino mass $m_{3/2} \sim 100$ GeV the constraint on the $\Omega_{DM}h^2$ constrains the reheating temperature at $T_{rh} \lesssim 10^{10}$ GeV. Whether or not these bounds are saturated specifies the percentage of gravitino at the dark matter abundance.

For a \itshape heavy gravitino \normalfont that its production cross section is dominated by the transverse $\pm \frac32$ helicity the gaugino soft mass $m_{\tilde{g}}$ is irrelevant.
Also, the contribution from the cascade decays of observable sector sparticles into gravitinos is completely negligible. Taking the gauge couplings at an average scale of $10^{10}$ GeV one takes the approximate relation for a \itshape heavy \normalfont gravitino
\begin{equation} \label{Omheavy-0}
\Omega_{3/2}h^2 \simeq 0.1 \frac{m_{3/2}}{100\, \text{GeV}} \frac{T_{rh}}{10^{10}\,\text{GeV}}
\end{equation}
and requiring $\Omega_{3/2}h^2 \lesssim 0.1$ one obtains
\begin{equation} \label{Theavy-9}
T_{rh} \lesssim 10^{10}\, \text{GeV} \times \frac{100 \, \text{GeV}}{m_{3/2}}
\end{equation}
for $m_{3/2} \lesssim m_\lambda, \tilde{m}$ i.e. the mass splitting in the observable sector supermultiplets. If this gravitino is stable then the problem shifts to the lifetime of the Next to the Lightest Supersymmetric Particle that it may be too long lived. Big Bang Nucleosynthesis requires the NLSP to be significantly heavier than the gravitino, as we discuss in the next subsection.

\subsection{NLSP Decay and Big Bag Nucleosynthesis}

A third mechanism for gravitino production is through the cascade decays of other supersymmetric particles. In the gauge mediation scenarios the gravitino is the LSP and all cascade decays will ultimately end in a gravitino. It may be possible the gravitino to account for the dominant dark matter component even if the $T_{rh}$ is very low but also large enough so that the Next to Lightest Supersymmetric Particle (NLSP) can equilibrate.  The NLSP freezes out as usual and if it is weakly interacting its relic density will be near the desired value. It decays to the gravitino LSP at 
\begin{equation}
\tau \sim \frac{M^2_P}{M^3_\text{weak}} \sim 10^5-10^8 \, \text{sec}.
\end{equation}  
This way the gravitino becomes dark matter with relic density 
\begin{equation}
\Omega_{3/2} =\frac{m_{3/2}}{m_\text{NLSP}} \Omega_\text{NLSP}.
\end{equation}
In the case that $m_{3/2}\simeq m_{\text{NLSP}}$ the gravitino dark matter inherits the desired relic density from WIMP decay. 

However, in the gauge mediated scenarios it can be $m_{3/2}\ll m_{\text{NLSP}}$ and the contribution to $\Omega_{3/2}$ from the NLSP decays is negligible. On the contrary, a fast (before BBN) decay of the NLSP is pursued. The calculation of the lifetime of the NLSP depends inversely on the ratio of the NLSP to the gravitino mass
\begin{equation}
\tau_{\,\text{NLSP}} = \left(5.9 \times 10^4 \text{sec}  \right) \left(\frac{m_{3/2}}{1 \text{GeV}} \right)^2 \left(\frac{100 \text{GeV}}{ m_\text{NLSP}}  \right)^5.
\end{equation}
The BBN commences at temperatures $T_\text{BBN} \sim 1$ MeV corresponding to the timescale $t\sim 0.2$sec and the $\tau_\text{NLSP}$ takes comparable values. When the NLSP decays during or after the BBN can alter its successful predictions if the relic abundance of the NLSP and/or the hadronic branching fraction in the NLSP decay is large enough \cite{Covi:2010au}. For $\Omega_\text{NLSP}h^2\ll 1$ or $m_{3/2}/m_\text{NLSP}\ll 1$ these decays are a negligile source of gravitino dark matter.
\\
\\
\itshape Big Bag Nucleosynthesis \normalfont
\\
The reaction rates, during Big Bag Nucleosynthesis, for generating the observed abundance of light elements $H^+$, $D^+$, $T^+$, $^3He^{++}$ are sensitive to the relativistic degrees of freedom during that time and to the presence of late decaying particles. The Standard Model is in reasonable accord with observation. We remark here, that there is the exception of the $^7 Li$ whose abundance is predicted to be larger by a factor of 2-5 when compared with observations. Late decaying particles are generally expected in the supersymmetric extensions of the Standard Model.  In the gravity mediation scenario the gravitino is the notorious late decaying sparticle. However, in the gauge mediation scheme the gravitino, in the presence of $R$-parity, is stable as the lightest supersymmetric particle (LSP). The late decaying particles should be found in the Next to the Lightest Supersymmetric Particle (NLSP). The identification of the NLSP is model dependent, nonetheless the bino, the stau or the sneutrino are certain candidates.

\section{Thermal Restoration of the $U(1)_R$ Symmetry}

\subsection{Exact $R$-Symmetry}
In the chapter 4 we stressed the importance of a global $U(1)_R$ symmetry for the supersymmetry breaking sector. $R$-symmetry  for generic superpotentials is a necessary condition for supersymmetry breaking in the true vacuum and a spontaneously broken $R$-symmetry a sufficient one. We expect $R$-symmetry to be violated by different sources that may restore supersymmetry but not in a nearby vacuum.  These sources may be $1/M_P$ suppressed dimension-five operators in the superpotential or a constant term that cancels the cosmological constant in the vacuum.

Hence, in the global supersymmetric limit the $U(1)_R$ symmetry can be an exact symmetry. In this thesis we have examined one minimal model of ordinary gauge mediation that is characterized by a spontaneously broken $R$-symmetry:
\begin{equation} \label{min-9}
W=FX+\lambda X\phi \bar{\phi}.
\end{equation}
\begin{equation} \label{KR-9}
K=|X|^2 - \frac{|X|^4}{\Lambda^2_1} +\frac{|X|^4}{\Lambda^4_2}.
\end{equation}
The superpotential is a "classical" ordinary gauge mediation superpotential and the K\"ahler potential can originate from an $R$-symmetric O'Raifeartaigh-like sector that breaks spontaneously the $R$-symmetry, as we saw at the section 4.4.3, \cite{Shih:2007av}:
\begin{equation} \label{Shih-9}
W=FX+kX\varphi_1\varphi_2+m_1\varphi_1\varphi_3+\frac12 m_2\varphi^2_2.
\end{equation}
The chiral superfields $\varphi_1$, $\varphi_2$ and $\varphi_3$ have $R$-charges $-1$, $1$ and $3$ respectively. They are heavy compared to the (\ref{min-9}) sector i.e. $m_1, \, m_2 \gg \lambda X$ and hence, integrated out from the low energy effective theory. In a thermalized universe it is the messenger fields $\phi$, $\bar{\phi}$ that control the thermal average value of the spurion $X$-field and thus, the thermal restoration or breaking of the $R$-symmetry.

The $R$-symmetry is a symmetry of the vacuum when $X=0$ and breaks spontaneously when $X\neq 0$. For the above theory (\ref{min-9}), (\ref{KR-9}) the $R$-symmetry is restored due to thermal effects at the temperature (\ref{T-S-2}), (\ref{gl-1})
\begin{equation} \label{TR-9}
T^{(R)}_X= \frac{4}{\sqrt{5}} \frac{1}{\lambda} \frac{F}{\Lambda_1}.
\end{equation}
The $\sqrt{5}$ at the denominator of (\ref{TR-9}) originates from the assumption of a minimal case of a $\bold{5}+\bold{\bar{5}}$ messenger sector i.e. the $\phi$, $\bar{\phi}$ messenger quarks and leptons form a single complete $SU(5)$ representation. We can add additional $SU(5)$ multiplets that couple to the spurion $X$ field preserving the gauge unification. However, the gauge coupling at the GUT scale increases and if we require it to remain perturbative then we may add only 1,2,3 or 4 $\bold{5}+\bold{\bar{5}}$ pairs or a single $\bold{10}+\bold{\bar{10}}$ pair or ($\bold{5}+\bold{\bar{5}}$)+($\bold{10}+\bold{\bar{10}}$) pairs to the particle content of the minimal $SU(5)$ GUT unless the messenger scale is large enough. According to (\ref{Nindex}) $N$ values as large as $5\times N_\phi \sim 50$ are allowed.
Hence, the square root at the denominator (\ref{TR-9}) increases respectively. Here we consider that 
\begin{equation} \label{TR2-9}
T^{(R)}_X = T_0 \equiv \frac{4}{\sqrt{N}}\frac{F} {\lambda \Lambda_1}.
\end{equation}
We recall that the minimum is at $|X| \sim \Lambda^2_2/\Lambda_1$. The $R$-symmetry breaks via a second order phase transition. The critical temperature (\ref{TR2-9}) can be quickly estimated from the fact that the spurion has negative squared mass $4F^2/\Lambda^2_1$ at the origin and receives thermal corrections $N\lambda^2 T^2/4$ from the messenger fields (\ref{V-S-T-2}).

\subsection{Approximate $R$-Symmetry}
The theory (\ref{min-9}), (\ref{KR-9}) exhibits an exact thermal restoration of the $U(1)_R$ symmetry. Nonetheless, theories like the (\ref{const-3}), (\ref{Mmag-3}) 
\begin{equation}
W=FX+\lambda X\phi\bar{\phi} + c, \quad \quad W=FX+\lambda X\phi\bar{\phi} - M\phi\bar{\phi} 
\end{equation}
that break the $U(1)_R$ explicitly are approximately $R$-symmetric at high tmperatures. Indeed, for high enough temperatures the $R$-violating terms $c$ and $M\phi\bar{\phi}$ are negligible and an approximate $R$-symmetry restoration takes place. This can be seen from the evolution of the thermal average value for the $R$-charged $X$ field that we recall here (\ref{S-min0}): 
\begin{equation} \label{X-min09}
X^{(c)}_{min}(T)=\frac{4\frac{c}{M^2_P} F -\frac{2Fc}{3\Lambda^2M^2_P}T^2}{8\frac{F^2}{\Lambda^2}+\frac{N}{2}\lambda^2T^2}, \quad \quad X^{(M)}_{min}(T)=\frac{\frac12 M\lambda T^2}{8 \frac{F^2}{\Lambda^2}+\frac{N}{2}\lambda^2 T^2}
\end{equation}

The $R$-symmetry breaking scale is the vev $\left \langle X \right\rangle \equiv X_0$. For the case of gravitational stabilization the vev is the $X^{(c)}_0 =c\Lambda^2/2FM^2_P$ and for the messenger mass case $X^{(M)}_0=0$. For the former case the $R$-symmetry breaking scale is apparently the $X_0$ ; for the later, where $W=FX+(\lambda X-M)\phi\bar{\phi}$, the $R$-symmetric point is not the $X=0$ but the $X=M/\lambda$. After the translation $X\rightarrow M/\lambda -X$ then $X^{(M)}_0 = M/\lambda$ which is here, the scale of $R$-symmetry breaking.

According to (\ref{X-min09}) the thermal average value tends to restore the $R$-symmetry. We can parametrize the degree of the $R$-symmetry breaking by defining the parameter $b_R$:
\begin{equation}
b_R(T)\equiv \frac{X(T)}{X_0}\,.
\end{equation}
Temperatures higher than the cut-off scale are not expected (for $\Lambda \gtrsim 10^{-4}M_P $) since a thermal equilibrium cannot be achieved. Thus, for the case of gravitational stabilization the second term at the numerator (\ref{X-min09}) is negligible. The parameter $b$ takes the universal (for both cases) form
\begin{equation} \label{b1-9}
b_R(T)=\frac{\left( 4 \frac{F}{\Lambda} \right)^2 }{ \left( 4 \frac{F}{\Lambda} \right)^2 + N \lambda^2 T^2}
\end{equation}
where $N$ is the number of messenger fields $\phi$ and $\bar{\phi}$ in the fundamental representation \cite{Giudice:1998bp}. We can now define the temperature $T_0$
\begin{equation}
T_0 \equiv \frac{4}{\sqrt{N}}\frac{F}{\lambda \Lambda}
\end{equation}
and recast the (\ref{b1-9}) to the simpler form
\begin{equation} \label{T0-9}
b_R(T) = \frac{1}{1+\left(\frac{T}{T_0} \right)^2}\,.
\end{equation}
Obviously, when $T\rightarrow 0$ the $R$-symmetry breaking scale takes its maximum value, i.e. the zero temperature one, and when $T \rightarrow \infty $ the $R$-symmetry is restored. In other words, the $b_R(T)$ parametrizes the $R$-symmetry breaking scale at finite temperature with respect to the zero temperature scale. Furthermore, solving the (\ref{b1-9}) with respect to the temperature we take
\begin{equation} \label{Tb-9}
T^2(b_R)\,=\,T^2_0 \, \frac{1-b_R}{b_R} = \left(\frac{4}{\sqrt{N}}\frac{F}{\lambda \Lambda} \right)^2 \frac{1-b_R}{b_R}\,.
\end{equation}
We note that $T_0=T(b_R=0.5)$. It is interesting here to re-derive the temperatures $T_X$ firstly given at the chapters 5 and 6. The $T_X$ corresponds to the temperature that the minimum at the $X$-direction crosses the (would-be at $T_{susy}$) tachyonic boundary $X=\sqrt{F/\lambda}$. Hence, 
\begin{equation}
X(T_X)= \sqrt{F/\lambda}=b_R(T_X)X_0
\end{equation}
which gives the following values for the parameter $b_R$:
\begin{equation}
b_R( T^{(c)}_X )=2\frac{F\, M^2_P}{c\Lambda^2}\sqrt{\frac{F}{\lambda}}\, , \quad \quad \quad b_R(T^{(M)}_X)=\frac{\sqrt{\lambda F}}{M}.
\end{equation}
Since $b_R( T^{(c)}_X ), \, b_R( T^{(M)}_X )\ll 1$, from (\ref{Tb-9}) we take
\begin{equation}
T^2_X \simeq \frac{8}{N}\frac{c}{\lambda M^2_P} \sqrt{\frac{F}{\lambda}}\, \quad \quad \text{and} \quad\quad T^2_X \simeq \frac{16}{N} \frac{F^2 M}{\lambda^2 \Lambda^2}\sqrt{\frac{F}{\lambda}}\, 
\end{equation}
which are the temperatures derived at chapters 5 and 6. We also note that $T_X>T_0$.
\\
\\
For the case of spontaneous breakdown of the $R$-symmetry, discussed in the previous subsection, the parameter $b_R(T)$ takes, approximately, the discrete values: 
\begin{equation}
b_R(T>T^{(R)}_X)=0
\quad \quad \text{and} \quad \quad
b_R(T<T^{(R)}_X)=1\, ,
\end{equation}
where $T^{(R)}_X = T_0$.

\section{Gravitino Thermal Production Revisited}

The crucial consequence of the thermal tendency to restore the $R$-symmetry is that the gravitino cross section from the scattering processes off thermal radiations, $\left\langle \sigma_{\tilde{G}} v\right\rangle$, becomes thermal dependent. Since the gaugino masses, $m_\lambda$, require an insertion of both the scalar and auxiliary components of $X$, while the scalars require only auxiliary components, the gauginos become lighter than the sfermions as $X(T)$ decreases by ${\cal O}\left(X(T)/X_0\right)$. In the models of ordinary gauge mediation, that we investigate here, there is no hierarchy between the sfermions and the gauginos at zero temperature, i.e. $m_\lambda \sim m_{\tilde{f}}$. Hence, due to the $R$-symmetry as the temperature increases and the $X(T)$ is driven to the origin the gauginos masses minimize compared to the zero temperature gaugino masses. We claim a scaling:
\begin{equation} \label{ratio-9}
\frac{m_\lambda(T)}{m_\lambda(T=0)} ={\cal O} \left(\frac{X(T)}{X_0}\right) = {\cal O}(b_R)\, .
\end{equation}
Here we will consider that: $m_\lambda(T)/m_\lambda(T=0)=b_R(T)$.

A connection with the $U(1)_R$ symmetry is straightforward. The supersymmetric Lagrangian contains the gauge interaction terms
\begin{equation} \nonumber
{\cal L}_{gauge}= \int d^4x d^2\theta \left\{\frac14 W^{\alpha}_1W_{1\, \alpha}+ \frac12 \text{tr}(W^{\alpha}_2W_{2\, \alpha}) + \frac12 \text{tr}(W^{\alpha}_3W_{3\, \alpha}) \right\} +\text{h.c}
\end{equation}
\begin{equation} \label{susyL-9}
+ \int d^4x d^2\theta d^2\bar{\theta} \left(\sum_{ij}\Phi^\dagger_ie^{g_jV_j}\Phi_i +\text{h.c} \right)
\end{equation}
where $V_j$ are the vector and $W_{j\alpha}$ the corresponding field strength superfields associated to $U(1)_Y$, $SU(2)$ and $SU(3)$ for $j=1,2,3$ respectively. The complete supersymmetric Lagrangian (without the soft terms) includes also the superpotential including the Yukawa interactions and the Higgs sector. The (\ref{susyL-9}) apart from being gauge invariat (section 2.2.1) it is also $R$-invariant. For instance,  the $W^\alpha W_\alpha$ has $R$ charge $+2$ and by a $U(1)_R$ rotarion it transforms to $e^{2i\alpha}W^\alpha W_\alpha$ which cancels out with the corresponding rotation at the Grassmannian variable $d^2\theta \rightarrow e^{-2i\alpha}d^2\theta$.  Introducing an explicit \itshape Majorana \normalfont gaugino mass in a supersymmertic manner extends the Lagrangian (\ref{susyL-9}) with the part
\begin{equation}
{\cal L}_{gaugino}= \int d^4x d^2\theta \left\{\frac14 2\theta\theta M_1W^{\alpha}_1W_{1\, \alpha}+ \frac12 2\theta\theta M_2\text{tr}(W^{\alpha}_2W_{2\, \alpha}) + \frac12 2\theta\theta M_3\text{tr}(W^{\alpha}_3W_{3\, \alpha}) \right\}
\end{equation}
which breaks the $R$-symmetry.
Thereby the radiatively generated at 1-loop gaugino masses by the operator $\int d^2 \theta \ln X W^\alpha W_\alpha+ \text{h.c.}$\footnote{This operator is generated after integrating the messengers out i.e. it is valid for temperatures and spurion mass $m_X$ lower than the messenger mass.}
\begin{equation}
m_\lambda=\frac{\alpha}{4\pi}\frac{\lambda F}{M_{mess}}= \frac{\alpha}{4\pi}\frac{ F}{X_0}
\end{equation}
should \itshape not \normalfont be expected  if the \itshape vacuum \normalfont of the theory is $R$-symmetric. Hence, when the $U(1)_R$ symmetry is a symmetry of the vacuum  then soft masses for majorana gauginos are prohibited.
\\
\\
The above arguments suggests a thermally sensitive cross section for the  production of gravitinos from scattering precesses off the thermal radiations. Considering the dominant QCD $2\rightarrow 2$ processes the (\ref{cross-9}) cross section becomes temperature dependent
\begin{equation} \label{crossT-9}
\sigma_{\tilde{G}}(p, T) \propto \frac{g^2}{M^2_P} \left(1+b^2_R(T)\frac{m^2_{\tilde{g}}}{3m^2_{3/2}} \right)
\end{equation}
where $m_{\tilde{g}}$ the gluino mass. We note that the gluino mass has already a dependence on the temperature due to the renormalization of the gauge coupling constants (\ref{run-6}):
\begin{equation}
m_{\tilde{g}}(T)=\frac{g^2_3(T)}{g^2_3(\mu)} m_{\tilde{g}}(\mu)
\end{equation}
where $\mu \simeq 100$ GeV. However, this running of the gaugino masses is negligible compared with the (\ref{ratio-9}) temperature dependence that we are considering here.

Another subtle point is whether the finite temperature effects enhance the gravitino mass or the Goldstino coupling. This was discussed at \cite{Leigh:1995jw} and \cite{Ellis:1995mr} where it was shown that thermal effects do \itshape not \normalfont give contribution to the gravitino production rate of the form $T^8/(m^2_{3/2}M^2_{P})$. 

Here, we are considering an opposite effect: at high temperatures the Goldstino generation rate may be suppressed due to the $R$-symmetry restoration. The formula  that gives the gravitino abundance $Y_{3/2}$ is given by (\ref{yield-9}).
According to our arguments the right hand side of the equation is not any more temperature independent but it reads:
\begin{equation} \label{}
\frac{dY_{3/2}}{dT}= -\frac{\left\langle \sigma_{\tilde{G}}(T) v\right\rangle n_{rad}}{HT}\,.
\end{equation}
For gravitino of mass $m_{3/2} < 100$ GeV we can neglect the yield of the helicity $\pm \frac32$ component for temperatures less than $10^{12}\,m^{-1}_{3/2}$ GeV$^2$, see (\ref{Theavy-9}). Focusing on the interactions of the helicity $\pm \frac12$ modes then the yield variable $Y_{3/2}$ is given by
\begin{equation} \label{int-9}
Y_{3/2}(T)-Y_{3/2}(T_{rh})=-\frac{g_*(T)}{g_*(T_{rh})} \left\{ \frac{n_{rad}\left\langle \sigma^{(1/2)}_{\tilde{G}} v\right\rangle}{HT}\right\} \int^T_{T_{rh}}dT\, b^2_R(T)
\end{equation}
where
\begin{equation}
\left\langle \sigma^{(1/2)}_{\tilde{G}}(T) v\right\rangle =\left\langle \sigma^{(1/2)}_{\tilde{G}}v \right\rangle b^2_R(T)
\end{equation}
as we can see from (\ref{crossT-9}). We remind the reader that the quantity in the brackets is temperature independent.

Another point that should be taken into consideration is whether the processes that involve chiral supermultiplets, i.e. quark and squarks, contribute to the production of the helicity $\pm \frac12$ component. If their contribution is non negligible then the suppression, due to the $R$-symmetry, of the Goldtino production may not be significant if the massive (not-suppressed by the $R$-symmetry) sfermions take over the gluino-trancated production process. We note, firstly, that the corresponding contribution to the Goldstino production rate is proportional to $m^4_{\tilde{q}}$ which is suppressed at high energies compared to the gluino contribution for dimensional reasons: the interactions of the gravitino to gauge supermultiplets are described by dimension-five operators and those to chiral supermultiplets by dimension-four ones. Hence, at high temperatures, $m_{\tilde{q}},\,m_{\tilde{g}}\ll T $, contributions involving the cubic Goldstino-quark-squark coupling are suppressed by $m^2_{\tilde{q}}/T^2$ relative to the gluino contributions. Secondly, at that high temperatures that we consider, the helicity $\pm \frac12 $ mode from cubic Goldstino-quark-squark interactions is subdominant compared to the helicity $\pm \frac32$ mode that is produced with a cross section $T^2/M^2_P$. Hence, we can focus only at those $2\rightarrow 2$ reactions where at least one of the three other external particles is a member of a (color)-gauge supermultiplet and ignore those with chiral supermultiplets altogether.

\subsection{Exact R-symmetry Thermal Restoration}

For models that exhibit exact thermal restoration of the $U(1)_R$ symmetry the $b_R(T)$ can be approximated by a step function: $b_R(T)=0$ for $T>T^{(R)}_X$ and $b_R(T)=1$ for $T<T^{(R)}_X$, where $T^{(R)}_X \simeq T_0$\footnote{This is indeed an approximation since a second order phase transition is not a discontinuous process; instead at the critical temperature $T_X$ there is no barrier and the transition occurs smoothly.}. Hence, the gravitino yield (\ref{int-9}) reads in this case:
\begin{equation} \label{intR-9}
Y_{3/2}(T)-Y_{3/2}(T_{rh})=-\frac{g_*(T)}{g_*(T_{rh})} \left\{ \frac{n_{rad}\left\langle \sigma^{(1/2)}_{\tilde{G}} v\right\rangle}{HT} \right\} \int^T_{T_{0}} dT\, b^2_R(T)
\end{equation}
\begin{equation}
\left. \simeq \frac{g_*(T)}{g_*(T_{0})} \left\{ \frac{n_{rad}\left\langle \sigma^{(1/2)}_{\tilde{G}} v\right\rangle}{HT} \right\}\right|_{T_{0}} T_0
\end{equation}
where we took into account that $T\ll T_0$. Also, $Y_{3/2}(T_{rh}) \sim 0$ since we consider that the dominant source of gravitino production are the scatterings in the thermal plasma and any pre-inflationary abundance was diluted by inflation. Therefore, for $T<1$ MeV, i.e. after nucleosynthesis, for a decoupled gravitino and for $T_{rh}>T_0$ the gravitino abundance is 
\begin{equation}
Y_{3/2} \simeq 1.1\times 10^{-10}\left(\frac{T_0}{10^{10}\, \text{GeV}} \right)\left(\frac{100\, \text{GeV}}{m_{3/2}} \right) \left(\frac{m_{\tilde{g}}(\mu)}{1\, \text{TeV}} \right)^2
\end{equation}
and the contribution to the $\Omega h^2$,
\begin{equation} \label{OR-9}
\Omega_{3/2}h^2 \simeq 0.2 \left(\frac{T_{0}}{10^{10}\,\text{GeV}} \right) \left(\frac{100\,\text{GeV}}{m_{3/2}} \right) \left(\frac{m_{\tilde{g}}(\mu)}{1\, \text{TeV}} \right)^2\, .
\end{equation}
The $T_0$ temperature is $T_0=4F/(\lambda \Lambda_1\sqrt{N})$ for the example (\ref{min-9}), (\ref{KR-9}). We can write it in terms of the gravitino mass
\begin{equation}
T_0= \frac{4}{\sqrt{N}}\frac{\sqrt{3}m_{3/2}M_P}{\lambda \Lambda_1}=4\frac{\sqrt{3} m_{3/2} }{\lambda \Lambda_1}\, 2.4\times 10^{18}\, \text{GeV}
\end{equation}
and the $\Omega h^2$ is recast to 
\begin{equation} 
\Omega_{3/2}h^2 = 0.2 \times \frac{16.6}{\sqrt{N}} \left(\frac{10^{10} \, \text{GeV}}{\lambda\Lambda_1} \right) \left(\frac{m_{\tilde{g}}(\mu)}{1\, \text{TeV}} \right)^2\, .
\end{equation}
This is a remarkable result. Firslty, the gravitino abundance does \itshape neither \normalfont depend on the gravitino mass \itshape nor \normalfont on the reheating temperature. Secondly, the quantities which control the yield are the supersymmetry breaking fundamental parameters $\lambda$ and $\Lambda_1$. For $\sqrt{N}={\cal O}(1-7)$, the gravitino does not overclose the universe when 
\begin{equation}
\lambda \Lambda_1 \gtrsim 10^{11} \, \text{GeV}.
\end{equation}
It can account for the dominant dark matter component when $\lambda\Lambda_1 \sim 10^{11}$; for example, when $\lambda=10^{-5}$ and $\Lambda_1=10^{16}$ GeV = ${\cal O}$(GUT) scale, the gravitino is the dark matter of the universe. It is interesting to note that these values of the parameters are the natural values for several models. A cut off of the order of the GUT scale is expected in many theories and, the fact that there is a physical cut off in the theory implies the smallness of the coupling since in the IR theory that appears it is expected to be suppressed. 

In addition, considering that the supersymmetry breaking local minimum must be thermally preferred we conclude that small values of the coupling $\lambda$ are favourable. However, decreasing the cut off and increasing the value of the coupling renders the supersymmetry preserving vacuum more attractive. Hence, we find an interesting window of values where the supersymmetry breaking vacuum is selected and the gravitino does not overclose the universe, or even accounts for the dominant dark matter component. This parameter space also specifies the gravitino mass range. 

We mention that the reheating temperature cannot be arbitrary high. Otherwise, the interactions of the helicity $\pm \frac32$ that are $T^2/M^2_P$ suppressed become significant. According to (\ref{Omheavy-0}), a gravitino with mass $1$ GeV constrains the reheating temperature to be $T_{rh} \lesssim 10^{12}$ GeV which is $10^4$ times relaxed relatively to the conventional bound (\ref{Tlight-9}).

\subsection{Approximate R-symmetry Thermal Restoration}

For models that break explicitly the $R$-symmetry like those we presented in the previous section the $b_R(T)$ is given by
\begin{equation} \label{}
b_R(T) = \frac{1}{1+\left(\frac{T}{T_0} \right)^2}\,.
\end{equation}
and the relevant part of the integral (\ref{int-9}) is
\begin{equation}
\left. \int^T_{T_{rh}} dT\, b^2_R(T)= \frac12 T_0 \left\{\frac{T_0 T}{T^2_0+T^2} +\text{Arctan}\left(\frac{T}{T_0}\right) \right\} \right|^{T}_{T_{rh}}\, .
\end{equation}
For, $T<1$ MeV, i.e. after nucleosynthesis we expect $T\ll T_0, \, T_{rh}$. Hence,
\begin{equation}
\int^T_{T_{rh}} dT\, b^2_R(T)\cong -\frac12 T_0 \left\{\frac{T_0 T_{rh}}{T^2_0+T^2_{rh}} +\text{Arctan}\left(\frac{T_{rh}}{T_0}\right) \right\}
\end{equation}
For reheating temperatures higher than the $T_0$, which are the cases that we are interested in, and especially for $T_{rh}\gg T_0$ the integral approximates to
\begin{equation}
\int^T_{T_{rh}} dT\, b^2_R(T)\simeq -\frac12 \, T_0 \text{Arctan}\left(\frac{T_{rh}}{T_0}\right)-\frac12 T_0\left(\frac{T_0}{T_{rh}}\right) 
\end{equation}
\begin{equation}
\simeq -\frac12 \, T_0 \text{Arctan}\left(\frac{T_{rh}}{T_0}\right)
\end{equation}
\begin{equation}
\equiv -\frac{\theta_{rh}}{2} \, T_0 
\end{equation}
where, $\theta_{rh} \equiv \text{Arctan}(T_{rh}/T_0)$ a coefficient that here is larger than one: $\pi/4<\theta_{rh} <\pi/2$.

For $T_{rh}=T_0$ the integral takes the value
\begin{equation}
\int^T_{T_{rh}} dT\, b^2_R(T)= -\frac{T_0}{4}-\frac{T_0}{2}\frac{\pi}{4}
\end{equation}
while for reheating temperatures lower than $T_0$, i.e. $T_{rh}<T_0$ the value is 
\begin{equation}
\int^T_{T_{rh}} dT\, b^2_R(T)\simeq -\frac12 \, T_{rh} -\frac{T_0}{2} \text{Arctan}\left(\frac{T_{rh}}{T_0}\right).
\end{equation}
Taking into account that
\begin{equation}
\text{Arctan} x=x-\frac{x^3}{3}+\frac{x^5}{5}-\frac{x^7}{7}+...
\end{equation}
the integral for temperatures for $T_{rh}<T_0$ converges to the $(-)T_{rh}$ value as expected.

The conclusion is that for the high reheating temperatures $T_{rh}>T_0$ the integral is  $- T_0 \theta_{rh}/2$ or roughly $-T_0/2$. Therefore the (\ref{int-9}) reads
\begin{equation} \label{}
Y_{3/2}(T)-Y_{3/2}(T_{rh})=-\frac{g_*(T)}{g_*(T_{rh})} \left\{ \frac{n_{rad}\left\langle \sigma^{(1/2)}_{\tilde{G}} v\right\rangle}{HT}\right\} \int^T_{T_{rh}}dT\, b^2_R(T) \Rightarrow
\end{equation}
\begin{equation} \label{}
\left. Y_{3/2}(T) \cong \frac{g_*(T)}{g_*(T_{0})} \left\{ \frac{n_{rad}\left\langle \sigma^{(1/2)}_{\tilde{G}} v\right\rangle}{HT}\right\}\right|_{T_{0}} \frac{T_0}{2}
\end{equation}
where we considered the coefficients $g_*$ to be dominated by the value given at the temperature $T_0$. The contribution of the gravitino abundance to $\Omega h^2$ is half times the (\ref{OR-9})
\begin{equation} \label{OgA-9}
\Omega_{3/2}h^2 \simeq 0.2 \left(\frac{T_{0}/2}{10^{10}\,\text{GeV}} \right) \left(\frac{100\,\text{GeV}}{m_{3/2}} \right) \left(\frac{m_{\tilde{g}}(\mu)}{1\, \text{TeV}} \right)^2\, .
\end{equation}
and substituting $T_0= 4F/(\sqrt{N}\lambda \Lambda_1)$ it is recast to 
\begin{equation} 
\Omega_{3/2}h^2 \simeq 0.1 \times \frac{16.6}{\sqrt{N}} \left(\frac{10^{10} \, \text{GeV}}{\lambda\Lambda} \right) \left(\frac{m_{\tilde{g}}(\mu)}{1\, \text{TeV}} \right)^2\, .
\end{equation}
 
We see that when the superpotential has an approximate $U(1)_R$ symmetry the result is basically the same with that of an exact $R$-symmetric superpotential. 
Hence, despite the fact that exact global symmetries are not expected and appear as accidental symmetries in the low energy effective theory the theories behave much like the exact $R$-symmetric theories in terms of the gravitino relic abundance. 

It is interesting to note that the all important temperature $T_0$ has \itshape no \normalfont dependence on the $R$-symmetry breaking parameters $c$ and $M$. Although they define the supersymmetry breaking vev $X_0$ they cancel out at the ratio $b_R(T)=X(T)/X_0$.

We mention, that in the case of approximate $R$-symmetric models the gravitino abundance is suppressed by a factor of two compared with the case of exact restoration of the $R$-symmetry. At first sight this seems a paradox for, the expectation may have been that the gravitino production is stronger suppressed when an exact restoration of the $R$-symmetry takes place than when the restoration is approximate. However, we believe that this small discrepancy originates from the fact that we simplified the thermal evolution of the $X(T)$ by assuming a step-like behaviour while the symmetry breaking occurs smoothly via a graduate increase of the mean value of the scalar field. The $T_0$ value of the integral (\ref{intR-9}) should be considered as the maximal and thus, the $Y_{3/2}$ bound in this case is a conservative one.

Finally, we would like to comment on the production of gravitinos from the thermal plasma due to the top Yukawa coupling, an effect considered at \cite{Rychkov:2007uq}. The production rate is enhanced by the additional term
\begin{equation}
\gamma_\text{top}=1.30 \frac{9\lambda_t T^6}{2M^2_P\pi^5}\left(1+\frac{A^2_t}{3m^2_{3/2}} \right)
\end{equation}
apart from the processes involving gauge supermultiplets, i.e. these that we have already analysed:
\begin{equation}
\gamma_V=\frac{T^6}{2(2\pi)^3M^2_P}\sum^3_{N=1}n_N\left(1+\frac{m^2_{\lambda_N}}{3m^2_{3/2}} \right)f_N
\end{equation}
where $f_N$ a factor which includes the gauge couplings. According to \cite{Rychkov:2007uq} the production processes that include the top quark Yukawa coupling enhance the gravitino production rate by almost $10\%$ or more if the $A_t$ is bigger than the gaugino massses. This effect can become very important when the production of the helicity $\pm \frac12$ component is suppressed by the vanishing gaugino masses. However, much like the gaugino masses, the $A$-terms require interactions which violate the  $U(1)_R$ symmetry and therefore, we expect to be suppresed as well at high temperatures.

\subsection{Conclusions and Discussion}

The standard paradigm in cosmology is that the dark matter is a weak interacting massive particle (WIMP). It is stable and it can annihilate to lighter observable, i.e. Standard Model, states. These interactions can maintain the dark matter in thermal equilibrium with the observable particles in the early universe. When the WIMP stops annihilating efficiently it freezes out and its relic abundance is given by the approximate formula
\begin{equation} \label{dmN-9}
\Omega_{WIMP}h^2 \simeq 0.1 \left(\frac{x_f}{10}\right) \left(\frac{1\times 10^{-26} \text{cm}\,\text{s}^{-1}}{\left\langle \sigma v \right\rangle}\right).
\end{equation}
where $x_f\equiv M_{DM}/T_f$ at the time of the freeze out. The cross section can be approximated by 
\begin{equation}
\left\langle \sigma v \right\rangle \sim \frac{\alpha^2}{M^2_{DM}} \sim \alpha^2\times \left(\frac{1\, \text{TeV}}{M_{dm}} \right)^2\times 10^{-26} \text{cm}\, \text{s}^{-1}
\end{equation}
where $\alpha=g^2/4\pi$. The reason that the neutralino dark matter is a compelling candidate is because their masses are predicted to be of the TeV order. The very attractive feature of the (\ref{dmN-9}) is that it is independent of the reheating temperature given that the reheating temperature is larger $\sim M_{dm}/10$, in order the WIMP to equilibrate. Therefore, it can be very high without the $\Omega_{WIMP}$ to be sensitive to it.  It is only sensitive to the mass  and the couplings of the WIMP; for the case of neutralino $g\sim 1 $ and $M_{DM}\sim M_{EW}$ making it a natural candidate.

A similar behaviour is suggested by the (\ref{OgA-9}) for the gravitino:
\begin{equation} \label{}
\Omega_{3/2}h^2 \simeq 0.1 \times \frac{16.6}{\sqrt{N}} \left(\frac{10^{10} \, \text{GeV}}{\lambda\Lambda} \right) \left(\frac{m_{\tilde{g}}(\mu)}{1\, \text{TeV}} \right)^2\, .
\end{equation}
The abundance depends on the product $\lambda\Lambda$ i.e. on fundamental quantities of supersymmetry breaking. A theory with a physical cut off $\Lambda$ related to the GUT scale and a coupling $\lambda \sim 10^{-5}$ which is naturally small in the sense that small Yukawa coupling values are expected in the IR macroscopic theory. Furthermore, the abundance is independent of the reheating temperature given that the reheating temperature is larger than $T_0$. The reheating temperature is bounded from above in order that only the helicity $\pm \frac32$ gravitino component not to be overproduced. In fact, the allowed window of the reheating temperatures is remarkably wide, see the figure 9.2.

Finally, an important but model dependent issue is the microscopic theory that implements the supersymmetry breaking. The thermal restoration of the $R$-symmetry implies that the origin where the messenger fields become massless is the attractive point of the free energy. Only the presence of hidden sector fields that may become massive at the origin could render an $R$-breaking minimum thermally favourable rather than the origin. Thereby, we expect that if such a hidden sector exists it must be nearly unpopulated. Moreover, a thermalized hidden sectror that couples directly to the Goldstino would lead to a large $dn_{3/2}/dt$. Hence, a thermalized hidden sector could give a rate of Goldstino production scaling like $T^8/F^2$ \cite{Leigh:1995jw}. Therefore, we expect an $R$-symmetry restoration to take place indeed at a high temperature enviroment of the early universe; otherwise the gravitino problem exacerbates.

We emphasize that the all important temperature is the $T_0=4F/(\lambda\Lambda\sqrt{N})$. Its order of maginitude can be understood easily by  recalling that the mass of the spurion $X$ (which at the tree-level is a flat direction) is $m_X \simeq 2F/\Lambda$ and its thermal mass is $\delta m_X\sim \sqrt{N}\lambda T$ i.e.
\begin{equation}
T_0 \sim \frac{m_X}{\sqrt{N}\lambda}
\end{equation}
The $m_X\sim F/\Lambda$ is of the order of the weak scale thereby, it is the smallness of the Yukawa coupling to the messenger fields that makes the $T_0$ large suggesting a GeV mass range gravitino for dark matter.

\begin{figure} 
\begin{tabular}{cc}

{(a)} \includegraphics [scale=.85, angle=0]{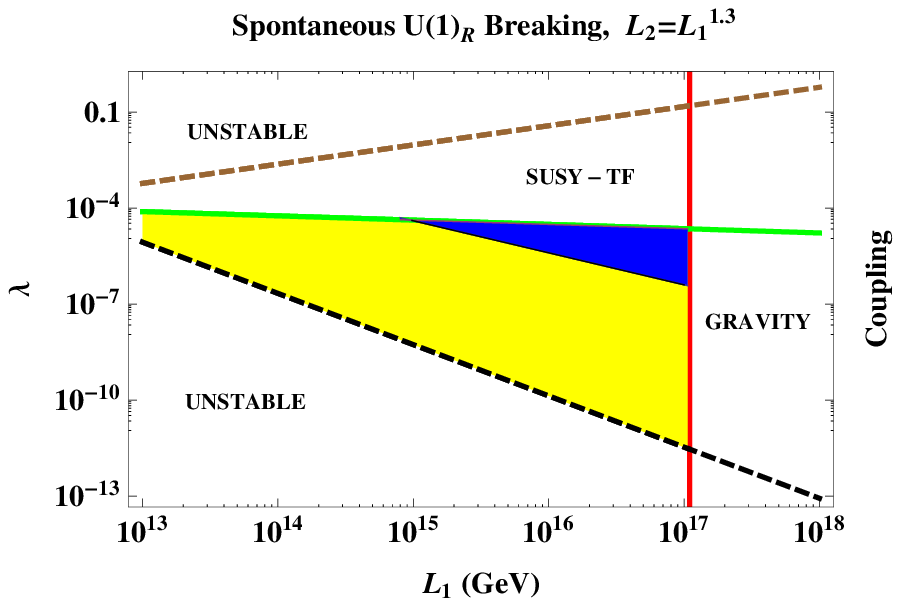} &
{(b)} \includegraphics [scale=.85, angle=0]{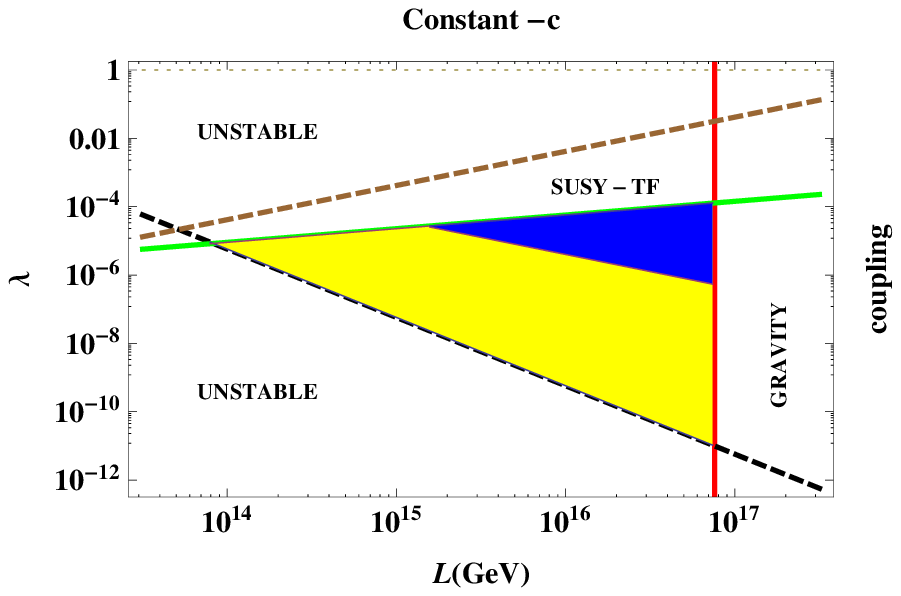}  \\
\end{tabular}
\begin{tabular}{cc}

{(c1)} \includegraphics [scale=.85, angle=0]{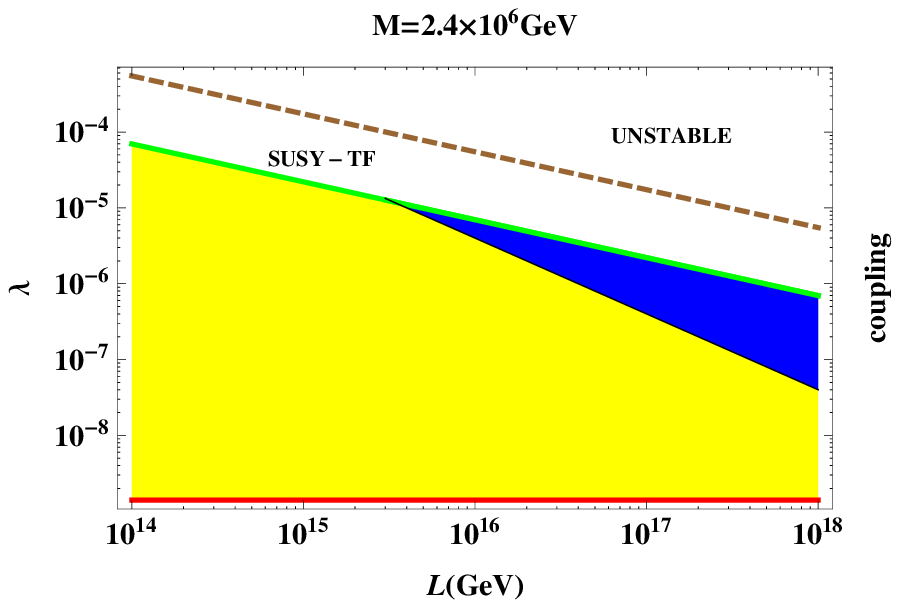} &
{(c2)} \includegraphics [scale=.85, angle=0]{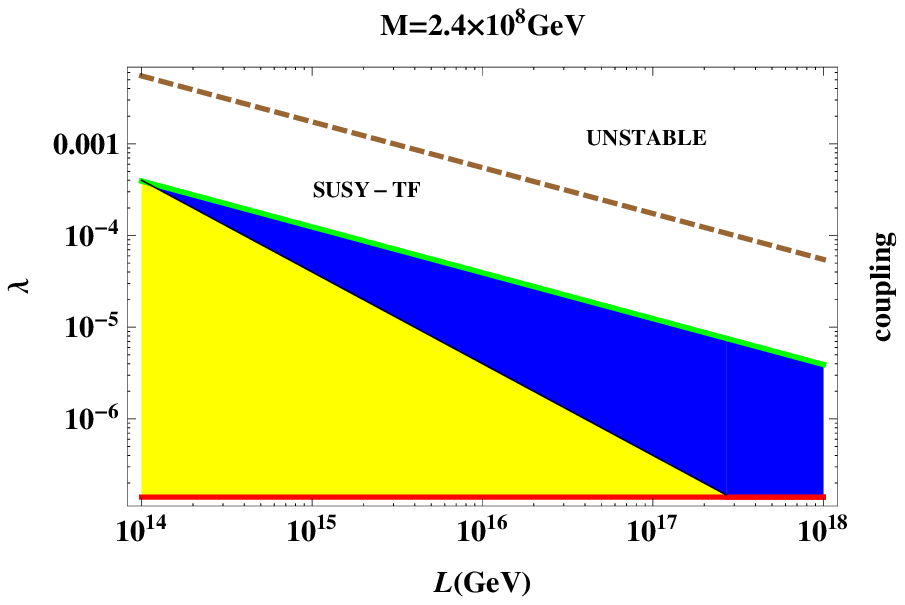} \\

\end{tabular}
\begin{tabular}{c}
{(c3)} \includegraphics [scale=.85, angle=0]{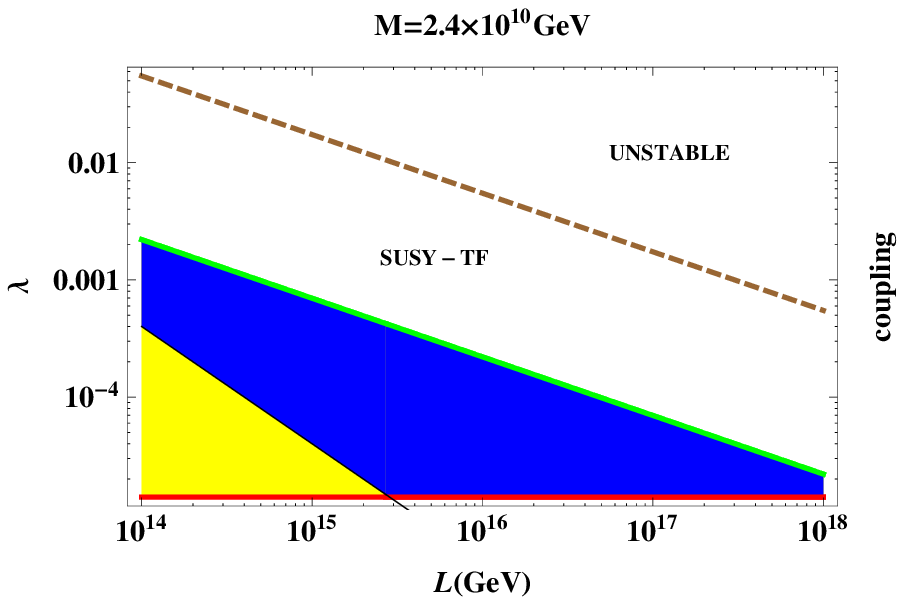} \\
\end{tabular}
\caption{\small{The figures show the region of parameter space where the supersymmetry breaking minimum is metastable (the region enclosed by the dashed lines),  thermally favourable (the yellow and the dark blue region) and where the gravitino thermal production does not overclose the universe (dark blue region) for the cases of spontaneous $U(1)_R$ breaking (a-panel), explicit $R$-breaking due to a constant $c$ at the superpotential (b-panel)  and due to an explicit messenger mass $M$ ((c)-panels). For the case of 6th order correction to the K\"ahler and spontaneous $R$-breaking we have considered the case where $\Lambda_2=\Lambda_1^{1.3}$ in Planck units. The red line separates gauge mediation, $m_{3/2}<0.1 m_{\tilde{g}}$, from gravity mediation.  The boundary line between the dark blue and the yellow region corresponds to the parameters that gravitino accounts for the dark matter i.e. $\Omega h^2 \sim 0.1$. We considered 700 GeV gaugino masses. In the dark blue region the gravitino abundance $\Omega h^2$ is less than $10\%$  and in the yellow it exceeds the observational bounds on the dark matter abundance i.e. it overcloses the universe. The $L$ stands for $\Lambda$. }}
\end{figure}

\begin{figure} 
\begin{center}
\begin{tabular}{c}
{(a)} \includegraphics [scale=.95, angle=0]{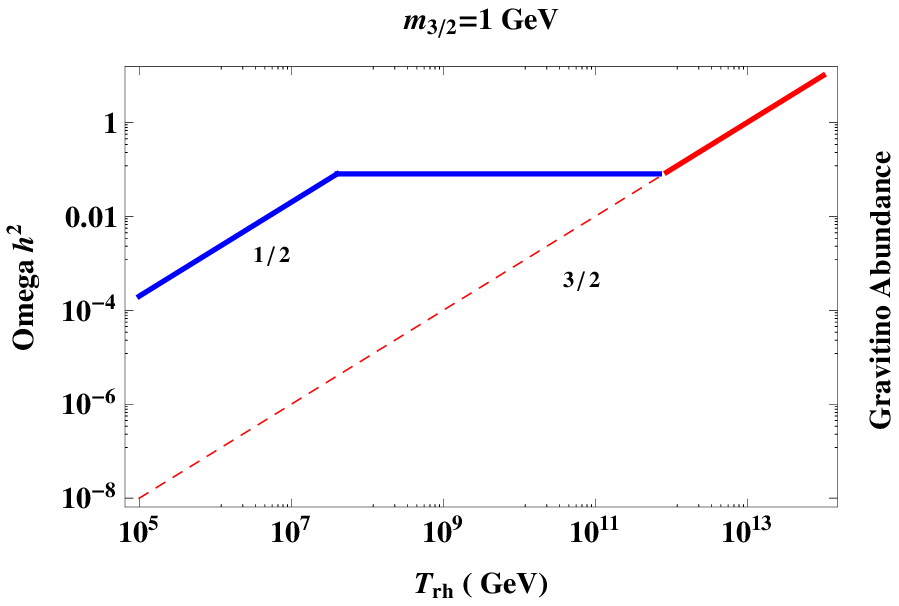} 
\end{tabular}
\begin{tabular}{c}
{(b)} \includegraphics [scale=.95 , angle=0]{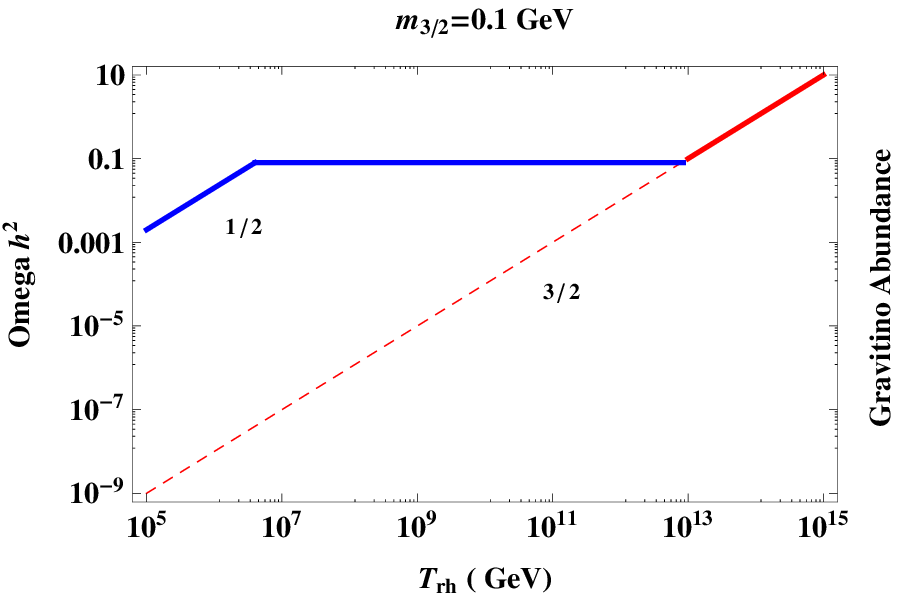} \\
\end{tabular}
\begin{tabular}{c}
{(c)} \includegraphics [scale=.95, angle=0]{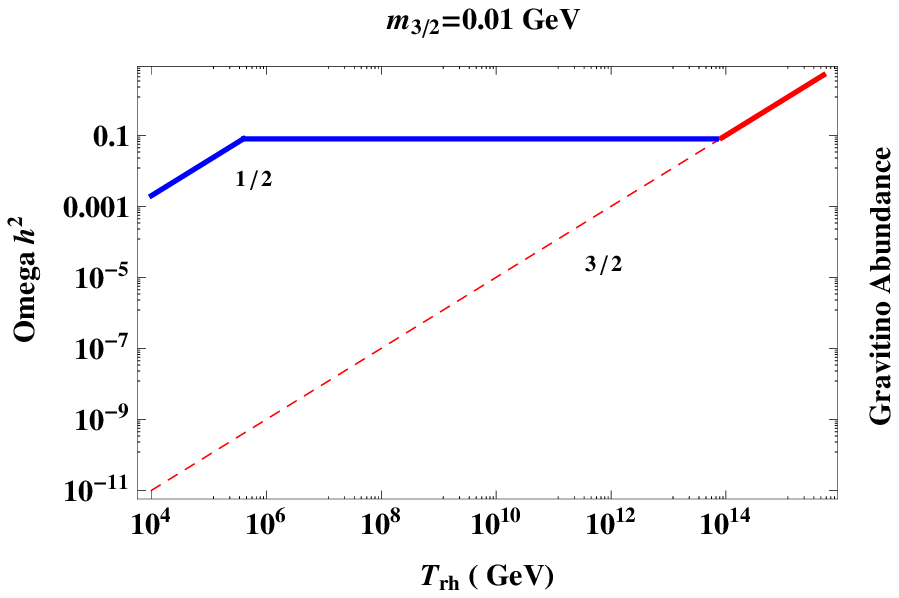} 
\end{tabular}
\caption{\small{The panels show the gravitino abundance (continuous line), for masses $m_{3/2}=1, 10^{-1}, 10^{-2}$ GeV from top to bottom, with respect to the reheating temperature of the universe. The hidden sector parameters where chosen such that the gravitino abundance is $\Omega_{3/2} h^2 = 0.1$ for reheating temperatures larger than $T_0$ and smaller than $10^{12}m^{-1}_{3/2}$ GeV$^2$; in particular, $\lambda \Lambda=5\times10^{10}$ GeV, $N=5$ and $m_{\tilde{g}}=700$ GeV. The figures demonstrate the insensitivity of the gravitino abundance to the reheating temperature. The red dashed line shows the abundance of the helicity $\pm \frac32$ gravitino component and  the blue line the $\pm \frac12$ component. The plateau corresponds to the $R$-symmetry restoration hence, to suppression of the $\pm \frac12$ component. This dependence of the $\Omega_{3/2}h^2$} on the reheating temperature is universal for all the three models.}
\end{center}
\end{figure}

\chapter{Summary}

We summarize here the main results and conclusions obtained in this thesis:
\\ 
\\
\itshape Ordinary Gauge Mediation  \normalfont

\begin{itemize}
	\item The spurion initial vev can be generic. The thermal effects can drive the spurion to the metastable vacuum for any initial value for the spurion inside the validity regime of the theory. 
\end{itemize}
\begin{itemize}
	\item There is no need the reheating temperature to be constrained from above. On the other hand, if the reheating temperature is not smaller than the messenger mass then the selection of the metastable vacuum can be realized non-thermally only for specific initial conditions.
\end{itemize}
\begin{itemize}
	\item The thermal selection of the metastable vacuum is realized for sufficiently weak coupling at the messenger superpotential. Although there is an upper bound on the Yukawa coupling the greatest part of the parameter space renders the metastable vacuum thermally favourable. Small Yukawa values are actually predicted in models without elementary singlets.
\end{itemize}
\begin{itemize}
	\item If the spurion is coupled with messenger like couplings with Standard Model singlets that introduce supersymmetric vacua then, the selection of the metastable vacuum requires the initial vev of the spurion to be around the metastable vacuum and the reheating temperature to be constrained from above.
\end{itemize}

\itshape Stringy Completion of  Supersymmetry Breaking Models   \normalfont
\begin{itemize}
	\item The constant $c$ in the supersymmetry breaking sector superpotential may be interpreted as the value of the overall modulus superpotential stabilized in an AdS minimum. Thereby the high of the barrier that protects the modulus from the runaway is related with the supersymmetry breaking scale. Large reheating temperatures can destabilize the minimum, nevertheless thermal selection of the supersymmetry breaking  metastable vacuum is possible without the destabilization of the modulus.
\end{itemize}
\begin{itemize}
	\item An alternative scenario is the stabilization of the volume modulus in a Minkowski vacuum with a barrier disconnected with the scale of supersymmetry breaking. A heavy modulus can then reheat the universe with a temperature large enough even to thermalize the messenger fields. 
\end{itemize}

\itshape R-Symmetry and Gravitino Abundance \normalfont
\begin{itemize}
	\item A long standing problem is the overproduction of gravitinos in the early universe. The fact that the supersymmetry breaking hidden sector is not rigid but dynamical and influenced by the thermal plasma radically alters the gravitino thermal production results. In particular, supersymmetry breaking models generally exhibit an $R$-symmetry which can be thermally restored resulting in a suppressed gravitino yield. Moreover, the gravitinos can account for the dark matter of the universe for reasonable values of the parameters.
\end{itemize}

\itshape In few words  \normalfont
\begin{itemize}
	\item The conclusion of this thesis can be encapsulated in the statement that a TeV supersymmetric universe can be generally thermally safe and high reheating temperatures are not problematic without the need to introduce  specially designed ingredients or unnecessarily complicated assumptions.
\end{itemize}

\appendix

\chapter{The Cosmological Moduli Problem}

We review here the cosmological moduli problem. In particular we present why the problem occurs and its connection with the supersymmetry breaking.

\section{Gravitational Relics}

Supersymmetry provides a natural solution to the large hierarchy between the weak scale and GUT or Planck scales. The presence of a GUT/Planck scale in the theory leads to the problem of quadratic divergences of the masses of the fundamental scalars; divergences which cancel in a supersymmetric theory.  It is also widely believed that string theory provides the consistent framework which lies behind low energy supersymmetry. These beyond the Standard Model particle physics theories include directions in field space which are flat in the supersymmetric limit and couple to light fields only through Planck scale suppressed interactions. If the potential for these flat directions is stabilized by the same physics responsible for susy breaking, a particle with mass of order of a TeV  and dangerously long lifetimes results. Possible examples of such particles are the dilaton of string theory, the massless gauge singlets of string compactifications, a Planck scale coupled singlet responsible for susy breaking, or a singlet field responsible for communicating susy breaking to the visible sector. These fields are often referred collectively as moduli. In the gravitational relics one must include, the supersymmetric partner of graviton, the spin-3/2 gravitino as well as the modulini fermionic fields. The estimation for the decay rate of these gravitational relics is at most 
\begin{equation} \label{Gamma}
\Gamma \sim \frac{m^3_\varphi}{8\pi M^2_P}
\end{equation} 
Even though the moduli are generic in string compactification their physical implications change considerably depending on the details of moduli stabilization and susy breaking.

\subsection{Incoherent Moduli}

A natural expected source of both scalar and fermionic relic moduli is thermal scattering in the plasma. For $T \sim M_P$ the moduli are in equilibrium with a thermal number density, $n_\varphi /s \sim 1/g_*$. A significant entropy release, such as from a standard inflationary scenario, would dilute this number density. Even so, moduli can be produced after the entropy release, i.e. after inlation, by thermal rescatterings in the reheated universe. Neglecting any initial number density, the Boltzmann equation for the number density of the incoherent moduli is
\begin{equation} \label{1}
\dot{n}_\varphi+3Hn_\varphi=\sum_{i\geq j} \left\langle \sigma v\right\rangle_{ij}n_i n_j
\end{equation} 
where $\left\langle \sigma v\right\rangle_{ij}$ is the thermally averaged cross section for the initial states $i$ and $j$. Typically, the $2 \rightarrow 2$ scattering cross section dominates and is of order $\alpha /M^{2}_{P}$ where $\alpha$ is a gauge coupling. Integrating (\ref{1}) from an initial tempreature $T_I$ (an inflaton reheating temprature for example, i.e. the maximum temrature of radiation dominated universe in thermal equilibrium) with this cross section gives an incoherent relic density of 
\begin{equation} \label{2}
Y_\varphi\equiv\frac{n_\varphi}{s} \sim \alpha\frac{T_{rh}}{M_P}.
\end{equation} 
For $T_I \ll M_P$ this is much smaller than a thermal number density and potentially quite insignificant compared with the coherent number density (see next).

\subsection{Gravitationally Produced Coherent Moduli}

Quantum fluctuations of moduli fields with $m<H$ produced at the last stages of inflation lead to the moduli problem even if initially there were no classical moduli fields \cite{Felder:1999wt}. These fluctuations have exponentially large wavelengths and for all practical purposes they have the same consequences as an homogeneous classical field of amplitude $\varphi_0=\sqrt{\left\langle (\delta\varphi)^2\right\rangle}$ where
\begin{equation} \label{3}
\varphi_0=\sqrt{\left\langle (\delta\varphi)^2\right\rangle}=\frac{1}{2\pi^2}\int dk k^2 |\varphi_k|^2.
\end{equation} 
The growth of the fluctuations for $m\ll H$ is given by 
\begin{equation} \label{3.5}
\frac{d\left\langle (\delta\varphi)^2\right\rangle}{dt}=\frac{H^3}{4\pi^2}\Rightarrow \left\langle (\delta\varphi)^2\right\rangle=\frac{H^3t}{4\pi^2}=\frac{H^2N}{4\pi^2}
\end{equation} 
where the integration was performed considering a constant Hubble parameter and the last equality came from the relation, $N\equiv Ht$, of the number of e-folds. Considering a simple quadratic potential of chaotic inflation model, $V(I)=\frac{1}{2}m^{2}_{I}I^2$. In this case one has 
\begin{equation} \label{4}
I(t)=I_0-\sqrt{\frac{2}{3}}M_P m_I t.
\end{equation} 
where $I_0$ the value of the inflaton field which gives 60 e-folds (i.e. initial value). The time-dependent Hubble parameter is given by 
\begin{equation} \label{5}
H=\frac{m_I}{\sqrt{6}M_P}I(t),
\end{equation} 
which yields 
\begin{equation} \label{6}
\varphi_0=\sqrt{\left\langle (\delta\varphi)^2\right\rangle}=\frac{m_I\varphi^{2}_{0}}{8\pi\sqrt{3}M^{2}_{P}}.
\end{equation} 
In order to have 60 e-folds of inflation in this model one needs $I_0 \sim 15 M_P$. This implies that a typical value of the (nearly) homogeneous scalar field $\varphi$ in a universe which experienced 60 e-folds of inflation in this model is given by 
\begin{equation} \label{7}
\varphi_0=\sqrt{\left\langle (\delta\varphi)^2\right\rangle} \sim 5m_I.
\end{equation}  
In large scale inflationary models one has $m_I \sim 5\times 10^{-6}M_P \sim 10^{13}$ GeV. The formula which connects the initial amplitude and the moduli number density is 
\begin{equation} \label{8}
\frac{n_\varphi}{s}\sim 10^{-2}\frac{\varphi^{2}_{0}}{\sqrt{m_\varphi M^{3}_{P}}} \sim \frac{\varphi^{2}_{0}T_{rh}}{3m_{\varphi}M^{2}_{P}}
\end{equation}  
which for the above $\varphi_0$, eq. (\ref{7}),  yields
\begin{equation} \label{9}
\frac{n_\varphi}{s} \sim 10^{-10}\frac{T_{rh}}{{m_\varphi}}.
\end{equation}

\subsection{Classical Coherent Moduli}

If a modulus has a nonzero vev, then its vev is generally expected to be of the order of $M_P$. In order to talk about a non-zero vev for any field there has to be a well defined origin, which will be defined as a point which is invariant ('fixed') under the group of symmetries under which the field transforms \cite{Lyth:1995ka}. For moduli the symmetries are relatively complicated, and are in general an infinite number of fixed points with a seperation of order $M_P$ (though only a finite number are physically distinct because the symmetry is a discrete gauge symmetry). The statement that the vev of some modulus is of order $M_P$ just means that it is not close to any particular fixed point.
  
The finite energy density in the early universe breaks susy \cite{Dine:1995uk}. In a thermal phase this is manifest through the disparate occupation numbers for bosons and fermions. In an inflationary phase in which a positive vacuum energy dominates, the inflaton $F$ or $D$ component is necessary nonzero, implying susy breaking. The same is true in the post-inflationary phase before reheating, when the inflaton oscillations dominate, and the time averaged vacuum energy is positive. In principle, susy breaking can be transmitted to flat directions by both renormalizable and non-renormalizable interactions. However, for large field values all fields which couple through renormalizable interactions gain a mass larger than any relevant scale of excitation. These states then effectively decouple and do not lift the flat directions. 

Non-renormalizable interactions can have important effects though. To illustrate this consider a term in the K\"ahler potential of the form 
\begin{equation} \label{10}
\delta K=\frac{\pm C^2}{M^{2}_{P}}I^{\dagger}I\varphi^\dagger\varphi
\end{equation}  
where $I$ is a field which dominates the energy density of the universe, $\varphi$ is a moduli field or generally any canonically normalized flat direction. No symmetry prevents such a term, which can be present directly at the Planck scale, or be generated by running to a lower scale. If $I$ dominates the energy density, then $\rho\cong\left\langle \int d^4\theta I^\dagger I \right\rangle$. The interaction ($\ref{10}$) gives an effective mass for $\varphi$ of 
\begin{equation} \label{11}
\delta {\cal L}=\pm C^2 \frac{\rho}{M^{2}_{P}}\varphi^\dagger\varphi
\end{equation}  
It is worth to note that a positive contribution to the K\"ahler potential gives a negative contribution to $m^2$. In a flat expanding background, $\rho=3H^2M^2_P$, so that $\Delta m^2_{\varphi}=\pm 3C^2H^2$. This is a generic result, independent of what specifically dominates the energy density. For $CH>m_{3/2}$, this source for the soft mass is more important than any hidden sector breaking. The general form of the induced potential along an exact flat direction is of the form 
\begin{equation} \label{12}
V(\varphi)=\pm C^2 H^2 M^2_P f(\varphi/M_P)
\end{equation}  
where $f$ is some function. The curvature is set by the Hubble constant, $V'' \sim H^2$ ($V''=\pm C^2 H^2$), and the scale for variations in the potential is $M_P$. Thus the contribution to the mass squared is proportional to $H^2$, but the absolute value and the sign of the coefficient $\pm C^2$ is unknown since it is determined by the counterterms which appear in a non-renormalizable theory.

During inflation the moduli evolve in the potential (\ref{12}) with $H \sim$ constant. Since the fields are parametrically close to critical damped, within few e-folds they are driven to a \itshape local \normalfont minimum of the potential. However, the form of the potential does not necessarily coincide with that after inflation, or from hidden sector susy breaking. In general the minima are seperated by ${\cal O}(M_P)$. Once $H \sim m_{3/2}$ the moduli start to oscillate freely about a true minimum with amplitude ${\cal O} (M_P)$. This gives a concrete realization of the initial conditions for the moduli problem by specifying the field vev for $H\geq m_{3/2}$. 

To illustrate the possible behaviour of the effective potential of the field $\varphi$ we can consider the simple model \cite{Linde:1996cx}
\begin{equation} \label{13}
V(\varphi)=\frac{1}{2}m^2_\varphi \varphi^2 +\frac{1}{2}C^2 H^2(\varphi-\varphi_0).
\end{equation}  
\begin{itemize}
	\item $C\ll 1$. In this case the moduli masses during inflation remain very small, and the motion of the field $\varphi$ towards the minimum of its effective potential will be very slow. Its quantum fluctuations will also not be suppressed. When $m^2_\varphi> C^2H^2$ if the field value is more than $10^{-10}M_P$ it will cause cosmological problems.
	\item $C\sim1$. In this case the field is critically damped and driven to the local minimum $\varphi_0 \sim {\cal O}(M_P)$. When $m^2_\varphi> C^2H^2$ the field starts oscillations around the low energy mimimum with an initial amplitude $\varphi_0 \sim {\cal O}(M_P)$ causing severe cosmological problems.  
	\item $C \ll 1$. In this case the behaviour of the solution changes dramatically. As Hubble parameter decreases, the minimum moves and drags with it the scalar field. As a result, the field $\varphi$ almost adiabatically moves to its new equilibrium value, and the amplitude of the oscillations about it is rather small. For example, for radiation dominated universe one needs $C \sim 30$ to reduce the amplitude of the oscillations by the factor $10^{-10}$.
\end{itemize}

If the minima coincide at early and late times, i.e. $\varphi_0=0$, the moduli are driven to the true minimum during inflation. One possibility under which the minima can coincide occurs if there are no K\"ahler potential couplings between the moduli and either the inflaton or hidden sectors. Another case that the minima might coincide is if there is a point of enhanced symmetry on moduli space \cite{Randall:1994fr, Dine:1995uk}. The main problem with this idea is the dilaton. It is not known that if such an enhanced symmetry exists for this field, and if it does, it is likely to lie at a point where the gauge coupling is extremely large. So, if symmetries are the solution of the moduli problem, the dilaton must be on a different footing than the other moduli. For example,the dilaton mass might arise from dynamics which does not break susy. 

\section{Cosmological Problems}

The gravitationally interacting particles decay very late. They are not part of the thermal equilibrium in the early universe plasma and their energy seems to be frozen until the expansion rate $H$ drops below the moduli mass. Then, they start rolling towards the minimum of their potential which can be considered quadratic close to the minimum. Indeed, at low energies i.e. $H<m_\varphi$, the mass term $1/2 m^2_{\varphi}\varphi^2$ dominates over terms like the Hubble induced masses and so, the potential is quadratic about the origin $\varphi=0$. The oscillations of the modulus can be considered as a boson condensation of zero-momentum particles. In other words the modulus behaves like nonrelativistic matter and thus scales like $\rho_{\varphi}\propto a^{-3}$. Taking into account that in the early universe the rest of the matter is expected to be relativistic scaling like $\rho_{rad}\propto a^{-4}$, the obvious conclusion is that the modulus will sooner or later dominate the universe (except if it is ultra-light as we will see). 
At this point the natural question is weather a given moduli field can decay before it dominates the universe. The decay rate, eq.(\ref{Gamma}), $\Gamma \sim m^3_\varphi/ M^2_P$, depends only on the mass of the modulus. It is straightforward to compare the lifetime of a modulus with the age of the universe today or at moments in the early universe when basic process took place. A modulus field will have not decay at present times if
\begin{equation}
\Gamma \leq H_0
\end{equation} 
that is if its mass is $m_{\varphi}\leq 20$ MeV. Therefore moduli e.g. of masses $1$ TeV, $1$ GeV have already decayed today while moduli of masses $1$ MeV, $1$ eV etc, are still stable. The second moment in the history of the universe which is one of the pillars of the modern cosmology is the so-called \itshape Big Bang Nucleosynthesis (BBN)\normalfont. Observationally it is verified that nucleosynthesis took place in the early universe and the abundances of the nuclei produced at these early moments put severe constraints on the energy content of that period. Firstly, the universe was dominated by radiation. A matter dominated universe would expand with a faster rate decreasing the period when the number of neutrons decay (speed-up effect). This is the period between the moment when temprature drops below the mass of the neutron, $m_n>T$ (at this period the electroweak interaction are still effective resulting in a suppressing Boltzmann exponent in the neutron relic number density) and the moment when the neutrons and protons form bound states i.e. nuclei. Making this period smaller we increase the ratio of neutrons over protons, $n/p$, and consequently we increase the helium, $^4 He$ abundance. The observational bound on the nonrelativistic matter at that period is 
\begin{equation} \label{m/rad}
\frac{\rho_m}{\rho_{rad}}<0.2 \, \, \, \, \, \, \, \, (95\%C.L.)
\end{equation} 
which can be translated as a bound on the coherent oscillating moduli fields (and not only moduli fields; inflaton or quintessence fields which oscillate or superheavy particles are also constrained). However, this is not the stringent observational bound on the moduli energy density. Even if the moduli energy density satisfies the bound (\ref{m/rad}) their subsequent decay can destroy the BBN products i.e. the observed nuclei. The reason is that at the moment of their decay their energy density stored in their oscillations (which until the decay moment steadily inrcreases over radiation) will be transformed to entropy, causing an 'entropy crisis' \cite{Coughlan:1983ci}\cite{Carroll:1998bd}. The energetic photos coming from the decay of the moduli will destroy $^4He$ nuclei overproducing $D+^3He$. This requires that moduli abundance at the time of the last reheating (last, in the sence that more than the inflation fields may have dominated the universe before, e.g. a curvaton field), i.e at the moment of maximum temprature of the radiation plasma, should satisfy 
\begin{equation} \label{n-phi}
Y_{\varphi}\equiv\frac{n_{\varphi}}{s}\leq 10^{-12}-10^{-15}.
\end{equation} 
The stringent yield bound $Y_{\varphi}\leq 10^{-15}$ is for a hadronic branching ration of the order ${\cal O}(1)$ while the bound $Y_{\varphi}\leq 10^{-12}$ is for hadronic branching ratio ${\cal O}(10^{-3})$. From eq. (\ref{8}) we can see that for $1$ TeV mass modulus, the stringent bound of (\ref{n-phi}) is equivalent to a constraint on the the initial modulus displacement $\varphi_{in}\leq 10^{-10}M_P$ . 

Another pillar of the modern cosmology is the presence of \itshape dark matter\normalfont. We know that dark matter exists in the universe and its nature is exotic in the sense that it is not composed by the standard particle physics model particles. The most popular candidates of dark matter are the supersymmetric particles and in particular the lightest supersymmetriic particle (LSP) in case that the R-parity is unbroken. Often it is considered as success of the (still non-verified) sypersymmetry that it predicts particles with the cosmological wanted mass spectrum of dark matter and that it predicts cross sections that can yield a dark matter of the observed abundance (neutralino dark matter). In order to have the correct abundance, the dark matter particles are assummed to have been in thermal equilibruim. In the case of neutralino dark matter particles they decouple at the temperature of 
\begin{equation} \label{T-DM}
T_{DM,\,f} = {\cal O}(1-10) \text{GeV}. 
\end{equation}
Thereby, after the decay of the moduli the LSP must be brought into thermal equilibrium so that it is not overproduced and have the correct ambundance. One needs a reheating temperature larger than the LSP decoupling temprature, eq. (\ref{T-DM}), constraining further the moduli masses and the energy densities. There are alternatives to LSP like the stable gravitino (studied in the last section), axions which don't impose these constraints on the reheating temperature and models which yield the dark matter ambundance by the moduli decay.

Another kind of constraint comes from the fact that our present universe is not matter-antimatter symmetric (which is also the reason that baryons survived). A \itshape baryogenesis/ leptogenesis \normalfont process should have taken place in the early universe following the inflaton decay. There are viable models of baryogenesis/leptogenesis like the thermal leptogenesis, the GUT-baryogenesis, the Affleck-Dine mechanism \cite{Dine:1995kz}, and the electroweak phase transitions baryogenesis (sphaleron mechanism) or even directly from the moduli decay \cite{Randall:1994fr}. Each baryogenesis process takes place at some temprature (e.g. GUT, ${\cal O}(10^{9})$ GeV or electroweak energy scale). However, most of the mechanisms is expected to work well at energies above the electroweak scale i.e. 100 GeV. Therefore the reheating temprature of the last moduli which decays must be of that order.

The bounds on the reheating temperatures are directly related to the moduli mass. The assumption is that the moduli will transform all their energy stored in the coherent oscillations into radiation when the age of the universe becomes comparable to their lifetime i.e.
\begin{equation} \label{decay}
\left\{t^{-1}_\text{universe}=H\right\}=\left\{\tau^{-1}=\Gamma_{tot}\approx \frac{m^3}{8\pi M^2_P}\right\}.
\end{equation}
This is a perturbative single body decay. The non-relativistic matter energy density will be transformed into radiation $\rho_{rad}=(2\pi^2/45) g_*(T_{rh})T^4_{rh}$ and from (\ref{decay}) the reheating temperature can be estimated as
\begin{equation} \label{Trh-A}
T_{rh}=\left(\alpha \frac{3\times45}{2\pi^2g_*(T_{rh})}\right)^{1/4}\sqrt{\Gamma_{tot}M_P}.
\end{equation}
In the above estimation of the reheating temperature we assumed that the ratio of the moduli energy density over the total energy density is $\rho_{\varphi}/\rho_{tot}=\alpha$. The usual and conservative assumption is that the modulus dominate the energy density of the universe at the moment of their decay, i.e $\alpha={\cal O}(1)$. Notice that the reheating temperature (\ref{Trh-A}) depends only on the total decay rate and not on the absolute value of the energy density of the moduli. 

As we said, and as we can see from the above formuli, a bound on the reheating temperature is a bound on the decay rate which is translated into a bound on the \itshape moduli mass \normalfont. If we want the decaying moduli not to spoil the BBN products we demand a reheating temperature above $T_{BBN} \geq 6$ MeV. This is equivalent to demanding a modili mass of the order of $m_\varphi \geq 10^2$ TeV. If we also want a reheating temperature of at least ${\cal O}(1)$ GeV able to yield a correct thermal abundance for the stable LSP, eg. neutralino dark matter, the moduli mass must be heavier than $m_\varphi \geq 10^5$ TeV. If we also consider an electroweak scale baryogenesis the moduli masses must be even heavier.

On the other hand, the stable moduli fields (those with a lifetime longer that the age of the universe) tend to overclose the universe. This can be avoided only if at the time of matter-radiation energy equivalence the moduli fields energy density is at most of the dark matter energy density:
\begin{equation} \label {light-mod}
 \rho_\varphi (t=t_{eq}) \leq \rho_{DM}\sim 0.3 \rho_{cr}.
\end{equation}
This is equivalent to a constraint on the moduli yield at that time $m_\varphi Y_\varphi<3eV$. Since matter-radiation equivalence the dominant part of the energy content of the universe has scaled like $a^{-3}=T^3$ and the (\ref{light-mod}) can be transformed to 
\begin{equation} \label{light-bound}
\rho_\varphi \cdot (T_0/T_I) \leq 0.3 \rho_{ct}\sim \rho_{cr}.
\end{equation}
The $T_I$ is the temperature that the oscillations start: $T_I\sim (m_\varphi M_P)^{1/2}$ and the moduli have energy at that time $\rho_\varphi\sim m^2_\varphi(\delta\varphi)^2$. $T_0$ is the present temperature of the universe. It is now straightforward to find the bound on the light modulus mass: 
\begin{equation}
m_\varphi<M_P\left(\frac{\rho_{cr}(t_0)M_P}{(\delta\varphi)^2T^3(t_0)}\right)^2
\end{equation}
where $t_0$ is the present time. For initial amplitude of oscillations $\delta\varphi\sim M_P$ we find that the moduli have to be lighter than 
\begin{equation} \label{lightmass}
m_\varphi<10^{-26} \text{eV}.
\end{equation}
It is useful to transform the eq. (\ref{light-bound}) bound on the mass to a bound on the initial displacement $\delta\varphi$ [16]:
\begin{equation}
\frac{\delta\varphi}{M_P}\leq 4\times10^{-9}\left(\frac{m_\varphi}{100\text{MeV}}\right)^{-1/4}.
\end{equation}
So a $m_\varphi\approx 10^{-26}$ eV moduli oversloses the universe for $\delta\varphi\approx M_P$, a $m_\varphi\approx 1$ eV for $\delta\varphi\approx 10^{-8} M_P$ and a $m_\varphi \approx 1$ MeV for $\delta\varphi\approx 10^{-10} M_P$. In these estimations we have neglected any entropy production that may took place since the moduli started oscillating. In this case the bounds on the moduli (both heavy and light) yield $Y_{\varphi}$ is relaxed to a value $Y_\varphi\rightarrow Y_\varphi/\Delta$ where $\Delta=S_{\text{after}}/S_{\text{before}}$. 

In summary, neglecting baryogenesis/leptogenesis the obvious goal is to look for a theory which gives either heavy moduli fields $m_\varphi >10^2-10^5$ TeV or ultralight $m_\varphi < 10^{-26}$ eV. The later masses are extremely light. Ultralight moduli fields are difficult to be generated because the supersymmetry breaking or no-renormalizable interaction terms induce a mass of $m_{3/2}$ scale. Moreover, they would cause the existance of a fifth force which is not observed by the up-to-date observations/experiments \cite{Carroll:1998bd, Adelberger:2003zx}.  Although light moduli are predicted in many theories they are not so light in order to satisfy the bound (\ref{lightmass}). The supersymmetry breaking induced moduli masses are around the $m_{3/2}$ scale. Taking into consideration only the BBN constraints a moduli of mass $m_{\varphi}\geq 10^2$ TeV seems to provide a resolution to the problem. Such heavy moduli masses can be realized e.g. in the flux compactification models \cite{Kachru:2003aw}. However, it was recently pointed out that heavier moduli overproduce gravitinos when they decay causing the so-called "moduli induced gravitino problem" \cite{Nakamura:2006uc, Endo:2006zj}.

\subsection{Moduli Induced Gravitino Problem}

In 2006 it was realized that the moduli fields can decay to gravitino with a branching ratio generically of ${\cal O}(0.01-1)$ \cite{Nakamura:2006uc, Endo:2006zj}. Consequently, the cosmological moduli problem cannot be solved simply by making the modulus mass heavier than $100$ TeV.  

The moduli fields can decay most efficiently into gauge bosons pairs and gaugino pairs. The decay into gravitinos pairs is computed to be of comparable magnitude (of course, a heavier than the gravtino modulus has been assumed). The two body decay of the moduli into Standard Model fermion pairs and sfermions can be shown to be suppressed by powers of the mass of the final states by using their equation of motion \cite{Moroi:1995fs}. The conclusion is that the branching ratio to gravitinos is
\begin {equation}
B_{3/2}\equiv Br(\varphi_{R,I}\rightarrow 2\psi_{3/2})=\frac{1}{54}\frac{12}{N_G}\frac{d^2_{3/2}}{d^2_g}\cong {\cal O} (1-0.01)
\end{equation}
where $N_G$ is the number of the gauge bosons and $d_g$, $d_{3/2}$ dimensionless constants of order unity. In order to see the cosmplogical problems \cite{205170} caused by the production of gravitinos one has to find the gravitino yield $Y_{3/2}$ which \\
1) if the LSP is the gravitino it must not exceed the dark matter abundance
\begin{equation}
Y_{3/2}\leq Y_{DM}
\end{equation}
2) or if $m_{3/2}>m_{LSP}$ it must satisfy the BBN constraint
\begin{equation}
Y_{3/2}\equiv \frac{n_{3/2}}{s}<10^{-12}-10^{-15}.
\end{equation}
A brancging ratio $B_{3/2}={\cal O}(1-10^{-2})$ violates these bounds. A theory with heavier than $10^2$ TeV gravitino, in order to avoid BBN problems, doesn't address the problem either. The reason is that the gravitino can decay to the LSP giving an, e.g. neutralino, abundance which exceeds the upperbound of the dark matter inferred by the observations. The neutralino LSPs produced this way are so abundant that they annihilate with each other, however, without being in a thermal equilibrium. The resulting neutralino yield, $Y_{LSP}$ is estimated to be unacceptable large both for wino and bino LSP.

A gravitino heavier than $10^{7}$ will produce LSPs that will get thermalized and the conventional computation of the relic abundance can be applied. However, with such a heavy gravitino, the resulting soft masses would be far above the electroweak scale, diminishing the very motivation of low energy supersymmetry. 

\section{Supersymmetry Breaking in the Early Universe due to the Inflaton Field}

We assume that in the early universe a standard inflation took place solving the homogeneity, isotropy and flatness problems and generating the initial curvature perturbations. During this quasi-de Sitter period vector bosons, fermions and heavy scalars were diluted. Saying heavy we mean scalars with mass heavier than the Hubble constant during inflation. On the other hand abundances of lighter scalar were not diluted since these fields cannot roll down due to the strong friction term and 'freeze' roughly in their pre-inflationary vev. After inflation except for the inflaton, $I$, there will be these light scalars and their classical super-Hubble long-wavelength quantum (initially) fluctuations.  

However, during inflation the finite vacuum energy density (non zero $F$ and/or $D$ components) breaks susy. For $H>m_{3/2}$ this breaking is dominant over the hidden sector inducing masses of the order of $H$ thereby making the fields heavy. Fields like \itshape flat directions \normalfont $\varphi_f$ (which can have non-Planck mass suppressed interactions) and moduli are not frozen as was generally assumed. As long as $H>m_{3/2}$ this source for the soft mass is the most important and persists also in the post inflationary epoch, i.e. during reheating and radiation dominated era. A well defined example of this effect is the behaviour of the flat directions $\varphi_f$; their relevant potential during inflation takes the form \cite{Dine:1995kz}
\begin{equation} \label{flatdir}
V(\varphi_f)=-cH^2_I|\varphi_f|^2+\left(\frac{a\lambda H_I\varphi_f^n}{nM^{n-3}}+h.c.\right)+|\lambda|^2\frac{|\varphi_f|^{2n-2}}{M^{2n-6}}
\end{equation}
where $c$ and $a$ are constants of ${\cal O}(1)$ and $M$ is some large mass such as the GUT or the Planck scale. We should mention that the flat direction is assumed to be stabilized even in the absence of supersymmetry breaking, by a high dimension operator in the superpotential. For $c<0$ the origin is the minimum and the average value of the flat direction evolves to $\varphi_f=0$ exponentially in time. For $c$ negative and $|c|\ll 1$, so $m^2 \ll H^2$, the origin  is still the minimum and in addition we have de-Sitter fluctuations with correlation length $l\approx H^{-1}e^{3H^2/2m^2}$. But for $c>0$ the potential has an unstable extremum at the origin. For $H_I \ll M_P$ energetically this limits $\varphi_f \ll M_P$. The only soft terms in (\ref{flatdir}) are therefore the lowest order ones, namely the mass and $A$ terms. The minimum of the potential (\ref{flatdir}) is given by  
\begin{equation}
|\varphi_{f \, 0}|=\left(\frac{\beta H_I M^{n-3}}{\lambda}\right)^{\frac{1}{n-2}}
\end{equation}
where $\beta$ is a numerical constant which depends on $a$, $c$, and $n$. The initial value of the flat direction, $\varphi_{f\,0}$, is parametrically between $H_I$ and $M_P$. For exapmle, with $H_I \sim 10^{13}$ GeV, $M/\lambda \sim M_P$, and $n=4$, $\varphi_{f\, 0} \sim 10^3 H_I$.

The \itshape string modulus \normalfont potential changes similarly during inflation. The finite inflationary energy induces soft potential , $V'' \sim H^2$ and the fields are driven to a local minimum within few e-foldings unless the induced mass happens to be numerically much less than $H$. If this last case, i.e. $V'' \ll H^2$, is the one that is realized then the \itshape couplings of the modulus \normalfont can drive it to the origin (plus the expected de-Sitter fluctuations). The case that the minima coincide at early and late times i.e. $V'' \sim H^2$, and the minimum is at the origin (this corresponds to $c<0$ for the flat directions) can happen if there is a point of enhanced symmetry on moduli space \cite{Dine:1995kz}. However, this idea does not apply for the dilaton. The moduli transform under some symmetry near such points. The lowest order invariants are therefore bilinears and the potential is necessarily an extremum at such points. So, it is possible that the potential is a minimum at both early and late times.

\end{document}